\documentclass[defaultstyle,11pt]{thesis}

\usepackage{latexsym}
\usepackage{graphicx}
\usepackage{rotating}
\usepackage[tbtags]{amsmath}
\usepackage{amssymb}
\usepackage{dcolumn}
\usepackage{bm}
\usepackage{url}

\title{Harmonic Gauge Perturbations of the Schwarzschild Metric}

\author{Mark~V.}{Berndtson}

\otherdegrees{B.A., University of Colorado, 1996}

\degree{Doctor of Philosophy}
	{Ph.D., Physics}

\dept{Department of}	
	{Physics}

\advisor{Prof.}
	{Neil Ashby}

\reader{David Bartlett}		

\degreeyear{2007}

\abstract{\OnePageChapter
The satellite observatory LISA will be capable of detecting gravitational 
waves from extreme mass ratio inspirals (EMRIs), such as a small black 
hole orbiting a supermassive black hole.  The gravitational effects of 
the much smaller mass can be treated as the perturbation of a known 
background metric, here the Schwarzschild metric.  The perturbed Einstein 
field equations form a system of ten coupled partial differential equations.  
We solve the equations in the harmonic gauge, also called 
the Lorentz gauge or Lorenz gauge.  Using separation of variables and 
Fourier transforms, we write the solutions in terms of six radial functions 
which satisfy decoupled ordinary differential equations.  We then use 
the solutions to calculate the gravitational self-force for circular 
orbits.  The self-force gives the first order perturbative corrections 
to the equations of motion.
}

\dedication[Dedication]{\centerline{To my family}} 

\acknowledgements{\OnePageChapter	
	I thank my advisor, Neil Ashby, for his advice and patience, 
	and the other committee members for their service.}

\LoFisShort

\LoTisShort

\begin{document}

\newcommand{\hz}{h_{0}}
\newcommand{\dhz}{h_{0}^{\prime}}
\newcommand{\ddhz}{h_{0}^{\prime\prime}}
\newcommand{\ho}{h_{1}}
\newcommand{\dho}{h_{1}^{\prime}}
\newcommand{\ddho}{h_{1}^{\prime\prime}}
\newcommand{\htw}{h_{2}}
\newcommand{\dhtw}{h_{2}^{\prime}}
\newcommand{\ddhtw}{h_{2}^{\prime\prime}}
\newcommand{\bhz}{H_{0}}
\newcommand{\dbhz}{H_{0}^{\prime}}
\newcommand{\ddbhz}{H_{0}^{\prime\prime}}
\newcommand{\bho}{H_{1}}
\newcommand{\dbho}{H_{1}^{\prime}}
\newcommand{\ddbho}{H_{1}^{\prime\prime}}
\newcommand{\bhtw}{H_{2}}
\newcommand{\dbhtw}{H_{2}^{\prime}}
\newcommand{\ddbhtw}{H_{2}^{\prime\prime}}
\newcommand{\bk}{K}
\newcommand{\dbk}{K^{\prime}}
\newcommand{\ddbk}{K^{\prime\prime}}
\newcommand{\bg}{G}
\newcommand{\dbg}{G^{\prime}}
\newcommand{\ddbg}{G^{\prime\prime}}
\newcommand{\ff}{\left(1-\frac{2 M}{r}\right)}
\newcommand{\ptw}{\psi_{2}}
\newcommand{\dptw}{\psi_{2}^{\prime}}
\newcommand{\ddptw}{\psi_{2}^{\prime\prime}}
\newcommand{\po}{\psi_{1}}
\newcommand{\dpo}{\psi_{1}^{\prime}}
\newcommand{\pz}{\psi_{0}}
\newcommand{\dpz}{\psi_{0}^{\prime}}
\newcommand{\pza}{\psi_{0a}}
\newcommand{\dpza}{\psi_{0a}^{\prime}}
\newcommand{\pzb}{\psi_{0b}}
\newcommand{\dpzb}{\psi_{0b}^{\prime}}
\newcommand{\pzc}{\psi_{0c}}
\newcommand{\dpzc}{\psi_{0c}^{\prime}}
\newcommand{\fmtwa}{M_{2a\!f}}
\newcommand{\dfmtwa}{M_{2a\!f}^{\prime}}
\newcommand{\pzbf}{\left(\pzb+\fmtwa\right)}
\newcommand{\dpzbf}{\left(\dpzb+\dfmtwa\right)}
\newcommand{\bmz}{M_{0}}
\newcommand{\dbmz}{M_{0}^{\prime}}
\newcommand{\ddbmz}{M_{0}^{\prime\prime}}
\newcommand{\bmo}{M_{1}}
\newcommand{\dbmo}{M_{1}^{\prime}}
\newcommand{\ddbmo}{M_{1}^{\prime\prime}}
\newcommand{\bmtw}{M_{2}}
\newcommand{\dbmtw}{M_{2}^{\prime}}
\newcommand{\ddbmtw}{M_{2}^{\prime\prime}}
\newcommand{\tbmz}{\widetilde{M}_{0}}
\newcommand{\tdbmz}{\widetilde{M}_{0}^{\prime}}
\newcommand{\tddbmz}{\widetilde{M}_{0}^{\prime\prime}}
\newcommand{\tbmo}{\widetilde{M}_{1}}
\newcommand{\tdbmo}{\widetilde{M}_{1}^{\prime}}
\newcommand{\tddbmo}{\widetilde{M}_{1}^{\prime\prime}}
\newcommand{\tbmtw}{\widetilde{M}_{2}}
\newcommand{\tdbmtw}{\widetilde{M}_{2}^{\prime}}
\newcommand{\tddbmtw}{\widetilde{M}_{2}^{\prime\prime}}
\newcommand{\bmtwa}{M_{2a}}
\newcommand{\dbmtwa}{M_{2a}^{\prime}}
\newcommand{\ddbmtwa}{M_{2a}^{\prime\prime}}
\newcommand{\fz}{f_{0}}
\newcommand{\dfz}{f_{0}^{\prime}}
\newcommand{\ddfz}{f_{0}^{\prime\prime}}
\newcommand{\fdz}{f_{d0}}
\newcommand{\dfdz}{f_{d0}^{\prime}}
\newcommand{\ddfdz}{f_{d0}^{\prime\prime}}
\newcommand{\dddfdz}{f_{d0}^{\prime\prime\prime}}
\newcommand{\ftw}{f_{2}}
\newcommand{\dftw}{f_{2}^{\prime}}
\newcommand{\ddftw}{f_{2}^{\prime\prime}}
\newcommand{\fdtw}{f_{d2}}
\newcommand{\dfdtw}{f_{d2}^{\prime}}
\newcommand{\ddfdtw}{f_{d2}^{\prime\prime}}
\newcommand{\nftw}{\tilde{f}_{2}}
\newcommand{\dnftw}{\tilde{f}_{2}^{\prime}}
\newcommand{\ddnftw}{\tilde{f}_{2}^{\prime\prime}}
\newcommand{\nfdtw}{\tilde{f}_{d2}}
\newcommand{\dnfdtw}{\tilde{f}_{d2}^{\prime}}
\newcommand{\ddnfdtw}{\tilde{f}_{d2}^{\prime\prime}}
\newcommand{\dddnfdtw}{\tilde{f}_{d2}^{\prime\prime\prime}}
\newcommand{\ddddnfdtw}{\tilde{f}_{d2}^{\prime\prime\prime\prime}}
\newcommand{\lnff}{\ln\!\left[1-\frac{2 M}{r}\right]}
\newcommand{\lnr}{\ln\!\left[\frac{2 M}{r}\right]}
\newcommand{\lnbff}{\ln\!\left[1-\frac{2 M}{R}\right]}
\newcommand{\lnbr}{\ln\!\left[\frac{2 M}{R}\right]}
\newcommand{\enbar}{\widetilde{E}}
\newcommand{\elbar}{\widetilde{L}}
\newcommand{\iom}{i\omega}
\newcommand{\deltar}{\delta(r-r^{\prime})}
\newcommand{\deltaom}{\delta^{2}(\Omega-\Omega^{\prime})}
\newcommand{\rdot}{\dot{r}^{\prime}}
\newcommand{\thdot}{\dot{\theta}^{\prime}}
\newcommand{\phdot}{\dot{\phi}^{\prime}}
\newcommand{\thetap}{\theta^{\prime}}
\newcommand{\phip}{\phi^{\prime}}
\newcommand{\ylmpi}{\overline{Y}_{lm}
\big(\textstyle{\frac{\pi}{2}},0\big)e^{-i m \phip}}
\newcommand{\dylmpi}{\frac{\partial\overline{Y}_{lm}\left(\frac{\pi}{2},
0\right)}{\partial\theta}e^{-i m \phip}}
\newcommand{\nylmpi}{\overline{Y}_{lm}\big(\textstyle{\frac{\pi}{2}},0\big)}
\newcommand{\ndylmpi}{\frac{\partial\overline{Y}_{lm}\left(\frac{\pi}{2},
0\right)}{\partial\theta}}
\newcommand{\costp}{\cos\big(\omega_{mk}\hat{t}-m\hat{\phi}\big)}
\newcommand{\sintp}{\sin\big(\omega_{mk}\hat{t}-m\hat{\phi}\big)}
\newcommand{\dphids}{\frac{d\phi^{\prime}}{d\tau}}
\newcommand{\axdot}{\big\lvert\frac{dr^{\prime}}{d\tau}\big\rvert}
\newcommand{\ardot}{\lvert\rdot\rvert}
\newcommand{\plog}{\text{PolyLog}\!\left[2,\frac{2 M}{r}\right]}
\newcommand{\mz}{m_{0}}
\newcommand{\hp}{h_{\scriptscriptstyle{+}}}
\newcommand{\hc}{h_{\times}}	
\chapter{\label{introchap}Introduction}
  
The satellite observatory LISA will be capable of detecting gravitational 
waves from extreme mass ratio inspirals (EMRIs).  These occur when 
a compact star, such as a black hole, neutron star, or white dwarf, is 
captured by a supermassive black hole \cite{lisamiss}.  
The evolution of EMRI orbits and their gravitational waveforms can be 
calculated using black hole perturbation theory.
An introduction to black hole perturbation theory, 
including the harmonic gauge, is in section \ref{sec:bhpt}.  
An outline of the remainder of the thesis is in section \ref{sec:sumth}.  

\section{\label{sec:bhpt}Black Hole Perturbation Theory}

The Schwarzschild metric is
\begin{equation}
\label{eq:schmet}
ds^{2}=g_{\mu\nu}dx^{\mu}dx^{\nu}=-\ff dt^{2}+\frac{1}{\ff}dr^{2}
+r^2 d\theta^{2}+r^2 \sin^{2}\theta d\phi^{2}\;,
\end{equation}
where the standard coordinates have been used.  
It is a solution of the Einstein field equations, which are
\begin{equation}
\label{eq:feqn}
R_{\mu\nu}-\tfrac{1}{2}g_{\mu\nu}R=8\pi\tfrac{G}{c^{4}} T_{\mu\nu}\;.
\end{equation}
The Schwarzschild metric is a vacuum solution, 
meaning $T_{\mu\nu}=0$.  In \eqref{eq:feqn}, the gravitational constant 
$G$ and speed of light $c$ are shown explicitly.  Generally though, 
we will use geometrized units, for which $G=c=1$.  We will use this and 
other notational conventions as described by Misner, Thorne and 
Wheeler \cite{mtw73}, including the $-$$+$$+$$+$ metric signature of 
\eqref{eq:schmet} and form of the field equations \eqref{eq:feqn}.  

Black hole perturbation theory for the Schwarzschild metric was 
formulated by Regge and Wheeler \cite{rw57} and extended by 
Zerilli \cite{zerp70}.  A summary of their method follows, which is 
taken mainly from their articles.  A small 
perturbation $h_{\mu\nu}$ is added to the background Schwarzschild 
metric $g_{\mu\nu}$.  Our physical problem involves a small mass $\mz$ 
orbiting a much larger black hole $M$, so $h_{\mu\nu}$ is proportional 
to the mass ratio $\mz/M$.  Accordingly, the total perturbed metric 
$\tilde{g}_{\mu\nu}$ is
\begin{equation}
\label{eq:tilmet}
\tilde{g}_{\mu\nu}=g_{\mu\nu}+h_{\mu\nu}+O((\mz/M)^{2})\;.  
\end{equation}
The inverse perturbed metric is
\begin{equation}
\label{eq:invtilmet}
\tilde{g}^{\mu\nu}=g^{\mu\nu}-h^{\mu\nu}+O((\mz/M)^{2})\;.
\end{equation}
The perturbed field equations are linear in $h_{\mu\nu}$:
\begin{multline}
\label{eq:pertfeqnfull}
-\left[h_{\mu\nu;\alpha}{}^{;\alpha}
+2 R^{\alpha}{}_{\!\mu}{}^{\beta}{}_{\!\nu}\,h_{\alpha\beta}
-(h_{\mu\alpha}{}^{;\alpha}{}_{;\nu}+h_{\nu\alpha}{}^{;\alpha}{}_{;\mu})
+h_{;\mu;\nu}-R^{\alpha}{}_{\nu}h_{\mu\alpha}
-R^{\alpha}{}_{\mu}h_{\nu\alpha}\right]
\\-g_{\mu\nu}(h_{\lambda\alpha}{}^{;\alpha;\lambda}
-h_{;\lambda}\!{}^{;\lambda})-h_{\mu\nu}R
+g_{\mu\nu}h_{\alpha\beta}R^{\alpha\beta}=16 \pi T_{\mu\nu}\;.
\end{multline}
The semicolons represent covariant differentiation with respect 
to the background metric $g_{\mu\nu}$.  For the Schwarzschild metric, 
the Ricci tensor $R_{\mu\nu}$ and Ricci scalar $R$ are zero, so the 
field equations simplify to
\begin{multline}
\label{eq:pertfeqn}
-\left[h_{\mu\nu;\alpha}{}^{;\alpha}
+2 R^{\alpha}{}_{\!\mu}{}^{\beta}{}_{\!\nu}\,h_{\alpha\beta}
-(h_{\mu\alpha}{}^{;\alpha}{}_{;\nu}+h_{\nu\alpha}{}^{;\alpha}{}_{;\mu})
+h_{;\mu;\nu}\right]\\-g_{\mu\nu}(h_{\lambda\alpha}{}^{;\alpha;\lambda}
-h_{;\lambda}\!{}^{;\lambda})=16 \pi T_{\mu\nu}\;.
\end{multline}
As discussed in \cite{mtw73}, the left side of the perturbed equations 
represents the propagation of a wave interacting with the background 
spacetime curvature.  The stress energy tensor $T_{\mu\nu}$ is the 
covariant form of
\begin{equation}
\label{eq:tmunuup}
T^{\mu\nu}=m_{0}\int_{\!-\infty}^{\infty}
\frac{\delta^{4}(x-z(\tau))}{\sqrt{-g}}\,{\frac{dz}{d\tau}}^{\!\mu}
{\frac{dz}{d\tau}}^{\!\nu}\,d\tau\;,
\end{equation}
where the delta function represents a point mass or test particle.  
The stress energy tensor divergence equation is
\begin{equation}
\label{eq:divtmunu}
T^{\mu\nu}\!{}_{;\nu}=0\;,
\end{equation}
which, when applied to \eqref{eq:tmunuup}, gives the background geodesic 
equations of motion \cite{det05}.  

Regge, Wheeler and Zerilli solved the perturbed field equations using 
separation of variables.  The angular dependence is contained in tensor 
harmonics, which are obtained from the familiar spherical harmonics.  
A Fourier transform separates the time dependence, leaving a set 
of radial ordinary differential equations to be solved.  The solution 
was simplified by making a particular choice of gauge, the so-called 
Regge-Wheeler gauge.  A gauge is a choice of coordinates.  A change of 
gauge is a small change of coordinates
\begin{equation}
\label{eq:xnew}
x^{\mu}_{\text{new}}=x^{\mu}_{\text{old}}+\xi^{\mu}\;,
\end{equation}
which causes the metric perturbation to change as
\begin{equation}
\label{eq:hnew}
h_{\mu\nu}^{\text{new}}=h_{\mu\nu}^{\text{old}}-\xi_{\mu;\nu}-\xi_{\nu;\mu}\;.
\end{equation}
Here, ``small'' means $O(\mz/M)$.  
Although the perturbation is gauge dependent, the perturbed field 
equations in \eqref{eq:pertfeqn} are gauge invariant \cite{det05}.

We will use the notation of Ashby \cite{ashby}, which is 
different from, and simpler than, Zerilli's notation \cite{zerp70}.  
Because the background metric is spherically symmetric, the perturbation 
can be split into odd and even parity parts.  This decomposition gives
\begin{equation}
\label{eq:hmunu}
h_{\mu\nu}(t,r,\theta,\phi)
=\sum_{l=0}^{\infty}\sum_{m=-l}^{l}\int_{\!-\infty}^{\infty} 
e^{-i\omega t} \left(h^{o,lm}_{\mu\nu}(\omega,r,\theta,\phi)
+h^{e,lm}_{\mu\nu}(\omega,r,\theta,\phi)\right) d\omega\;.
\end{equation}
The odd parity terms are
\begin{equation}
\label{eq:ohmunu}
h^{o,lm}_{\mu\nu}(\omega,r,\theta,\phi)
=\begin{pmatrix}
0 & 0 & h_{0}^{lm}(\omega,r) 
\csc\theta\frac{\partial Y_{lm}(\theta,\phi)}{\partial\phi} 
& -h_{0}^{lm}(\omega,r)\sin\theta\frac{\partial 
Y_{lm}(\theta,\phi)}{\partial\theta}
\\ * & 0 & h_{1}^{lm}(\omega,r) \csc\theta
\frac{\partial Y_{lm}(\theta,\phi)}{\partial\phi} 
& -h_{1}^{lm}(\omega,r)\sin\theta\frac{\partial Y_{lm}(\theta,\phi)}
{\partial\theta}
\\ * & * & -h_{2}^{lm}(\omega,r)X_{lm}(\theta,\phi) 
& h_{2}^{lm}(\omega,r)\sin\theta W_{lm}(\theta,\phi)
\\ * & * & *  & h_{2}^{lm}(\omega,r)\sin^{2}\theta 
X_{lm}(\theta,\phi)
\end{pmatrix}\;,
\end{equation}
where
\begin{equation}
\label{eq:wlm}
W_{lm}(\theta,\phi)
=\frac{\partial^{2}Y_{lm}(\theta,\phi)}{\partial\theta^{2}}
-\cot\theta\frac{\partial Y_{lm}(\theta,\phi)}{\partial\theta}
-\frac{1}{\sin^{2}\theta}\frac{\partial^{2}
Y_{lm}(\theta,\phi)}{\partial\phi^{2}}\;,
\end{equation}
\begin{equation}
\label{eq:xlm}
X_{lm}(\theta,\phi)
=\frac{2}{\sin\theta}\frac{\partial}{\partial\phi}\!
\left(\frac{\partial Y_{lm}(\theta,\phi)}{\partial\theta}
-\cot\theta\, Y_{lm}(\theta,\phi)\right)\;.
\end{equation}
The even parity part is 
\begin{multline}
\label{eq:ehmunu}
h^{e,lm}_{\mu\nu}(\omega,r,\theta,\phi)=
\\\begin{pmatrix}
\ff H^{lm}_{0}(\omega,r)Y_{lm} 
& H^{lm}_{1}(\omega,r)Y_{lm}  & h^{lm}_{0}(\omega,r) 
\frac{\partial Y_{lm}}{\partial\theta} & h^{lm}_{0}(\omega,r)
\frac{\partial Y_{lm}}{\partial\phi}
\\ * & \frac{H^{lm}_{2}(\omega,r)Y_{lm}}{\ff} 
&  h^{lm}_{1}(\omega,r) \frac{\partial Y_{lm}}{\partial\theta} 
& h^{lm}_{1}(\omega,r)\frac{\partial Y_{lm}}{\partial\phi}
\\ * & * & \begin{subarray}{l} \displaystyle{r^{2}}
\scriptstyle{\big(K^{lm}(\omega,r) Y_{lm}}
\\\,\,\,\,\,\,+G^{lm}(\omega,r) W_{lm} \big)  
\end{subarray}  & r^{2}\sin\theta\, G^{lm}(\omega,r) X_{lm}
\\ * & * & *  &\begin{subarray}{l}
\displaystyle{ r^{2}\sin^{2}\theta }
\scriptstyle{\big(K^{lm}(\omega,r) Y_{lm}}
\\\,\,\,\,\quad\quad-G^{lm}(\omega,r) W_{lm} \big)\end{subarray}
\end{pmatrix}\;.
\end{multline}
Asterisks represent symmetric components.  The angular functions are 
the tensor harmonics.  The Regge-Wheeler gauge is defined by setting 
four radial factors equal to zero: the odd parity $\htw$ and the even 
parity $\hz$, $\ho$ and $\bg$.  The trace $h$ of the metric perturbation is
\begin{equation}
\label{eq:nhtrace}
h(t,r,\theta,\phi)
=\sum_{l=0}^{\infty}\sum_{m=-l}^{l} Y_{lm}(\theta,\phi)
\int_{\!-\infty}^{\infty}e^{-i\omega t}h^{lm}(\omega,r) 
d\omega\;,
\end{equation}
where
\begin{equation}
\label{eq:rhtrace}
h^{lm}(\omega,r)=-\bhz^{lm}(\omega,r)+\bhtw^{lm}(\omega,r)
+2 \bk^{lm}(\omega,r)\;.
\end{equation}
Similarly, the stress energy tensor may be decomposed in terms of 
tensor harmonics and Fourier transforms.  The covariant components are
\begin{equation}
\label{eq:tmunu}
T_{\mu\nu}(t,r,\theta,\phi)
=\sum_{l=0}^{\infty}\sum_{m=-l}^{l}\int_{\!-\infty}^{\infty}
e^{-i\omega t} \left(T^{o,lm}_{\mu\nu}(\omega,r,\theta,\phi)
+T^{e,lm}_{\mu\nu}(\omega,r,\theta,\phi)\right)d\omega\;,
\end{equation}
where
\begin{equation}
\label{eq:otmunu}
T^{o,lm}_{\mu\nu}(\omega,r,\theta,\phi)
=\begin{pmatrix}
0 & 0 & So_{02}^{lm}(\omega,r) \csc\theta
\frac{\partial Y_{lm}}{\partial\phi} & -So_{02}^{lm}(\omega,r)
\sin\theta\frac{\partial Y_{lm}}{\partial\theta}
\\ * & 0 & So_{12}^{lm}(\omega,r) \csc\theta
\frac{\partial Y_{lm}}{\partial\phi} & -So_{12}^{lm}(\omega,r)
\sin\theta\frac{\partial Y_{lm}}{\partial\theta}
\\ * & * & -So_{22}^{lm}(\omega,r)X_{lm} 
& So_{22}^{lm}(\omega,r)\sin\theta W_{lm}
\\ * & * & *  & So_{22}^{lm}(\omega,r)\sin^{2}\theta X_{lm}
\end{pmatrix}\;,
\end{equation}
and
\begin{multline}
\label{eq:etmunu}
T^{e,lm}_{\mu\nu}(\omega,r,\theta,\phi)=
\\\begin{pmatrix}
Se^{lm}_{00}(\omega,r)Y_{lm} 
& Se^{lm}_{01}(\omega,r)Y_{lm}  & Se^{lm}_{02}(\omega,r) 
\frac{\partial Y_{lm}}{\partial\theta} & Se^{lm}_{02}(\omega,r)
\frac{\partial Y_{lm}}{\partial\phi}
\\ * & Se^{lm}_{11}(\omega,r)Y_{lm} &  Se^{lm}_{12}(\omega,r) 
\frac{\partial Y_{lm}}{\partial\theta} & Se^{lm}_{12}(\omega,r)
\frac{\partial Y_{lm}}{\partial\phi}
\\ * & * & \begin{subarray}{l} Ue^{lm}_{22}(\omega,r) Y_{lm}
\\\,\,\,\,\,\,+Se^{lm}_{22}(\omega,r) W_{lm}   \end{subarray}  
& Se^{lm}_{22}(\omega,r) \sin\theta X_{lm}
\\ * & * & *  &\begin{subarray}{l}\displaystyle{\sin^{2}\theta}
\scriptstyle{ \big(Ue^{lm}_{22}(\omega,r) Y_{lm}}
\\\quad\quad-Se^{lm}_{22}(\omega,r) W_{lm} \big) \end{subarray} 
\end{pmatrix}\;.
\end{multline}
The trace of the stress energy tensor is
\begin{equation}
\label{eq:trtmunu}
T=g^{\mu\nu}T_{\mu\nu}\;,
\end{equation}
and its multipole decomposition is
\begin{equation}
\label{eq:ntrtmunu}
T(t,r,\theta,\phi)
=\sum_{l=0}^{\infty}\sum_{m=-l}^{l}Y_{lm}(\theta ,\phi)
\int_{\!-\infty}^{\infty}e^{-\iom  t}\,T^{lm}(\omega,r)d\omega\;,
\end{equation}
where
\begin{equation}
\label{eq:rtrtmunu}
T^{lm}(\omega,r)=\frac{r }{2 M-r}Se_{00}^{lm}(\omega,r)
+\left(1-\frac{2 M}{r}\right) Se_{11}^{lm}(\omega,r)
+\frac{2 }{r^2}Ue_{22}^{lm}(\omega,r)\;.
\end{equation}

The gauge change vector $\xi_{\mu}$ is 
\begin{multline}
\label{eq:chimu}
\xi_{\mu}(t,r,\theta,\phi)
=\sum_{l=0}^{\infty}\sum_{m=-l}^{l}\bigg\{\bigg[\int_{\!-\infty}^{\infty} 
e^{-i\omega t} \left(\xi^{o,lm}_{\mu}(\omega,r,\theta,\phi)\right.
\\\left.+\xi^{e,lm}_{\mu}(\omega,r,\theta,\phi)\right) d\omega\bigg]
+\delta_{l0}\delta_{\mu t}C_{0}\ff t \,Y_{00}(\theta,\phi)
\\+\delta_{l 1}C_{1}\,t\,r^2 \left[\delta_{\mu\theta}\csc\theta
\frac{\partial Y_{1m}(\theta,\phi)}{\partial\phi} 
-\delta_{\mu\phi}\sin\theta\frac{\partial Y_{1m}(\theta,\phi)}
{\partial\theta}\right]\bigg\}\;,
\end{multline}
where $\delta_{a b}$ is the Kronecker delta and where
\begin{equation}
\label{eq:ochimu}
\xi^{o,lm}_{\mu}(\omega,r,\theta,\phi)
=\begin{pmatrix}
0 , 0 , Z^{lm}(\omega,r) \csc\theta
\frac{\partial Y_{lm}}{\partial\phi} , 
-Z^{lm}(\omega,r)\sin\theta\frac{\partial Y_{lm}}{\partial\theta}
\end{pmatrix}\;,
\end{equation}
\begin{equation}
\label{eq:echimu}
\xi^{e,lm}_{\mu}(\omega,r,\theta,\phi)
=\begin{pmatrix}
M^{lm}_{0}(\omega,r)Y_{lm} , 
M^{lm}_{1}(\omega,r)Y_{lm}  , M^{lm}_{2}(\omega,r) 
\frac{\partial Y_{lm}}{\partial\theta} , M^{lm}_{2}(\omega,r)
\frac{\partial Y_{lm}}{\partial\phi}
\end{pmatrix}\;.
\end{equation}
Because a gauge change is a small change of coordinates, the quantities 
$Z^{lm}(\omega,r)$ and $M^{lm}_{i}(\omega,r)$ are of order $\frac{m_{0}}{M}$, 
just as the perturbation $h_{\mu\nu}$ is.  The remaining two terms in 
$\xi_{\mu}$ \eqref{eq:chimu} are also order $\frac{m_{0}}{M}$ and are 
discussed in subsections \ref{sec:zevpareqz} and \ref{sec:zoddpareq}, respectively.

Regge and Wheeler showed that the odd parity components could be expressed 
in terms of a single radial scalar function, the Regge-Wheeler function, 
which satisfies a second order ordinary differential equation.  
Zerilli did the same for the even parity components; his function is 
called the Zerilli function.  Later, Moncrief showed that these two 
functions are gauge invariant \cite{monc74a}.

The Regge-Wheeler gauge is comparatively simple, but we will work 
in a different gauge, the harmonic gauge.  The reason for this choice 
is that the equations for the gravitational self-force were derived 
in the harmonic gauge \cite{mst97}, \cite{qwald97}.  
The basic harmonic gauge field equations are given in \cite{mtw73}, 
which uses the term Lorentz gauge.  The spelling Lorenz is also used in 
the literature \cite{poisslrr}.  We first define the trace-reversed 
metric
\begin{equation}
\label{eq:hbar}
\overline{h}_{\mu\nu}=h_{\mu\nu}-{\frac{1}{2}g_{\mu\nu}h}\;,
\end{equation}
where the trace $h$ is
\begin{equation}
\label{eq:htrace}
h=g^{\alpha\beta}h_{\alpha\beta}\;.
\end{equation}
The harmonic gauge is defined by the condition
\begin{equation}
\label{eq:divheqn}
\overline{h}_{\mu\nu}{}^{;\nu}=0\;.
\end{equation}
Using \eqref{eq:hbar} and \eqref{eq:divheqn}, the field equations 
\eqref{eq:pertfeqn} simplify to
\begin{equation}
\label{eq:hpertfeqn}
{\overline{h}_{\mu\nu;\alpha}}{}^{;\alpha}
+2 R^{\alpha}{}_{\!\mu}{}^{\beta}{}_{\!\nu}
\,\overline{h}_{\alpha\beta}=-16 \pi T_{\mu\nu}\;.
\end{equation}
Adding \eqref{eq:pertfeqn} and \eqref{eq:hpertfeqn} gives
\begin{equation}
\label{eq:hzterms}
\overline{h}_{\mu\alpha}{}^{;\alpha}{}_{;\nu}
+\overline{h}_{\nu\alpha}{}^{;\alpha}{}_{;\mu}
-g_{\mu\nu}\overline{h}_{\lambda\alpha}{}^{;\alpha;\lambda}=0\;,
\end{equation}
which are the terms eliminated in going from \eqref{eq:pertfeqn} to
\eqref{eq:hpertfeqn}.  The harmonic gauge condition \eqref{eq:divheqn} 
and field equations \eqref{eq:hpertfeqn} are preserved after a gauge 
change which satisfies
\begin{equation}
\label{eq:divchi}
\xi_{\mu;\nu}{}^{;\nu}=0\;.
\end{equation}
The harmonic gauge equations above apply when the background metric 
describes a curved spacetime, as the Schwarzschild metric does.  

Before discussing further the harmonic gauge for the Schwarzschild metric, 
it is helpful to review a much simpler example, the plane wave.  The 
analysis below of the plane wave is taken from Weinberg's chapter on 
gravitational radiation \cite{wein72}, supplemented by \cite{schutz90}.  
The plane wave is a perturbation $h_{\mu\nu}$ of the flat space metric 
$\eta_{\mu\nu}$.  The perturbed metric $\tilde{g}_{\mu\nu}$ is 
\begin{equation}
\label{eq:pwpert}
\tilde{g}_{\mu\nu}=\eta_{\mu\nu}+h_{\mu\nu}\;.
\end{equation}
We can choose the gauge so that
\begin{equation}
\label{eq:pwdivheqn}
\frac{\partial}{\partial x^{\mu}}h^{\mu}{}_{\nu}
=\frac{1}{2}\frac{\partial}{\partial x^{\nu}}h^{\mu}{}_{\mu}\;.
\end{equation}
In this gauge, the homogeneous field equations are
\begin{equation}
\label{eq:pwhpertfeqn}
\square\,h_{\mu\nu}=0\;,
\end{equation}
where $\square$ is the flat spacetime D'Alembertian operator.  
The gauge \eqref{eq:pwdivheqn} and field equations \eqref{eq:pwhpertfeqn} 
are preserved by gauge changes $\xi_{\nu}$ which satisfy
\begin{equation}
\label{eq:pwdivchi}
\square\,\xi_{\nu}=0\;.
\end{equation}
Equations \eqref{eq:pwdivheqn}-\eqref{eq:pwdivchi} are the flat 
spacetime background metric equivalents of the curved background spacetime 
equations \eqref{eq:divheqn}-\eqref{eq:hpertfeqn}, \eqref{eq:divchi}.  
Moreover, we can derive equations \eqref{eq:pwdivheqn}-\eqref{eq:pwdivchi} 
from the curved spacetime expressions, although Weinberg derives his 
equations \textit{ab initio}.  

Weinberg shows that the solution to the field equations 
\eqref{eq:pwhpertfeqn} is a wave
\begin{equation}
\label{eq:pwsol}
h_{\mu\nu}(x)=e_{\mu\nu}\exp(i k_{\lambda}x^{\lambda})
+\overline{e}_{\mu\nu}\exp(-i k_{\lambda}x^{\lambda})\;,
\end{equation}
where $e_{\mu\nu}$ is the symmetric polarization tensor.  Here and 
elsewhere in this thesis, an overbar usually represents complex conjugation; 
however, this notational rule does not apply to $\overline{h}_{\mu\nu}$ 
\eqref{eq:hbar} and its trace, $\overline{h}$.  
Substituting $h_{\mu\nu}$ \eqref{eq:pwsol} into the field equations 
\eqref{eq:pwhpertfeqn} and gauge definition \eqref{eq:pwdivheqn} yields
\begin{equation}
\label{eq:kpwsol}
k_{\mu}k^{\mu}=0\;,\,k_{\mu}e^{\mu}{}_{\nu}
=\tfrac{1}{2}k_{\nu}e^{\mu}{}_{\mu}\;,
\end{equation}
respectively.  A symmetric $4\times 4$ matrix has at most ten independent 
components.  The gauge definition \eqref{eq:pwdivheqn} represents four 
constraints, which reduce the number of independent components, or 
polarizations, to six.  The gauge transformation vector appropriate 
to $h_{\mu\nu}$ \eqref{eq:pwsol} is
\begin{equation}
\label{eq:plxi}
\xi^{\mu}(x)=i B^{\mu}\exp(i k_{\lambda}x^{\lambda})
-i\overline{B}^{\mu}\exp(-i k_{\lambda}x^{\lambda})\;,
\end{equation}
where $B^{\mu}$ is a constant vector (equation (9.14) of \cite{schutz90}).  
A change of gauge modifies the polarization tensors as
\begin{equation}
\label{eq:deltemn}
e_{\mu\nu}^{\text{new}}=e_{\mu\nu}^{\text{old}}+k_{\mu}\xi_{\nu}+k_{\nu}\xi_{\mu}\;.
\end{equation}
The gauge transformation vector \eqref{eq:plxi} satisfies \eqref{eq:pwdivchi}.

The solution \eqref{eq:pwsol} is a plane wave traveling in the 
positive $z$-direction if the wave vector $k^{\lambda}$ is
\begin{equation}
\label{eq:zks}
k^{x}=k^{y}=0\;,\,k^{z}=k^{t}>0\;.
\end{equation}
For such a plane wave, the six independent polarizations can be written 
as the following linear combinations of $e_{\mu\nu}$:
\begin{equation}
\label{eq:allpol}
e_{\pm}=e_{xx}\mp i e_{xy}\;,\,
f_{\pm}=e_{zx}\mp i e_{zy}\;,\,
e_{tt}\;,\,e_{zz}\;.
\end{equation}
The remaining components are not independent and depend on the six above, 
by symmetry and by the following equations:
\begin{equation}
\label{eq:fourpol}
e_{tx}=-e_{zx}\;,\,e_{ty}=-e_{zy}\;,\,
e_{tz}=-\tfrac{1}{2}(e_{tt}+e_{zz})\;,\,e_{yy}=-e_{xx}\;.
\end{equation}
Weinberg shows the components $e_{\pm}$ are gauge invariant, because they 
can not be removed by a coordinate transformation.  However, $f_{\pm}$, 
$e_{tt}$ and $e_{zz}$ can be eliminated by a coordinate transformation which 
satisfies \eqref{eq:pwdivchi} and preserves the gauge 
condition~\eqref{eq:pwdivheqn}.  

The six independent components \eqref{eq:allpol} behave differently when 
a rotation is made about the $z$-axis, the direction of wave propagation.  
Specifically, Weinberg shows
\begin{equation}
\label{eq:pwrot}
e_{\pm}^{\prime}=\exp(\pm 2 i\theta)e_{\pm}\;,\,
f_{\pm}^{\prime}=\exp(\pm i\theta)f_{\pm}\;,\,
e_{tt}^{\prime}=e_{tt}\;,\,e_{zz}^{\prime}=e_{zz}\;.
\end{equation}
where the primes denote polarizations after rotation through an angle 
$\theta$.  A plane wave $\psi$ has helicity $h$ if 
\begin{equation}
\label{eq:heldef}
\psi^{\prime}=\exp(i h\theta)\psi\;.
\end{equation}
The rotation results \eqref{eq:pwrot} show that the gravitational 
plane wave $h_{\mu\nu}$ \eqref{eq:pwsol} can be decomposed into six pieces: 
two of helicity~$\pm 2$, two of helicity~$\pm 1$ and two of helicity~$0$.  
The trace of the polarization tensor is $e_{zz}-e_{tt}$, so the zero helicity 
pieces are related to the trace of $h_{\mu\nu}$.  Only the helicity~$\pm 2$ 
components are physically meaningful, because they alone can not be 
removed by a coordinate transformation.

Schwarzschild metric perturbation theory has an analogue to the different 
helicity functions of the plane wave example.  
The generalized Regge-Wheeler equation is
\begin{equation}
\label{eq:grweqn}
\frac{d^2 \psi_s(r_{*})}{d r_{*}^2}
+\omega^2 \psi_s(r_{*})-V_{sl}(r) \psi_s(r_{*})=S_{slm}(\omega,r_{*})
\;,\;\;s=0,1,2\;,
\end{equation}
where the potential $V$ is
\begin{equation}
\label{eq:vgrweqn}
V_{sl}(r)=\ff\left(\frac{2(\lambda+1)}{r^2}
+(1-s^{2})\frac{2M}{r^3}\right)\;.
\end{equation}
The generalized Regge-Wheeler equation is discussed in \cite{hs00}, 
\cite{leav86}, \cite{leopois97}.  It represents a wave interacting 
with an effective potential that results from the background spacetime 
curvature.  The parameter $s$ is the spin or spin weight, and it 
corresponds to the different helicities of the plane wave example.  
The case $\psi_2$ is equal to the Regge-Wheeler function that 
was derived in the Regge-Wheeler gauge.  
Described in \cite{rw57} and \cite{zerp70}, the coordinate $r_{*}$ is 
\begin{equation}
\label{eq:rstar}
dr_{*}=\frac{dr}{\ff}\;,\;
r_{*}=r+2 M \ln\!\left(\frac{r}{2M}-1\right)\;,
\end{equation}
so that
\begin{equation}
\label{eq:drstar}
\frac{\!\!\!\!\!d }{dr_{*}}=\ff\frac{\!\!\!d }{dr}
\end{equation}
and
\begin{equation}
\label{eq:ddrstar}
\frac{\!\!\!d^2 }{d r_{*}^2}=
\left(1-\frac{2M}{r}\right)^2 \frac{\!\!\!d^2 }{dr^2}+
\left(1-\frac{2M}{r}\right)\frac{2M}{r^2} \frac{\!\!\!d }{dr}\;.
\end{equation}
Because $2 M<r<\infty$, we have $-\infty<r_{*}<\infty$.  Due to 
this relation, $r_{*}$ is called the ``tortoise'' coordinate \cite{mtw73}.  
Throughout this thesis, the parameter $\lambda$ is defined 
as \cite{zerp70}
\begin{equation}
\label{eq:lam}
\lambda=\frac{1}{2}(l-1)(l+2)\;,
\end{equation}
which implies
\begin{equation}
\label{eq:ltolam}
2(\lambda+1)=l(l+1)\;.
\end{equation}
The generalized Regge-Wheeler equation will often be abbreviated as
\begin{equation}
\label{eq:grweqnop}
\mathcal{L}_{s}\psi_{s}=S_{s}\;,
\end{equation}
where the operator $\mathcal{L}_{s}$ is defined by the left side of 
\eqref{eq:grweqn}.  The source $S_{s}$ is constructed from the radial 
coefficients of the stress energy tensor \eqref{eq:tmunu}-\eqref{eq:etmunu}.

The solutions for the harmonic gauge are described in terms of generalized 
Regge-Wheeler functions.  Specifically, the odd parity solutions are 
written in terms of two generalized Regge-Wheeler functions, one with 
$s=2$ and the other with $s=1$.  The even parity solutions are written 
in terms of two functions with $s=0$, one with $s=1$ and one with $s=2$.  
The even parity spin $2$ function is actually the Zerilli function, but 
can be related to the spin $2$ Regge-Wheeler function by differential 
operators.  The even and odd parity spin $2$ functions are gauge 
invariant.  These six functions correspond to the six different helicity 
states of the plane wave example.  

\section{\label{sec:sumth}Summary of Thesis}

Most of this thesis discusses the solution of the field 
equations in the harmonic gauge \eqref{eq:hpertfeqn}.  
Chapters~\ref{oddpar} and~\ref{evpar} summarize the derivation of the 
odd and even parity solutions, respectively.  The solutions are 
expressions for the radial coefficients (such as $\hz$ and $\bhz$) of 
the angular functions in the odd \eqref{eq:ohmunu} and even 
\eqref{eq:ehmunu} parity metric perturbations $h_{\mu\nu}$.  
The cases of non-zero and zero angular frequency are handled 
separately.  Both inhomogeneous and homogeneous solutions are 
covered.  Some of the even parity solutions are listed separately in the 
appendices.  With a few cited exceptions, the solutions derived 
in these two chapters are new and constitute the main research 
results of this thesis.  

Chapter~\ref{eqmochap} covers the equations of motion.  Both the 
background geodesic and gravitational self-force equations are presented.  
Chapter~\ref{tmunuchap} discusses the stress energy tensor for a point 
mass, its multipole decomposition and its Fourier transforms for 
circular and elliptic orbits.  The chapter includes formulae for the 
radial coefficients (such as $So_{22}$ and $Se_{22}$) of the angular 
functions of the stress energy components $T_{\mu\nu}$ in equations 
\eqref{eq:otmunu} and \eqref{eq:etmunu}.  Chapter~\ref{rweqnchap} 
has homogeneous and inhomogeneous solutions for the generalized 
Regge-Wheeler and Zerilli equations.  Chapter~\ref{radchap} explains 
how to use the harmonic gauge solutions to calculate gravitational 
waveforms and energy and angular momentum fluxes.  Chapter~\ref{numchap} 
contains results of numerical calculations, mainly of 
the gravitational self-force for circular orbits.  Chapter~\ref{concchap} 
is a brief conclusion.  

\textit{Mathematica} was used extensively for the derivations.  Some 
results are attributed to unpublished work of Neil Ashby, which is 
cited as reference \cite{ashby}.  Among other things, he rederived 
and corrected the published solutions for the Regge-Wheeler gauge.  
He also provided \textit{Mathematica} tools for simplifying expressions 
(including derivatives of angular functions) and for deriving recursion 
relations for infinite series.  Quantities related to the background 
Schwarzschild metric, such as Christoffel symbols and Riemann 
curvature tensor components, were calculated with 
\textit{Mathematica}-based software written by him.		
\chapter{\label{oddpar}Odd Parity Solutions}

This chapter contains the derivation of the odd parity field equations 
and their solutions in the harmonic gauge.  We will use separation of 
variables to reduce the perturbed field equations to a system of three 
coupled ordinary differential equations, with independent variable $r$.  
The resulting equations are solved in terms of generalized 
Regge-Wheeler functions, with $s=2$ and 
$s=1$.  Section~\ref{sec:nzoddpar} describes the non-zero frequency 
solutions, with the cases of spherical harmonic index $l\ge2$ and $l=1$ 
handled separately.  Section~\ref{sec:zoddpar} does the same for 
zero frequency.  Section~\ref{sec:hoddpar} concludes with a discussion 
of homogeneous solutions.

\section{\label{sec:nzoddpar}Non-Zero Frequency Solutions}

Subsection~\ref{sec:nzoddparge} covers the case $l\ge2$.  
Subsection~\ref{sec:nzoddpareq} shows how the solutions for $l=1$ can 
be obtained from the solutions for $l\ge2$.

\subsection{\label{sec:nzoddparge}Solutions for $l \ge 2$}

The first step is to derive the radial field equations, using separation 
of variables.  The method of derivation is similar to that used by Regge, 
Wheeler and Zerilli for the Regge-Wheeler gauge \cite{rw57}, \cite{zerp70}.  
We substitute the odd 
parity metric perturbations from (\ref{eq:hmunu})-(\ref{eq:ohmunu}) and 
stress energy tensor components from (\ref{eq:tmunu})-(\ref{eq:otmunu}) 
into the harmonic gauge field equations (\ref{eq:hpertfeqn}), which are a 
system of coupled partial differential equations.  Because the equations are 
Fourier transforms, the time dependence is 
in the exponentials $e^{-i\omega t}$, which divide off.  Partial 
derivatives with respect to time become factors of angular frequency, 
using the rule $\frac{\!\!\partial}{\partial t}\to-\iom$.  The angular 
variables $\theta$ and $\phi$ are contained entirely in the tensor 
harmonics and their derivatives.  After simplifying the 
angular derivatives, the tensor harmonics also separate off, and  
we are left with the following radial factors to the odd parity field 
equations: 
\begin{multline}
\ff^2 \ddhz+\frac{(-8 M^2+4(2+\lambda)M r-r^2 
(2+2 \lambda+(\iom)^2 r^2))}{r^4}\hz\\+\frac{2\iom M(2 M-r)}{r^3}\ho
=-16\pi\ff So_{02}\;,\label{eq:ddhzeqn}
\end{multline}
\begin{multline}
\ff^2 \ddho+\frac{4 M}{r^2}\ff\dho+\frac{2\iom M}
{2 M r-r^2}\hz+\frac{\lambda(8 M-4 r)}{r^4}\htw
\\+\frac{-16 M^2+4(5+\lambda)M r-r^2 (6+2 \lambda+(\iom)^2 r^2)}
{r^4}\ho=-16\pi\ff So_{12}\;,\label{eq:ddhoeqn}
\end{multline}
\begin{multline}
\ff^2 \ddhtw-\frac{2(2 M-r)(3 M-r)}{r^3}\dhtw
-\frac{2(r-2 M)^2}{r^3}\ho
\\+\frac{16 M^2+4(-3+\lambda)M r-r^2 (-2+2 \lambda+(\iom)^2 r^2)}{r^4}\htw
\\=-16\pi\ff So_{22}\;.\label{eq:ddhtweqn}
\end{multline}
Primes signify differentiation with respect to $r$.  Generally, 
functional dependence on $l,m,\omega$ and $r$ will be suppressed, so that, 
for example, $\hz$ abbreviates $\hz^{lm}(\omega,r)$ and $So_{02}$ represents 
$So^{lm}_{02}(\omega,r)$.  However, the system of equations 
(\ref{eq:ddhzeqn})-(\ref{eq:ddhtweqn}) must be solved for each combination of 
indices $l$, $m$ and $\omega$.  Equation (\ref{eq:ddhzeqn}) comes from 
the $t\phi$ 
component of the field equations (multiplied by a factor of $-(1-2M/r)$).  
Similarly, (\ref{eq:ddhoeqn}) and (\ref{eq:ddhtweqn}) are obtained from the 
$r\phi$ and $\phi\phi$ components, respectively.  Because there are only 
three odd parity radial functions ($\hz$, $\ho$ and $\htw$), there are 
only three odd parity radial field equations.  The remaining odd parity  
components of the field equations are either zero or have the same radial 
factors as (\ref{eq:ddhzeqn})-(\ref{eq:ddhtweqn}).   

Each of the three radial field equations can be written in the form
\begin{equation}
\ff^{2}h^{\prime\prime}_{i}-(\iom)^{2}h_{i}+\text{other terms}\;,\quad i=0,1, 2\;,
\end{equation}
or, alternatively,
\begin{equation}
\label{eq:waveqn}
\frac{d^2h_{i}}{dr_*^2}-(-\iom)^{2}h_{i}+\text{other terms}\;,
\end{equation}
where the ``other terms'' are at most first order derivatives.  
The inverse Fourier transform of (\ref{eq:waveqn}) is
\begin{equation}
\frac{\partial^2h_{i}}{\partial r_*^2}
-\frac{\partial^2h_{i}}{\partial t^2}+\text{other terms}\;,
\end{equation}
which is a wave equation.  In the time domain, the field equations are 
a hyperbolic system of partial differential equations.  The system is 
a well-posed initial value problem and can be solved numerically 
in the time domain \cite{bl05}, \cite{mst97}.  The field equations above 
agree with those derived by Barack and Lousto \cite{bl05}, 
who used different notation and worked in the time domain.  

Related equations are also separable.  The harmonic gauge condition 
$\overline{h}_{\mu\nu}{}^{;\nu}=0$ leads to the radial equation
\begin{equation}
\ff \dho-\frac{\iom r}{2 M-r}\hz-\frac{2(M-r)}{r^2}\ho
+\frac{2\lambda}{r^2}\htw=0\;.\label{eq:diveqnh}
\end{equation}
The stress energy tensor divergence equation, $T_{\mu\nu}\!{}^{;\nu}=0$, is 
\cite{ashby}
\begin{equation}
\label{eq:divseqn}
\ff So_{12}^{\prime}-\frac{\iom r}{2 M-r}So_{02}
-\frac{2(M-r)}{r^2}So_{12}+\frac{2 \lambda}{r^2}So_{22}=0\;.
\end{equation}
For a change of gauge described by equations \eqref{eq:hnew} and 
\eqref{eq:chimu}-\eqref{eq:ochimu}, the radial perturbation factors 
transform as
\begin{equation}
\hz^{\text{new}}=\hz^{\text{old}}+\iom Z\;,\label{eq:hznew}
\end{equation}
\begin{equation}
\ho^{\text{new}}=\ho^{\text{old}}+\frac{2}{r}Z-Z^{\prime}\;,\label{eq:honew}
\end{equation}
\begin{equation}
\htw^{\text{new}}=\htw^{\text{old}}+Z\;,\label{eq:htwnew}
\end{equation}
where the function $Z$ is order $\frac{m_{0}}{M}$ \cite{ashby}, \cite{rw57}, 
\cite{zerp70}.  To preserve the harmonic gauge, a change of gauge must 
satisfy $\xi_{\mu;\nu}{}^{;\nu}=0$, and the associated radial equation is
\begin{equation}
\frac{d^2 Z}{dr_*^2}+\omega^2 Z-\ff\frac{2(\lambda+1)}{r^2}Z=0\;.
\label{eq:divzeqn}
\end{equation}
This is the homogeneous generalized Regge-Wheeler 
equation, with $s=1$.  Odd parity gauge changes which preserve the 
harmonic gauge are implemented by adding homogeneous spin~1 solutions 
to the metric perturbations.  

To derive another first order equation, differentiate \eqref{eq:diveqnh} with 
respect to $r$ and use \eqref{eq:ddhoeqn} and \eqref{eq:diveqnh} 
to eliminate $\ddho$ and $\dho$.  This gives
\begin{multline}
-\iom\dhz-\frac{2 \lambda}{r^2}\ff\dhtw+\frac{2 \iom}{r}
\hz+\left(-(\iom)^2+\frac{\lambda(4 M-2 r)}{r^3}\right)\ho
\\+\frac{4 \lambda}{r^3}\ff\htw=-16 \pi\ff So_{12}\;.\label{eq:dhzeqn}
\end{multline}
Unlike the harmonic gauge condition \eqref{eq:diveqnh}, 
which applies only in the harmonic gauge, 
equation \eqref{eq:dhzeqn} is gauge invariant and applies in any gauge.  
As shown by \eqref{eq:hznew}-\eqref{eq:htwnew}, 
the radial functions are gauge dependent; however, a gauge change to one 
(say $\hz$) is canceled by changes to the remaining functions, leaving 
equation \eqref{eq:dhzeqn} the same in the new gauge.  The stress energy 
tensor term $So_{12}$ is coordinate dependent; however, it is order 
$\frac{m_{0}}{M}$, so any changes to it 
are order $\left(\frac{m_{0}}{M}\right)^{2}$.  Thus, \eqref{eq:dhzeqn} 
is gauge invariant to linear order in $\frac{m_{0}}{M}$.  In contrast, 
a gauge change applied to \eqref{eq:diveqnh} results in additional terms 
involving $Z$ and its derivatives, unless the change satisfies 
\eqref{eq:divzeqn} and preserves the harmonic gauge.  Equation 
\eqref{eq:diveqnh} is invariant only under changes which preserve 
the harmonic gauge, but equation \eqref{eq:dhzeqn} is invariant under 
arbitrary gauge changes.  Another way of deriving \eqref{eq:dhzeqn} is to 
substitute the odd parity metric perturbations into \eqref{eq:pertfeqn}, 
the general perturbation field equations applicable to any gauge.  Equation 
\eqref{eq:dhzeqn} is obtained from the $r\phi$ component, multiplied 
by $-\ff$.  Because \eqref{eq:pertfeqn} is gauge invariant, 
so is \eqref{eq:dhzeqn}.  

Regge and Wheeler showed that, in the Regge-Wheeler gauge, the odd parity 
perturbations can be written in terms of a single scalar function  
which satisfies a wave equation called the Regge-Wheeler equation \cite{rw57}.  
In the notation of this thesis, this function is the odd parity $\ptw$, 
and the Regge-Wheeler equation is
\begin{equation}\frac{d^2 \ptw}{dr_*^2}+\omega^2 \ptw
-\ff\left(\frac{2(\lambda+1)}{r^2}-\frac{6M}{r^3}\right)\ptw
=S_{2}\;.\label{eq:ptweqn}
\end{equation}
Equation \eqref{eq:ptweqn} is the generalized Regge-Wheeler equation from 
\eqref{eq:grweqn}, with $s=2$.  Subsequently, Moncrief proved that the 
Regge-Wheeler function $\ptw$ is gauge invariant  
\cite{monc74a}-\cite{monc74d}.  He deduced a formula for $\ptw$ in terms 
of the metric perturbations.  Adjusting for differences in notation between 
his paper and this thesis, his expression is
\begin{equation}
\label{eq:spin2func}
\ptw=\ff\left(\frac{\ho}{r}-\frac{2 \htw}{r^2}
+\frac{\dhtw}{r}\right)\;.
\end{equation}
Although the radial perturbation functions are gauge dependent, $\ptw$ 
is gauge invariant.  Moncrief showed that any gauge change in $\ho$ 
would be offset by changes to $\htw$ and $\dhtw$, as follows.  
From equation \eqref{eq:honew}, the term containing $\ho$ changes by 
$\frac{2 Z}{r^2}-\frac{Z^{\prime}}{r}$.  Using \eqref{eq:htwnew}, the terms 
with  $\htw$ and $\dhtw$ change by $-\frac{2 Z}{r^2}+\frac{Z^{\prime}}{r}$.  
The changes cancel each other, leaving $\ptw$ invariant.  Moncrief did not 
derive his result in the harmonic gauge, but his work can be used 
here, because \eqref{eq:spin2func} is gauge invariant.

In equation \eqref{eq:ptweqn}, the quantity $S_{2}$ is a source term 
constructed from the radial components of the stress energy tensor.  
To find $S_{2}$, substitute \eqref{eq:spin2func} into the Regge-Wheeler 
equation and simplify with the other harmonic gauge equations, obtaining 
\begin{equation}S_{2}=-\frac{16\pi}{r}\ff^2 So_{12}
+\frac{32\pi (6 M^2 -5 M r+r^2)}{r^4}So_{22}-\frac{16\pi}{r}
\ff^2 So_{22}^{\prime}\;.\label{eq:ptwsource}
\end{equation}
Taking into account differences in notation, equation \eqref{eq:ptwsource} 
agrees with Zerilli's result in the Regge-Wheeler gauge \cite{zerp70}, as 
corrected by others \cite{ashby}, \cite{sago03}.

Solutions for $\ho$ and $\hz$ can be written terms of $\htw$ 
and $\ptw$.  By solving \eqref{eq:spin2func} for $\ho$, we find
\begin{equation}
\ho=\frac{2}{r}\htw+\frac{r^2}{r-2 M}\ptw-\dhtw\;.
\label{eq:nh1sol}
\end{equation}
We then use \eqref{eq:nh1sol} and its radial derivative to eliminate $\ho$ 
and $\dho$ from \eqref{eq:diveqnh} and solve for $\hz$ to get
\begin{equation}
\hz=\frac{(2 M-r)}{\iom}\dptw-\ff\frac{\ptw}{\iom }
+\iom \htw - \ff\frac{16\pi}{\iom}So_{22}\;.\label{eq:nh0sol}
\end{equation}
Substituting for $\ho$ in the field equation \eqref{eq:ddhtweqn}, we have
\begin{equation}
\mathcal{L}_{1}\htw=2 \ff \ptw -16 \pi \ff So_{22}\;,
\label{eq:nddhtweqn}
\end{equation}
where the operator $\mathcal{L}_{1}$ was defined in \eqref{eq:grweqnop}.  
The left side is the generalized Regge-Wheeler equation, with $s=1$.  To 
complete the odd parity solutions, we will solve \eqref{eq:nddhtweqn} 
for $\htw$ and substitute the result into \eqref{eq:nh0sol} and 
\eqref{eq:nh1sol}.

The form of \eqref{eq:nddhtweqn} suggests the following trial solution
\begin{equation}
\htw^{\text{try}}=\frac{e_{d}}{C}\dptw+\frac{\po}{C}
+\frac{e_{r}}{C}\ptw+16 \pi\left(f_{02}So_{02}+f_{12}So_{12}+f_{22}So_{22}\right)\;,
\end{equation}
where $C$ is a constant and the other quantities are functions of $r$.  
To find the unknowns, we insert $\htw^{\text{try}}$ into the left side of 
\eqref{eq:nddhtweqn} and obtain
\begin{equation}
\label{eq:ddpsi1}
\mathcal{L}_{1}\psi_{1}=\ptw,\dptw\text{ terms}+\text{source terms}\;.
\end{equation}
The ``source terms'' are complicated expressions involving $So_{02}$, 
$So_{12}$ and $So_{22}$, and their first and second radial derivatives.  
The ``$\ptw,\dptw$ terms'' are terms proportional to either $\ptw$ 
or $\dptw$.  In order that $\po$  does not couple to $\ptw$, we set 
the coefficients of $\ptw$ and $\dptw$ equal to zero, which produces 
two second order differential equations for $e_{r}$ and $e_{d}$.  The 
coefficient of $\ptw$ gives
\begin{multline}
\label{eq:ereqn}
\frac{(-2 M+r)^2}{r^2}e_{r}^{\prime\prime}
+\frac{2 M (-2 M+r)}{r^3}e_{r}^{\prime}+\frac{6 M (2 M-r)}{r^4}e_{r}
\\+\frac{\left(-48 M^3+4 (17+2 \lambda) M^2 r+4 (1+\lambda) r^3
+2 M r^2 \left(-15-6 \lambda+2 (\iom)^{2}r^2\right)\right)}{(2 M-r) r^5}e_{d}
\\+\frac{2 \left(12 M^2-2 (5+2 \lambda) M r+r^2 \left(2+2 \lambda
+(\iom)^{2} r^2\right)\right)}{r^4}e_{d}^{\prime}=\frac{2 C (r-2 M)}{r}\;,
\end{multline}
and the coefficient of $\dptw$ yields
\begin{equation}
\label{eq:edeqn}
\frac{(-2 M+r)^2}{r^2}e_{d}^{\prime\prime}
+\frac{2 M (2 M-r)}{r^3}e_{d}^{\prime}
+\frac{2 (-2 M+r)^2}{r^2}e_{r}^{\prime}
+\frac{2 M (4 M-r)}{r^4}e_{d}=0\;.
\end{equation}

Equations \eqref{eq:ereqn} and \eqref{eq:edeqn} can be solved by 
substituting series trial solutions, namely,
\begin{equation}
e_{r}^{\text{try}}=\sum_{n=-3}^{n=4}\frac{a_n}{r^n}\;,\,
e_{d}^{\text{try}}=\sum_{n=-3}^{n=4}\frac{b_n}{r^n}\;.
\end{equation}
Doing so leads to
\begin{multline}
e_{r}=C_{r}-\frac{2 C M}{(\iom)^{2} r}
+\frac{-\frac{3 C_{d} M}{2}-\frac{3C M^2}{(\iom)^{2}}}{r^2}
\\-\frac{(\iom)^{2} (2 C_{d} (1+\lambda)+3 C_{r} M)
+C (M+2 \lambda M)}{2 (\iom)^{4} r^3}+O(r^{-4})
\end{multline}
and
\begin{multline}
e_{d}=\frac{C r}{(\iom)^{2}}+C_{d}
+\frac{-2 C_{d} M-\frac{4 C M^2}{(\iom)^{2}}}{r}
-\frac{(\iom)^{2} (2 C_{d} (1+\lambda)+3 C_{r} M)
+C (M+2 \lambda M)}{2 (\iom)^{4} r^2}
\\+\frac{M \left(4 C (4+3 \lambda) M+(\iom)^{2} (C_{d} (11
+8 \lambda)+6 C_{r} M)\right)}{2 (\iom)^{4} r^3}+O(r^{-4})\;.
\end{multline}
The series terminate if 
\begin{equation}
C_{r}=\frac{C (3+2 \lambda)}{3 (\iom)^{2}}\;,\,
C_{d}= -\frac{2 C M}{(\iom)^{2}}\;,
\end{equation}
which gives
\begin{equation}
e_{r}=\frac{C (-6 M+(3+2 \lambda) r)}{3 (\iom)^{2} r}\;,\,
e_{d}=\frac{C (-2 M+r)}{(\iom)^{2}}\;.
\end{equation}
These solutions can be verified by substitution into \eqref{eq:ereqn} 
and \eqref{eq:edeqn}.  If we choose $C=(\iom)^2$, we have
\begin{multline}
\htw^{\text{try}}=\frac{(-2 M+r) }{(\iom)^{2}}\dptw
+\frac{\po}{(\iom)^{2}}+\frac{(-6 M+(3+2 \lambda) r) }{3 (\iom)^{2}r}\ptw
\\+16 \pi \left(f_{02}So_{02}+f_{12}So_{12}+f_{22}So_{22}\right)\;.
\end{multline}

As mentioned previously, the ``source terms'' in equation \eqref{eq:ddpsi1} 
include second radial derivatives of $So_{02}$, $So_{02}$ and $So_{02}$.  
The second derivatives are eliminated if $f_{02}=f_{12}=0$ and 
\begin{equation}
\label{eq:oddfrules}
f_{22}=\frac{r-2 M}{(\iom)^{2}r}\;.
\end{equation}
Substituting these results into $\htw^{\text{try}}$, we finally obtain 
the solution to equation \eqref{eq:nddhtweqn},
\begin{equation}
\label{eq:h2sol}
\htw=\frac{1}{(\iom)^2}\left[(r-2 M)\dptw+\po
+\frac{-6 M+(3+2\lambda)r}{3 r}\ptw+16\pi\ff So_{22}\right]\;,
\end{equation}
and its radial derivative,
\begin{multline}
\label{eq:dh2sol}
\dhtw=\frac{1}{(\iom)^2}\left[\frac{2(-6 M
+(3+\lambda)r)}{3r}\dptw+\dpo\right.
\\+\frac{-8 M^2+4(2+\lambda)Mr-r^2(2+2\lambda
+(\iom)^2 r^2)}{(2 M-r)r^2}\ptw\\\left.-16\pi\ff So_{12}
+\frac{32\pi}{r}\ff So_{22}\right]\;.
\end{multline}

Equation \eqref{eq:spin2func} defines $\ptw$ in terms of odd parity 
radial functions.  An analogous expression for $\po$ is derived by 
solving \eqref{eq:h2sol} for $\po$ and using \eqref{eq:nh0sol} and 
\eqref{eq:nh1sol} to eliminate $\dptw$ and $\ptw$, respectively.  
The result is
\begin{equation}
\label{eq:spin1func}
\po=\iom\hz+\ff\frac{2\lambda}{3 r}
\left(-\ho+\frac{2}{r}\htw-\dhtw\right)\;.
\end{equation}
Substituting \eqref{eq:spin1func} into the spin~$1$ generalized Regge-Wheeler 
equation gives the differential equation for $\po$,
\begin{multline}
\label{eq:poeqn}
\mathcal{L}_{1}\po=\frac{32 (3+\lambda) \pi (r-2 M)^2 }{3 r^3}So_{12}
+\frac{32 \lambda \pi (r-2 M) }{3 r^3}So_{22}
\\+\frac{16 \pi (r-2 M)^3 }{r^3}So_{12}^{\prime}
+\frac{32 \lambda \pi (r-2 M)^2 }{3 r^3}So_{22}^{\prime}\;.
\end{multline}
Equation \eqref{eq:divseqn} may be used to eliminate $So_{12}^{\prime}$.  
Although \eqref{eq:spin2func} is gauge invariant, equation 
\eqref{eq:spin1func} is not:  it is valid only in the harmonic gauge.  
If a gauge change is made which satisfies \eqref{eq:divzeqn} and 
thereby preserves the harmonic gauge, $\po$ changes by a homogeneous 
spin~$1$ solution (in other words, a homogeneous solution of 
\eqref{eq:poeqn}).  
Stated differently, 
\begin{equation}
\label{eq:nspin1func}
\po^{\text{new}}=\po^{\text{old}}+\po^{\text{hom}}
=\iom\hz^{\text{new}}+\ff\frac{2\lambda}{3 r}
\left(-\ho^{\text{new}}+\frac{2}{r}\htw^{\text{new}}
-\frac{d\htw^{\text{new}}}{dr}\right)\;,
\end{equation}
where $\po^{\text{hom}}$ refers to the homogeneous solution.  
Even though $\po$ has changed by $\po^{\text{hom}}$, 
the right side of \eqref{eq:nspin1func} has the same form as 
\eqref{eq:spin1func}.  In this limited sense, the formula for $\po$ in 
\eqref{eq:spin1func} is invariant under gauge changes which 
preserve the harmonic gauge.  Nevertheless, the behavior of $\po$ is 
different from that of $\ptw$, because $\ptw^{\text{new}}=\ptw^{\text{old}}$ 
after any change of gauge.

The above solutions for $\htw$ and $\dhtw$ are substituted into 
\eqref{eq:nh0sol} and \eqref{eq:nh1sol} to obtain
\begin{equation}
\label{eq:h0sol}
\hz=\frac{1}{\iom}\left(\po+\frac{2\lambda}{3}\ptw\right)
\end{equation}
and
\begin{equation}
\label{eq:h1sol}
\ho=\frac{1}{(\iom)^2}\left[-\frac{2\lambda}{3}\dptw
+\frac{2}{r}\po-\frac{2\lambda}{3 r}\ptw+16\pi\ff So_{12}-\dpo\right]\;,
\end{equation}
and their radial derivatives,
\begin{equation}
\label{eq:dh0sol}
\dhz=\frac{1}{\iom}\left(\dpo+\frac{2\lambda}{3}\dptw\right)
\end{equation}
and
\begin{multline}
\label{eq:dh1sol}
\dho=\frac{1}{(\iom)^2}\left[\frac{2\lambda(-4 M+r)}
{3(2 M-r)r}\dptw+\frac{-8 M^2+4(3+\lambda)Mr-r^2(4+2\lambda
+(\iom^2)r^2)}{r^2(-2 M+r)^2}\po\right.
\\-\frac{2\lambda(8 M^2-2(3+2\lambda)M r+r^2(1+2\lambda+(\iom)^2 r^2)}
{3 r^2(-2 M+r)^2}\ptw\\\left.+\frac{32\pi(M-r)}{r}So_{12}
-\frac{32\pi\lambda}{r^2}So_{22}+\frac{2(-M+r)}{r(-2 M+r)}\dpo\right]\;.
\end{multline}
One may verify by substitution that these solutions satisfy the field 
equations, as well as the harmonic gauge condition, 
the first order equation \eqref{eq:dhzeqn}, and the definitions of 
$\ptw$ and $\po$.  

The solutions also can be checked by transforming from the harmonic 
gauge (``H'') to the Regge-Wheeler gauge (``RW''). By definition, 
$\htw^{\text{RW}}=0$.  Applying  equation \eqref{eq:htwnew}, we set 
$Z=-\htw^{H}$ and substitute into equations \eqref{eq:hznew} and 
\eqref{eq:honew} to obtain
\begin{equation}
\label{eq:rwhz}
\hz^{\text{RW}}=\frac{(2 M-r)}{\iom}\dptw-\ff\frac{\ptw}{\iom }
- \ff\frac{16\pi}{\iom}So_{22}\;,
\end{equation}
and
\begin{equation}
\label{eq:rwho}
\ho^{\text{RW}}=\frac{r^2}{r-2 M}\ptw\;.
\end{equation}
Adjusting for differences in notation, the Regge-Wheeler gauge solutions 
agree with those obtained by Zerilli and others 
\cite{ashby}, \cite{sago03}, \cite{zerp70}.  
Equation \eqref{eq:rwho} can be solved for 
$\ptw$ to give the Regge-Wheeler expression for $\ptw$.  In the limit 
$\htw\to 0$, Moncrief's formula, equation \eqref{eq:spin2func}, 
reduces to theirs.

The preceding paragraph suggests another way of deriving the harmonic 
gauge solutions.  Instead of working throughout in the harmonic gauge, 
we could have started in the Regge-Wheeler gauge (where the solutions 
are known) and looked for the gauge transformation vector that would 
take us from the Regge-Wheeler gauge to the harmonic gauge.  To do so, 
we would need to derive the radial function $Z$ in \eqref{eq:ochimu} 
that we would then substitute into \eqref{eq:hznew}-\eqref{eq:htwnew}, 
with the superscript ``old'' referring to the Regge-Wheeler gauge and 
``new'' being the harmonic gauge.  This approach was begun in 
\cite{sago03}, where a differential equation for the gauge 
transformation vector was derived.  The equation is similar to 
\eqref{eq:nddhtweqn} above, although the derivation in \cite{sago03} 
is very different.  However, the authors of 
\cite{sago03} did not solve their gauge transformation equation (they put 
it aside for ``future study'') and did not complete the 
derivation of the harmonic gauge solutions.  

Because equation \eqref{eq:dhzeqn} is gauge invariant, the formula for 
$\ptw$ in \eqref{eq:spin2func} is not unique.  We can solve 
\eqref{eq:dhzeqn} for $\dhtw$ and substitute for $\dhtw$ in 
\eqref{eq:spin2func}.  The result is 
\begin{equation}
\label{eq:jtspin2funcps}
\ptw=-\iom\ptw^{\text{JT}}+\frac{8\pi(r-2 M)}{\lambda}So_{12}\;,
\end{equation}
where
\begin{equation}
\label{eq:jtspin2func}
\ptw^{\text{JT}}=\frac{1}{\lambda}\left(-\hz+\frac{\iom r}{2}\ho
+\frac{r}{2}\dhz\right)\;.
\end{equation}
The superscript ``JT'' stands for Jhingan and Tanaka, who derived 
$\ptw^{\text{JT}}$ in the Regge-Wheeler gauge using a different method and 
notation and showed that it is gauge invariant \cite{jt03}.  We will use the 
superscript ``JT'' because they discuss this form extensively, 
even though they acknowledge it was derived earlier by others.  
In \eqref{eq:jtspin2func}, the factor of $-\iom$ indicates that 
$\ptw$ is the time derivative of $\ptw^{\text{JT}}$, and Jhingan and Tanaka 
constructed $\ptw^{\text{JT}}$ so that it would be the Fourier transform of a 
time integral of $\ptw$.  Applying equations \eqref{eq:hznew} and 
\eqref{eq:honew}, a gauge change in $\ho$ is canceled by corresponding 
gauge changes to $\hz$ and $\dhz$, leaving $\ptw^{\text{JT}}$ invariant.  After 
substituting $\ptw^{\text{JT}}$ into the Regge-Wheeler equation and simplifying, 
we obtain a source term given by
\begin{equation}
\label{eq:jtptweqn}
\mathcal{L}_{2}\ptw^{\text{JT}}=-\frac{8\pi(r-2 M)}
{\lambda}\left(\iom So_{12}+So_{02}^{\prime}\right)\;.
\end{equation}
Adjusting for differences in notation, the source term agrees with that in 
\cite{jt03}.  Using \eqref{eq:jtspin2funcps}, the field equation solutions 
can be rewritten in terms of $\ptw^{\text{JT}}$, if desired.

This completes the derivation of the odd parity non-zero frequency solutions, 
for $l\ge 2$.  We started with the harmonic gauge field equations given by 
\eqref{eq:hpertfeqn}, which are partial differential equations.  Using 
separation of variables, that system was reduced to the three radial 
field equations in \eqref{eq:ddhzeqn}-\eqref{eq:ddhtweqn}, a system of 
coupled ordinary differential equations.  The solutions to 
the radial equations are written in 
terms of $\ptw$ and $\po$, each of which satisfies its own decoupled 
second order differential equation.  To calculate $\hz$, $\ho$ and $\htw$, 
we would first solve the decoupled equations for $\ptw$ and $\po$ and then 
substitute the results into the solutions \eqref{eq:h0sol}, \eqref{eq:h1sol} 
and \eqref{eq:h2sol}.  In this manner, we have simplified the problem from 
a system of coupled partial differential equations to two decoupled 
ordinary differential equations.

\subsection{\label{sec:nzoddpareq}Solutions for $l = 1$}

Although subsection~\ref{sec:nzoddparge} assumed that $l\ge 2$, most of the 
results derived there can be used for $l=1$.  One difference is that 
$\htw$ is no longer present.  In equation \eqref{eq:ohmunu} for the odd 
parity metric perturbations, the angular functions $W$ and $X$ are zero 
\cite{zerp70}, so $\htw$ does not exist for $l=1$.  Similarly, $So_{22}$ is 
non-existent.  Further, there are only two field equations, \eqref{eq:ddhzeqn} and 
\eqref{eq:ddhoeqn}, because \eqref{eq:ddhtweqn} is the radial coefficient of 
$W$ or  $X$.  To show that $W$ and $X$ are zero, substitute the spherical 
harmonics for $l=1$ \cite{arfken85} into the definitions of $W$ \eqref{eq:wlm} 
and $X$ \eqref{eq:xlm}.

If $l=1$, then  $\lambda=0$, from the definition of $\lambda$ \eqref{eq:lam}.  
In the remaining field equations, any terms involving $\htw$ have a factor 
of $\lambda$, so such terms are zero.  Other, 
related equations --- \eqref{eq:diveqnh}-\eqref{eq:honew} 
and \eqref{eq:divzeqn}-\eqref{eq:dhzeqn} --- also still 
apply, with the substitution $\lambda\to 0$ to ensure that terms with 
$\htw$ and $So_{22}$ are zero.  

Finally, the solutions for $\hz$ and $\ho$ apply, again with 
$\lambda\to 0$.  This means the solutions do not depend on $\ptw$, so 
$\ptw$ is not defined for $l=1$.  The solutions now depend on 
$\po$, as given by \eqref{eq:spin1func} and \eqref{eq:poeqn}.

\section{\label{sec:zoddpar}Zero Frequency Solutions}

The zero-frequency equations and solutions have $\omega=0$.   
Factors of $\omega$ are due to time derivatives of the 
factor $e^{-\iom t}$.  Because this exponential contains the time dependence 
of the metric perturbation, solutions with $\omega=0$ are time independent 
solutions.  
Below, the cases $l\ge 2$ and $l=1$ are discussed separately.

\subsection{\label{sec:zoddparge}Solutions for $l \ge 2$}

We can use the field and related equations from 
subsection~\ref{sec:nzoddparge}, after substituting $\omega=0$.  
However, $\hz$ no longer couples 
to $\ho$ and $\htw$ through the radial field equations.  Instead, $\hz$ 
has a separate field equation, \eqref{eq:ddhzeqn}, while $\ho$ and $\htw$ 
solve a coupled system, \eqref{eq:ddhoeqn} and \eqref{eq:ddhtweqn}.  
This decoupling of $\hz$ also appears in other equations, namely, the 
harmonic gauge condition \eqref{eq:diveqnh} and the gauge invariant 
first order equation \eqref{eq:dhzeqn}.  Neither contains $\hz$ if 
$\omega=0$.  Further, the source for \eqref{eq:ddhzeqn}, $So_{02}$, 
no longer couples to $So_{12}$ and $So_{22}$ through the stress energy 
tensor divergence equation \eqref{eq:divseqn}.

We will solve for $\ho$ and $\htw$ first.  Using \eqref{eq:dhzeqn} 
to eliminate $\ho$ from \eqref{eq:ddhtweqn}, we have
\begin{multline}
\ff^{2}\ddhtw+\ff \frac{2 M}{r^2}\dhtw+
\frac{2 (1+\lambda) (2 M-r)}{r^3}\htw
\\=16\pi\ff \left[\frac{(r-2 M)}{\lambda}So_{12}-So_{22}\right]\;.
\end{multline}
This is the generalized Regge-Wheeler equation, with $s=1$.  Accordingly, 
\begin{equation}
\label{eq:h2sol0}
\htw=\po\;,\,\dhtw=\dpo\;,
\end{equation}
where
\begin{equation}\mathcal{L}_{1}\po
=16\pi\ff \left[\frac{(r-2 M)}{\lambda}So_{12}-So_{22}\right]\;.
\end{equation}
Here, $\po$ is different from $\po$ in subsection~\ref{sec:nzoddparge}, 
even though the same notation is used for both functions.  
Substituting \eqref{eq:h2sol0} into \eqref{eq:dhzeqn} gives
\begin{equation}
\label{eq:h1sol0}
\ho=\frac{2}{r}\po-\dpo+\frac{8\pi r^2}{\lambda}So_{12}\;,
\end{equation}
and the radial derivative is
\begin{equation}
\dho=-\frac{2(-2 M+(2+\lambda)r)}{r^2(r-2 M)}\po
+\frac{2(r-M)}{r(r-2 M)}\dpo+\frac{16\pi r(r-M)}{\lambda(2 M-r)}So_{12}\;.
\end{equation}
By substitution, the reader may verify that \eqref{eq:h2sol0} and 
\eqref{eq:h1sol0} are solutions to the zero frequency field and 
related equations that involve $\ho$ and $\htw$.  The two solutions 
depend on $\po$, but not $\ptw$.  This means the definition of $\ptw$ in 
\eqref{eq:spin2func} is not valid for zero frequency, as that definition is 
in terms of $\ho$ and $\htw$.

The next step is to solve the remaining field equation, 
\eqref{eq:ddhzeqn}, for $\hz$.  From \eqref{eq:hznew}, $\hz$ is gauge 
invariant when $\omega=0$, and the solution will be in terms of the 
spin~$2$ Regge-Wheeler function, which is also gauge invariant.  Using 
\eqref{eq:jtspin2func}, we set $\ptw=\ptw^{\text{JT}}$, so that
\begin{equation}
\label{eq:spin2func0}
\ptw=\frac{1}{\lambda}\left(-\hz+\frac{r}{2}\dhz\right)\;,
\end{equation}
and
\begin{equation}
\mathcal{L}_{2}\ptw=-\frac{8\pi(r-2 M)}{\lambda}So_{02}^{\prime}\;.
\end{equation}
After differentiating \eqref{eq:spin2func0} and using \eqref{eq:ddhzeqn} to 
eliminate $\ddhz$, we find that
\begin{equation}
\label{eq:dspin2func0}
\dptw=-\frac{(-2 M+r+\lambda r)}{\lambda(2 M r-r^2)}\hz
-\frac{8\pi r^2 }{\lambda (-2 M+r)}So_{02}
-\frac{1}{2 \lambda}\dhz\;.
\end{equation}
Equations \eqref{eq:spin2func0} and \eqref{eq:dspin2func0} can be solved for 
$\hz$ and $\dhz$, giving
\begin{equation}
\hz=(r-2 M)\dptw+\ff\ptw+\frac{8\pi r^2}{\lambda}So_{02}\;,
\end{equation}
and
\begin{equation}
\dhz=2\ff\dptw+\frac{2(-2 M+r+\lambda r)}{r^2}\ptw
+\frac{16\pi r}{\lambda}So_{02}\;.
\end{equation}
For zero frequency and $l\ge 2$, the metric perturbation can be written in 
terms of spin~$2$ and spin~$1$ generalized Regge-Wheeler functions, 
just as for non-zero frequency modes.

\subsection{\label{sec:zoddpareq}Solutions for $l = 1$}

For $l=1$, the function $\htw$ is not present, as explained 
in subsection~\ref{sec:nzoddpareq}.  From \eqref{eq:ddhzeqn} 
and \eqref{eq:ddhoeqn}, the field equations become
\begin{equation}
\label{eq:ddhzeqnz1}
\ff^2 \ddhz-\frac{2 (r-2 M)^2 }{r^4}\hz=-16\pi\ff So_{02}\;,
\end{equation}
\begin{multline}
\label{eq:ddhoeqnz1}
\ff^2 \ddho+\frac{4 M}{r^2}\ff\dho
\\-\frac{2 \left(8 M^2-10 M r+3 r^2\right)}{r^4}\ho
=-16\pi\ff So_{12}\;.
\end{multline}
The harmonic gauge condition simplifies to
\begin{equation}
\label{eq:diveqnh1}
\ff\dho-\frac{2 (M-r)}{r^2}\ho=0\;.
\end{equation}
For the non-zero frequency mode having $l=1$, we used the solutions 
for $l\ge 2$, but we can not do that here.  The zero frequency solutions 
for $l\ge 2$ have factors of $\frac{1}{\lambda}$, but $\lambda=0$ 
if $l=1$.

The rules for a change of gauge are somewhat different from 
\eqref{eq:hznew}-\eqref{eq:honew} and are discussed in \cite{vish70}.  
Rewritten in terms of $\xi^{o}_{1}(t,r)$ instead of $Z(\omega,r)$, the 
rules become
\begin{equation}
\label{eq:hznew1}
\hz^{\text{new}}=\hz^{\text{old}}-\frac{\partial \xi^{o}_{1}(t,r)}{\partial t}\;,
\end{equation}
\begin{equation}
\label{eq:honew1}
\ho^{\text{new}}=\ho^{\text{old}}+\frac{2}{r}\xi^{o}_{1}(t,r)
-\frac{\partial \xi^{o}_{1}(t,r)}{\partial r}\;,
\end{equation}
where the replacement $\iom\to-\frac{\!\!\partial}{\partial t}$ 
has been used in \eqref{eq:hznew}.  The symbol $\xi^{o}_{1}$ is short 
for $\xi^{\text{odd}}_{l=1}$. Referring to \eqref{eq:chimu}-\eqref{eq:ochimu}, 
the odd parity gauge change vector for $l=1$ is
\begin{equation}
\begin{pmatrix}
0 , 0 , \xi^{o}_{1}(t,r) \csc\theta
\frac{\partial Y_{1m}}{\partial\phi} , 
-\xi^{o}_{1}(t,r)\sin\theta\frac{\partial Y_{1m}}{\partial\theta}
\end{pmatrix}\;.
\end{equation}
The rules are modified in order to allow a gauge change 
of the form
\begin{equation}
\label{eq:z1t}
\xi^{o}_{1}(t,r)=C_{1}\,t\,r^2\;,
\end{equation}
where $C_{1}$ is a constant.  
Although written in terms of $t$, this 
is actually a change in the coordinate $\phi$ \cite{vish70}.  The 
modification is necessary in order to allow a gauge change in $\hz$.  
If we substitute \eqref{eq:z1t} into \eqref{eq:hznew1} and 
\eqref{eq:honew1}, then $\hz$ changes by $-C_{1}\, r^{2}$, but $\ho$ 
is unaltered \cite{vish70}.  Because $\xi^{o}_{1}(t,r)$ is only linear 
in $t$, $\hz$ remains time independent.  This form of gauge change 
could not have been used for other modes.  
For non-zero frequency modes, the gauge vector time dependence is in 
the factor $e^{-i\omega t}$.  
For zero frequency modes with $l\ge 2$, 
the new form would cause $\htw$ to grow linearly with time, as 
\eqref{eq:htwnew} shows.  As a result, the zero frequency $\hz$ is 
effectively gauge invariant for $l\ge 2$ (by \eqref{eq:hznew}), but gauge 
dependent for $l=1$.

From \eqref{eq:divzeqn}, a gauge change which preserves the harmonic 
gauge must be a solution of
\begin{equation}
\label{eq:divzeqn1}
\frac{\partial^2 \xi^{o}_{1}(t,r)}{\partial r_*^2}
-\frac{\partial^2 \xi^{o}_{1}(t,r)}{\partial t^2}
-\ff\frac{2}{r^2}\xi^{o}_{1}(t,r)=0\;,
\end{equation}
which is the generalized Regge-Wheeler equation with $s=1$, written 
in the time domain.  The most general zero frequency 
gauge change which satisfies \eqref{eq:divzeqn1} and leaves the 
perturbation time independent is
\begin{equation}
\label{eq:z1g}
\xi^{o}_{1}(t,r)=C_{1}\,t\,r^2
+C_{2}\,\frac{2 M^2+2 M r+r^2\ln\left[1-\frac{2 M}{r}\right]}{8 M^3}
+C_{3}\,r^{2}\;.
\end{equation}
If $\xi^{o}_{1}(t,r)$ were non-linear in $t$, or had some other time dependence, 
$\hz^{\text{new}}$ and $\ho^{\text{new}}$ would be time dependent.  The first term of 
\eqref{eq:z1g} changes only $\hz$; the second, only $\ho$.  
The last term, $C_{3}\,r^{2}$, changes neither $\hz$ nor $\ho$, so we 
will disregard it.  

For $l=1$, the zero and non-zero frequency gauge changes rules may 
be combined to give
\begin{equation}
\xi^{o}_{1}(t,r)=C_{1}\, t \,r^{2}
+\int_{\!-\infty}^{\infty}e^{-\iom t} Z(\omega, r)d\omega\;.
\end{equation} 
This satisfies \eqref{eq:divzeqn1}, provided that $Z(\omega, r)$ 
is a solution of \eqref{eq:divzeqn} with $\lambda=0$.  
For non-zero frequency, $Z(\omega, r)$ 
is the function $Z$ referred to in section~\ref{sec:nzoddpar}.  
For zero frequency, $Z(\omega=0, r)$ is equal to the last two 
terms of \eqref{eq:z1g}.  

We will solve the field equations first for $\ho$ and then for $\hz$.  
If we use \eqref{eq:diveqnh1} and its radial derivative
to eliminate $\ddho$ and then $\dho$ from \eqref{eq:ddhoeqnz1}, 
we find that the left side of \eqref{eq:ddhoeqnz1} is reduced to zero.  
This means $So_{12}=0$.  More directly, the same result can be obtained 
from \eqref{eq:dhzeqn}, by substituting $\lambda=0$ and $\omega=0$.  
Because $So_{12}=0$, the function $\ho$ is a homogeneous solution of 
\eqref{eq:ddhoeqnz1} and \eqref{eq:diveqnh1}.  Accordingly, 
\begin{equation}
\label{eq:h1sol01}
\ho=\frac{C}{r(r-2 M)}\;,
\end{equation}
where $C$ is a constant.  If C is non-zero, we can nevertheless zero 
out $\ho$ by a gauge transformation described in \eqref{eq:z1g}, with 
$C_{1}=C_{3}=0$ and $C_{2}=C$.  Because $\ho$ is entirely gauge dependent, 
we can set $C=0$ in \eqref{eq:h1sol01}, so $\ho=0$ for $l=1$ and $\omega=0$.

The other radial function, $\hz$, represents the orbital angular momentum 
of the small mass \cite{zerp70}.  
Equation \eqref{eq:ddhzeqnz1} has homogeneous solutions of $\frac{2 M}{r}$ 
and $(\frac{r}{2 M})^2$.  From them, we can construct an inhomogeneous 
solution using variation of parameters \cite{mandw}, or a Green's 
function \cite{jack75}.  Either method produces
\begin{equation}
\label{eq:h0sol01}
\hz=\frac{1}{r}\,\int_{\!2 M}^{r}\frac{16\pi 
{r^{\prime}}^{3}So_{02}(r^{\prime})}{3 (r^{\prime}-2 M)}dr^{\prime}
+r^2\,\int^{\infty}_{r}\frac{16\pi So_{02}(r^{\prime})}
{3 (r^{\prime}-2 M)}dr^{\prime}\;,
\end{equation}
\begin{equation}
\dhz=-\frac{1}{r^2}\,\int_{\!2 M}^{r}
\frac{16\pi {r^{\prime}}^{3}So_{02}(r^{\prime})}{3 (r^{\prime}-2 M)}dr^{\prime}
+2 r\,\int^{\infty}_{r}\frac{16\pi So_{02}(r^{\prime})}
{3 (r^{\prime}-2 M)}dr^{\prime}\;,
\end{equation}
where $So_{02}(r^{\prime})$ is short for $So_{02}(\omega=0,r^{\prime})$.  
Alternatively, the solution for $\hz$ may be written in terms of a spin~$1$ 
Regge-Wheeler function and its radial derivative; however, the resulting 
expressions are more complicated than those given above.  The limits 
$2 M$ and $\infty$ are generic and should be replaced by limits 
appropriate to the orbital motion.  For example, an elliptic orbit 
would have limits of $r_{\text{min}}$ (periastron) instead of $2 M$ and 
$r_{\text{max}}$ (apastron) instead of $\infty$.

We can analytically evaluate the integrals in $\hz$ \eqref{eq:h0sol01} 
for a circular orbit of radius $R$, with the stress energy tensor 
expressions in Chapter \ref{tmunuchap}.  Doing so gives
\begin{equation}
\label{eq:h0sol01cir}
\hz=-\mz\frac{4\elbar r^2}{R^3}\sqrt{\frac{\pi }{3}}\,\theta(R-r)
-\mz\frac{4\elbar}{r}\sqrt{\frac{\pi}{3}}\,\theta(r-R)\;,
\end{equation}
where we have used \cite{arfken85}
\begin{equation}
\label{eq:dy10}
\frac{\partial Y_{10}(\theta,\phi)}{\partial\theta}
=-\sqrt{\frac{3}{4\pi}}\,\sin \theta\;.
\end{equation}
The factor $\elbar$ is the orbital angular momentum per unit mass 
\eqref{eq:endef}, \eqref{eq:cendef}.  To find 
the perturbation, we substitute $\hz$ \eqref{eq:h0sol01cir} and 
$\ho=0$ into $h^{o,lm}_{\mu\nu}$ \eqref{eq:ohmunu}, getting
\begin{equation}
\label{eq:hmunu10}
h_{t\phi}^{10}=h_{\phi t}^{10}
=\left(-\mz\frac{2\elbar r^{2}}{R^{3}}\sin^{2}\theta\right)\theta(R-r)
+\left(-\mz\frac{2\elbar}{r}\sin^{2}\theta\right)\theta(r-R)\;.
\end{equation}
The other perturbation components are zero for this mode, and the 
superscript~``10'' is short for $l=1,m=0$.  The solution \eqref{eq:hmunu10} 
was also calculated by Detweiler and Poisson \cite{dp04}, following 
Zerilli \cite{zerp70}.  They note the solution for $r>R$ resembles the 
linearized Kerr metric.  Specifically, the $r^{-1}$ term 
has the same form as the $g_{t\phi}$ component of the Kerr metric 
linearized to $O(a)$, with $\mz\elbar/M$ taking the place of the Kerr 
angular momentum parameter~$a$.  In contrast, the solution for $r<R$ 
differs from the background metric only by a gauge transformation 
\cite{dp04}.  As discussed in \cite{zerp70} and \cite{dp04}, the 
solution for this multipole represents the orbital angular momentum 
of the mass $\mz$ and the change in angular momentum that occurs at 
the orbital radius.

\section{\label{sec:hoddpar}Homogeneous Solutions}

This section discusses homogeneous solutions of the odd parity 
field equations.  We will start with the non-zero frequency case, 
with $l\ge 2$.

The radial field equations in \eqref{eq:ddhzeqn}-\eqref{eq:ddhtweqn} 
form a system of three second order linear ordinary differential 
equations.  From the theory of differential equations, such a system 
may be reformulated as an equivalent system of six first 
order equations, and the new system will have at most six linearly 
independent homogeneous solutions \cite{zill93}.  Accordingly, one might 
expect that the field equations could also have six homogeneous 
solutions.  However, consider the following system of four first 
order homogeneous linear differential equations: 
\begin{equation}
\label{eq:sys4a}
\dhtw-d_{2}=0\;,
\end{equation}
\begin{multline}
\label{eq:sys4b}
\ff^2 d_{2}^{\,\prime}-\frac{2(2 M-r)(3 M-r)}{r^3}d_{2}
-\frac{2(r-2 M)^2}{r^3}\ho
\\+\frac{16 M^2+4(-3+\lambda)M r-r^2(-2+2\lambda+(\iom)^2 r^2)}{r^4}\htw
=0\;,
\end{multline}
\begin{equation}
\label{eq:sys4c}
\ff \dho-\frac{\iom r}{2 M-r}\hz-\frac{2(M-r)}{r^2}\ho
+\frac{2\lambda}{r^2}\htw=0\;,
\end{equation}
\begin{multline}
\label{eq:sys4d}
-\iom\dhz-\frac{2 \lambda}{r^2}\ff d_{2}+\frac{2 \iom}{r}\hz
+\left(-(\iom)^2+\frac{\lambda(4 M-2 r)}{r^3}\right)\ho
\\+\frac{4 \lambda}{r^3}\ff\htw=0\;.
\end{multline}
Equation \eqref{eq:sys4b} is the homogeneous form of \eqref{eq:ddhtweqn}, 
rewritten in terms of $d_{2}$ \eqref{eq:sys4a}.  Similarly, \eqref{eq:sys4c} is 
the harmonic gauge condition, \eqref{eq:diveqnh}, and \eqref{eq:sys4d} 
is from \eqref{eq:dhzeqn}.  Thus, the system of four is derived from 
the field equations and the harmonic gauge condition.  Moreover, it is 
possible to show that the homogeneous form of the field equations can 
be derived from \eqref{eq:sys4a}-\eqref{eq:sys4d}.  Each system 
implies the other, so the system of four and the homogeneous field 
equations are equivalent.  The system of four 
has only four linearly independent solutions. Because these four equations 
are equivalent to the homogeneous form of the field equations, the 
odd parity field equations also have only four linearly independent 
homogeneous solutions.  This reduction from six to four evidently 
occurs because of the harmonic gauge condition, which is a differentiable 
constraint.  

Solutions to a system of linear differential equations are written 
in the form of column vectors.  If a matrix formed from the solution 
vectors has a non-zero determinant, the solutions are linearly independent 
\cite{zill93}.  

We need to find four linearly independent solution vectors 
for the system \eqref{eq:sys4a}-\eqref{eq:sys4d}.  As explained in 
Chapter \ref{rweqnchap}, the non-zero frequency generalized Regge-Wheeler 
equation has two linearly independent homogeneous solutions:  
$\psi^{\text{in}}_{s}$, which represents an ingoing wave near the event 
horizon, and $\psi^{\text{out}}_{s}$, which represents an outgoing wave at 
large $r$.  This suggests that we form one solution vector from each of 
$\po^{\text{in}}$, $\po^{\text{out}}$, $\ptw^{\text{in}}$ and $\ptw^{\text{out}}$.  
The first solution vector is
\begin{equation}
\label{eq:bigx1}
\bm{X}_{1}^{T}=\left(\frac{\po^{\text{in}}}{\iom}\;,\,
\frac{1}{(\iom)^2}\left[\frac{2}{r}\po^{\text{in}}
-(\po^{\text{in}})^{\prime}\right]\;,\,
\frac{\po^{\text{in}}}{(\iom)^2}\;,\,
\frac{(\po^{\text{in}})^{\prime}}{(\iom)^2}\right)\;.
\end{equation}
The $T$ is for transpose, because $\bm{X}_{1}$ is a column vector, 
rather than the row vector displayed above.  The first component of 
\eqref{eq:bigx1} comes from the $\po$ term of $\hz$ \eqref{eq:h0sol}, 
the second component from the $\po$ terms of $\ho$ \eqref{eq:h1sol}, 
the third from $\htw$ \eqref{eq:h2sol} and the fourth from 
$d_{2}=\dhtw$ \eqref{eq:dh2sol}.  The second solution, $\bm{X}_{2}$, 
is obtained from $\bm{X}_{1}$ by replacing $\po^{\text{in}}$ with 
$\po^{\text{out}}$.  The third vector is 
\begin{multline}
\label{eq:bigx3}
\bm{X}_{3}^{T}=\left(\frac{2\lambda}{3\iom}\ptw^{\text{in}}\;,\,
-\frac{2\lambda}{3(\iom)^2}\left[(\ptw^{\text{in}})^{\prime}
+\frac{\ptw^{\text{in}}}{r}\right]\;,\,
\frac{1}{(\iom)^2}\bigg[(r-2 M)(\ptw^{\text{in}})^{\prime}\right.\\\left.
+\frac{-6 M+(3+2\lambda)r}{3 r}\ptw^{\text{in}}\bigg]\;,\,
\frac{1}{(\iom)^2}\left[A\,(\ptw^{\text{in}})^{\prime}
+B\,\ptw^{\text{in}}\right]\right)\;,
\end{multline}
where
\begin{equation}
\label{eq:bigx3a}
A=\frac{2(-6 M+(3+\lambda)r)}{3r}\;,
\end{equation}
\begin{equation}
\label{eq:bigx3b}
B=\frac{-8 M^2+4(2+\lambda)Mr-r^2(2+2\lambda+(\iom)^2 r^2)}{(2 M-r)r^2}\;.
\end{equation}
The fourth solution, $\bm{X}_{4}$, is obtained from $\bm{X}_{3}$ by 
replacing $\ptw^{\text{in}}$ with $\ptw^{\text{out}}$.  The function
\begin{equation}
\label{eq:detfirst}
\frac{r^4 W_{1}W_{2}}{(\iom)^{5}(r-2 M)^2}
\end{equation}
is the determinant of the matrix whose four columns are $\bm{X}_{1}$ 
through $\bm{X}_{4}$.  In equation \eqref{eq:detfirst}, the factors 
$W_{1}$ and $W_{2}$ are the Wronskians $W_{s}$ \eqref{eq:grwwr} of the 
generalized Regge-Wheeler homogeneous solutions.  Because the determinant 
\eqref{eq:detfirst} is non-zero, the solution vectors $\bm{X}_{1}$ through 
$\bm{X}_{4}$ are linearly independent.  From the theory of ordinary 
differential equations, any homogeneous solution to the system 
\eqref{eq:sys4a}-\eqref{eq:sys4d} (and, by extension, the field 
equations \eqref{eq:ddhzeqn}-\eqref{eq:ddhtweqn}) can be written 
as a linear combination of the solution vectors described above.  
Accordingly, homogeneous solutions of the odd parity field equations 
are formed from combinations of the generalized Regge-Wheeler 
functions $\ptw$ and $\po$.  This result is to be expected, given the form 
of the inhomogeneous solutions derived in subsection~\ref{sec:nzoddparge}.

Gauge transformations which preserve the harmonic gauge are implemented 
by adding homogeneous spin $1$ solutions \eqref{eq:divzeqn}.  Because 
of this gauge freedom, the $\po$ contributions to the homogeneous 
solutions above can be removed by a gauge transformation which preserves 
the harmonic gauge.  However, the $\ptw$ solutions can not be so 
removed.  This is because the spin $2$ Regge-Wheeler functions are 
gauge invariant.  To modify the two solution vectors attributable to 
$\ptw$ ($\bm{X}_{3}$, $\bm{X}_{4}$), we would have to transform to a 
different gauge, such as the Regge-Wheeler gauge.  Also, a harmonic 
gauge preserving change adds only homogeneous solutions, so it can 
not remove the inhomogeneous $\po$ contributions derived in 
subsection \ref{sec:nzoddparge}.  

The homogeneous solutions above assumed $l\ge 2$.  For $l=1$, the 
radial functions $\htw$ and $\dhtw$ are not present, as explained 
in subsection \ref{sec:nzoddpareq}.  We are left with a two equation 
system that is composed of \eqref{eq:sys4c}-\eqref{eq:sys4d}, modified 
by the substitution $\lambda=0$.  Using similar arguments to the 
$l\ge 2$ case, we can show that the $l=1$ homogeneous solutions 
can be written solely in terms of $\po^{\text{in}}$ and $\po^{\text{out}}$.  
Specifically, the solution vectors are the first two components of 
$\bm{X}_{1}$ and $\bm{X}_{2}$.  Moreover, these homogeneous solutions 
can be removed by a gauge transformation which preserves the harmonic 
gauge.

The homogeneous solutions above are for non-zero frequency.  For 
zero frequency and $l\ge2$, we use the following first order system 
instead:
\begin{equation}
\label{eq:sysa0}
\dhz-d_{0}=0\;,
\end{equation}
\begin{equation}
\label{eq:sysb0}
\ff^2 d_{0}^{\,\prime}+\frac{(-8 M^2+4(2+\lambda)M r-r^2 
(2+2 \lambda))}{r^4}\hz=0\;,
\end{equation}
\begin{equation}
\label{eq:sysc0}
\ff \dho-\frac{2(M-r)}{r^2}\ho
+\frac{2\lambda}{r^2}\htw=0\;,
\end{equation}
\begin{equation}
\label{eq:sysd0}
-\frac{2 \lambda}{r^2}\ff\dhtw+\frac{\lambda(4 M-2 r)}{r^3}\ho
+\frac{4 \lambda}{r^3}\ff\htw=0\;.
\end{equation}
We use these four first order equations because of the decoupling 
of the zero frequency $\hz$ from $\ho$ and $\htw$.  If we had wished, 
we could have used the non-zero frequency equivalents of 
\eqref{eq:sysa0}-\eqref{eq:sysd0} instead of 
\eqref{eq:sys4a}-\eqref{eq:sys4d} in our discussion of the non-zero 
frequency case.  Equation \eqref{eq:sysb0} is the homogeneous form of 
the field equation \eqref{eq:ddhzeqn}, rewritten in terms of $d_{0}$ 
\eqref{eq:sysa0}.  Equation \eqref{eq:sysc0} is the zero frequency 
harmonic gauge condition \eqref{eq:diveqnh}, and equation \eqref{eq:sysd0} 
is the zero frequency form of \eqref{eq:dhzeqn}.  
It is possible to show that the system \eqref{eq:sysa0}-\eqref{eq:sysd0} 
is equivalent to the zero frequency field equations, in the same way 
that \eqref{eq:sys4a}-\eqref{eq:sys4d} are equivalent to the non-zero 
frequency field equations.  Accordingly, the zero frequency field 
equations have only four linearly independent homogeneous solutions.  

As discussed in Chapter \ref{rweqnchap}, there are two linearly 
independent zero frequency homogeneous solutions of the generalized 
Regge-Wheeler equation.  One solution, $\psi^{\text{in}}_{s}$, is 
finite as $r\to 2 M$, but diverges like $r^{l+1}$ as $r\to\infty$.  The 
other solution, $\psi^{\text{out}}_{s}$, is bounded for large $r$, but diverges 
logarithmically near the event horizon.  With arguments similar to the 
non-zero frequency case, we can show that all zero frequency homogeneous 
solutions can be written as combinations of $\psi^{\text{in}}_{1}$, 
$\psi^{\text{out}}_{1}$, $\psi^{\text{in}}_{2}$ and $\psi^{\text{out}}_{2}$.  
Further, the $\psi^{\text{in}}_{1}$ and $\psi^{\text{out}}_{1}$ solutions 
can be removed by means of a gauge transformation which preserves 
the harmonic gauge.  Although the spin~$2$ solutions remain, they 
are divergent, either at the horizon or for large $r$.  To prevent 
an unphysical divergence, we set them equal to zero by choice of 
integration constants.  A similar conclusion was reached by Vishveshwara 
\cite{vish70}, although he did not work in the harmonic gauge.

For $l=1$, $\htw$ is no longer present, and the homogeneous solution 
for $\ho$ is zero, as discussed in subsection \ref{sec:zoddpareq}.  
This leaves the single second order differential equation for $\hz$, 
which is written as two first order equations in 
\eqref{eq:sysa0}-\eqref{eq:sysb0}.  For reference, the second order 
equation is \eqref{eq:ddhzeqnz1}
\begin{equation}
\label{eq:ddhzeqnz1h}
\ddhz-\frac{2}{r^2}\hz=0\;,
\end{equation}
which has homogeneous solutions of the form
\begin{equation}
\label{eq:ddhzeqnz1hom}
C^{\text{in}}r^{2}+\frac{C^{\text{out}}}{r}\;.
\end{equation}
The $r^{2}$ solution can be removed by means of a transformation which 
preserves the harmonic gauge and does not change $\ho$.  To do so, 
simply set $C_{1}=C^{\text{in}}$ and $C_{2}=C_{3}=0$ in the gauge change vector 
$\xi^{o}_{1}(t,r)$ \eqref{eq:z1g}.  However, the $r^{-1}$ solution 
can not be so removed by \eqref{eq:z1g}.  Moreover, if we try to remove 
this solution by means of a gauge transformation which does not preserve 
the harmonic gauge, then $\ho$ will grow linearly with time, based on 
\eqref{eq:hznew1}-\eqref{eq:honew1}.  In other words, the $r^{-1}$ 
solution is not a purely gauge perturbation.  It will be zero only 
if we set $C^{\text{out}}=0$ on physical grounds.

Homogeneous solutions of \eqref{eq:ddhzeqnz1h} were studied by Vishveshwara 
\cite{vish70}, who did not work in the harmonic gauge specifically.  
He showed that the $r^{-1}$ solution gives
\begin{equation}
\label{eq:vishh03}
h_{t\phi}=\frac{c}{r}\sin^{2}\theta\;,
\end{equation}
which he observed is a rotational perturbation.  Vishveshwara further 
demonstrated that this solution can be made regular near the event 
horizon and elsewhere in Kruskal coordinates, following a gauge 
transformation which is equivalent to adding a specific $r^{2}$ solution 
described in \cite{vish70}.  

The rotational perturbation mentioned by Vishveshwara would describe 
the slow rotation of the central mass $M$.  Because we are assuming 
that the central mass is not rotating, we set $C^{\text{out}}=0$ in 
\eqref{eq:ddhzeqnz1hom}.  Orbital angular momentum of the small mass 
$\mz$ is described by the inhomogeneous perturbation \eqref{eq:h0sol01}, 
not by a homogeneous solution.

The work above shows that the physically meaningful odd parity 
homogeneous solutions to the harmonic gauge field equations are 
constructed from the non-zero frequency $\ptw$, which is gauge invariant.  
Other possible homogeneous solutions either can be removed by means 
of a gauge transformation which preserves the harmonic gauge, or 
do not satisfy the applicable boundary conditions. 		
\chapter{\label{evpar}Even Parity  Solutions}

Although more complicated, the calculation of the even parity solutions 
is similar to the odd parity derivation.  First, the field and related 
equations are obtained using separation of variables.  Next, the seven 
radial field equations are solved in terms of solutions to decoupled 
equations.  The spin~2 Regge-Wheeler function is replaced by a related 
spin~2 function, the solution to Zerilli's equation.  There are three 
generalized Regge-Wheeler functions:  one with $s=1$ and two with $s=0$.  
Non-zero frequency solutions are in section~\ref{sec:nzevpar}, 
zero frequency solutions are in section~\ref{sec:zevpar}, 
homogeneous solutions are in section~\ref{sec:hevpar}, and an interim 
summary of results is in section~\ref{sec:intsum}.

\section{\label{sec:nzevpar}Non-Zero Frequency Solutions}

Subsection~\ref{sec:nzevparge} describes solutions for $l\ge 2$.  
Subsection~\ref{sec:nzevpareq} explains how the solutions for $l=1$ 
and $l=0$ can be derived from the $l\ge 2$ solutions.

\subsection{\label{sec:nzevparge}Solutions for $l \ge 2$}

Using separation of variables, we derive the seven radial field 
equations:
\begin{multline}
\label{eq:ddbh0eqn}
\frac{(-2 M+r)^2 }{r^2}\ddbhz+\frac{2 (M-r) (2 M-r)}{r^3}\dbhz
-\frac{4 \iom M}{r^2}\bho+\frac{2 M (3 M-2 r) }{r^4}\bhtw
\\+\frac{\left(-2 M^2+4 (1+\lambda) M r-r^2 \left(2+2 \lambda
+(\iom)^2 r^2\right)\right) }{r^4}\bhz
+\frac{4 M (-2 M+r) }{r^4}\bk
\\=-8 \pi Se_{00}-\frac{8 \pi (-2 M+r)^2 }{r^2}Se_{11}
-\frac{16 \pi (-2 M+r) }{r^3}Ue_{22}\;,
\end{multline}

\begin{multline}
\label{eq:ddbh2eqn}
\begin{aligned}\frac{(-2 M+r)^2 }{r^2}\ddbhtw
&+\frac{2 (M-r) (2 M-r)}{r^3}\dbhtw+\frac{2 M (3 M-2 r) }{r^4}\bhz
-\frac{4 \iom M }{r^2}\bho\\&+\frac{8 (1+\lambda) (-2 M+r)^2 }{r^5}\ho
+\frac{4 (2 M-r) (3 M-r) }{r^4}\bk
\\&+\frac{\left(-18 M^2+4 (5+\lambda) M r-r^2 \left(6+2 \lambda
+(\iom)^2 r^2\right)\right) }{r^4}\bhtw\end{aligned}
\\=-8 \pi Se_{00}-\frac{8 \pi (-2 M+r)^2 }{r^2}Se_{11}
-\frac{16 \pi (2 M-r) }{r^3}Ue_{22}\;,
\end{multline}
\begin{multline}
\label{eq:ddkeqn}
\frac{(-2 M+r)^2 }{r^2}\ddbk+\frac{2 (M-r) (2 M-r)}{r^3}\dbk
+\frac{2 M (-2 M+r) }{r^4}\bhz-\frac{4 (1+\lambda) (-2 M+r)^2 }{r^5}\ho
\\+\frac{2 (2 M-r) (3 M-r) }{r^4}\bhtw+\frac{\left(-16 M^2+4 (4+\lambda) M r
-r^2 \left(4+2 \lambda+(\iom)^2 r^2\right)\right) }{r^4}\bk
\\=-8 \pi Se_{00}+\frac{8 \pi (-2 M+r)^2 }{r^2}Se_{11}\;,
\end{multline}
\begin{multline}
\label{eq:ddbh1eqn}
\frac{(-2 M+r)^2 }{r^2}\ddbho
+\frac{2 \left(2 M^2-3 M r+r^2\right) }{r^3}\dbho
-\frac{4 (1+\lambda) (2 M-r) }{r^4}\hz\\-\frac{2 \iom M }{r^2}\bhz+
\frac{\left(-4 M^2+4 (2+\lambda) M r-r^2 \left(4+2 \lambda
+(\iom)^2 r^2\right)\right) }{r^4}\bho\\-\frac{2 \iom M }{r^2}\bhtw
=-\frac{16 \pi (-2 M+r) }{r}Se_{01}\;,
\end{multline}
\begin{multline}
\label{eq:ddh0eqn}
\frac{(-2 M+r)^2 }{r^2}\ddhz+\frac{\left(-8 M^2+4
 (2+\lambda) M r-r^2 \left(2+2 \lambda+(\iom)^2 r^2\right)\right) }{r^4}\hz
\\+\frac{2 \iom M(2 M-r) }{r^3}\ho
+\frac{2 (-2 M+r)^2 }{r^3}\bho=-\frac{16 \pi (-2 M+r) }{r}Se_{02}\;,
\end{multline}
\begin{multline}
\frac{(-2 M+r)^2 }{r^2}\ddho+\frac{4 M (-2 M+r) }{r^3}\dho
+\frac{4 \lambda (-2 M+r) }{r^2}\bg\\+\frac{2 \iom M }{2 M r-r^2}\hz+
\frac{\left(-16 M^2+4 (5+\lambda) M r-r^2 \left(6+2 \lambda
+(\iom)^2 r^2\right)\right) }{r^4}\ho
\\+\frac{2 (-2 M+r) }{r^2}\bhtw+\frac{(4 M-2 r) }{r^2}\bk
=-\frac{16 \pi (-2 M+r) }{r}Se_{12}\;,
\end{multline}
\begin{multline}
\label{eq:ddgeqn}
\frac{(-2 M+r)^2 }{r^2}\ddbg+\frac{2 (M-r) (2 M-r)}{r^3}\dbg
+\left(-(\iom)^2+\frac{\lambda (4 M-2 r)}{r^3}\right) \bg
\\+\frac{2 (-2 M+r)^2 }{r^5}\ho=-\frac{16 \pi (-2 M+r) }{r^3}Se_{22}\;.
\end{multline}
Equations \eqref{eq:ddbh0eqn}, \eqref{eq:ddbh2eqn} and \eqref{eq:ddkeqn} 
are formed by combining the $tt$, $rr$ and $\theta\theta$ components of 
\eqref{eq:hpertfeqn} so that each equation contains the second derivative 
of only one radial function.  The $\theta\theta$ and $\phi\phi$ components 
of the even parity metric perturbation in \eqref{eq:ehmunu} each has two 
angular functions, $Y_{lm}(\theta,\phi)$ and $W_{lm}(\theta,\phi)$.  This 
structure carries over to the field equations, and the $\theta\theta$ 
component used above is the coefficient of the 
$Y_{lm}(\theta,\phi)$ term.  The remaining four field equations come from 
the $tr$, $t\theta$, $r\theta$ and $\theta\theta$ components, respectively, 
and here the $\theta\theta$ component is the coefficient of the 
$W_{lm}(\theta,\phi)$ term.  Other components of \eqref{eq:hpertfeqn} 
duplicate the equations listed above.  As is the case for odd parity, the 
even parity field equations form a hyperbolic system of partial differential 
equations when written in the time domain.  Taking into account differences 
in notation, the field equations agree with those in \cite{bl05}.

The harmonic gauge condition \eqref{eq:divheqn} gives three radial 
equations,
\begin{equation}
\label{eq:divheqn0}
\left(1-\frac{2 M}{r}\right) \dbho
-\frac{2 (1+\lambda) }{r^2}\hz+\frac{1}{2} \iom \bhz
-\frac{2 (M-r) }{r^2}\bho+\frac{1}{2} \iom \bhtw+\iom\bk=0\;,
\end{equation}
\begin{equation}
\label{eq:divheqn1}
\frac{\dbhz}{2}+\frac{\dbhtw}{2}-\dbk-\frac{M}{2 M r-r^2}\bhz
-\frac{2 (1+\lambda) }{r^2}\ho-\frac{\iom r }{2 M-r}\bho
+\frac{(3 M-2 r) }{2 M r-r^2}\bhtw-\frac{2 }{r}\bk=0\;,
\end{equation}
\begin{equation}
\label{eq:divheqn2}
\left(1-\frac{2 M}{r}\right) \dho-2 \lambda \bg
-\frac{\iom r }{2 M-r}\hz+\frac{\bhz}{2}-\frac{2 (M-r) }{r^2}\ho
-\frac{\bhtw}{2}=0\;.
\end{equation}
Equations \eqref{eq:divheqn0} and \eqref{eq:divheqn1} are from the $t$ 
and $r$ components of \eqref{eq:divheqn}, respectively, while 
\eqref{eq:divheqn2} can be obtained from either the $\theta$ or $\phi$ 
component.

The stress energy tensor divergence equation \eqref{eq:divtmunu} also 
generates three radial equations,
\begin{equation}
\label{eq:divt0}
\left(1-\frac{2
M}{r}\right) Se_{01}^{\prime}-\frac{\iom r }{2 M-r}Se_{00}
-\frac{2 (M-r)}{r^2}Se_{01}-\frac{2 (1+\lambda ) }{r^2}Se_{02}=0\;,
\end{equation}
\begin{multline}
\label{eq:divt1}
\left(1-\frac{2 M}{r}\right)Se_{11}^{\prime}
+\frac{M }{(-2 M+r)^2}Se_{00}-\frac{\iom r }{2 M-r}Se_{01}
-\frac{(M-2 r) }{r^2}Se_{11}\\-\frac{2 (1+\lambda
) }{r^2}Se_{12}-\frac{2 }{r^3}Ue_{22}=0\;,
\end{multline}
\begin{equation}
\label{eq:divt2}
\left(1-\frac{2
M}{r}\right) Se_{12}^{\prime}-\frac{\iom r }{2 M-r}Se_{02}
-\frac{2 (M-r) }{r^2}Se_{12}-\frac{2 \lambda  }{r^2}Se{22}
+\frac{1}{r^2}Ue_{22}=0\;.
\end{equation}
Equations \eqref{eq:divt0}, \eqref{eq:divt1} and \eqref{eq:divt2} are 
from the $t$, $r$ and $\theta$ (or $\phi$) components of 
\eqref{eq:divtmunu}, respectively.

Applying \eqref{eq:hnew}, a gauge change alters the radial perturbation 
functions by
\begin{equation}
\label{eq:bh0new}
\bhz^{\text{new}}=\bhz^{\text{old}}+\frac{2\iom}{\left(1-\frac{2 M}{r}\right)}\bmz
+\frac{2 M}{r^2}\bmo\;,
\end{equation}
\begin{equation}
\bho^{\text{new}}=\bho^{\text{old}}-\frac{2 M}{(2 M-r) r}\bmz+\iom \bmo-\dbmz\;,
\end{equation}
\begin{equation}
\bhtw^{\text{new}}=\bhtw^{\text{old}}-\frac{2 M}{r^2}\bmo-2\ff\dbmo\;,
\end{equation}
\begin{equation}
\label{eq:bknew}
\bk^{\text{new}}=\bk^{\text{old}}+\frac{2(2 M-r)}{r^2}\bmo
+\frac{2(1+\lambda)}{r^2}\bmtw\;,
\end{equation}
\begin{equation}
\label{eq:evhznew}
\hz^{\text{new}}=\hz^{\text{old}}-\bmz+\iom \bmtw\;,
\end{equation}
\begin{equation}
\ho^{\text{new}}=\ho^{\text{old}}-\bmo+\frac{2}{r}\bmtw-\dbmtw\;,
\end{equation}
\begin{equation}
\label{eq:gnew}
\bg^{\text{new}}=\bg^{\text{old}}-\frac{\bmtw}{r^2}\;,
\end{equation}
where $\bmz$, $\bmo$ and $\bmtw$ are defined 
in \eqref{eq:echimu} \cite{ashby}, \cite{rw57}, \cite{zerp70}.  
From \eqref{eq:divchi}, a gauge transformation that preserves the harmonic 
gauge must satisfy a system of three coupled differential equations,
\begin{multline}
\label{eq:ddm0eqn}
\left(1-\frac{2 M}{r}\right)^2 \ddbmz
+\frac{2(-2 M+r)^2 }{r^3}\dbmz+\frac{2 \iom M (2 M-r) }{r^3}\bmo
\\+\frac{\left(4 (1+\lambda) M-r \left(2+2 \lambda
+(\iom)^2 r^2\right)\right) }{r^3}\bmz=0\;,
\end{multline}
\begin{multline}
\label{eq:ddm1eqn}
\left(1-\frac{2 M}{r}\right)^2 \ddbmo
+\frac{2 (-2 M+r) }{r^2}\dbmo+\frac{2 \iom M }{2 M r-r^2}\bmz
-\frac{4 (1+\lambda) (2 M-r) }{r^4}\bmtw
\\+\frac{\left(-8 M^2+4 (3+\lambda) M r-r^2 \left(4+2 \lambda
+(\iom)^2 r^2\right)\right)}{r^4}\bmo=0\;,
\end{multline}
\begin{multline}
\label{eq:ddm2eqn}
\left(1-\frac{2 M}{r}\right)^2 \ddbmtw
+\frac{2 M (-2 M+r) }{r^3}\dbmtw+\frac{2 (-2 M+r)^2 }{r^3}\bmo
\\+\frac{\left(4 (1+\lambda) M-r \left(2+2 \lambda
+(\iom)^2 r^2\right)\right) }{r^3}\bmtw=0\;.
\end{multline}
Equations \eqref{eq:ddm0eqn}, \eqref{eq:ddm1eqn} and \eqref{eq:ddm2eqn} 
are from the $t$, $r$ and $\theta$ (or $\phi$) components of 
\eqref{eq:divchi}, respectively.
Later, we will show that solutions to this system can be written 
in terms of homogeneous solutions of the generalized Regge-Wheeler 
equation, with $s=1$ or $s=0$.  

By differentiating the harmonic gauge conditions in 
\eqref{eq:divheqn0}-\eqref{eq:divheqn2}, we can eliminate second 
derivatives from the field equations to get four additional 
first order equations,
\begin{multline}
\label{eq:ndh0eqns}
\left(1-\frac{2 M}{r}\right) \dbhz
+\frac{2 \lambda (1+\lambda) }{r}\bg
+\frac{2 (1+\lambda) \left(M^2-M r+(\iom)^2 r^4\right) }{\iom (2 M-r)r^4}\hz
-\frac{(1+\lambda) }{r}\bhz\\+\frac{(1+\lambda) (M-r) }{r^3}\ho
+\frac{\left(-(1+\lambda) M+r \left(1+\lambda
+2 (\iom)^2 r^2\right)\right) }{\iom r^3}\bho+\frac{M }{r^2}\bhtw
\\+\frac{\left(-3 M^2+2 (2+\lambda) M r-r^2 \left(1+\lambda
+(\iom)^2 r^2\right)\right) }{(2 M-r) r^2}\bk
+\frac{(1+\lambda) (M-r) }{\iom r^3}\dhz
\\=-\frac{8 \pi (M-r) }{\iom r}Se_{01}
-8 \pi (-2 M+r) Se_{11}\;,
\end{multline}
\begin{multline}
\label{eq:ndkeqns}
\left(1-\frac{2 M}{r}\right) \dbk
+\frac{2 (1+\lambda) M }{\iom r^4}\hz-\frac{(1+\lambda) (2 M-r) }{r^3}\ho
-\frac{(1+\lambda) (2 M-r) }{\iom r^3}\bho
\\+\frac{(2 M-r) }{r^2}\bhtw+\frac{(-3 M+r) }{r^2}\bk
+\frac{(1+\lambda) (2 M-r) }{\iom r^3}\dhz
=-\frac{8 \pi (2 M-r) }{\iom r}Se_{01}\;,
\end{multline}
\begin{multline}
\label{eq:ndgeqns}
\left(1-\frac{2 M}{r}\right)\lambda\dbg
+\frac{\lambda (1+\lambda) }{r}\bg
+\frac{\left(-(1+\lambda) M^2+2 (\iom)^2 M r^3
+(\iom)^2 \lambda r^4\right) }{\iom(2 M-r) r^4}\hz
\\-\frac{(3 M+\lambda r) }{2 r^2}\bhz-\frac{\left(M-3 \lambda M
+2 \lambda r+(\iom)^2 r^3\right) }{2 r^3}\ho+\frac{\left(M+\lambda M
+(\iom)^2 r^3\right) }{2 \iom r^3}\bho
\\+\frac{\left(3 M^2-M r+2 \lambda M r-\lambda r^2
-(\iom)^2 r^4\right) }{4 M r^2-2 r^3}\bk-\frac{\left(M+\lambda M+(\iom)^2
r^3\right) }{2 \iom r^3}\dhz
\\=\frac{4 M \pi }{\iom r}Se_{01}-4 \pi (-2 M+r) Se_{11}
-\frac{8 \pi (-2 M+r) Se_{12}}{r}\;,
\end{multline}
\begin{multline}
\label{eq:ndh2eqns}
\left(1-\frac{2 M}{r}\right)\dbhtw
-\frac{2 \lambda (1+\lambda)}{r}\bg
-\frac{2 (1+\lambda) \left(-3 M^2+M r
+(\iom)^2 r^4\right) }{\iom  (2 M-r) r^4}\hz
\\+\frac{(2 M+r+\lambda r) }{r^2}\bhz
+\frac{(1+\lambda) (3 M-r) }{r^3}\ho
-\frac{(1+\lambda) (3 M-r) }{\iom  r^3}\bho
\\+\frac{(-3 M+2 r) }{r^2}\bhtw
+\frac{\left(7 M^2-2 (5+\lambda) M r+r^2 \left(3+\lambda+(\iom)^2
r^2\right)\right)}{(2 M-r) r^2}\bk
\\+\frac{(1+\lambda) (3 M-r) }{\iom r^3}\dhz
=\frac{8 \pi (r-3 M)}{\iom  r}Se_{01}+8 \pi (r-2 M)Se_{11}\;.
\end{multline}
Alternatively, these four equations may be derived 
by manipulating the radial equations which can be extracted 
from \eqref{eq:hzterms}, after using the field equations to 
eliminate second derivatives with respect to $r$.  Three of the equations, 
\eqref{eq:ndh0eqns}, \eqref{eq:ndkeqns} and \eqref{eq:ndgeqns}, 
are gauge invariant to linear order, just as odd parity equation 
\eqref{eq:dhzeqn} is.  The three also can be obtained from the gauge 
invariant general perturbation field equations in \eqref{eq:pertfeqn}.  
In the Regge-Wheeler gauge, equation \eqref{eq:ndgeqns} 
simplifies to the so-called ``algebraic relation'' used by Regge, 
Wheeler and Zerilli to solve the field equations in their 
gauge \cite{rw57}, \cite{zerp70}.  

The first step in solving the field equations is to derive an even parity gauge 
invariant function, as was done for the odd parity case.  To do so, we write 
a trial solution in the form
\begin{multline}
\label{eq:evptwtry}
\ptw^{\text{try}}=f_{1}\bhz+f_{2}\bho+f_{3}\bhtw+f_{4}\bk+f_{5}\hz+f_{6}\ho+f_{7}\bg
\\+f_{d1}\dbhz+f_{d2}\dbho+f_{d3}\dbhtw+f_{d4}\dbk+f_{d5}\dhz
+f_{d6}\dho+f_{d7}\dbg\;,
\end{multline}
and then use \eqref{eq:bh0new}-\eqref{eq:gnew} to find a combination 
of the radial functions which leaves $\ptw^{\text{try}}$ invariant.  One 
combination, which will be called $\ptw$, is
\begin{multline}
\label{eq:evspin2func}
\ptw=2 r \bg+\frac{2 M }{\iom r(3 M+\lambda r)}\hz
+\frac{(2 M-r) }{3 M+\lambda r}\ho+\frac{(-2 M+r) }{\iom (3 M+\lambda r)}\bho
\\+\frac{r^2 }{3 M+\lambda r}\bk+\frac{(2 M-r) }{\iom(3 M+\lambda r)}\dhz\;.
\end{multline}
A gauge change in one radial function, say $G$, is canceled by changes 
in the remaining radial functions.  In the Regge-Wheeler gauge, we have 
$G=\hz=\ho=\dhz=0$.  With these substitutions,
\begin{equation}
\label{eq:evspin2funcrw}
\ptw=\frac{(-2 M+r)}{\iom(3 M+\lambda r)}\bho^{\text{RW}}
+\frac{r^2 }{3 M+\lambda r}\bk^{\text{RW}}\;,
\end{equation}
where the superscript ``RW'' signifies that $\bho$ and $\bk$ are 
computed in the Regge-Wheeler gauge.  Zerilli derived expressions for 
$\bho^{\text{RW}}$ and $\bk^{\text{RW}}$ \cite{zerp70}, which have been corrected by 
others \cite{ashby}, \cite{sago03}.  The corrected expressions can be 
solved for $\ptw$ \cite{ashby}, and the resulting ``Zerilli form'' 
agrees with \eqref{eq:evspin2funcrw}.  Solutions in the Regge-Wheeler 
gauge are provided below, in equations \eqref{eq:rwh0sol}-\eqref{eq:rwksol}.

The odd parity gauge invariant function, $\ptw$, is a solution of the 
generalized Regge-Wheeler equation, with $s=2$.  The even parity $\ptw$ 
is the solution of a related equation, the Zerilli equation \cite{zerp70}, 
which is
\begin{equation}
\label{eq:zereqn}
\mathcal{L}_{\!Z}\ptw=\frac{d^2 \ptw}{dr_*^2}+\omega^{2}\ptw
+\frac{2 (2 M-r) \left(9 M^3+9 \lambda M^2 r+3 \lambda^2 M r^2
+\lambda^2 (1+\lambda) r^3\right)}{r^4 (3 M+\lambda r)^2}\ptw=S_{2}\;.
\end{equation}
As will be shown in Chapter~\ref{rweqnchap}, homogeneous solutions 
of the Zerilli equation can be written in terms of homogeneous solutions 
of the Regge-Wheeler equation, using differential operators.  This means 
the Zerilli function is also a spin~$2$ function, which justifies using 
the notation $\ptw$ for it as well.  In fact, Jhingan and Tanaka showed 
that, in the Regge-Wheeler gauge, the even parity metric perturbation 
can be written in terms of Regge-Wheeler rather than Zerilli functions, 
although it is somewhat more complicated to do so \cite{jt03}.  
Presumably, the same could be done for other gauges.

The source term, $S_{2}$, in \eqref{eq:zereqn} is calculated by 
substituting $\ptw$ into the Zerilli equation and simplifying with 
the field and related equations.  The result is
\begin{multline}
\label{eq:zereqns1}
\mathcal{L}_{\!Z}\ptw=-\frac{8 \pi r^2 }{3 M+\lambda r}Se_{00}
-\frac{16 \lambda \pi (-2 M+r)^2 }{\iom (3 M+\lambda r)^2}Se_{01}
+\frac{16 \pi (-2 M+r)^2 }{r (3 M+\lambda r)}Se_{12}
\\+\frac{32 M \pi (2 M-r) (3 M-(3+\lambda) r)}
{\iom r^2 (3 M+\lambda r)^2}Se_{02}
+\frac{8 \pi (-2 M+r)^2 }{3 M+\lambda r}Se_{11}
\\+\frac{32 \pi (2 M-r) }{r^2}Se_{22}
+\frac{16 \pi(-2 M+r)^2}{\iom r (3 M+\lambda r)}Se_{02}^{\prime}\;.
\end{multline}
Using \eqref{eq:divt0}, we can eliminate $Se_{00}$ and rewrite the source as
\begin{multline}
\label{eq:zereqns2}
\mathcal{L}_{\!Z}\ptw=\frac{16 M \pi (2 M-r)(3 M-(3+\lambda)r)
 }{\iom r (3 M+\lambda r)^2}Se_{01}+\frac{8 \pi (-2 M+r)^2 }{3 M
+\lambda r}Se_{11}\\+\frac{16 \pi (2 M-r)\left(6 M^2
+(-3+\lambda) M r+\lambda (1+\lambda) r^2\right) }{\iom r^2 (3 M
+\lambda r)^2}Se_{02}+\frac{16 \pi (-2 M+r)^2 }{r (3 M+\lambda r)}Se_{12}
\\+\frac{32\pi (2 M-r) }{r^2}Se_{22}+\frac{8 \pi (-2 M+r)^2 }
{\iom (3 M+\lambda r)}Se_{01}^{\prime}+\frac{16 \pi (-2 M+r)^2 }{\iom r 
(3 M+\lambda r)}Se_{02}^{\prime}\;.
\end{multline}
Although \eqref{eq:zereqns1} and \eqref{eq:zereqns2} are equal,  
the latter expression agrees with Zerilli's form in a different notation 
\cite{zerp70}, as corrected by others \cite{ashby}, \cite{sago03}.

Because equations \eqref{eq:ndh0eqns}-\eqref{eq:ndgeqns} are gauge 
invariant, the definition of $\ptw$ in \eqref{eq:evspin2func} is not 
unique.  In particular, we can solve \eqref{eq:ndgeqns} for $\dhz$ 
and substitute the result into \eqref{eq:evspin2func} to get
\begin{multline}
\label{eq:step1}
\ptw=\bigg\{2 r \bg+\frac{(4 M-2 r) }{3 M+\lambda r}\ho
+\frac{r (-2 M+r) }{(1+\lambda) (3 M+\lambda r)}\bhtw
+\frac{r }{1+\lambda}\bk
\\+\frac{(2 M-r) r^2 }{(1+\lambda) (3 M+\lambda r)}\dbk\bigg\}
-\frac{8\pi  (2 M-r) r^2 }{(1+\lambda)(3 \iom M+\iom \lambda r)}Se_{01}\;.
\end{multline}
The part in curly brackets is gauge invariant.  It is Moncrief's form of $\ptw$, 
although he used different notation (including for the radial functions) 
and derived his result by other means \cite{monc74a}, \cite{monc74e}.  
Accordingly, we can define
\begin{multline}
\label{eq:evspin2funcmon}
\ptw^{\text{Mon}}=2 r \bg+\frac{(4 M-2 r) }{3 M+\lambda r}\ho
+\frac{r (-2 M+r) }{(1+\lambda) (3 M+\lambda r)}\bhtw
\\+\frac{r}{1+\lambda}\bk+\frac{(2 M-r) r^2}
{(1+\lambda)(3 M+\lambda r)}\dbk\;,
\end{multline}
so that
\begin{equation}
\ptw^{\text{Mon}}=\frac{8 \pi (2 M-r) r^2 }{\iom (1+\lambda)
 (3 M+\lambda r)}Se_{01}+\ptw\;.
\end{equation}
Substituting \eqref{eq:evspin2funcmon} into the Zerilli equation gives
\begin{multline}
\label{eq:moneqns}
\mathcal{L}_{\!Z}\ptw^{\text{Mon}}
=-\frac{8 \pi r \left(24 M^2+(-9+7 \lambda) M r+(-1+\lambda) 
\lambda r^2\right)}{(1+\lambda) (3 M+\lambda r)^2}Se_{00}
\\+\frac{8 \iom \pi r^2 (-2 M+r) }{(1+\lambda) (3 M+\lambda r)}Se_{01}
+\frac{8 \pi (-2 M+r)^2}{3 M+\lambda r}Se_{11}
+\frac{16 \pi (-2 M+r)^2 }{r (3 M+\lambda r)}Se_{12}
\\+\frac{32 \pi (2 M-r) }{r^2}Se_{22}
-\frac{8 \pi(2 M-r) r^2}{(1+\lambda) (3 M+\lambda r)}Se_{00}^{\prime}\;.
\end{multline}
For non-zero frequency modes, we will use $\ptw$ as given by 
\eqref{eq:evspin2func}, because it simplifies to the Zerilli form 
in the Regge-Wheeler gauge.  

We can use the definition of $\ptw$ to simplify the field equations 
by writing $\bhz$, $\bho$, $\bhtw$ and $\bk$ in terms of $\ptw$, 
$\hz$, $\ho$ and $\bg$.  The radial derivative of $\ptw$ is
\begin{multline}
\label{eq:devspin2func}
\dptw=\frac{6 M }{3 M+\lambda r}\bg
+\frac{2 M \left(6 M^2+3 \lambda M r
+\lambda (1+\lambda) r^2\right) }{\iom (2 M-r) r^2 (3 M+\lambda r)^2}\hz
\\-\frac{\left(12 M^2+9 \lambda M r+(-1+\lambda) \lambda r^2\right) }
{r (3 M+\lambda r)^2}\ho-\frac{\left(6 M^2+3 \lambda M r+\lambda 
(1+\lambda) r^2\right) }{\iom  r (3 M+\lambda r)^2}\bho
\\+\frac{r \left(-3 M^2-3 \lambda M r+\lambda r^2\right) }
{(2 M-r) (3 M+\lambda r)^2}\bk
+\frac{8 \pi  r^2 }{3 \iom  M+\iom  \lambda r}Se_{01}
+\frac{16 \pi  r }{3 \iom  M+\iom \lambda r}Se_{02}
\\+2 r \dbg+\frac{\left(6 M^2+3 \lambda M r+\lambda 
(1+\lambda) r^2\right) }{\iom  r (3 M+\lambda r)^2}\dhz\;,
\end{multline}
which is also gauge invariant.  We solve \eqref{eq:evspin2func} 
and \eqref{eq:devspin2func} for $\bho$ and $\bk$ to obtain
\begin{multline}
\label{eq:newho}
\bho=-\iom  r \dptw+\frac{2 \iom  (3 M-r) r }{2 M-r}\bg
+\frac{2 M }{2 M r-r^2}\hz-\iom  \ho
\\+\frac{\iom  \left(-3 M^2-3 \lambda M r+\lambda r^2\right) }
{(2 M-r) (3 M+\lambda r)}\ptw+\frac{8 \pi  r^3 }{3M+\lambda r}Se_{01}
\\+\frac{16 \pi  r^2 }{3 M+\lambda r}Se_{02}+2 \iom  r^2 \dbg+\dhz
\end{multline}
\begin{multline}
\label{eq:newk}
\bk=\left(1-\frac{2 M}{r}\right) \dptw
-2 (1+\lambda) \bg+\frac{2 (-2 M+r) }{r^2}\ho
\\+\frac{\left(6 M^2+3 \lambda M r+\lambda (1+\lambda) r^2\right) }
{r^2 (3 M+\lambda r)}\ptw
+\frac{8 \pi  (2 M-r) r }{\iom  (3 M+\lambda r)}Se_{01}
\\+\frac{(32 M \pi -16 \pi  r) }{3 \iom M+\iom  \lambda r}Se_{02}
+(4 M-2 r) \dbg\;.
\end{multline}
Equation \eqref{eq:ndkeqns} can be solved for $\bhtw$.  After using 
\eqref{eq:newho} and \eqref{eq:newk} to eliminate $\bho$, $\bk$ and 
$\dbk$, we have
\begin{equation}
\label{eq:newhtw}
\begin{split}
\bhtw=&\frac{\left(-3 M^2-3 \lambda M r
+\lambda r^2\right) }{r (3 M+\lambda r)}\dptw
+\frac{2 \left(-4 \lambda M+2 \lambda
r+(\iom)^2 r^3\right) }{2 M-r}\bg+\frac{(-6 M+4 r) }{r^2}\ho
\\&+\frac{1}{(2 M-r) r^2 (3 M+\lambda r)^2}\left[18 M^4+9 
(-1+2 \lambda) M^3 r+\lambda M r^3 \left(-\lambda+2 \lambda^2
\right.\right.\\&\left.\left.-6 (\iom)^2 r^2\right)-\lambda^2 r^4 
\left(1+\lambda+(\iom)^2 r^2\right)-3 M^2 r^2 \left(3 \lambda
-2 \lambda^2+3 (\iom)^2 r^2\right)\right]\ptw
\\&+\frac{8 \lambda \pi  (2 M-r) r^2 }{\iom  (3 M+\lambda r)^2}Se_{01}
+\frac{16 \pi  \left(3 M^2+3 \lambda M r-\lambda
r^2\right) }{\iom  (3 M+\lambda r)^2}Se_{02}
\\&-\frac{8 \pi  (2 M-r) r^2 }{3 M+\lambda r}Se_{11}
-\frac{16 \pi  (2 M-r) r }{3 M+\lambda r}Se_{12}
+2 M \dbg+\left(2-\frac{4 M}{r}\right) \dho\;.
\end{split}
\end{equation}
Similarly, \eqref{eq:divheqn0} can be solved for $\bhz$.  We use 
\eqref{eq:newho}, \eqref{eq:newk} and \eqref{eq:newhtw} to eliminate 
$\bho$, $\bk$, $\bhtw$ and $\dbho$, which gives
\begin{equation}
\label{eq:newhz}
\begin{split}
\bhz=&\frac{\left(-3 M^2-3 \lambda M r+\lambda r^2\right) }
{r (3 M+\lambda r)}\dptw+\frac{2 (\iom)^2 r^3 }{2 M-r}\bg
-\frac{2\iom  r \hz}{-2 M+r}-\frac{2 M }{r^2}\ho
\\&+\frac{1}{(2 M-r) r^2 (3 M+\lambda r)^2}\left[18 M^4+9 (-1+2 \lambda) 
M^3 r+\lambda M r^3 \left(-\lambda+2 \lambda^2\right.\right.
\\&\left.\left.-6 (\iom)^2 r^2\right)-\lambda^2 r^4 \left(1+\lambda
+(\iom)^2 r^2\right)-3 M^2 r^2 \left(3 \lambda-2 \lambda^2+3 
(\iom)^2 r^2\right)\right]\ptw
\\&+\frac{8 \lambda \pi  (2 M-r) r^2 }{\iom  (3 M+\lambda r)^2}Se_{01}
+\frac{16 \pi  \left(3 M^2+3 \lambda M r-\lambda
r^2\right) }{\iom  (3 M+\lambda r)^2}Se_{02}
\\&+\frac{8 \pi  r^2 (-2 M+r) }{3 M+\lambda r}Se_{11}
+\frac{16 \pi  r (-2 M+r) }{3 M+\lambda r}Se_{12}+2 M \dbg\;.
\end{split}
\end{equation}
The expressions for $\bhz$, $\bho$, $\bhtw$ and $\bk$ solve four of the 
field equations, specifically \eqref{eq:ddbh0eqn}-\eqref{eq:ddbh1eqn}.  
In the remaining three field equations, the results above can be used 
to eliminate $\bho$, $\bhtw$, and $\bk$, so that 
\eqref{eq:ddh0eqn}-\eqref{eq:ddgeqn} become
\begin{multline}
\label{eq:nddh0eqn}
\frac{(-2 M+r)^2 }{r^2}\ddhz+\frac{2 (-2 M+r)^2}{r^3}\dhz
+\frac{4 \iom  (2 M-r) (3 M-r) }{r^2}\bg+\frac{4 \iom  (-2 M+r)^2 }{r}\dbg
\\+\frac{\left(4 (1+\lambda) M-r \left(2+2
\lambda+(\iom)^2 r^2\right)\right) }{r^3}\hz
-\frac{2 \iom  \left(2 M^2-3 M r+r^2\right) }{r^3}\ho
\\=\frac{2 \iom  (-2 M+r)^2 }{r^2}\dptw
+\frac{2 \iom  (2 M-r) \left(3 M^2+3 \lambda M r-\lambda r^2\right) }
{r^3(3 M+\lambda r)}\ptw
\\-\frac{16 \pi  (-2 M+r)^2 }{3 M+\lambda r}Se_{01}
-\frac{16 \pi  (2 M-r) (M-(2+\lambda) r) }{r (3 M+\lambda r)}Se_{02}\;,
\end{multline}
\begin{multline}
\label{eq:nddh1eqn}
\frac{(-2 M+r)^2 }{r^2}\ddho
+\frac{4\left(2 M^2-3 M r+r^2\right) }{r^3}\dho
-\frac{4 \left(2 M-r+(\iom)^2 r^3\right) }{r^2}\bg
+\frac{2 \iom  M \hz}{2 M r-r^2}
\\+\frac{\left(-8 M^2+4 (2+\lambda)M r-r^2 \left(2+2 
\lambda+(\iom)^2 r^2\right)\right) }{r^4}\ho
+\frac{4 \left(2 M^2-3 M r+r^2\right) }{r^2}\dbg
\\=\frac{2 M (2 M-r) (3 M-(3+\lambda) r) }{r^3 (3 M+\lambda r)}\dptw
-\frac{2  }{r^4 (3 M+\lambda r)^2}\left[18 M^4+3 (-3+4 \lambda) M^3 r
\right.\\\left.+(\iom)^2 \lambda^2 r^6-3 \lambda M r^3 \left(1+\lambda
-2(\iom)^2 r^2\right)+M^2 \left(6\lambda^2 r^2+9 (\iom)^2 r^4\right)\right]\ptw
\\-\frac{48 M \pi  (-2 M+r)^2 }{\iom  r (3 M+\lambda r)^2}Se_{01}
-\frac{32 M \pi(2 M-r)(3 M-(3+\lambda) r)}{\iom r^2(3 M+\lambda r)^2}Se_{02}
\\-\frac{16 \pi  (-2 M+r)^2 }{3 M+\lambda r}Se_{11}
-\frac{16 \pi  (2 M-r) (M-(2+\lambda) r) }{r (3 M+\lambda r)}Se_{12}\;,
\end{multline}
\begin{multline}
\label{eq:nddgeqn}
\frac{(-2 M+r)^2 }{r^2}\ddbg
+\frac{2(M-r) (2 M-r) }{r^3}\dbg+\left(-(\iom)^2
+\frac{\lambda (4 M-2 r)}{r^3}\right) \bg
\\+\frac{2 (-2 M+r)^2 }{r^5}\ho
=-\frac{16 \pi  (-2 M+r) }{r^3}Se_{22}\;.
\end{multline}
Equation \eqref{eq:nddgeqn} is actually the same as \eqref{eq:ddgeqn}, 
but is reprinted here for convenience.  To summarize, we have used the 
definition of $\ptw$ to reduce the number of unsolved field equations 
from seven to three and the number of unknown radial perturbation 
functions from seven to three, namely, $\hz$, $\ho$ and $\bg$.

We can rewrite the unknown radial functions as
\begin{equation}
\label{eq:mhz}
\hz=-\tbmz+\iom\tbmtw\;,
\end{equation}
\begin{equation}
\label{eq:mho}
\ho=-\tbmo+\frac{2}{r}\tbmtw-\tdbmtw\;,
\end{equation}
\begin{equation}
\label{eq:mbg}
\bg=-\frac{\tbmtw}{r^2}\;.
\end{equation}
This can be done because $\hz$, $\ho$ and $\bg$ are zero in the 
Regge-Wheeler gauge. If we were to transform from that gauge to 
the harmonic gauge, we would apply \eqref{eq:evhznew}-\eqref{eq:gnew} 
with the ``old'' quantities set equal to zero and the ``new'' 
functions representing the harmonic gauge radial factors.  The tildes 
distinguish this particular gauge transformation from others.
Substituting \eqref{eq:mhz}-\eqref{eq:mbg} into equations 
\eqref{eq:nddh0eqn}-\eqref{eq:nddgeqn}, we find
\begin{multline}
\label{eq:iddm0eqn}
\frac{(-2 M+r)^2 }{r^2}\tddbmz+\frac{2 (-2 M+r)^2 }{r^3}\tdbmz
+\frac{2 \iom  M (2 M-r) }{r^3}\tbmo
\\+\frac{\left(4 (1+\lambda) M-r \left(2+2 \lambda
+(\iom)^2 r^2\right)\right)}{r^3}\tbmz
=-\frac{2 \iom  (-2 M+r)^2 }{r^2}\dptw
\\-\frac{2 \iom  (2 M-r) \left(3 M^2+3 \lambda M r
-\lambda r^2\right) }{r^3 (3 M+\lambda r)}\ptw
+\frac{16 \pi  (-2 M+r)^2 }{3 M+\lambda r}Se_{01}
\\+\frac{16 \pi  (2 M-r) (M-(2+\lambda) r) }{r (3 M+\lambda r)}Se_{02}
+16\iom\pi\ff Se_{22}\;,
\end{multline}
\begin{multline}
\label{eq:iddm1eqn}
\frac{(-2 M+r)^2 }{r^2}\tddbmo
+\frac{2 (-2 M+r) }{r^2}\tdbmo+\frac{2 \iom  M }{2 M r-r^2}\tbmz
\\-\frac{4 (1+\lambda) (2 M-r)}{r^4}\tbmtw
+\frac{\left(-8 M^2+4 (3+\lambda) M r-r^2 \left(4+2 \lambda
+(\iom)^2 r^2\right)\right) }{r^4}\tbmo
\\=-\frac{2 M (2 M-r) (3 M-(3+\lambda) r) }{r^3 (3 M+\lambda r)}\dptw
+\frac{2  }{r^4 (3 M+\lambda r)^2}\left[18 M^4+3 (-3+4 \lambda) M^3 r
\right.\\\left.+(\iom)^2 \lambda^2 r^6-3 \lambda M r^3 
\left(1+\lambda-2 (\iom)^2 r^2\right)+M^2 \left(6\lambda^2 r^2
+9 (\iom)^2 r^4\right)\right]\ptw
\\+\frac{48 M \pi  (-2 M+r)^2 }{\iom  r (3 M+\lambda r)^2}Se_{01}
+\frac{32 M \pi(2 M-r) (3 M-(3+\lambda) r) }
{\iom r^2 (3 M+\lambda r)^2}Se_{02}
\\+\frac{16 \pi  (-2 M+r)^2}{3 M+\lambda r} Se_{11}
+\frac{16 \pi  (2 M-r) (M-(2+\lambda) r) }{r (3 M+\lambda r)}Se_{12}
\\-\frac{32 \pi  (-2 M+r) }{r^2}Se_{22}
-\frac{16 \pi (-2 M+r)}{r} Se_{22}^{\prime}\;,
\end{multline}
\begin{multline}
\label{eq:iddm2eqn}
\frac{(-2 M+r)^2 }{r^2}\tddbmtw
+\frac{2 M (-2 M+r) }{r^3}\tdbmtw+\frac{2 (-2 M+r)^2 }{r^3}\tbmo
\\+\frac{\left(4 (1+\lambda) M-r \left(2+2 \lambda
+(\iom)^2 r^2\right)\right) }{r^3}\tbmtw
=\frac{16 \pi  (-2 M+r) }{r}Se_{22}\;.
\end{multline}
The left hand sides of \eqref{eq:iddm0eqn}-\eqref{eq:iddm2eqn} are the 
same as \eqref{eq:ddm0eqn}-\eqref{eq:ddm2eqn}, with the substitution 
$\widetilde{M}_{i}\to M_{i}$, $i=0,1,2$.  Stated differently, a gauge change 
which preserves the harmonic gauge is a homogeneous solution of 
\eqref{eq:iddm0eqn}-\eqref{eq:iddm2eqn}.  

To complete the even parity solutions, we will 
solve \eqref{eq:iddm0eqn}-\eqref{eq:iddm2eqn} for 
$\tbmz$, $\tbmo$ and $\tbmtw$, substitute the results into 
\eqref{eq:mhz}-\eqref{eq:mbg} to find $\hz$, $\ho$ and $\bg$, and 
then substitute \eqref{eq:mhz}-\eqref{eq:mbg} into 
\eqref{eq:newho}-\eqref{eq:newhz} to obtain $\bhz$, $\bho$, $\bhtw$ 
and $\bk$.  Before doing so, it is worth revisiting the plane wave 
example from Weinberg \cite{wein72}, which is discussed in 
section~\ref{sec:bhpt}.  The plane 
wave can be decomposed into six pieces of different helicities, with only 
the $\pm 2$ components being gauge invariant.  
The odd and even parity $\ptw$ functions, which are gauge invariant, 
are Schwarzschild metric analogues of the helicity $\pm 2$ parts.  
Counting these two plus the odd parity function $\po$, we have three 
pieces unaccounted for so far, a deficiency which is rectified below.  
We will see that the even parity solutions contain three additional 
generalized Regge-Wheeler functions:  one with $s=1$ and two with $s=0$.  
The two with $s=0$ have  different source terms and participate in the 
metric perturbation in different ways.  

We start by deriving one of the $s=0$ functions.  From \eqref{eq:rhtrace}, 
the radial component of the perturbation trace is
\begin{equation}
\label{eq:homr}
h(\omega,r)=-\bhz+\bhtw+2 \bk\;,
\end{equation}
where the indices $l,m$ have been omitted for simplicity.  
To obtain a differential equation for the trace, subtract 
\eqref{eq:ddbh0eqn} from \eqref{eq:ddbh2eqn}, add twice 
\eqref{eq:ddkeqn} to the difference, and rewrite the result 
in terms of $h(\omega,r)$, all of which gives  
\begin{multline}
\label{eq:ddhomreqn}
\frac{(-2 M+r)^2 }{r^2}h^{\prime\prime}(\omega,r)
+\frac{2 \left(2 M^2-3 M r+r^2\right) }{r^3}h^{\prime}(\omega,r)
\\+\frac{\left(4 (1+\lambda) M-r \left(2+2 \lambda+(\iom)^2
r^2\right)\right) }{r^3}h(\omega,r)=-16 \pi  Se_{00}
\\+\frac{16 \pi  (-2 M+r)^2}{r^2} Se_{11}
-\frac{32 \pi  (2 M-r) }{r^3}Ue_{22}\;.
\end{multline}
Defining $\pz$ as
\begin{equation}
\label{eq:spin0func}
\pz= r\,h(\omega,r)=r (-\bhz+\bhtw+2 \bk)
\end{equation}
and substituting for $h(\omega,r)$ in \eqref{eq:ddhomreqn} yields
\begin{equation}
\label{eq:pzeqn}
\mathcal{L}_{0}\pz=S_{0}=-16 \pi r Se_{00}
+\frac{16 \pi (-2 M+r)^2 }{r}Se_{11}
+\frac{32 \pi (-2 M+r) }{r^2}Ue_{22}\;.
\end{equation}
This is the generalized Regge-Wheeler equation, with $s=0$.  The 
operator $\mathcal{L}_{0}$ is defined by \eqref{eq:grweqnop}.  The 
right side of \eqref{eq:pzeqn} may be simplified to
\begin{equation}
S_{0}=16\pi(r-2 M)T(\omega,r)\;.
\end{equation}
Here, $T(\omega,r)$ is the radial component of the trace of the 
stress energy tensor from \eqref{eq:rtrtmunu}.  Alternatively, 
we could have taken the trace of the harmonic gauge field 
equations, \eqref{eq:hpertfeqn}, to obtain 
\begin{equation}
\label{eq:trfeqn}
{\overline{h}_{;\alpha}}{}^{;\alpha}=-{h_{;\alpha}}{}^{;\alpha}
=-16\pi g^{\mu\nu}T_{\mu\nu}=-16\pi T\;.
\end{equation}
Applying separation of variables to the equation 
${h_{;\alpha}}{}^{;\alpha}=16\pi T$ gives \eqref{eq:ddhomreqn}.  
In the plane wave example, the zero helicity functions are related to the 
trace of the perturbation.  The Schwarzschild metric analogue of this is the 
relation \eqref{eq:spin0func} between $\pz$ and the radial component of the 
trace.

The derivation of the even parity $\po$ function follows.  We can 
write a trial solution in the form
\begin{multline}
\label{eq:potry}
\po^{\text{try}}=\tilde{\alpha}(r)\bhz+\tilde{\beta}(r)\bho
+\tilde{\gamma}(r)\bhtw+\tilde{\delta}(r)\bk+\tilde{\epsilon}(r)\bg
\\+\tilde{\mu}(r)\hz+\tilde{\nu}(r)\ho
+\tilde{\lambda}(r)\dbhz+\tilde{\omega}(r)\dbhtw
+\tilde{\rho}(r)\dbg+\tilde{\psi}(r)\dhz\;,
\end{multline}
where tildes distinguish the Greek-lettered functions 
used here from similarly 
labeled quantities found elsewhere in this thesis.  The trial solution has 
only four first derivative terms because the other three radial function 
derivatives can be eliminated by solving the three harmonic gauge 
conditions for them.  Applying equations \eqref{eq:newho}-\eqref{eq:newhz} 
and \eqref{eq:mhz}-\eqref{eq:mbg}, the radial perturbation functions 
can be rewritten in terms of $\ptw$, $\tbmz$, $\tbmo$, $\tbmtw$, radial 
coefficients of the stress energy tensor, and their derivatives.  The 
resulting expression for $\po^{\text{try}}$ is substituted into the 
generalized Regge-Wheeler equation with $s=1$.  After simplifying, there 
are two groups of terms:  (1) eight terms proportional to $\ptw$, $\tbmz$, 
$\tbmo$, $\tbmtw$, and their first radial derivatives, and (2) terms 
proportional to the stress energy tensor coefficients and their 
derivatives.  The terms in the first group are set equal to zero, forming 
a system of eight coupled ordinary differential equations.  The system 
must be solved to obtain the Greek-lettered functions in $\po^{\text{try}}$.  
Lengthy calculations produce
\begin{multline}
\label{eq:evspin1func}
\po=\frac{2 \lambda r }{3 \iom}\bg
-\frac{4 M }{3 (\iom)^2 r^2}\hz+\frac{r \bhz}{2 \iom}
-\frac{(-2 M+r) }{3 \iom r}\ho
-\frac{2 (-2 M+r) }{3 (\iom)^2 r}\bho
\\-\frac{r }{2 \iom}\bhtw-\frac{2 r }{3\iom}\bk
+\frac{2 (-2 M+r) }{3 (\iom)^2 r}\dhz\;.
\end{multline}
The terms in the second group become the source for the $\po$ 
differential equation, giving
\begin{equation}
\begin{split}
\label{eq:evpoeqnl}
\mathcal{L}_{1}\po=&\frac{16 \pi(2 M-r)\left(2 (3+\lambda) M
-3 (\iom)^2 r^3\right)}{3 (\iom)^2 r (3 M+\lambda r)}Se_{01}
+\frac{16 \pi}{3(\iom)^2 r^3 (3 M+\lambda r)} \left[24 M^3\right.
\\&-4 (9+4 \lambda) M^2 r+M r^2 \left(12+4 \lambda-4 \lambda^2
-3 (\iom)^2 r^2\right)+r^3 \left(2 \lambda^2+6(\iom)^2 r^2
\right.\\&\left.\left.+\lambda \left(2+3 (\iom)^2 r^2\right)\right)
\right]Se_{02}-\frac{16 \pi (-2 M+r)^2 (6 M+(-6+\lambda) r) }{3 \iom r 
(3 M+\lambda r)}Se_{11}
\\&-\frac{16 \pi (2 M-r) ((-3+4 \lambda) M+(6+\lambda) r) }
{3 \iom r (3 M+\lambda r)}Se_{12}+\frac{64 \lambda (3+2
\lambda) \pi (2 M-r) }{3 \iom r (3 M+\lambda r)}Se_{22}
\\&-\frac{16 \pi (-2 M+r)^2 (6 M+\lambda r) }{3 (\iom)^2 r 
(3 M+\lambda r)}Se_{01}^{\prime}-\frac{32 \pi (-2 M+r)^2 }
{3 (\iom)^2 r^2}Se_{02}^{\prime}\\&-\frac{16 \pi (2 M-r)^3}{\iom 
(3 M+\lambda r)} Se_{11}^{\prime}-\frac{16 \pi (-2 M+r)^2 
(M-(2+\lambda) r) }{\iom r (3 M+\lambda r)}Se_{12}^{\prime}\;.
\end{split}
\end{equation}
Using \eqref{eq:divt0}-\eqref{eq:divt2}, the source simplifies to
\begin{multline}
\label{eq:evpoeqn}
\mathcal{L}_{1}\po=\frac{16 \pi r }{3 \iom}Se_{00}
+\frac{32 \pi (-2 M+r)^2 }{3 (\iom)^2 r^2}Se_{01}
+\frac{64 M \pi(2 M-r)}{3 (\iom)^2 r^3} Se_{02}
\\-\frac{16 \pi (-2 M+r)^2 }{3 \iom r}Se_{11}
+\frac{16 \pi (-2 M+r)^2 }{3 \iom r^2}Se_{12}
+\frac{32 \lambda\pi (2 M-r) }{3 \iom r^2}Se_{22}
\\+\frac{16 \pi (2 M-r) }{\iom r^2}Ue_{22}
-\frac{32 \pi (-2 M+r)^2}{3 (\iom)^2 r^2}Se_{02}^{\prime}\;.
\end{multline}
This derivation has been brief, but the reader may verify by 
substitution that \eqref{eq:evspin1func} is the solution 
to \eqref{eq:evpoeqn}.  Because of the various first order 
differential equations given previously, the definition of $\po$ 
in \eqref{eq:evspin1func} is not unique, just as 
we have seen that there is not a single expression for $\ptw$.  

In order to obtain the seven radial metric perturbation 
functions, we need to find $\tbmz$, $\tbmo$ and 
$\tbmtw$, as explained previously.  The formulae for $\pz$ and 
$\po$ help us to do so.  
After substituting \eqref{eq:newho}-\eqref{eq:newhz} and 
\eqref{eq:mhz}-\eqref{eq:mbg} into \eqref{eq:spin0func} and 
\eqref{eq:evspin1func}, we have
\begin{multline}
\label{eq:pzm}
\pz=2 \left(1-\frac{2 M}{r}\right) r \dptw
-\frac{2 \iom  r^2 }{-2 M+r}\tbmz
+\left(-4+\frac{4 M}{r}\right) \tbmo+\frac{4(1+\lambda) }{r}\tbmtw
\\+\frac{2 \left(6 M^2+3 \lambda M r+\lambda 
(1+\lambda) r^2\right) }{r (3 M+\lambda r)}\ptw
+\frac{16 \pi  (2 M-r) r^2 }{\iom(3 M+\lambda r)}Se_{01}
\\-\frac{32 \pi  r (-2 M+r) }{\iom  (3 M+\lambda r)}Se_{02}
-32 \pi  r Se_{22}+(4 M-2 r) \tdbmo\;,
\end{multline}
and
\begin{multline}
\label{eq:pom}
\po=\frac{r^2 }{-2 M+r}\tbmz+\frac{1}{\iom }\tbmo
-\frac{2 (1+\lambda) }{\iom  r}\tbmtw
-\frac{2(3 M+\lambda r) }{3 \iom  r}\ptw
\\+\frac{16 \pi  r }{\iom }Se_{22}+\frac{(-2 M+r) }{\iom }\tdbmo
+\frac{(-2 M+r) }{\iom  r}\tdbmtw\;.
\end{multline}
Equations \eqref{eq:pzm}, \eqref{eq:pom}, and their first radial 
derivatives can be solved for four unknowns:  $\tbmz$, $\tdbmz$, 
$\tbmo$ and $\tdbmo$.  The resulting lengthy expressions will contain 
$\tbmtw$ and $\tdbmtw$, which remain undetermined.  However, the 
expression for $\tbmo$ can be substituted into \eqref{eq:iddm2eqn}.  
This gives a single second order differential equation for $\tbmtw$, 
which becomes, after substituting $\tbmtw=\bmtwa/r$,
\begin{multline}
\label{eq:ddm2aeqn}
\mathcal{L}_{0}\bmtwa=-\frac{2 (-2 M+r)^2 }{r}\dptw
+\left(1-\frac{2 M}{r}\right) \pz
-\iom  \left(-2+\frac{4 M}{r}\right) \po
\\-\frac{2 \lambda (2 M-r) (3 M-(3+\lambda) r) }{3 r (3 M+\lambda r)}\ptw
+\frac{16\pi  r (-2 M+r)^2 }{\iom  (3 M+\lambda r)}Se_{01}
\\+\frac{32 \pi  (-2 M+r)^2 }{\iom  (3 M+\lambda r)}Se_{02}
+16 \pi  (-2 M+r) Se_{22}\;.
\end{multline}
Once this last equation is solved for $\bmtwa$, we can backtrack to 
obtain $\tbmtw$ and then $\tbmz$ and $\tbmo$, as well as their 
derivatives.  

We solve \eqref{eq:ddm2aeqn} in a manner similar to \eqref{eq:nddhtweqn}, 
the odd parity differential equation for $\htw$.  The result is
\begin{multline}
\label{eq:m2asol}
\bmtwa=\frac{\lambda (3+2\lambda)(2 M-r)r}{6 (\iom)^2(3 M+\lambda r)}\dptw
+\pza+\fz\pz+\fdz\dpz
-\frac{1}{6 (\iom)^2 (3 M+\lambda r)^2}\big[4 \lambda^3 r^2
\\+\lambda^4 r^2+27 (\iom)^2 M^2 r^2+9\lambda M\left(M+2(\iom)^2 r^3\right)
+3 \lambda^2 \left(M^2+M r+r^2+(\iom)^2 r^4\right)\big]\ptw
\\+\frac{4 \pi r \left(-48 M^3+15 (1-2 \lambda) M^2 r+(7-6 \lambda)
\lambda M r^2+\lambda(1+2 \lambda)r^3\right)}
{(\iom)^3(3 M+\lambda r)^2}Se_{01}\\+\frac{2(1+\lambda)}{\iom}\po
-\frac{8 \pi(2 M-r) \left(12 M^2+8\lambda M r+\lambda(1
+2\lambda)r^2\right)}{(\iom)^3 (3 M+\lambda r)^2}Se_{02}
+\frac{(-2 M+r)}{\iom}\dpo
\\-\frac{4 \pi (2 M-r) r^2 (8 M+(-1+2 \lambda) r)}
{(\iom)^2 (3 M+\lambda r)}Se_{11}
-\frac{8 \pi r (-2 M+r)^2}{(\iom)^2 (3 M+\lambda r)}Se_{12}\;.
\end{multline}
From \eqref{eq:ddm2aeqn}, $\bmtwa$ satisfies a generalized Regge-Wheeler 
equation with $s=0$, so $\pza$ must as well.  The subscript ``$a$'' 
distinguishes this second $s=0$ function from $\pz$.  The factors 
$\fz$ and $\fdz$ are solutions to the following two coupled differential 
equations
\begin{multline}
\label{eq:nzf0eqn}
\frac{(-2 M+r)^2 }{r^2}\ddfz
+\frac{2 M (-2 M+r) }{r^3}\dfz
\\+\frac{2 \left(8 M^3+2 (-3+2 \lambda) M^2 r+2 (1+\lambda) r^3
+M r^2 \left(-3-6 \lambda+2 (\iom)^2 r^2\right)\right)
}{(2 M-r)r^5}\fdz
\\+\frac{\left(-8 M^2-4 M (r+2 \lambda r)+2 r^2 
\left(2+2 \lambda+(\iom)^2 r^2\right)\right) }{r^4}\dfdz
=1-\frac{2 M}{r}\;,
\end{multline}
\begin{equation}
\label{eq:nzfd0eqn}
\frac{(-2 M+r)^2 }{r^2}\ddfdz
+\frac{2 M (2 M-r) }{r^3}\dfdz
+\frac{2 (-2 M+r)^2 }{r^2}\dfz
-\frac{4 M (M-r) }{r^4}\fdz=0\;,
\end{equation}
which do not have elementary solutions and are solved numerically.  

We can derive a formula for $\pza$ in terms of the radial perturbation 
functions by starting with a trial solution for $\pza$ in the form of 
the right side of \eqref{eq:potry}, then by using 
\eqref{eq:newho}-\eqref{eq:newhz} and 
\eqref{eq:mhz}-\eqref{eq:mbg} to rewrite the trial solution in terms of 
$\tbmz$, $\tbmo$ and $\tbmtw$, and finally by applying \eqref{eq:m2asol} 
and the rules following \eqref{eq:pom} to substitute for the three 
$\widetilde{M}_{i}$.  These manipulations produce a expression for $\pza$ in 
which the Greek-lettered functions are coefficients of complicated terms 
containing $\pz$, $\po$, $\ptw$, $\pza$ itself, the radial factors of the 
stress energy tensor, and derivatives of the foregoing.  The last step is 
to solve algebraically for the Greek-lettered functions  so 
that the expression reduces to $\pza$.  This procedure gives
\begin{multline}
\label{eq:evspin0afunc}
\pza=\frac{\lambda (1+\lambda) r }{(\iom)^2}\bg
+\frac{(1+\lambda) \left(-M+2 (\iom)^2 r^3\right) }{(\iom)^3(2 M-r) r}\hz
+\frac{(-5-4 \lambda) (2 M-r) r}{2 (\iom)^2(2 M-r)}\bhz
\\-\frac{(1+\lambda)r}{2 (\iom)^2 r}\ho
+\frac{(2 M-r)\left(1+\lambda+2 (\iom)^2 r^2\right)}{2 (\iom)^3 (2 M-r)}\bho
+\frac{(1+\lambda) (2 M-r) r}{(\iom)^2 (2 M-r)}\bhtw
\\+\frac{r \left((11+10 \lambda) M-r \left(5+5 \lambda
+2 (\iom)^2 r^2\right)\right)}{2 (\iom)^2 (2 M-r)}\bk
-\frac{(1+\lambda) }{2 (\iom)^3}\dhz
-\fz (r(-\bhz\\+\bhtw+2 \bk))-\fdz\left(\frac{(4 M-r)}{2 M-r}\bhz
+\frac{4 (1+\lambda)}{r}\ho+\frac{2 \iom r^2}{2 M-r}\bho
\right.\\\left.+\frac{(-8 M+5 r)}{2 M-r}\bhtw+2 \bk-2 r\dbhtw\right)\;.
\end{multline}
Like $\po$ and $\ptw$, this definition of $\pza$ is not unique, because 
of the first order differential equations derived earlier.  
By substituting \eqref{eq:evspin0afunc} into the generalized 
Regge-Wheeler equation for $s=0$, we find that
\begin{equation}
\mathcal{L}_{0}\pza=S_{0b}+S_{0c}\;,
\end{equation}
where
\begin{multline}
\begin{aligned}
S_{0b}=&\frac{4 \pi r \left((-5-6 \lambda)M+r\left(2
+3 \lambda+2 (\iom)^2 r^2\right)\right) }{(\iom)^2 (2 M-r)}Se_{00}
-\frac{8 \pi}{(\iom)^3 r}\left[r+\lambda r-2 (\iom)^2 r^3
\right.\\&\left.+M \left(-3-2 \lambda+8 (\iom)^2 r^2\right)\right]Se_{01}
-\frac{16 (1+\lambda) \pi \left(M-2 (\iom)^2 r^3\right)}{(\iom)^3 r^2}Se_{02}
\\&+\frac{4 \pi (2 M-r) \left((17+14 \lambda) M-r \left(8+7 \lambda
+2 (\iom)^2 r^2\right)\right)}{(\iom)^2 r}Se_{11}
\\&-\frac{8 (1+\lambda) \pi (2 M-r)}{(\iom)^2 r}Se_{12}
+\frac{16 \lambda (1+\lambda) \pi (2 M-r)}{(\iom)^2 r^2}Se_{22}
\end{aligned}
\\-\frac{8 (7+6 \lambda) \pi (2 M-r)}{(\iom)^2 r^2}Ue_{22}
-\frac{8 (1+\lambda) \pi (2 M-r)}{(\iom)^3 r}Se_{02}^{\prime}\;,
\end{multline}
and
\begin{multline}
S_{0c}=\frac{32 \pi\left((2 M-r) \fz
+\fdz+2 (2 M-r) \dfdz\right)}{r^2}Ue_{22}
-\frac{16 \pi (-2 M+r)^2 }{r^2}\left[r \fz+\fdz
\right.\\\left.+2 r \dfdz\right]Se_{11}
+\frac{16 \pi\left((2 M-r) r \fz+(6 M-r) \fdz
+2 (2 M-r) r \dfdz\right)}{2 M-r}Se_{00}
\\+16 \pi \fdz\left(r  Se_{00}^{\prime}
-\frac{(-2 M+r)^2}{r}Se_{11}^{\prime}
+\frac{2(2 M-r)}{r^2}Ue_{22}^{\prime}\right)\;.
\end{multline}
The functions $\pz$ and $\pza$ are Schwarzschild metric equivalents 
of the two helicity $0$ pieces for the plane wave.  

Equations \eqref{eq:nzf0eqn} and \eqref{eq:nzfd0eqn} are difficult to 
solve numerically, particularly as the spherical harmonic index $l$ increases.  
An alternative is to reformulate the problem so that we do not have to 
calculate $\fz$ and $\fdz$.  We start by breaking $\pza$ into two pieces 
to get
\begin{equation}
\label{eq:pzasplit}
\pza=\pzb+\pzc\;,
\end{equation}
where
\begin{equation}
\label{eq:ddpzb}
\mathcal{L}_{0}\pzb=S_{0b}
\end{equation}
and
\begin{equation}
\mathcal{L}_{0}\pzc=S_{0c}\;.
\end{equation}
We then define
\begin{equation}
\label{eq:fm2a}
\fmtwa=\pzc+\fz\pz+\fdz\dpz\;,
\end{equation}
and applying the differential operator $\mathcal{L}_{0}$ to $\fmtwa$ gives
\begin{equation}
\label{eq:ddfmtwa}
\mathcal{L}_{0}\fmtwa=\ff\pz\;.
\end{equation}
A method for solving \eqref{eq:ddfmtwa} numerically is explained at 
the end of section~\ref{sec:irweqn}, where the difficulties in solving 
\eqref{eq:nzf0eqn} and \eqref{eq:nzfd0eqn} are also discussed.  
The radial derivative of $\fmtwa$ is
\begin{multline}
\label{eq:dfm2a}
\dfmtwa=\dpzc
-\frac{16 \pi r^3 \fdz}{(-2 M+r)^2}Se_{00}
+16 \pi r \fdz Se_{11}+\frac{32 \pi \fdz}{-2 M+r}Ue_{22}
\\+\left(\frac{\left(-4 M^2-2 M (r+2 \lambda r)+r^2 \left(2+2 \lambda
+(\iom)^2 r^2\right)\right) \fdz}{r^2 (-2 M+r)^2}+\dfz\right)\pz
\\+\left(\fz+\frac{2 M \fdz}{2 M r-r^2}
+\dfdz\right) \dpz\;.
\end{multline}

An examination of \eqref{eq:pzasplit} and \eqref{eq:fm2a} shows that
\begin{equation}
\label{eq:ftonof}
\pza+\fz\pz+\fdz\dpz=\pzb+\fmtwa\;,
\end{equation}
and this equality can be used to rewrite the first 
line of $\bmtwa$ \eqref{eq:m2asol} 
in terms of $\pzb$ and $\fmtwa$.  The differential equations for 
$\pzb$ and $\fmtwa$ (see \eqref{eq:ddpzb} and \eqref{eq:ddfmtwa}) 
do not depend on $\fz$ and $\fdz$, so it is not necessary to solve 
\eqref{eq:nzf0eqn} and \eqref{eq:nzfd0eqn} for them.  Using the 
relations mentioned after equation \eqref{eq:pom}, we can find 
$\tbmtw$ and then $\tbmz$ and $\tbmo$.  The three 
$\widetilde{M}_{i}$, $i=0,1,2$, are
\begin{equation}
\begin{split}
\label{eq:m0sol}
\tbmz=&-\frac{\lambda (-2 M+r)^2 \dptw}
{2 \iom  r (3 M+\lambda r)}
-\frac{(-2 M+r) \pz}{2\iom  r^2}
-\frac{\iom  \pzbf}{r}-\frac{2 (1+\lambda) \po}{r}
\\&+\frac{1}{6 \iom  r^2 (3 M+\lambda r)^2}\left[2 \lambda^3 (6 M-r) r^2
+\lambda^4 r^3-27 (\iom)^2 M^2 r^3+9 \lambda M \left(4 M^2-M r
\right.\right.\\&\left.\left.-2 (\iom)^2 r^4\right)
-3 \lambda^2 r \left(-9 M^2-M r+r^2+(\iom)^2 r^4\right)\right] \ptw
+\frac{4 \pi}{(\iom)^2(3 M\!+\!\lambda r)^2}\left[48 M^3\right.\\&\left.
+(-15+38 \lambda)M^2 r+3 \lambda (-5+2 \lambda) M r^2+(1-2 \lambda) 
\lambda r^3\right] Se_{01}
\!-\!\frac{8 \pi \! (-2 M\!+\!r)^2 Se_{12}}{\iom (3 M+\lambda r)}
\\&+\frac{8 \lambda \pi  (-2 M+r)^2 Se_{02}}{(\iom)^2 (3 M+\lambda r)^2}
+\frac{4 \pi  (2 M-r) r (4 M+r+2 \lambda r) Se_{11}}{\iom  (3 M+\lambda r)}
+\frac{(-2 M+r) \dpz}{2 \iom  r}\;,
\end{split}
\end{equation}

\begin{equation}
\begin{split}
\label{eq:m1sol}
\tbmo=&\frac{\left(\lambda^2 (3 M-4 r)
-\lambda^3 r+9 (\iom)^2 M r^2+3 \lambda \left(M-r+(\iom)^2 
r^3\right)\right) \dptw}{6(\iom)^2 r (3 M+\lambda r)}
+\frac{r \pz}{4 M-2 r}\\&-\frac{\pzbf}{r^2}
-\frac{1}{2 (\iom)^2 r^2 (3 M+\lambda r)^2}\left[\lambda^4 r^2
+9 (\iom)^2 M^2 r^2+2 \lambda^3 r (M+r)\right.\\&\left.
+\lambda^2 \left(3 M^2+2 M r+r^2\right)+3 \lambda M \left(M+(\iom)^2
r^3\right)\right] \ptw+\frac{4 \pi  r}{(\iom)^3 (2 M-r) (3 M+\lambda r)^2}
\\&\times\left[-M^2 \left(9+4 \lambda+4 \lambda^2-12 (\iom)^2 r^2\right)
+\lambda (1+2 \lambda) r^2 \left(-1+(\iom)^2 r^2\right)+M r 
\left(4 \lambda^2\right.\right.\\&\left.\left.
+3 (\iom)^2 r^2+2 \lambda \left(-1+5 (\iom)^2 r^2\right)\right)\right] Se_{01}
-\frac{8 \pi}{(\iom)^3 (2 M-r) r (3 M+\lambda r)^2}\left[-48 (1
\right.\\&\left.+\lambda) M^3+\lambda r^3 \left(1+3 \lambda+2 \lambda^2
-(\iom)^2 r^2\right)+3 M^2 r \left(5-5 \lambda-10 \lambda^2+2 (\iom)^2 
r^2\right)\right.\\&\left.+M r^2 \left(\lambda^2-6 \lambda^3-3 (\iom)^2 r^2
+\lambda\left(7+2 (\iom)^2 r^2\right)\right)\right] Se_{02}+\frac{\dpzbf}{r}
\\&-\frac{4 \pi  r (M-2 \lambda M+r+2 \lambda r) Se_{11}}
{(\iom)^2 (3 M+\lambda r)}+\frac{8 (1+\lambda) \pi 
(8 M+(-1+2 \lambda) r) Se_{12}}{(\iom)^2 (3 M+\lambda r)}
\\&+\frac{2 (1+\lambda) \dpo}{\iom  r}\;,
\end{split}
\end{equation}

\begin{equation}
\begin{split}
\label{eq:m2sol}
\tbmtw=&\frac{\lambda (3+2 \lambda) (2 M-r) \dptw}
{6 (\iom)^2 (3 M+\lambda r)}+\frac{\pzbf}{r}+\frac{2(1+\lambda) \po}{\iom  r}
-\frac{1}{6 (\iom)^2 r (3 M+\lambda r)^2}\\&\times\left[4 \lambda^3 r^2
+\lambda^4 r^2+27 (\iom)^2 M^2 r^2+9 \lambda M \left(M+2 (\iom)^2 r^3\right)
+3 \lambda^2 \left(M^2+M r+r^2\right.\right.\\&\left.\left.
+(\iom)^2 r^4\right)\right] \ptw
+\frac{4 \pi}{(\iom)^3(3 M+\lambda r)^2}\left[-48 M^3+15 (1-2 \lambda)M^2 r
+(7-6 \lambda) \lambda M r^2\right.
\\&\left.+\lambda (1+2 \lambda) r^3\right] Se_{01}
-\frac{8 \pi  (2 M-r) \left(12 M^2+8 \lambda M r+\lambda 
(1+2 \lambda) r^2\right) Se_{02}}{(\iom)^3 r (3 M+\lambda r)^2}
\\&-\frac{4 \pi  (2 M-r) r (8 M+(-1+2 \lambda) r) Se_{11}}
{(\iom)^2 (3 M+\lambda r)}-\frac{8 \pi  (-2 M+r)^2 Se_{12}}{(\iom)^2
(3 M+\lambda r)}+\frac{(-2 M+r) \dpo}{\iom  r}\;.
\end{split}
\end{equation}
One can verify by substitution that \eqref{eq:m0sol}-\eqref{eq:m2sol} 
solve the system \eqref{eq:iddm0eqn}-\eqref{eq:iddm2eqn}.  

We now can show that a gauge change which preserves the harmonic gauge 
can be expressed in terms of homogeneous solutions of the 
generalized Regge-Wheeler equation with $s=0,1$.  As noted previously, 
homogeneous solutions of \eqref{eq:iddm0eqn}-\eqref{eq:iddm2eqn} are 
also solutions of \eqref{eq:ddm0eqn}-\eqref{eq:ddm2eqn}, which are the 
homogeneous differential equations that define a gauge change which 
preserves the harmonic gauge.  To find homogeneous solutions to 
\eqref{eq:iddm0eqn}-\eqref{eq:iddm2eqn}, take the $\widetilde{M}_{i}$ 
and set terms with $\ptw$, $\dptw$ and the radial components of the 
stress energy tensor equal to zero.  Also, replace $\pzb$, $\fmtwa$ 
and their derivatives with $\pza$, $\pz$ and $\dpz$, using 
\eqref{eq:dfm2a} and \eqref{eq:ftonof}.  These steps lead to
\begin{multline}
\label{eq:hm0sol}
\tbmz^{h}=-\frac{\left(-2 M+r+2 (\iom)^2 r \fz\right)}{2 \iom  r^2}\pz
-\frac{\iom  }{r}\pza\\-\frac{2 (1+\lambda)}{r}\po
+\frac{\left(-2 M+r-2 (\iom)^2 \fdz\right) }{2 \iom  r}\dpz\;,
\end{multline}
\begin{multline}
\label{eq:hm1sol}
\tbmo^{h}=-\frac{\pza}{r^2}
+\frac{1}{2 r^3 (-2 M+r)^2}\left[-2 r (-2 M+r)^2 \fz+\left(-8 M^2
-4 M (r+2 \lambda r)\right.\right.\\\left.\left.+2 r^2 \left(2+2\lambda
+(\iom)^2 r^2\right)\right) \fdz+(2 M-r) r^2 
\left(r^2+(4 M-2 r)\dfz\right)\right]\pz 
\\+\frac{\left((2 M-r) \fz+\fdz+(2 M-r) \dfdz\right) }{(2 M-r) r}\dpz
+\frac{\dpza}{r}+\frac{2 (1+\lambda)}{\iom  r}\dpo\;,
\end{multline}

\begin{equation}
\label{eq:hm2sol}
\tbmtw^{h}=\frac{\fz }{r}\pz+\frac{\pza}{r}
+\frac{2 (1+\lambda) }{\iom  r}\po+\frac{\fdz }{r}\dpz
+\frac{(-2 M+r) }{\iom  r}\dpo\;,
\end{equation}
where the superscript ``$h$'' stands for ``homogeneous''.  Although not 
specifically indicated, the functions $\pz$, $\pza$ and $\po$ are also 
homogeneous solutions of their respective differential equations.  
The reader may verify by substitution that 
\eqref{eq:hm0sol}-\eqref{eq:hm2sol} 
satisfy \eqref{eq:ddm0eqn}-\eqref{eq:ddm2eqn}, with the replacement 
$M_{i}=\widetilde{M}_{i}^{h}$, $i=0,1,2$.  

A harmonic gauge preserving change does not have to involve all three 
generalized Regge-Wheeler functions.  For example, if we set 
$\po=\pza=0$ in the $\widetilde{M}_{i}^{h}$, then the gauge change 
will involve only $\pz$.  Applying the gauge change rules 
in \eqref{eq:bh0new}-\eqref{eq:gnew}, such a gauge change alters 
$\pz$ in \eqref{eq:spin0func} by a homogeneous spin~$0$ solution, but 
leaves $\po$ and $\pza$ the same.  Similarly, a gauge change involving 
$\po$ (or $\pza$) adds a homogeneous spin~$1$ (or spin~$0$) solution 
to $\po$ \eqref{eq:evspin1func} (or $\pza$ \eqref{eq:evspin0afunc}), 
as the case may be.  This behavior is similar to what we saw in 
Chapter~\ref{oddpar}:  an odd parity gauge change which preserves 
the harmonic gauge modifies the odd parity $\po$ by a homogeneous 
spin~$1$ solution, as discussed in the text above at \eqref{eq:nspin1func}.  
Note that, even though  $\pz$ and $\pza$ share the same homogeneous 
differential equation, they generate different gauge change vectors, 
because they participate in the $\widetilde{M}_{i}^{h}$ in linearly 
independent ways.  

To complete the derivation of the solutions, substitute the 
$\widetilde{M}_{i}$ from \eqref{eq:m0sol}-\eqref{eq:m2sol} 
into \eqref{eq:mhz}-\eqref{eq:mbg} to find 
$\hz$, $\ho$ and $\bg$, which are then used to obtain $\bhz$, 
$\bho$, $\bhtw$ and $\bk$ from \eqref{eq:newho}-\eqref{eq:newhz}.  
The seven solutions and their radial derivatives are set forth 
in Appendix \ref{appa}.  The hardy reader may verify the 
solutions by substitution into the field and related equations.  The 
solutions are written in terms of $\pzb$ and $\fmtwa$, but this is 
only for numerical convenience.  Using \eqref{eq:dfm2a} and 
\eqref{eq:ftonof}, the solutions can be restated in terms of $\pz$ 
and $\pza$, which, along with $\ptw$ and $\po$, are the elemental 
constituents.

As is the case for odd parity, another way to check the even parity solutions 
is to transform from the harmonic gauge to the Regge-Wheeler gauge.  
If we set $M_{i}=-\widetilde{M}_{i}$ in the gauge transformation 
equations \eqref{eq:bh0new}-\eqref{eq:gnew}, we obtain
\begin{equation}
\hz^{\text{RW}}=\ho^{\text{RW}}=\bg^{\text{RW}}=0\;,
\end{equation}

\begin{equation}
\begin{split}
\label{eq:rwh0sol}
\bhz^{\text{RW}}=&\frac{\left(-3 M^2-3 \lambda M r
+\lambda r^2\right)\dptw}{r (3 M+\lambda r)}
-\frac{1}{(2 M-r) r^2 (3 M+\lambda r)^2}\left[-18 M^4+9 (1\!-\!2 \lambda) M^3 r
\right.\\&+\lambda^2 r^4 \left(1+\lambda+(\iom)^2 r^2\right)
+3 M^2 r^2 \left(3 \lambda-2 \lambda^2+3(\iom)^2 r^2\right)
+\lambda M r^3 \left(\lambda-2 \lambda^2\right.\\&\left.\left.
+6 (\iom)^2 r^2\right)\right]\ptw
-\frac{8 \lambda \pi  r^2 (-2 M+r) Se_{01}}{\iom(3 M+\lambda r)^2}
+\frac{16 \pi  \left(3 M^2+3 \lambda M r-\lambda
r^2\right) Se_{02}}{\iom  (3 M+\lambda r)^2}
\\&-\frac{8 \pi  (2 M-r) r^2 Se_{11}}{3 M+\lambda r}
-\frac{16 \pi  (2 M-r) r Se_{12}}{3 M+\lambda r}\;,
\end{split}
\end{equation}

\begin{equation}
\bho^{\text{RW}}=-\iom  r \dptw+\frac{\iom  \left(-3 M^2-3 
\lambda M r+\lambda r^2\right) \ptw}{(2 M-r) (3 M+\lambda r)}
+\frac{8 \pi  r^3 Se_{01}}{3 M+\lambda r}
+\frac{16 \pi  r^2 Se_{02}}{3 M+\lambda r}\;,
\end{equation}

\begin{equation}
\begin{split}
\bhtw^{\text{RW}}=&\frac{\left(-3 M^2-3\lambda M r+\lambda r^2\right) \dptw}
{r (3 M+\lambda r)}-\frac{1}{(2 M-r) r^2 (3 M+\lambda r)^2}
\left[-18 M^4+9 (1\!-\!2 \lambda) M^3 r\right.\\&+\lambda^2 r^4 
\left(1+\lambda+(\iom)^2 r^2\right)+3 M^2 r^2 \left(3 \lambda-2 \lambda^2
+3(\iom)^2 r^2\right)+\lambda M r^3 \left(\lambda-2 \lambda^2
\right.\\&\left.\left.+6 (\iom)^2 r^2\right)\right] \ptw
-\frac{8 \lambda \pi  r^2 (-2 M+r) Se_{01}}{\iom  (3 M+\lambda r)^2}
+\frac{16 \pi  \left(3 M^2+3 \lambda M r-\lambda
r^2\right) Se_{02}}{\iom  (3 M+\lambda r)^2}
\\&-\frac{8 \pi  (2 M-r) r^2 Se_{11}}{3 M+\lambda r}
-\frac{16 \pi  (2 M-r) r Se_{12}}{3 M+\lambda r}-32 \pi Se_{22}\;,
\end{split}
\end{equation}
\begin{multline}
\label{eq:rwksol}
\bk^{\text{RW}}=\left(1-\frac{2 M}{r}\right) \dptw
+\frac{\left(6 M^2+3 \lambda M r+\lambda (1+\lambda) r^2\right) \ptw}
{r^2(3 M+\lambda r)}
\\+\frac{8 \pi  (2 M-r) r Se_{01}}{\iom  (3 M+\lambda r)}
+\frac{(32 M \pi -16 \pi  r) Se_{02}}{3 \iom M+\iom  \lambda r}\;.
\end{multline}
The Regge-Wheeler solutions above agree with those obtained by 
Zerilli \cite{zerp70}, as corrected by others 
\cite{ashby}, \cite{sago03}.  

If we wished, we could have begun in the Regge-Wheeler gauge and derived 
the gauge transformation vectors from the Regge-Wheeler gauge to the 
harmonic gauge.  
This method was attempted in \cite{sago03}, which contains scalar 
differential equations involving Teukolsky functions with spin~$\pm 1$.  
The authors of \cite{sago03} did not solve their gauge transformation 
equations and did not obtain the gauge transformation vectors and 
harmonic gauge solutions.  

This concludes the derivation of the even parity non-zero frequency 
solutions, for $l\ge 2$.  
The seven radial metric perturbation factors can be solved 
in terms of four functions of various spins:  
the Zerilli function (the even parity version of the $s=2$ Regge-Wheeler 
function) and three generalized Regge-Wheeler 
functions (one with $s=1$ and two with $s=0$).  
The spin~$2$ function is gauge invariant.  
The two spin~$0$ functions have different sources and participate 
in the metric perturbation in different ways.  Although the solutions 
in Appendix \ref{appa} are written in terms of $\pzb$ and $\fmtwa$, 
these two quantities are actually composed of the two spin~$0$ functions.  
Like the odd parity case, the more complicated even parity problem 
can be reduced to solving a set of decoupled ordinary differential 
equations.  

\subsection{\label{sec:nzevpareq}Solutions for $l=0, 1$}

Non-zero frequency solutions for $l=1$ and $l = 0$ can be obtained from 
the results for $l\ge 2$.  From the definition of $\lambda$ \eqref{eq:lam}, 
we have $\lambda=0$ for $l=1$ and $\lambda=-1$ for $l=0$.  

For $l=1$, the radial function $\bg$ is not 
present, because the angular functions associated with it in 
the even parity metric perturbation \eqref{eq:ehmunu} are zero for 
$l\le 1$.  This is analogous to the odd parity case, where $\htw$ was not 
present for the same reason.  Similarly, the corresponding stress energy 
tensor coefficient, $Se_{22}$, is not present, and the $\bg$ field 
equation \eqref{eq:ddgeqn} 
does not exist.  The remaining field equations, together with the 
harmonic gauge conditions \eqref{eq:divheqn0}-\eqref{eq:divheqn2} and 
stress energy divergence equations \eqref{eq:divt0}-\eqref{eq:divt2}, 
still apply; however, terms containing $\bg$ and $Se_{22}$ are zero, 
because they have coefficients of $\lambda$.  
Except for $\bg$, the solutions in Appendix \ref{appa} are still 
applicable.  In the remaining six solutions, terms with $\ptw$ and $\dptw$ 
each have a factor of $\lambda$, so such terms are zero.  
For $l=1$, the solutions are constructed from $\pz$, $\pza$ and 
$\po$, which have the same definitions and sources as for $l\ge 2$.  
However, source terms with $Se_{22}$ are zero, because each has a factor of 
$\lambda$.  Since the radial perturbation functions do not contain 
$\ptw$, the definition of $\ptw$ in \eqref{eq:evspin2func} is no longer  
valid, so $\ptw$ is not defined for $l=1$.

Also for $l=1$, 
gauge changes which preserve the harmonic gauge are still specified by 
the differential equations \eqref{eq:ddm0eqn}-\eqref{eq:ddm2eqn}, with 
the radial factors given by the $\widetilde{M}_{i}^{h}$ in 
\eqref{eq:hm0sol}-\eqref{eq:hm2sol}.  However, the inhomogeneous equations 
\eqref{eq:iddm0eqn}-\eqref{eq:iddm2eqn} for the $\widetilde{M}_{i}$ are 
no longer applicable, because they were only intermediate 
steps in obtaining the solutions for $l\ge 2$ and their derivation 
presupposed the existence of $\ptw$.  It follows that the 
$\widetilde{M}_{i}$ set forth in \eqref{eq:m0sol}-\eqref{eq:m2sol} 
do not apply for $l=1$.  Moreover, negating the $\widetilde{M}_{i}$ 
gives the gauge transformation from the harmonic gauge to the Regge-Wheeler 
gauge.  The resulting Regge-Wheeler gauge expressions 
\eqref{eq:rwh0sol}-\eqref{eq:rwksol} are applicable only for 
$l\ge 2$ \cite{zerp70}, so one would not expect that the $\widetilde{M}_{i}$ 
would apply for $l=1$.

For $l=0$, the radial functions $\hz$, $\ho$ and $\bg$ are not present, 
because their associated angular functions are zero.  The relevant 
spherical harmonic is a constant ($Y_{00}=\frac{1}{\sqrt{4\pi}}$, 
\cite{arfken85}), and their associated angular functions are composed of 
spherical harmonic derivatives.  There are only four radial perturbation 
functions:  $\bhz$, $\bho$, $\bhtw$ and $\bk$.  Likewise, there are 
only four field equations, \eqref{eq:ddbh0eqn}-\eqref{eq:ddbh1eqn}, 
and four stress energy tensor components, $Se_{00}$, $Se_{01}$, $Se_{11}$ 
and $Ue_{22}$.  In the remaining field equations, terms with $\hz$, $\ho$ 
or $\bg$ are zero, because those terms have factors of $\lambda+1$.  
The gauge transformation vector $\xi_{\mu}$ \eqref{eq:echimu} does 
not have $\theta$ and $\phi$ components, so $M_{2}$ is not present.  
Vector equations, which have only one free index, have the same 
angular functions as $\xi_{\mu}$, so their $\theta$ and $\phi$ components 
are zero also.  For example, equation \eqref{eq:divheqn2} is from 
the $\theta$ (or $\phi$) component of the harmonic gauge condition, 
$\overline{h}_{\mu\nu}{}^{;\nu}=0$, so \eqref{eq:divheqn2} does not 
exist for $l=0$.  The other two conditions, \eqref{eq:divheqn0} and 
\eqref{eq:divheqn1}, are still applicable, but the substitution 
$\lambda=-1$ ensures that terms containing $\hz$ and $\ho$ vanish.  
Other vector equations are treated similarly, so equations 
\eqref{eq:divt2} (from $T_{\theta\nu}\!{}^{;\nu}=0$) and 
\eqref{eq:ddm2eqn} (from $\xi_{\theta;\nu}{}^{;\nu}=0$) also are not present.  
Notwithstanding these differences, the solutions in Appendix \ref{appa} 
for $\bhz$, $\bho$, $\bhtw$ and $\bk$ may still be used, after setting 
$\lambda=-1$.  With this substitution, terms having $\ptw$, $\po$, 
$Se_{02}$ and $Se_{12}$ are zero.  The solutions depend on $\pz$ and 
$\pza$, and they are given by the same expressions and differential 
equations as for $l\ge 2$, with $\lambda=-1$.  
Because the radial perturbation functions do not depend on $\po$ and 
$\ptw$, these two are not defined for $l=0$.

The solutions in Appendix \ref{appa} have factors of $3 M+\lambda r$ 
in the denominator of some terms.  For $l=0$, we have $\lambda=-1$, 
so these denominators are zero at $r=3 M$.  We need to check that there 
will not be division by zero.  Some of the denominators are in terms 
which have factors of $\lambda+1$, such as terms having $\ptw$ 
or $Se_{12}$, but these terms are identically zero because 
$\lambda+1=0$ for all $r$ when $l=0$.  However, the other denominators are 
in coefficients of $Se_{01}$ and $Se_{11}$, and these coefficients do not 
have factors of $\lambda+1$ and are not zero.  It turns out that when 
the coefficients of $Se_{01}$ and $Se_{11}$ are actually calculated 
using $\lambda=-1$, the numerators have factors of $3 M-r$ which cancel 
any troublesome denominator factors, thereby avoiding division by zero when 
$r=3 M$.  These denominators are not a problem for $l\ge 1$, because then 
$\lambda\ge 0$ and $3 M+\lambda r \ne 0$ always.   

The radial factors $\widetilde{M}_{0}^{h}$ and $\widetilde{M}_{1}^{h}$ still 
describe gauge changes which preserve the harmonic gauge, but 
$\widetilde{M}_{2}^{h}$ is not applicable for $l=0$, because it comes 
from the $\theta$ (or $\phi$) component of $\xi_{\mu}$.  Inspection of 
\eqref{eq:hm0sol} and \eqref{eq:hm1sol} shows that terms containing 
$\po$ and $\dpo$ have factors of $\lambda+1$, so such terms are zero.  
For $l=0$, harmonic gauge preserving changes consist of adding only 
homogeneous $s=0$ solutions of the generalized Regge-Wheeler equation.  
The two spin~$0$ functions, $\pz$ and $\pza$, represent different 
gauge changes, because they participate in the gauge change vectors in 
linearly independent ways.  Also, using the same reasoning as for $l=1$, 
the $\widetilde{M}_{i}$ do not apply for $l=0$.

The reader may verify by substitution that the $l=1$ and $l=0$ solutions, as 
constructed above, satisfy the relevant field and related equations.  

\section{\label{sec:zevpar}Zero Frequency Solutions}

The even parity zero frequency solutions are derived separately from the 
non-zero frequency solutions, just as was done for the odd parity 
results.  The cases $l\ge 2$, $l=1$ and $l=0$ are covered in 
subsections~\ref{sec:zevparge}, \ref{sec:zevpareqo} and~\ref{sec:zevpareqz}, 
respectively.  Subsection~\ref{sec:zeveqn} shows how to solve two systems 
of equations which are related to the $l\ge 2$ solutions.

\subsection{\label{sec:zevparge}Solutions for $l \ge 2$}

For zero frequency, we substitute $\omega=0$ into the non-zero frequency 
field equations \eqref{eq:ddbh0eqn}-\eqref{eq:ddgeqn}, the harmonic gauge 
conditions \eqref{eq:divheqn0}-\eqref{eq:divheqn2}, the stress energy tensor 
divergence equations \eqref{eq:divt0}-\eqref{eq:divt2}, the gauge 
transformation formulae \eqref{eq:bh0new}-\eqref{eq:gnew}, and the equations 
for gauge changes which preserve the harmonic gauge 
\eqref{eq:ddm0eqn}-\eqref{eq:ddm2eqn}.  One consequence of this substitution 
is that equations involving $\bho$, $\hz$, $Se_{01}$, $Se_{02}$ and $\bmz$ 
decouple from the remaining equations, so these radial functions are solved 
for separately.  

For non-zero frequency, we derived four first order differential equations 
\eqref{eq:ndh0eqns}-\eqref{eq:ndh2eqns}, which were in addition to the 
three harmonic gauge conditions.  Using similar methods, we find 
four additional equations for zero frequency,
\begin{multline}
\label{eq:ndh0eqns0}
\left(1-\frac{2 M}{r}\right) \dbhz
+\frac{2 \lambda (1+\lambda)}{r}\bg-\frac{(1+\lambda) }{r}\bhz
+\frac{2 (1+\lambda) (M-r) }{r^3}\ho\\+\frac{\bhtw}{r}+\frac{\lambda }{r}\bk
+\left(-1+\frac{M}{r}\right)\dbk=-8 \pi (-2 M+r) Se_{11}\;,
\end{multline}
\begin{multline}
\label{eq:ndhzeqns0}
\frac{(1+\lambda)(2 M-r)}{r}\dhz+\frac{2(1+\lambda)M}{r^2}\hz
-\frac{(1+\lambda) (2 M-r) }{r}\bho\\=-8 \pi (2 M-r) r Se_{01}\;,
\end{multline}
\begin{multline}
\label{eq:ndhgeqns0}
\lambda\left(1-\frac{2 M}{r}\right) \dbg
+\frac{\lambda (1+\lambda) }{r}\bg-\frac{(3 M+\lambda r) }{2 r^2}\bhz
-\frac{(M-\lambda M+\lambda r) }{r^3}\ho\\+\frac{M}{2 r^2}\bhtw
+\frac{\lambda }{2 r}\bk-\frac{M}{2 r}\dbk
=-4 \pi (-2 M+r) Se_{11}-\frac{8 \pi (-2 M+r)}{r} Se_{12}\;,
\end{multline}
\begin{multline}
\label{eq:ndh2eqns0}
\left(1-\frac{2 M}{r}\right) \dbhtw
-\frac{2 \lambda (1+\lambda) }{r}\bg+\frac{(2 M+r+\lambda r)}{r^2}\bhz
\\+\frac{2 (1+\lambda) (3 M-r) }{r^3}\ho+\frac{3(-2 M+r) }{r^2}\bhtw
+\frac{(8 M-(4+\lambda) r) }{r^2}\bk\\+\left(-1+\frac{3 M}{r}\right)
\dbk=-8 \pi (2 M-r) Se_{11}\;.
\end{multline}
Equations \eqref{eq:ndh0eqns0}, \eqref{eq:ndhzeqns0} and 
\eqref{eq:ndhgeqns0} are gauge invariant to linear order.  

Alternatively, we can derive \eqref{eq:ndh0eqns0}-\eqref{eq:ndh2eqns0} 
from the non-zero frequency 
equivalents \eqref{eq:ndh0eqns}-\eqref{eq:ndh2eqns}.  We solve 
\eqref{eq:ndkeqns} for $\dhz$ and use the result to eliminate $\dhz$ from 
\eqref{eq:ndh0eqns}, \eqref{eq:ndgeqns} and \eqref{eq:ndh2eqns}, which 
leads to
\begin{multline}
\label{eq:neweqn1}
\left(1-\frac{2 M}{r}\right) \dbhz+\left(-1+\frac{M}{r}\right)\dbk
+\frac{2 \lambda (1+\lambda) }{r}\bg
\\+\frac{2 \iom(1+\lambda) }{2 M-r}\hz-\frac{(1+\lambda) }{r}\bhz
+\frac{2(1+\lambda) (M-r) }{r^3}\ho+2\iom\bho
\\+\frac{\bhtw}{r}-\frac{\left(-2 \lambda M+\lambda r
+(\iom)^{2} r^3\right) }{2 M r-r^2}\bk=-8 \pi (-2 M+r) Se_{11}\;,
\end{multline}
\begin{multline}
\label{eq:neweqn2}
\left(\lambda-\frac{2 \lambda M}{r}\right)\dbg
-\frac{\left(M+\lambda M+\iom ^2 r^3\right) }{2 r+2 \lambda r}\dbk
+\frac{(3 \iom  M+\iom  \lambda r) }{2 M r-r^2}\hz
\\+\frac{\lambda (1+\lambda)}{r}\bg
-\frac{\left(M-\lambda M+\lambda r+(\iom)^{2} r^3\right) }{r^3}\ho
+\frac{\left(M+\lambda M+(\iom)^{2} r^3\right)}{2 (1+\lambda) r^2}\bhtw
\\-\frac{(3 M+\lambda r)}{2 r^2}\bhz
+\frac{\left(\lambda^2 (2 M-r)-3 \iom ^2 M r^2+\lambda
\left(2 M-r \left(1+\iom ^2 r^2\right)\right)\right)}
{2 (1+\lambda) (2 M-r) r}\bk
\\=-\frac{4 \iom  \pi r^2}{1+\lambda}Se_{01}
-4 \pi (-2 M+r) Se_{11}-\frac{8 \pi (-2 M+r) }{r}Se_{12}\;,
\end{multline}
\begin{multline}
\label{eq:neweqn3}
\left(1-\frac{2 M}{r}\right)\dbhtw
+\left(-1+\frac{3 M}{r}\right)\dbk
-\frac{2 \lambda (1+\lambda) }{r}\bg
-\frac{2 \iom  (1+\lambda) }{2 M-r}\hz
\\+\frac{\left(16 M^2-2 (8+\lambda) M r+r^2 
\left(4+\lambda+(\iom)^{2} r^2\right)\right) }{(2 M-r) r^2}\bk
+\frac{3 (-2 M+r) }{r^2}\bhtw
\\+\frac{(2 M+r+\lambda r) }{r^2}\bhz
+\frac{2(1+\lambda) (3 M-r) }{r^3}\ho
=-8 \pi (2 M-r) Se_{11}\;.
\end{multline}
Further, multiplying \eqref{eq:ndkeqns} by $\iom r^{2}$ gives
\begin{multline}
\label{eq:neweqn4}
\frac{(1+\lambda) (2 M-r)}{r}\dhz
+\iom  r (-2 M+r) \dbk
+\frac{2 (1+\lambda) M }{r^2}\hz
\\-\frac{\iom  (1+\lambda) (2 M-r) }{r}\ho
-\frac{(1+\lambda) (2 M-r) }{r}\bho
+\iom (2 M-r) \bhtw
\\+\iom  (-3 M+r) \bk=-8 \pi (2 M-r) r Se_{01}\;.
\end{multline}
Equations \eqref{eq:neweqn1}, \eqref{eq:neweqn2} and \eqref{eq:neweqn4} 
are gauge invariant to linear order, but \eqref{eq:neweqn3} is 
invariant only for changes which preserve the harmonic gauge.  
If we set $\omega=0$ in \eqref{eq:neweqn1}-\eqref{eq:neweqn4}, we 
obtain \eqref{eq:ndh0eqns0}-\eqref{eq:ndh2eqns0}.  

The solutions for $\bho$ and $\hz$ are derived as follows.  
From \eqref{eq:ddbh1eqn} and \eqref{eq:ddh0eqn}, the relevant field 
equations for zero frequency are
\begin{multline}
\label{eq:ddbh1eqn0}
\frac{(-2 M+r)^2 }{r^2}\ddbho
+\frac{2 \left(2 M^2-3 M r+r^2\right) }{r^3}\dbho
-\frac{4 (1+\lambda) (2 M-r) }{r^4}\hz
\\+\frac{\left(-4 M^2+4 (2+\lambda) M r-r^2 \left(4+2 \lambda
\right)\right) }{r^4}\bho=-\frac{16 \pi (-2 M+r) }{r}Se_{01}\;,
\end{multline}
\begin{multline}
\label{eq:ddh0eqn0}
\frac{(-2 M+r)^2 }{r^2}\ddhz+\frac{\left(-8 M^2+4
(2+\lambda) M r-r^2 \left(2+2 \lambda\right)\right) }{r^4}\hz
\\+\frac{2 (-2 M+r)^2 }{r^3}\bho=-\frac{16 \pi (-2 M+r) }{r}Se_{02}\;.
\end{multline}
The applicable harmonic gauge condition \eqref{eq:divheqn0} becomes
\begin{equation}
\label{eq:zdivheqn0}
\left(1-\frac{2 M}{r}\right) \dbho
-\frac{2(1+\lambda)}{r^2}\hz-\frac{2 (M-r) }{r^2}\bho=0\;.
\end{equation}
After solving \eqref{eq:ndhzeqns0} for $\bho$ and substituting the result 
into \eqref{eq:ddh0eqn0}, we have
\begin{multline}
\label{eq:nddh0eqn0}
\frac{(-2 M+r)^2 }{r^2}\ddhz+\frac{2 (-2 M+r)^2 }{r^3}\dhz
+\frac{2 (1+\lambda) (2 M-r)}{r^3}\hz
\\=-\frac{16 \pi (-2 M+r)^2}{(1+\lambda) r}Se_{01}
-\frac{16 \pi (-2 M+r)}{r}Se_{02}\;.
\end{multline}
We then insert a trial solution
\begin{equation}
\label{eq:h0try}
\hz^{\text{try}}=\tilde{\alpha}(r)\po+\tilde{\beta}(r)\dpo
\end{equation}
into the homogeneous form of \eqref{eq:nddh0eqn0} and solve for 
$\tilde{\alpha}$ and $\tilde{\beta}$ using the series solution method 
that was used to solve \eqref{eq:nddhtweqn}.  
Differentiating and simplifying the result gives
\begin{equation}
\label{eq:po0}
\po=-r^2 \dhz\;,
\end{equation}
which, after substitution into the generalized Regge-Wheeler equation 
for $s=1$, produces
\begin{equation}
\label{eq:po0eqn}
\mathcal{L}_{1}\po=\frac{16\pi(-2 M+r)^2}{1+\lambda}Se_{01}
-64 \pi (2 M-r) Se_{02}-16 \pi (2 M-r) r Se_{02}^{\prime}\;.
\end{equation}
Alternatively, we could have found $\po$ in the same way as for 
non-zero frequency, but it would be more complicated to do so.  

To find $\hz$, differentiate both sides of \eqref{eq:po0} with respect 
to $r$, use \eqref{eq:ddh0eqn0} and \eqref{eq:ndhzeqns0} to eliminate 
$\ddhz$ and $\dhz$, respectively, and then solve the resulting expression for 
\begin{equation}
\label{eq:h0sol0}
\hz=\frac{8 \pi r^2 (-2 M+r) Se_{01}}{(1+\lambda)^2}
+\frac{8 \pi r^2 Se_{02}}{1+\lambda}+\frac{(2M-r) \dpo}{2 r+2 \lambda r}\;.
\end{equation}
Unlike $\hz^{\text{try}}$, the solution $\hz$ has terms containing $Se_{01}$ and 
$Se_{02}$, and this is because $\hz^{\text{try}}$ is only a homogeneous 
solution of \eqref{eq:nddh0eqn0}.  
The radial derivative of \eqref{eq:h0sol0} is, after simplification,
\begin{equation}
\label{eq:dh0sol0}
\dhz=-\frac{\po}{r^2}\;,
\end{equation}
which agrees with the definition of $\po$ \eqref{eq:po0}.  
We now can substitute $\hz$ and $\dhz$ into the first order equation 
\eqref{eq:ndhzeqns0}  and solve for 
\begin{equation}
\label{eq:bh1sol0}
\bho=-\frac{\po}{r^2}+\frac{8\pi r(-2 M+r+\lambda r)Se_{01}}
{(1+\lambda)^2}+\frac{16 M \pi r Se_{02}}{(1+\lambda) (2 M-r)}
+\frac{M \dpo}{(1+\lambda) r^2}\;.
\end{equation}
The derivative is
\begin{multline}
\label{eq:dbh1sol0}
\dbho=\frac{2 (M-r) \po}{(2 M-r) r^3}
+\frac{16 M \pi (2 M+(-1+\lambda) r) Se_{01}}{(1+\lambda)^2(2 M-r)}
\\+\frac{16 \pi \left(-2 M^2-2 \lambda M r
+(1+\lambda) r^2\right) Se_{02}}{(1+\lambda) (-2 M+r)^2}
\\+\frac{\left(-2 M^2-2 \lambda M r+(1+\lambda) r^2\right)\dpo}
{(1+\lambda) (2 M-r) r^3}\;.
\end{multline}
For zero frequency, only $\bho$ and $\hz$ depend on the spin~$1$ 
generalized Regge-Wheeler function; the other radial metric perturbations 
do not.

The remaining five radial metric perturbation factors -- $\bhz$, $\bhtw$, 
$\ho$, $\bk$ and $\bg$ -- are obtained in a manner similar to the 
non-zero frequency derivation.  Due to the similarities, only key 
intermediate results are described below.  

For a gauge invariant function, we can 
not use the definition of $\ptw$ in \eqref{eq:evspin2func}, because it 
has factors of $\omega$ in the denominator of some terms.  However, we 
can use the alternative Moncrief form, $\ptw^{\text{Mon}}$ 
\eqref{eq:evspin2funcmon}, which is also gauge invariant.  
For zero frequency, $\ptw^{\text{Mon}}$ becomes
\begin{multline}
\label{eq:evspin2func0}
\ptw=2 r \bg+\frac{(4 M-2 r) }{3 M+\lambda r}\ho
+\frac{r (-2 M+r) }{(1+\lambda) (3 M+\lambda r)}\bhtw
\\+\frac{r }{1+\lambda}\bk
+\frac{(2 M-r)r^2 }{(1+\lambda)(3 M+\lambda r)}\dbk\;,
\end{multline}
which, following \eqref{eq:moneqns}, satisfies a Zerilli-type 
differential equation
\begin{multline}
\label{eq:moneqns0}
\mathcal{L}_{\!Z}\ptw=-\frac{8 \pi r \left(24 M^2
+(-9+7 \lambda) M r+(-1+\lambda) \lambda r^2\right) }
{(1+\lambda) (3 M+\lambda r)^2}Se_{00}\\+\frac{8 \pi (-2 M+r)^2 }
{3 M+\lambda r}Se_{11}+\frac{16 \pi (-2 M+r)^2 }{r (3 M+\lambda r)}Se_{12}
\\+\frac{32 \pi (2 M-r)}{r^2} Se_{22}-\frac{8 \pi(2 M-r) r^2 }
{(1+\lambda) (3 M+\lambda r)}Se_{00}^{\prime}\;.
\end{multline}
Using $\ptw$ and its radial derivative, $\dptw$, we can write the zero 
frequency $\bhz$, $\bhtw$ and $\bk$ in terms of $\ptw$, $\ho$, $\bg$ 
and their derivatives, just as was done for non-zero frequency 
in \eqref{eq:newk}-\eqref{eq:newhz}.  Further, the non-zero 
frequency definition of $\pz$ \eqref{eq:spin0func} and its associated 
generalized Regge-Wheeler differential equation \eqref{eq:pzeqn} still apply.  

With these results, we can reduce the problem to the solution of a 
single second order differential equation of the generalized 
Regge-Wheeler form
\begin{equation}
\begin{split}
\label{eq:m2aeqn0}
\mathcal{L}_{0}\bmtwa=&\frac{(-2 M+r)^2}{2 (1+\lambda) r}\dpz
-\frac{(-2 M+r)^2 ((6+5 \lambda) M
+\lambda (1+\lambda) r) }{(1+\lambda) r (3 M+\lambda r)}\dptw
\\&-\frac{(-2 M+r)^2 }{2 (1+\lambda) r^2}\pz
-\frac{\lambda (-2 M+r)^2 \left(3 M^2+6 (1+\lambda) M r+\lambda 
(1+\lambda) r^2\right)}{(1+\lambda) r^2 (3 M+\lambda r)^2}\ptw
\\&-\frac{8 \pi (2 M-r) r^3 ((6+5 \lambda) M+\lambda (1+\lambda) r)}
{(1+\lambda)^2 (3 M+\lambda r)^2}Se_{00}
-\frac{8\pi (2 M-r)^3 r }{(1+\lambda) (3 M+\lambda r)}Se_{11}
\\&+\frac{16 \pi (-2 M+r)^2(M+r+\lambda r)}
{(1+\lambda)(3 M+\lambda r)}Se_{12}-16 \pi (2 M-r) Se_{22}\;,
\end{split}
\end{equation}
where $\bmtwa$ is related to $\bg$ by
\begin{equation}
\label{eq:m2atog}
\bmtwa=-r^{3} \bg\;.
\end{equation}
The derivation of \eqref{eq:m2aeqn0} resembles that of its non-zero 
frequency counterpart, equation \eqref{eq:ddm2aeqn}, but is simpler because 
the zero frequency $\po$ couples only to $\hz$ and $\bho$.  The 
solution of \eqref{eq:m2aeqn0} is
\begin{multline}
\label{eq:m2asol0}
\bmtwa=\ftw \ptw+\fdtw\dptw+\fz \pz+\fdz \dpz\\+\pza
+\frac{8 \pi r^4 \fdtw }{(1+\lambda)(2 M-r) (3 M+\lambda r)}Se_{00}\;.
\end{multline}
Its derivative is
\begin{equation}
\begin{split}
\dbmtwa=&\frac{8 \pi r (2 (3 M+\lambda r)\fdz+r\fdtw)}{3 M+\lambda r}Se_{11}
+\frac{16\pi r \fdtw }{3 M+\lambda r}Se_{12}+\frac{32 \pi \fdtw }
{2 M-r}Se_{22}\\&+\frac{32 \pi \fdz }{-2 M+r}Ue_{22}
+\left(-\frac{2(M+r+\lambda r) \fdz}{(2 M-r) r^2}
+\dfz\right)\pz\\&+ \left(-\frac{2 \left(9 M^3+9 \lambda M^2 r
+3 \lambda^2 M r^2+\lambda^2 (1+\lambda) r^3\right) \fdtw}{(2 M-r) r^2
(3 M+\lambda r)^2}+\dftw\right)\ptw\\&+\left(\ftw+\frac{2 M \fdtw}
{2 M r-r^2}+\dfdtw\right)\dptw+\left(\fz+\frac{2 M \fdz}{2 M r-r^2}
+\dfdz\right) \dpz+\dpza
\\&-\frac{8 \pi r^3}{(1+\lambda) (-2 M+r)^2 (3 M+\lambda r)^2}
\left[2 (1+\lambda) (3 M+\lambda r)^2 \fdz\right.
\\&\left.+r \left(\lambda (M+r+\lambda r) \fdtw-(2 M-r) (3 M+\lambda r) 
\dfdtw\right)\right] Se_{00}\;.
\end{split}
\end{equation}

The quantities $\fz$, $\fdz$, $\ftw$ and $\fdtw$ are functions of $r$ 
and are solutions of two systems of differential equations.  
To find $\fz$ and $\fdz$, we must solve 
\begin{multline}
\label{eq:ddf0eqn}
\frac{(-2 M+r)^2 }{r^2}\ddfz+\frac{2 M (-2 M+r) }{r^3}\dfz
-\frac{4 (2 M-r) (M+r+\lambda r) }{r^4}\dfdz
\\+\frac{2 \left(4 M^2+(-1+2 \lambda) M r-2 (1+\lambda) r^2\right) }{r^5}\fdz
=-\frac{(2 M-r)^2}{2 (1+\lambda) r^2}\;,
\end{multline}
\begin{equation}
\label{eq:ddfd0eqn}
\frac{(-2 M+r)^2 }{r^2}\ddfdz+\frac{2 M (2 M-r) }{r^3}\dfdz
+\frac{2 (-2 M+r)^2 }{r^2}\dfz-\frac{4 M (M-r) }{r^4}\fdz
=\frac{(2 M-r)^2}{2 (1+\lambda) r}\;.
\end{equation}
The differential equations for $\ftw$ and $\fdtw$ are
\begin{multline}
\label{eq:ddf2eqn}
\frac{(-2 M+r)^2 }{r^2}\ddfdtw+\frac{2 M (2 M-r) }{r^3}\dfdtw
+\frac{2 (-2 M+r)^2 }{r^2}\dftw\\-\frac{2 M \left(18 M^3-36 M^2 r-3 
\left(-3+6 \lambda+2 \lambda^2\right) M r^2+2\lambda(3+\lambda)r^3\right) }
{r^4 (3 M+\lambda r)^2}\fdtw\\=-\frac{(2 M-r)^2 ((6+5 \lambda) M+\lambda 
(1+\lambda) r)}{(1+\lambda) r (3 M+\lambda r)}\;
\end{multline}
\begin{multline}
\label{eq:ddfd2eqn}
\frac{(-2 M+r)^2 }{r^2}\ddftw+\frac{2 M (-2 M+r) }{r^3}\dftw
\\\begin{aligned}&-\frac{4 (2 M-r) \left(9 M^3+9 \lambda M^2 r+3 \lambda^2 
M r^2+\lambda^2 (1+\lambda) r^3\right) }{r^4 (3 M+\lambda r)^2}\dfdtw
\\&+\frac{2(3+2 \lambda) M (2 M-r) (3 M+2 \lambda r) }{r^3 (3 M+\lambda r)^2}
\ftw+\frac{2}{r^5 (3 M+\lambda r)^3}\\&\times\left[108 M^5+9 (-9+14 \lambda)
M^4 r+9 \lambda (-11+6 \lambda) M^3 r^2\right.\\&\left.
+3\lambda^2 (-17+2\lambda)M^2 r^3+\lambda^3 (-7+2 \lambda)
M r^4-2 \lambda^3 (1+\lambda) r^5\right]\fdtw\end{aligned}
\\=-\frac{\lambda (2 M-r)^2 \left(3 M^2+6(1+\lambda)M r
+\lambda (1+\lambda)r^2\right)}{(1+\lambda) r^2 (3 M+\lambda r)^2}\;.
\end{multline}
These two systems do not have simple analytic solutions.  We will solve them 
in subsection \ref{sec:zeveqn}.  

The function $\pza$ is a solution of the generalized Regge-Wheeler equation 
with $s=0$.  In terms of the radial metric perturbation functions, it is
\begin{equation}
\begin{split}
\label{eq:spin0afunc0}
\pza=&\left(-r^3-2 r \ftw-\frac{6 M }
{3 M+\lambda r}\fdtw\right)\bg+(r \fz+\fdz) \bhz+\left(\frac{2 (-2 M+r) }
{3 M+\lambda r}\ftw\right.\\&\left.+\frac{\left(6 M^2+6 \lambda M r-2 
\lambda r^2\right)}{r (3 M+\lambda r)^2}\fdtw\right)\ho+(-r \fz-\fdz)\bhtw
+\left(\frac{(2 M-r) r }{(1+\lambda) (3 M+\lambda r)}\ftw
\right.\\&\left.+\frac{\left(6 M^2+3 \lambda M r+\lambda (1+\lambda) 
r^2\right) }{(1+\lambda)(3 M+\lambda r)^2}\fdtw\right) \bhtw-2 (r \fz+\fdz) \bk
+\left(-\frac{r }{1+\lambda}\ftw\right.\\&\left.-\frac{3 M }{(1+\lambda)
(3 M+\lambda r)}\fdtw\right) \bk-2 r \fdtw \dbg+r \fdz\dbhz-r \fdz \dbhtw
-2 r \fdz \dbk\\&+\left(\frac{r^2 (-2 M+r)}{(1+\lambda) (3 M+\lambda r)}\ftw
-\frac{r \left(6 M^2+3 \lambda M r+\lambda (1+\lambda) r^2\right) }
{(1+\lambda) (3 M+\lambda r)^2}\fdtw\right) \dbk\;.
\end{split}
\end{equation}
The differential equation for $\pza$ is
\begin{equation}
\mathcal{L}_{0}\pza=S_{0a}= S_{0b}+S_{0c}+S_{0d}.
\end{equation}
The source terms $S_{0b}$, $S_{0c}$ and $S_{0d}$ can be found by substituting 
$\pza$ \eqref{eq:spin0afunc0} into the spin~$0$ generalized Regge-Wheeler 
equation and simplifying.  
Doing so gives
\begin{multline}
S_{0b}=\frac{8 \pi r (-2 M+r)^3 }
{(1+\lambda) (3 M+\lambda r)}Se_{11}+\frac{16 \pi (-2 M+r)^2 
(M+r+\lambda r) }{(1+\lambda)(3 M+\lambda r)}Se_{12}
\\+16 \pi (-2 M+r) Se_{22}\;,
\end{multline}
\vspace*{-20pt}
\begin{equation}
\begin{split}
S_{0c}=&-\frac{32 \pi  \left((2 M-r) \ftw
+\fdtw+2 (2 M-r) \dfdtw\right)}{r^2}Se_{22}
\\&-\frac{16 \pi (-2 M+r)^2  \left(r (3 M+\lambda r) 
\ftw+3 M \fdtw+2 r (3 M+\lambda r) \dfdtw\right)}{r^2(3 M+\lambda r)^2}
Se_{12}\\&-\frac{8 \pi (-2 M+r)^2\left(r(3 M+\lambda r)\ftw+(6 M+\lambda r) 
\fdtw+2 r (3 M+\lambda r)\dfdtw\right)}{r (3 M+\lambda r)^2} Se_{11}
\\&+\frac{8 \pi Se_{00}}{(1+\lambda) (2 M-r) (3 M+\lambda r)^3}
\left[(2 M-r) r \left(72 M^3+9 (-3+5 \lambda) M^2 r
\right.\right.\\&\left.+2 \lambda(-6+5 \lambda) M r^2+(-1+\lambda)
\lambda^2 r^3\right)\ftw+\left(72 M^4+48\lambda M^3 r
+6\lambda(6+5 \lambda)\right.\\&\left.\times M^2 r^2+\lambda\left(-9
-\lambda+2 \lambda^2\right) M r^3+\lambda^2 (1+\lambda)r^4\right)\fdtw
+2(2 M-r)r\\&\left.\times (3 M+\lambda r) \left((2 M-r) r (3 M+\lambda r)\dftw
+\left(6 M^2+3 \lambda M r+\lambda (1+\lambda)r^2\right)\dfdtw\right)\right]
\\&+\frac{8 \pi r \left((2 M-r) r (3 M+\lambda r) \ftw+\left(6 M^2
+3 \lambda M r+\lambda (1+\lambda) r^2\right) \fdtw\right)}
{(1+\lambda) (3 M+\lambda r)^2}Se_{00}^{\prime}
\\&-\frac{8 \pi (-2 M+r)^2 \fdtw }{3 M+\lambda r}Se_{11}^{\prime}
-\frac{16 \pi (-2 M+r)^2 \fdtw }{r (3 M+\lambda r)}Se_{12}^{\prime}
-\frac{32 \pi (2 M-r) \fdtw }{r^2}Se_{22}^{\prime}\;,
\end{split}
\end{equation}
\begin{equation}
\begin{split}
\label{eq:s0d}
S_{0d}=&\frac{32 \pi\left((2 M-r) \fz+\fdz+2 (2 M-r) \dfdz\right)}{r^2}Ue_{22}
\\&-\frac{16 \pi (-2 M+r)^2 \left(r \fz+\fdz+2 r \dfdz\right)}{r^2}Se_{11}
\\&+\frac{16 \pi  \left((2 M-r) r \fz+(6 M-r) \fdz+2 (2 M-r) r \dfdz\right)}
{2 M-r}Se_{00}
\\&+16 \pi\fdz\left(r Se_{00}^{\prime}
-\frac{(-2 M+r)^2}{r}Se_{11}^{\prime}
+\frac{2(2 M-r)}{r^2}Ue_{22}^{\prime}\right)\;.
\end{split}
\end{equation}
The derivation of $\bmtwa$ and $\pza$ is similar to the non-zero frequency 
analysis for them.

We use the solution for $\bmtwa$ to find the remaining radial metric 
perturbation factors.  The calculation is similar to that of the 
non-zero frequency derivation, for which we backtracked from $\bmtwa$ 
through intermediate steps.  The resulting solutions for $\bhz$, 
$\bhtw$, $\bk$, $\ho$ and $\bg$ are in Appendix \ref{appb}, and they may be 
verified by substitution into the zero frequency field equations.  

For non-zero frequency, we derived the three radial factors 
\eqref{eq:hm0sol}-\eqref{eq:hm2sol} for a change of gauge which  
preserves the harmonic gauge.  Using similar notation and derivations, 
the zero frequency equivalents for $l\ge 2$ are
\begin{equation}
\label{eq:hm0sol0}
\tbmz^{h}=\frac{(-2 M+r) }{2 (1+\lambda) r}\dpo\;,
\end{equation}
\vspace*{-20pt}
\begin{equation}
\label{eq:hm1sol0}
\begin{split}
\tbmo^{h}=&-\frac{\pza}{r^2}+\frac{1}{4 (1+\lambda) (2 M-r) r^3}
\left[-4 (1+\lambda) (2 M-r) r \fz\right.\\&\left.
-8 (1+\lambda) (M+r+\lambda r) \fdz+(2 M-r) r^2 
\left(r+4 (1+\lambda) \dfz\right)\right]\pz+\frac{\dpza}{r}
\\&+\frac{\left(4 (1+\lambda) (2 M-r) \fz+4 (1+\lambda) \fdz
-(2 M-r) \left(r^2-4 (1+\lambda) \right)\right)
}{4 (1+\lambda) (2 M-r) r}\dpz\;,
\end{split}
\end{equation}
\begin{equation}
\label{eq:hm2sol0}
\tbmtw^{h}=\frac{\fz }{r}\pz+\frac{\pza}{r}+\frac{\fdz }{r}\dpz\;.
\end{equation}
Here, $\po$, $\pz$ and $\pza$ are homogeneous solutions of the generalized 
Regge-Wheeler equation for $s=1$ or $s=0$.  Using the substitution 
$M_{i}\to \widetilde{M}_{i}^{h}$ ($i=0,1,2$), the reader may verify 
that equations \eqref{eq:hm0sol0}-\eqref{eq:hm2sol0} satisfy the zero 
frequency forms of the harmonic gauge preservation equations 
\eqref{eq:ddm0eqn}-\eqref{eq:ddm2eqn}.  For $l\ge 2$, a zero frequency 
gauge change which preserves the harmonic gauge is made by adding 
homogeneous solutions of the spin~$1$ or spin~$0$ generalized Regge-Wheeler 
equation, just as is done for the non-zero frequency case.

The seven even parity zero frequency solutions for $l\ge 2$ are written 
in terms of four functions:  three generalized Regge-Wheeler functions 
(two with $s=0$ and one with $s=1$), and the Zerilli-Moncrief function, 
which is the even parity equivalent of the $s=2$ Regge-Wheeler function.  
This is the same structure as the corresponding non-zero 
frequency solution set.  

\subsection{\label{sec:zeveqn}Solution of Two Systems of Equations}

In this subsection, we will solve the two systems of differential equations 
in \eqref{eq:ddf0eqn}-\eqref{eq:ddfd0eqn} and 
\eqref{eq:ddf2eqn}-\eqref{eq:ddfd2eqn}.  The solutions will be in the 
form of infinite series.  

We will start with the equations for $\fz$ and $\fdz$.  It is helpful to 
rewrite this system as a single third order differential equation.  To do so, 
solve \eqref{eq:ddfd0eqn} for $\dfz$ and use this result and 
its derivative to eliminate $\ddfz$ and $\dfz$ from \eqref{eq:ddf0eqn}.  
This procedure leads to
\begin{multline}
\label{eq:dddfd0eqnr}
\frac{(-2 M+r)^2}{r^2}\dddfdz
+\frac{4 \left(M^2+4 (1+\lambda) M r-2 (1+\lambda) r^2\right)}{r^4}\dfdz
\\-\frac{8 (M-r) \left(M^2+2 (1+\lambda ) M r
-(1+\lambda)r^2\right)}{(2 M-r) r^5}\fdz
=\frac{(2 M-r)(4 M-3 r)}{2 r^2 (1+\lambda)}\;.
\end{multline}
To find an expression for $\fz$ in terms of $\fdz$, define
\begin{equation}
\label{eq:ftil}
\fdz=\ff \tilde{f}_{d0}\;,
\end{equation}
substitute into \eqref{eq:ddfd0eqn}, and simplify to obtain
\begin{equation}
\label{eq:ddfd0aeqn}
\left[\ff\tilde{f}_{d0}^{\prime}\right]^{\prime}
+2 \dfz=\frac{r}{2(1+\lambda)}\;,
\end{equation}
which can be integrated with respect to $r$ to give
\begin{equation}
\label{eq:tryf0rule}
\fz=\frac{1}{2}\left[\frac{r^2}{4(1+\lambda)}
-\ff\tilde{f}_{d0}^{\prime}\right]+C\;.
\end{equation}
After solving \eqref{eq:ftil} for $\tilde{f}_{d0}$ and differentiating, we 
rewrite \eqref{eq:tryf0rule} as
\begin{equation}
\label{eq:f0rule}
\fz=\frac{r^2}{8(1+\lambda)}
-\frac{M \fdz}{2 M r-r^2}
-\frac{\dfdz}{2}+C\;.
\end{equation}
The first term does not apply when $\fz$ and $\fdz$ are 
homogeneous solutions of \eqref{eq:ddf0eqn}-\eqref{eq:ddfd0eqn}.  
The constant of integration $C$ may be set to zero as explained 
below, in the discussion following \eqref{eq:c4sol}.

In terms of the dimensionless quantities $x=\frac{r}{2 M}$ and 
$\fdz(x)=\frac{1}{(2 M)^{3}}\fdz(r)$, equation 
\eqref{eq:dddfd0eqnr} is
\begin{multline}
\label{eq:dddfd0eqnx}
\frac{(x-1)^2}{x^2}\frac{d^{3}\fdz(x)}{dx^{3}}
+\frac{\left(1+8 (1+\lambda) x
-8 (1+\lambda) x^2\right)}{x^4}\frac{d\,\fdz(x)}{dx}
\\+\frac{(2 x-1) \left(-1-4 (1+\lambda) x
+4 (1+\lambda) x^2\right)} {(x-1) x^5}\fdz(x)
=\frac{(x-1)(3 x-2)}{2 x^{2}(1+\lambda)}\;.
\end{multline}
An inhomogeneous solution of \eqref{eq:dddfd0eqnx} is
\begin{equation}
\label{eq:fd0inf}
\fdz^{\infty}(x)=\sum_{n=-3}^{\infty}\frac{a_{n}}{x^{n}}
-\ln x \sum_{n=-1}^{\infty}\frac{b_{n}}{x^{n}}\;,
\end{equation}
where the recursion relations for $a_{n}$ and $b_{n}$ are
\begin{equation}
\begin{split}
\label{eq:fd0an}
a_{n}=&\frac{1}{(1+n) (n-2 l) (2+2 l+n)}\left[(-2+n)^3a_{n-3}
+\left(l(l+1)(-6+4 n)\right.\right.
\\&\left.+n \left(-7+9 n-3 n^2\right)\right)a_{n-2}
+\left(l(l+1)(2-8 n)+3 n \left(-1+n^2\right)\right)a_{n-1}
\\&+3 (-2+n)^2 b_{n-3}+\left(-7+4 l+4 l^2+18 n-9 n^2\right)b_{n-2}
\\&\left.-\left(3+8 l+8 l^2-9 n^2\right)b_{n-1}
-\left(2-4 l-4 l^2+6 n+3 n^2\right)b_{n}\right]\;,
\end{split}
\end{equation}
\begin{multline}
\label{eq:fd0bn}
b_{n}=\frac{1}{(1+n) (n-2 l) (2+2 l+n)}\left[(-2+n)^3b_{n-3}
+\left(l(l+1)(-6+4 n)\right.\right.\\\left.\left.
+n \left(-7+9 n-3 n^2\right)\right)b_{n-2}
-\left(l(l+1)(-2+8 n)-3 n \left(-1+n^2\right)\right)b_{n-1}\right]\;.
\end{multline}
It is simpler to use the spherical harmonic index $l$ instead of $\lambda$ 
here.  The initial values for $a_{n}$ are
\begin{equation}
a_{n}=0,\;n\le-4\;,
\end{equation}
\begin{equation}
a_{-3}=-\frac{3}{2 l (1+l) (2 l-1) (3+2 l)}\;,
\end{equation}
\begin{equation}
a_{-2}=\frac{3+5 l+5 l^2}{4 l^2 (1+l)^2 (2 l-1) (3+2 l)}\;,
\end{equation}
\begin{equation}
a_{-1}=0\;,
\end{equation}
\begin{equation}
a_{0}=\frac{-3-11 l-5 l^2+12 l^3+6 l^4}
{8 l^2 (1+l)^2 (2 l-1) (1+2 l)^2 (3+2 l)}\;,
\end{equation}
and for $b_{n}$ are
\begin{equation}
b_{n}=0,\;n\le-2\;,
\end{equation}
\begin{equation}
b_{-1}=\frac{1}{2 (2 l-1) (1+2 l)^2 (3+2 l)}\;,
\end{equation}
\begin{equation}
b_{0}=-\frac{1}{4 (2 l-1) (1+2 l)^2 (3+2 l)}\;.
\end{equation}
The expressions for $a_{n}$ and $b_{n}$ have a factor of $n-2 l$ in the 
denominator, so it would appear that they are singular when $n=2 l$.  
However, the coefficients $a_{2 l}$ and $b_{2 l}$ are actually finite 
for specific values of $l$.  Evidently, when calculated, $a_{2 l}$ 
and $b_{2 l}$ end up having a factor of $n-2 l$ in the numerator 
which cancels the factor of $n-2 l$ in the denominator, removing the 
singularity.  The superscript ``$\infty$'' attached to $\fdz^{\infty}$ 
signifies that the  series \eqref{eq:fd0inf} is suitable for larger $r$, 
rather than for $r$ close to $2 M$, where its convergence is much slower.  
We can derive a second inhomogeneous series solution for analysis
near the event horizon of the form
\begin{equation}
\fdz^{2 M}(X)=\sum_{n=2}^{\infty} d_{n}X^{n}\;,
\end{equation}
where $X=1-\frac{1}{x}=1-\frac{2 M}{r}$ and where the $d_{n}$ have 
a multiterm recursion relation like $a_{n}$ and $b_{n}$ above.  
This series converges slowly for larger $r$.

The series $\fdz^{\infty}$ and $\fdz^{2 M}$ are not equal.  
If they are both evaluated at an intermediate point, say $r=4 M$, 
the calculated values do not agree.  This disparity would cause the 
metric perturbations to be discontinuous and is not physical.  To make 
the two series match, we need to add homogeneous solutions of 
\eqref{eq:dddfd0eqnr} to each, because two inhomogeneous 
solutions of a linear ordinary differential equation may 
differ only by a homogeneous solution \cite{zill93}.  
For \eqref{eq:dddfd0eqnr}, homogeneous solutions have the form
\begin{equation}
\label{eq:hfdo}
\fdz^{h}=\ff\pz^{a}\pz^{b}\;,
\end{equation}
where $\pz^{a}$ and $\pz^{b}$ are any two homogeneous solutions of the 
zero frequency generalized Regge-Wheeler equation with $s=0$.  
Based on this result and $\fz$ from \eqref{eq:f0rule}, we have 
\begin{equation}
\label{eq:hf0fd0}
\fz^{h}=-\frac{1}{2}\ff\left(\pz^{a}\pz^{b}\right)^{\prime},\;
\fdz^{h}=\ff\pz^{a}\pz^{b}\;,
\end{equation}
as homogeneous solutions of the system 
\eqref{eq:ddf0eqn}-\eqref{eq:ddfd0eqn}.  Equations \eqref{eq:hfdo} 
and \eqref{eq:hf0fd0} may be verified by substitution.  
In section~\ref{sec:zrweqn}, we will derive two linearly independent 
homogeneous Regge-Wheeler solutions:  $\psi_{0}^{\text{in}}$ \eqref{eq:psinz}, 
a polynomial which is bounded as $r\to 2 M$, but diverges like $r^{l+1}$ 
as $r\to\infty$; and $\psi_{0}^{\text{out}}$ \eqref{eq:psoutz}, an infinite 
series which is bounded as $r\to\infty$, but diverges 
logarithmically (like $\ln\left[1-\frac{2 M}{r}\right]$) as $r\to 2 M$. 
 
The series with $b_{n}$ in \eqref{eq:fd0inf} is a homogeneous solution 
of \eqref{eq:dddfd0eqnx}, because there is not a logarithm term ($\ln x$) 
on the right side of \eqref{eq:dddfd0eqnx}.  Specifically, 
\begin{equation}
\label{eq:fd0bser}
\sum_{n=-1}^{\infty}\frac{b_{n}}{x^{n}}=\frac{\left(1-\frac{1}{x}\right)
\pz^{\text{in}}\pz^{\text{out}}}{2(1+2 l)^2 (2 l-1) (3+2 l)}\;.
\end{equation}
The series with $a_{n}$ does not have a simple form like this.  

To incorporate homogeneous solutions, we may define
\begin{equation}
\label{eq:fd0outh}
\fdz^{\text{out}}=\fdz^{\infty}
+c_{3}\left(1-\frac{1}{x}\right)\psi^{\text{out}}_{0}\psi^{\text{out}}_{0}
\end{equation}
and
\begin{equation}
\label{eq:fd0inh}
\fdz^{\text{in}}=\fdz^{2 M}+c_{1}X\psi^{\text{in}}_{0}\psi^{\text{in}}_{0}
+c_{2}X\psi^{\text{out}}_{0}\psi^{\text{in}}_{0}.
\end{equation}
To find the constants, we evaluate $\fdz^{\text{out}}$, $\fdz^{\text{in}}$ and their 
first and second derivatives at an intermediate point $x_{0}$ and 
solve the system
\begin{equation}
\label{eq:fdzsys}
\fdz^{\text{out}}=\fdz^{\text{in}}\;,
\frac{d \fdz^{\text{out}}}{dx}
=\frac{d \fdz^{\text{in}}}{dx}\;,
\frac{d^{2} \fdz^{\text{out}}}{dx^{2}}
=\frac{d^{2} \fdz^{\text{in}}}{dx^{2}}\;,
\end{equation}
for $c_{1}$, $c_{2}$ and $c_{3}$.  There are as many equations 
as unknowns, so \eqref{eq:fdzsys} has a unique solution.  
From \eqref{eq:dddfd0eqnx}, 
continuity of $\fdz$ and its first and second derivatives 
implies continuity of its all higher order derivatives.  Applying 
\eqref{eq:f0rule} and \eqref{eq:ddf0eqn}-\eqref{eq:ddfd0eqn} 
shows the continuity of $\fz$ and its derivatives of all orders.  
Accordingly, there will not be unphysical discontinuities in the metric 
perturbation functions due to $\fz$ and $\fdz$.

Because of continuity, we need only one of 
the two solutions \eqref{eq:fd0outh} and \eqref{eq:fd0inh} in order 
to calculate the metric perturbations numerically.  The series 
$\fdz^{2 M}$ converges much more slowly than $\fdz^{\infty}$, so it is 
better to use $\fdz^{\text{out}}$.  Numerical values of $c_{3}$ for different 
values of spherical harmonic index $l$ are set forth in 
Table~\ref{tab:rotmsoc} on the following page.  
Solving the system \eqref{eq:fdzsys} is numerically difficult because 
the different terms in $\fdz^{\text{in}}$ and $\fdz^{\text{out}}$ may vary by many 
orders of magnitude, particularly as $l$ increases.  Based on 
\eqref{eq:psinz} and \eqref{eq:psoutz}, $(\pz^{\text{in}})^{2}$ increases 
as $\left(\frac{r}{2 M}\right)^{2(l+1)}$, while $(\pz^{\text{out}})^{2}$ 
decreases as $\left(\frac{2 M}{r}\right)^{2 l}$.  
To minimize this source of error, the calculations for the table 
were done with \textit{Mathematica} using arbitrary precision 
arithmetic, that is, arithmetic with fractions of integers rather 
than finite precision decimal numbers.  However, there is still error 
due to truncation of the infinite series for $\fdz^{\infty}$, $\fdz^{2 M}$ 
and $\pz^{\text{out}}$, which are not exact.  As part of this method, the 
coefficients $a_{n}$ and $b_{n}$ were calculated for $l$ in general 
and then specific values of $l$ were substituted.  This made it possible 
to calculate $a_{2 l}$ and $b_{2l}$, despite the factors of $n-2 l$ 
in the denominator.  The intermediate matching 
point used was $x_{0}=\frac{r_{0}}{2 M}$, where $r_{0}=4 M$.  A 
larger value of $x_{0}$ would require significantly 
more terms in $\fdz^{2 M}$, the series that converges most slowly, but a 
smaller value would require more terms in $\fdz^{\infty}$.  

\newcolumntype{d}{D{.}{.}{26}}
\begin{sidewaystable}
\caption[Numerical Values of the Constant $c_{3}$ in 
Equation \eqref{eq:fd0outh}]
{\label{tab:rotmsoc}
Numerical values of the constant $c_{3}$ in equation \eqref{eq:fd0outh}, for 
selected values of the spherical harmonic index $l$.  The 
heading ``$100/200/500$'' means the calculation used $100$ terms 
in the series for $\fdz^{\infty}$, $200$ terms in the series for 
$\pz^{\text{out}}$ and $500$ terms in the series for $\fdz^{2 M}$.  
The heading ``$70/100/300$'' is interpreted similarly, but the figures 
in its column are less accurate, because the series have a larger 
truncation error.  All terms were used for $\pz^{\text{in}}$, which is a 
polynomial, not an infinite series.  The column 
labeled ``$c_{3}$'' was computed using the 
analytic formula \eqref{eq:fd0c3}.  Comparing the first and second columns 
gives a conservative upper bound on the series truncation error of the 
figures in the first column.  Discrepancies between the first and third columns 
are much less than the truncation error.}
\begin{center}
\begin{tabular}{|r|d|d|d|}
\hline
\multicolumn{1}{|c|}{$l$} & \multicolumn{1}{c|}{$100/200/500$}
 & \multicolumn{1}{c|}{$70/100/300$} & \multicolumn{1}{c|}{$c_{3}$}\\
\hline
\rule[2mm]{0mm}{2mm}
2 & -1.3227513227513227513\times 10^{-6}
 & -1.3227513227513227516\times 10^{-6}
 & -1.3227513227513227513\times 10^{-6}\\
4 & -4.5446001606243480771\times 10^{-10}
 & -4.544600160624347\times 10^{-10}
 & -4.5446001606243480771\times 10^{-10}\\
6 & -4.0387974132551233772\times 10^{-13}
 & -4.0387974132547\times 10^{-13}
 & -4.0387974132551233772\times 10^{-13}\\
8 & -5.3896590808114810814\times 10^{-16}
 & -5.3896590805\times 10^{-16}
 & -5.3896590808114810815\times 10^{-16}\\
10 & -9.0484374716258353447\times 10^{-19}
 & -9.0484370\times 10^{-19}
 & -9.0484374716258354477\times 10^{-19}\\
12 & -1.7617119604491278512\times 10^{-21}
 & -1.7616\times 10^{-21}
 & -1.7617119604491441344\times 10^{-21}\\
14 & -3.8048919205023529200\times 10^{-24}
 & -3.6\times 10^{-24}
 & -3.8048919205342432413\times 10^{-24}\\
\hline
\end{tabular}
\end{center}
\end{sidewaystable}

An analytic formula for $c_{3}$ is
\begin{equation}
\label{eq:fd0c3}
c_{3}=-\frac{[\Gamma(1+l)]^4}{8 (2 l-1) (1+2 l) (3+2 l)[\Gamma(2+2 l)]^2}\;.
\end{equation}
This expression was obtained by experimenting with constants appearing in 
related equations, \eqref{eq:fd0bser}, \eqref{eq:psioutofx} 
and \eqref{eq:psinzx}.  Table~\ref{tab:rotmsoc} has numerical values for 
this form of $c_{3}$ as well.

Additionally, we can solve \eqref{eq:dddfd0eqnr} for specific values of $l$ 
by constructing an inhomogeneous solution from homogeneous solutions 
of \eqref{eq:dddfd0eqnr}.  From 3.1.1 of \cite{odebook}, an 
inhomogeneous third order linear differential equation given by
\begin{equation}
\label{eq:genthrd}
f_{3}(x)\frac{d^{3} y}{dx^{3}}+f_{2}(x)\frac{d^{2} y}{dx^{2}}
+f_{1}(x)\frac{d y}{dx}+f_{0}(x)y=g(x)
\end{equation}
has the solution
\begin{equation}
\label{eq:genthrdy}
y(x)=C_{1}y_{1}+C_{2}y_{2}+C_{3}\left(y_{2}\int y_{1}\, \psi\, dx
-y_{1}\int y_{2}\, \psi\, dx\right)\;.
\end{equation}
Here, $y_{1}$ and $y_{2}$ are two linearly independent homogeneous solutions of 
\eqref{eq:genthrd}, and 
\begin{equation}
\label{eq:genthrdp}
\psi= \Delta^{-2} e^{-F}\left(1
+\frac{1}{C_{3}}\int \frac{g}{f_{3}}\Delta e^{F} dx\right)\;,
\end{equation}
where
\begin{equation}
F=\int\frac{f_{2}}{f_{3}}dx\;,\;\Delta=y_{1}\frac{d y_{2}}{dx}
-\frac{d y_{1}}{dx}y_{2}\;.
\end{equation}
For \eqref{eq:dddfd0eqnr}, we have $f_{2}=0$, so $F=0$.  
Applying \eqref{eq:genthrdy} for $l=2$ leads to
\begin{equation}
\begin{split}
\label{eq:fd02}
(2 M)^{3}\fdz^{\text{out}}(x)=&\frac{(2 M-r) r \left(656 M^3
-1764 M^2 r+2187 M r^2-729 r^3\right)}{25200 M^2}
+(2 M-r) r \\&\times\left(2 M^2-6 M r+3 r^2\right)
\Bigg\{\frac{ \left(146 M^2-198 M r+39 r^2\right) \lnff}{16800 M^3}
\\&+\frac{\left(-6 M^2+6 M r+\left(2 M^2-6 M r+3 r^2\right)
\lnff\right)\lnr}{420 M^3}
\\&+\frac{\left(2 M^2-6 M r+3 r^2\right)}{1680 M^3}
\left(4\,\plog-\lnff^2\right)\Bigg\}
\\&+c_{3}(2 M)^{3}\ff(\pz^{\text{out}})^{2}\;,
\end{split}
\end{equation}
where $\fdz^{\text{out}}$ is the dimensionless quantity 
defined by \eqref{eq:fd0outh} and where
\begin{equation}
\label{eq:pzoutl2}
(\pz^{\text{out}})^{2}=\left[-\frac{15 r}{2 M^{3}}\left(6 M(r-M)
+\left(2 M^2-6 M r+3 r^2\right) \lnff\right)\right]^{2},\text{ for }l=2.
\end{equation}
The first four lines of the right side of \eqref{eq:fd02} are an analytic 
expression for the series $\fdz^{\infty}$ \eqref{eq:fd0inf}, expressed in 
terms of $r$ instead of $x$ and multiplied by $(2 M)^{3}$ to give the 
correct dimensions.   
The expression ``$\text{PolyLog}[2,z]$'' is the \textit{Mathematica} 
notation for the dilogarithm function $Li_{2}(z)$, which 
\textit{Mathematica} defines as \cite{wolfbk}
\begin{equation}
\label{eq:polylog}
Li_{2}(z)=\int^{0}_{z}\frac{\ln[1-t]}{t}dt\;.
\end{equation}
By substituting \eqref{eq:fd02} into the expression for $\fz$ 
\eqref{eq:f0rule}, we may calculate $\fz^{\text{out}}$.  

When we substitute 
$\fz^{\text{out}}$ and $\fdz^{\text{out}}$ into $\bmtwa$ \eqref{eq:m2asol0}, 
we discover 
that $\bmtwa$ (and therefore $\bg$) will diverge like $\lnff$ as $r\to 2 M$, 
unless $c_{3}$ has the numerical value given by \eqref{eq:fd0c3}.  
To see this, set
\begin{equation}
\label{eq:fd0terms}
\fz\pz+\fdz\dpz=\fz^{\text{out}}\pz^{\text{in}}+\fdz^{\text{out}}(\pz^{\text{in}})^{\prime}\;.
\end{equation}
The left side is from \eqref{eq:m2asol0}, and
\begin{equation}
\pz^{\text{in}}=C^{\text{in}}\left[\left(\frac{r}{2 M}\right)^{3}
-\left(\frac{r}{2 M}\right)^{2}
+\frac{r}{12 M}\right]\;,\text{ for }l=2.
\end{equation}
We use $\pz^{\text{in}}$ rather than $\pz^{\text{out}}$ in \eqref{eq:fd0terms}, 
because $\pz^{\text{in}}$ is bounded as 
$r\to 2M$.  The constant $C^{\text{in}}$ is the amplitude of the ingoing 
solution.  After a brief calculation, we find that \eqref{eq:fd0terms} 
is exactly equal to
\begin{multline}
\label{eq:nfd0terms}
\frac{r C^{\text{in}}}{100800 M^3}\bigg\{-94 M^4-90720000\,M^3 c_{3}(r-M)
+2 M^3 r+149 M^2 r^2-500 M r^3
\\+300 r^4-20 (1+756000\, c_{3}) M^2 \left(2 M^2-6 M r+3 r^2\right)\lnff
\\+40 M^2 \left(2 M^2-6 M r+3 r^2\right) \lnr\bigg\}\;.
\end{multline}
To prevent a logarithmic divergence as $r\to 2 M$, we require that 
$c_{3}=-\frac{1}{756000}$, which agrees with \eqref{eq:fd0c3}.  The 
terms proportional to $c_{3}$ are a constant multiple of $\pz^{\text{out}}$, 
as given by \eqref{eq:pzoutl2}.  The remaining terms are the contribution 
of $\fdz^{\infty}$ to $\bmtwa$.  Each of the two types of terms diverges 
logarithmically, but the divergences cancel if $c_{3}=-\frac{1}{756000}$.  
The possibility of the logarithmic divergence is hidden in the series 
expansion for $\fdz^{\infty}$, because $\lnff$ can be expanded as a power 
series.  
This example is only for $l=2$, but similar results presumably also hold 
for larger $l$.  
The system \eqref{eq:fdzsys} has a unique solution, and 
it is physically necessary both to have continuity of $\fdz$ and 
to avoid a logarithmic divergence.  

Equation \eqref{eq:genthrdy} shows that an inhomogeneous solution of 
\eqref{eq:dddfd0eqnr} may be constructed from homogeneous solutions.  
The homogeneous solutions of \eqref{eq:dddfd0eqnr} 
have products of $\pz^{\text{in}}$ and $\pz^{\text{out}}$, 
which are formed from hypergeometric functions as shown in 
Chapter~\ref{rweqnchap}.  It follows that the inhomogeneous solutions 
\eqref{eq:fd0outh} and \eqref{eq:fd0inh} are themselves related to the 
hypergeometric functions.  Also as explained in Chapter~\ref{rweqnchap}, 
a hypergeometric series of the form ${}_{2}F_{1}(a,b;c;z)$ will converge 
quickly, provided $\lvert z\rvert\le\frac{1}{2}$.  This suggests that 
the series $\fdz^{\infty}$ in \eqref{eq:fd0outh} will converge efficiently 
as long as $\frac{1}{x}=\frac{2 M}{r}\lesssim\frac{1}{2}$, a requirement 
which generally will be met.  The series $\fdz^{\infty}$ usually converges 
in fewer than $100$ terms, if double precision arithmetic is used.  
On the other hand, if $r\ge 4 M$, then $X=1-\frac{2 M}{r}\ge \frac{1}{2}$ 
and $\fdz^{2 M}$ converges slowly.  For this reason, it is better to use 
$\fdz^{\text{out}}$ \eqref{eq:fd0outh} than $\fdz^{\text{in}}$ \eqref{eq:fd0inh} to 
calculate the metric perturbations numerically.  

Another issue concerns the number of constants.  Equation 
\eqref{eq:dddfd0eqnr} is a linear third order ordinary differential 
equation, so it has three linearly independent homogeneous solutions 
which are used in \eqref{eq:fd0outh}-\eqref{eq:fd0inh} and which 
are associated with the three constants $c_{1}$, $c_{2}$ and $c_{3}$.  
The third order equation is derived from the two equation system 
\eqref{eq:ddf0eqn}-\eqref{eq:ddfd0eqn}.  
By inspection, the system has an additional homogeneous solution 
given by
\begin{equation}
\label{eq:c4sol}
\fz^{h}=c_{4}\;,\;\fdz^{h}=0\;,
\end{equation}
where $c_{4}$ is equivalent to the constant of integration $C$ in the 
expression for $\fz$ \eqref{eq:f0rule}.  Thus, the system has 
a total of four homogeneous solutions and four constants of 
integration.  However, the system is only 
a mathematical tool used to solve the equation for $\bmtwa$ 
\eqref{eq:m2aeqn0}, which is a second order differential equation 
with only two homogeneous solutions ($\pz^{\text{out}}$ and $\pz^{\text{in}}$)
and therefore two constants of integration.  
This method of solution leads to two additional constants.  
It turns out that two of the constants, $c_{2}$ and $c_{4}$, affect 
$\fz$ and $\fdz$, but not $\bmtwa$.  
The functions $\fz$ and $\fdz$ 
appear in two places in $\bmtwa$:  (1) in the terms $\fz\pz+\fdz\dpz$ 
and (2) in the source for $\pza$ \eqref{eq:s0d}.  Adding the 
homogeneous solution corresponding to $c_{2}$ changes both (1) 
and (2), but the changes cancel, leaving $\bmtwa$ unaltered.  
The same is true for $c_{4}$.  To prove these results, it is necessary 
to write out $\pz$ and $\pza$ in integral form, using 
the inhomogeneous solution 
\eqref{eq:retgrwsol}.  Because $c_{4}$ does not affect $\bmtwa$, 
it will not affect the metric perturbations and may be set equal to 
zero, which is why the integration constant $C$ in \eqref{eq:f0rule} 
may be disregarded.  Even though the value of $c_{2}$ does 
not affect the metric perturbations, it is not arbitrary and is 
determined when we solve the system \eqref{eq:fdzsys}.  
On the other hand, the homogeneous solutions associated with $c_{1}$ 
and $c_{3}$ do affect $\bmtwa$.  Using equations \eqref{eq:retgrwsol} 
and \eqref{eq:zwron}, we can show that adding the $c_{1}$ homogeneous 
solution to $\fdz$ and $\fz$ is equivalent to adding a constant 
multiple of $\pz^{\text{in}}$ to $\bmtwa$, while adding the $c_{3}$ 
solution results in adding a constant multiple of $\pz^{\text{out}}$ 
to $\bmtwa$.  The lengthy calculations required to prove these results 
will not be described here, but the example given above for $l=2$ 
is a specific application of these principles to the $c_{3}$ 
homogeneous solution.

To summarize, we can calculate $\fz$ and $\fdz$ as follows.  Compute 
$\fdz$ using the formula for $\fdz^{\text{out}}$ \eqref{eq:fd0outh} and  
the values of $c_{3}$ in Table~\ref{tab:rotmsoc}.  Find $\fz$ by 
applying \eqref{eq:f0rule}, without the constant $C$.  The derivatives 
$\dfdz$ and $\dfz$ are obtained by differentiating \eqref{eq:fd0outh} 
and \eqref{eq:f0rule}.  For dimensions, $\fdz(r)=(2 M)^{3}\fdz(x)$ 
and $\fz(r)=(2 M)^{2}\fz(x)$.  

Solutions for the other two equation system, which consists of
\eqref{eq:ddf2eqn}-\eqref{eq:ddfd2eqn}, are derived in a similar manner.  
We can simplify the equations somewhat by defining
$\nftw$ and $\nfdtw$ such that
\begin{multline}
\label{eq:nftwdef}
\ftw=\frac{3 \left(-18 M^3+9 M^2 r+3 \lambda (1+\lambda) M r^2
+\lambda^2 (1+\lambda) r^3\right)}
{2 \left(\lambda+\lambda^2\right)^2 r^2 (3 M+\lambda r)}\nftw
\\-\frac{9 M \left(-18 M^3+9 M^2 r+6 \lambda (1+\lambda) M r^2
+2 \lambda^2 (1+\lambda) r^3\right)}
{2 \left(\lambda+\lambda^2\right)^2 r^3 (3 M+\lambda r)^2}\nfdtw\;,
\end{multline}
\begin{multline}
\label{eq:nfdtwdef}
\fdtw=\frac{9 M (2 M-r)}
{2 \left(\lambda+\lambda^2\right)^2 r}\nftw
\\+\frac{3\left(-18 M^3+9 M^2 r+3 \lambda (1+\lambda) M r^2
+\lambda^2(1+\lambda) r^3\right)}
{2 \left(\lambda+\lambda^2\right)^2 r^2 (3 M+\lambda r)}\nfdtw\;.
\end{multline}
These definitions originate as follows.  Two terms in $\bmtwa$ 
\eqref{eq:m2asol0} are
\begin{equation}
\label{eq:f2fd2}
\ftw\ptw+\fdtw\dptw\;,
\end{equation}
where $\ptw=\psi_{Z}$ is either an inhomogeneous or homogeneous solution 
of the Zerilli equation.  As explained in Chapter~\ref{rweqnchap}, 
homogeneous solutions of the Zerilli equation are related to homogeneous 
solutions of the spin~$2$ Regge-Wheeler equation by differential 
operators \eqref{eq:oddevout}-\eqref{eq:oddevin}.  A modified form of 
these relations for zero frequency is
\begin{equation}
\label{eq:rwtoz}
\psi_{Z}^{}=\frac{2}{3}\left[\left(\lambda+\lambda^2+\frac{9 M^2(r-2 M)}
{r^2(3 M+\lambda r)}\right)\psi_{\text{RW}}^{}
+3 M \left(1-\frac{2 M}{r}\right)\psi_{\text{RW}}^{\prime}\right]
\end{equation}
where $\psi_{Z}^{}$ and $\psi_{\text{RW}}^{}$ are homogeneous solutions only.  
Using \eqref{eq:rwtoz}, the definitions 
\eqref{eq:nftwdef}-\eqref{eq:nfdtwdef} are derived so that 
\eqref{eq:f2fd2} becomes
\begin{equation}
\label{eq:nf2fd2}
\nftw\psi_{\text{RW}}^{}+\nfdtw\psi_{\text{RW}}^{\prime}\;.
\end{equation}
After substituting \eqref{eq:nftwdef} and \eqref{eq:nfdtwdef} into the 
system \eqref{eq:ddf2eqn}-\eqref{eq:ddfd2eqn} and simplifying, we have
\begin{multline}
\label{eq:nddf2eqn}
\frac{(-2 M+r)^2 }{r^2}\ddnftw
+\frac{2 M (-2 M+r) }{r^3}\dnftw
+\frac{4 (-2 M+r) (-3 M+r+\lambda r) }{r^4}\dnfdtw
\\+\frac{8 M (2 M-r)}{r^4}\nftw
+\frac{\left(-24 M^2+2(11+2\lambda)M r-4(1+\lambda)r^2\right)}{r^5}\nfdtw
\\=-\frac{2 (2 M-r)^{2} \left(-12 M^2+6 (1+\lambda) M r
+\lambda (1+\lambda) r^2\right)}{3 r^4}\;,
\end{multline}
\begin{multline}
\label{eq:nddfd2eqn}
\frac{(-2 M+r)^2}{r^2}\ddnfdtw
-\frac{2 M (-2 M+r)}{r^3}\dnfdtw
+\frac{2 (-2 M+r)^2 }{r^2}\dnftw
+\frac{4 M (3 M-r)}{r^4}\nfdtw
\\=-\frac{2 (2 M-r)^{2} \left(-12 M^2+2 (3+\lambda) M r
+\lambda (1+\lambda) r^2\right)}{3 r^3}\;.
\end{multline}

We can eliminate $\nftw$ and its derivatives from the new system to 
obtain a single fourth order differential equation for $\nfdtw$,
\begin{equation}
\begin{split}
\label{eq:d4eqnr}
\frac{(-2 M+r)^2}{r^2}&\ddddnfdtw
+\frac{(4 M-3 r) (2 M-r)}{r^3}\dddnfdtw
+\frac{4 \left(-7 M^2+4 (2+\lambda)M r
-2 (1+\lambda) r^2\right)}{r^4}\ddnfdtw
\\&-\frac{4 M \left(14 M^2+(-13+4 \lambda) M r
-2 (-2+\lambda) r^2\right)}{r^5 (-2 M+r)}\dnfdtw
\\&+\frac{8 M \left(18 M^3-11 M^2 r+2 (-1+\lambda) M r^2
-(-2+\lambda) r^3\right)}{r^6 (-2 M+r)^2}\nfdtw
\\&=\frac{2 (2 M-r) \left(32 M^3+4 (2+\lambda) M^2 r
-8 (1+\lambda) M r^2-\lambda (1+\lambda) r^3\right)}{r^5}\;,
\end{split}
\end{equation}
and a relation between $\nftw$ and $\nfdtw$,
\begin{multline}
\label{eq:f2rule}
\nftw=-\frac{\left(-3 M^3+(9+2 \lambda) M^2 r
-3 (2+\lambda)M r^2+(1+\lambda) r^3\right)}{2 M r(-2 M+r)^2}\nfdtw
\\+\frac{\left(-11 M^2+2 (5+2 \lambda) M r
-2 (1+\lambda) r^2\right)}{4 M (2 M-r)}\dnfdtw
+\frac{(2 M-r) r^2 }{16 M}\dddnfdtw
\\+\frac{96 M^3-4 (18+7 \lambda) M^2 r-4 \left(-3-2 \lambda
+\lambda^2\right) M r^2+\lambda(1+\lambda) r^3}{24 M}\;.
\end{multline}
In terms of $l$ instead of $\lambda$ and $x=\frac{r}{2 M}$, 
equation \eqref{eq:d4eqnx} is
\begin{equation}
\begin{split}
\label{eq:d4eqnx}
\frac{(-1+x)^4}{x^4}&\frac{d^{4}\nfdtw(x)}{dx^{4}}
+\frac{(-1+x)^3 (-2+3 x)}{x^5}\frac{d^{3}\nfdtw(x)}{dx^{3}}
\\-&\frac{(-1+x)^2 \left(7-4 \left(2+l+l^2\right) x+4 l (1+l) 
x^2\right) }{x^6}\frac{d^{2}\nfdtw(x)}{dx^{2}}
\\+&\frac{(-1+x) \left(-7+\left(17-2 l-2 l^2\right)
x+2 \left(-6+l+l^2\right) x^2\right) }{x^7}\frac{d\nfdtw(x)}{dx}
\\+&\frac{\left(9-11 x+2 \left(-4+l+l^2\right) x^2-2 
\left(-6+l+l^2\right) x^3\right) }{x^8}\nfdtw(x)
\\=&\frac{(-1+x)^3 \left(-16-2 \left(2+l+l^2\right) x
+8 l (1+l) x^2+l \left(-2-l+2 l^2+l^3\right) x^3\right)}{2 x^7}\;.
\end{split}
\end{equation}
The function $\nfdtw(x)$ is dimensionless, with 
$\nfdtw(r)=(2 M)^{3}\nfdtw(x)$.

We solve \eqref{eq:d4eqnx} in the same way that we solved the 
third order equation for $\fdz$ \eqref{eq:dddfd0eqnx}.  An 
inhomogeneous solution of \eqref{eq:d4eqnx} is
\begin{equation}
\nfdtw^{\infty}(x)=\sum_{n=-3}^{\infty}\frac{a_{n}}{x^{n}}
-\ln x \sum_{n=-1}^{\infty}\frac{b_{n}}{x^{n}}\;.
\end{equation}
The coefficients $a_{n}$ and $b_{n}$ are defined by the recursion relations
\begin{multline}
a_{n}=\frac{1}{n (1+n) (n-2 l) (2+2 l+n)}\Big[
-\left(5-6 n+n^2\right)^2 a_{n-4}
\\-\left(-5+37 n-76 n^2+33 n^3-4 n^4
+l(l+1) \left(30-22 n+4 n^2\right)\right) a_{n-3}
\\-\left(4+22 n+16 n^2-27 n^3+6 n^4
-2 l (l+1)\left(15-20 n+6 n^2\right)\right) a_{n-2}
\\-n\left(1+12 n+3 n^2-4 n^3+2 l(l+1)(-7+6 n)\right) a_{n-1}
\\-4 (-3+n) \left(5-6 n+n^2\right)b_{n-4}
\\-\left(37-152 n+99 n^2-16 n^3+l(l+1)(-22+8 n)\right)b_{n-3}
\\-\left(22+32 n-81 n^2+24 n^3-8 l(l+1)(-5+3 n)\right)b_{n-2}
\\-\left(1+24 n+9 n^2-16 n^3+2 l(l+1)(-7+12 n)\right)b_{n-1}
\\+\left(l(l+1)(4+8 n)-n \left(4+9 n+4 n^2\right)\right)b_{n}\Big]\;,
\end{multline}
\begin{multline}
b_{n}=\frac{1}{n (1+n) (n-2 l) (2+2 l+n)}\Big[-\left(5-6 n+n^2\right)^2 b_{n-4}
-\left(-11+2 \left(20+l+l^2\right)\right.\\\left.\times (-3+n)
+\left(5+4 l+4 l^2\right) (-3+n)^2-15 (-3+n)^3-4 (-3+n)^4\right)b_{n-3}
\\+\left(-4-22 n-16 n^2+27 n^3-6 n^4+2 l(l+1)\left(15-20 n+6 n^2\right)
\right)b_{n-2}
\\-n\left(1+12 n+3 n^2-4 n^3+2 l(l+1)(-7+6 n)\right)b_{n-1}\Big]\;.
\end{multline}
Again, there are denominator factors of $n-2 l$, but $a_{2 l}$ and 
$b_{2l}$ are finite when calculated for specific values of $l$.  
The initial values for $a_{n}$ are
\begin{equation}
a_{n}=0,\;n\le-4\;,
\end{equation}
\begin{equation}
a_{-3}=-\frac{(-1+l) l (1+l) (2+l)}{12 (2 l-1) (3+2 l)}\;,
\end{equation}
\begin{equation}
a_{-2}=-\frac{-42+49 l+50 l^2+2 l^3+l^4}{24 (2 l-1) (3+2 l)}\;,
\end{equation}
\begin{equation}
a_{-1}=0\;,
\end{equation}
\begin{equation}
a_{0}=\frac{3 \left(-24-82 l-15 l^2+165 l^3+162 l^4+101 l^5
+43 l^6+8 l^7+2 l^8\right)}{16 l (1+l) (2 l-1) (1+2 l)^2 (3+2 l)}\;,
\end{equation}
\begin{multline}
a_{1}=\frac{1}{16 l (1+l)(2 l-1)^3 (1+2 l)^2 (3+2 l)^3}\big[1296+1278 l
\\-11103 l^2-10206 l^3+26927 l^4+25891 l^5-13277 l^6
\\-20738 l^7-7550 l^8-1445 l^9-157 l^{10}+72 l^{11}+12 l^{12}\big]\;,
\end{multline}
\begin{multline}
a_{2}=\frac{1}{96 l(1+l) (2 l-1)^3 (1+2 l)^2 (3+2 l)^3}\big[-2268-4464 l
\\+14619 l^2+29036 l^3-12274 l^4-42771 l^5-30469 l^6
\\-10830 l^7+1323 l^8+2995 l^9+907 l^{10}+168 l^{11}+28 l^{12}\big]\;,
\end{multline}
\begin{multline}
a_{3}=\frac{(l-1) (2+l)}{384 l (1+l) (2 l-3)^2 (2 l-1)^3 
(1+2 l)^2(3+2 l)^3 (5+2 l)^2}\big[291600+489240 l\\-2014200 l^2-3380832 l^3
+2712054 l^4+5973457 l^5+2401081 l^6\\-1062286 l^7-1876168 l^8-822579 l^9
+100593 l^{10}+162368 l^{11}\\+41864 l^{12}+6832 l^{13}+976 l^{14}\big]\;,
\end{multline}
and for $b_{n}$ are
\begin{equation}
b_{n}=0,\;n\le-2\;,
\end{equation}
\begin{equation}
b_{-1}=\frac{(-1+l) (2+l) \left(2+6 l+7 l^2+2 l^3+l^4\right)}
{4 (2 l-1) (1+2 l)^2 (3+2 l)}\;,
\end{equation}
\begin{equation}
b_{0}=-\frac{(-1+l) (2+l) \left(10+20 l+21 l^2+2 l^3+l^4\right)}
{8 (2 l-1) (1+2 l)^2 (3+2 l)}\;,
\end{equation}
\begin{equation}
b_{1}=-\frac{(-1+l) (2+l) \left(18-12 l-35 l^2-45 l^3-20 l^4+3 l^5+l^6\right)}
{8 (2 l-1)^2 (1+2 l)^2 (3+2 l)^2}\;,
\end{equation}
\begin{equation}
b_{2}=-\frac{(-1+l)^2 l (1+l) (2+l)^2 \left(-7+l+l^2\right)}
{16 (2 l-1)^2 (1+2 l)^2 (3+2 l)^2}\;,
\end{equation}
\begin{equation}
b_{3}=-\frac{5 (-2+l) (-1+l)^3 l (1+l) (2+l)^3 (3+l)}
{32 (2 l-3) (2 l-1)^2 (1+2 l)^2 (3+2 l)^2 (5+2 l)}\;.
\end{equation}
A second inhomogeneous solution is 
\begin{equation}
\nfdtw^{2 M}(X)=\sum_{n=3}^{\infty} d_{n}X^{n}\;, 
\end{equation}
The series $\nfdtw^{\infty}$ and $\nfdtw^{2M}$ are not equal.  They differ 
by homogeneous solutions of \eqref{eq:d4eqnx}, which have the form
\begin{equation}
\label{eq:nfd2h}
\nfdtw^{h}=\left(1-\frac{1}{x}\right)\psi^{a}_{0}\psi^{b}_{\text{RW}}\;.
\end{equation}
Here, $\psi^{a}_{0}$ and $\psi^{b}_{\text{RW}}$ are homogeneous solutions of the 
generalized Regge-Wheeler equations for $s=0$ and $s=2$, respectively.  
The subscript ``RW'' is used instead of ``$2$'' because, for even parity, 
we have used $\ptw$ to refer to solutions of the Zerilli equation, not 
the Regge-Wheeler equation.  In contrast, homogeneous solutions of the 
two equation system \eqref{eq:ddf2eqn}-\eqref{eq:ddfd2eqn} are given by
\begin{equation}
\ftw^{h}=-\left(1-\frac{2 M}{r}\right)\psi^{a}_{0}
\left(\psi^{b}_{2}\right)^{\prime},
\fdtw^{h}=\left(1-\frac{2 M}{r}\right)\psi^{a}_{0}\psi^{b}_{2}\;,
\end{equation}
where $\ptw^{b}$ is an ``in'' or ``out'' homogeneous solution of 
the Zerilli equation.  

To add homogeneous solutions, define
\begin{equation}
\label{eq:fd2outh}
\nfdtw^{\text{out}}=\nfdtw^{\infty}
+c_{3}\left(1-\frac{1}{x}\right)\psi^{\text{out}}_{0}\psi^{\text{out}}_{\text{RW}}\;,
\end{equation}
\begin{equation}
\label{eq:fd2inh}
\nfdtw^{\text{in}}=\nfdtw^{2 M}+c_{1}X\psi^{\text{in}}_{0}\psi^{\text{in}}_{\text{RW}}
+c_{2}X\psi^{\text{out}}_{0}\psi^{\text{in}}_{\text{RW}}
+c_{4}X\psi^{\text{in}}_{0}\psi^{\text{out}}_{\text{RW}}.
\end{equation}
Using \textit{Mathematica} as before, we find 
the constants by solving the four equation system
\begin{equation}
\label{eq:fdtwsys}
\nfdtw^{\text{out}}=\nfdtw^{\text{in}}\;,
\frac{d \nfdtw^{\text{out}}}{dx}
=\frac{d \nfdtw^{\text{in}}}{dx}\;,
\frac{d^{2} \nfdtw^{\text{out}}}{dx^{2}}
=\frac{d^{2} \nfdtw^{\text{in}}}{dx^{2}}\;,
\frac{d^{3} \nfdtw^{\text{out}}}{dx^{3}}
=\frac{d^{3} \nfdtw^{\text{in}}}{dx^{3}}\;.
\end{equation}
Table~\ref{tab:rotrlmc} gives numerical values of $c_{3}$ on the following 
page.  The constant $c_{3}$ may also be calculated using the expression
\begin{equation}
\label{eq:fd2c3}
c_{3}=-\frac{\left(-2+l+l^2\right) \left(\left(2+3 l+l^2\right)^2 
[\Gamma(1+l)]^4+(l-1)^2 l^2 [\Gamma(l-1)]^2[\Gamma(3+l)]^2\right)}
{32(1+2 l)^2\left(-3+4 l+4 l^2\right)\Gamma(1+2 l) \Gamma(2+2 l)}\;,
\end{equation}
which is obtained in a manner similar to \eqref{eq:fd0c3}.  

\begin{sidewaystable}
\caption[Numerical Values of the Constant $c_{3}$ in 
Equation \eqref{eq:fd2outh}]
{\label{tab:rotrlmc}
Numerical values of the constant $c_{3}$ in equation \eqref{eq:fd2outh}, for 
selected values of the spherical harmonic index $l$.  The 
heading ``$150/200/700$'' means the calculation used $150$ terms 
in the series for $\nfdtw^{\infty}$, $200$ terms in the series for 
$\pz^{\text{out}}$ and $\psi_{\text{RW}}^{\text{out}}$, and $700$ terms in the series 
for $\nfdtw^{2 M}$.  The heading ``$125/175/600$'' is interpreted 
similarly, but the figures in its column are less accurate, because 
the series have a larger truncation error.  All terms were used 
for $\pz^{\text{in}}$ and $\psi_{\text{RW}}^{\text{in}}$, which are polynomials, 
not infinite series.  The column labeled ``$c_{3}$'' was computed using the 
analytic formula \eqref{eq:fd2c3}.  The figures in the columns may be compared 
in the same way as for Table \ref{tab:rotmsoc}.}
\begin{center}
\begin{tabular}{|r|d|d|d|}
\hline
\multicolumn{1}{|c|}{$l$} & \multicolumn{1}{c|}{$150/200/700$}
 & \multicolumn{1}{c|}{$125/175/600$} & \multicolumn{1}{c|}{$c_{3}$}\\
\hline
\rule[2mm]{0mm}{2mm}
2 & -3.8095238095238095238\times 10^{-4}
 & -3.8095238095238095238\times 10^{-4}
 & -3.8095238095238095238\times 10^{-4}\\
4 & -3.6811261301057219425\times 10^{-6}
 & -3.6811261301057219425\times 10^{-6}
 & -3.6811261301057219425\times 10^{-6}\\
6 & -2.5331337375936133822\times 10^{-8}
 & -2.5331337375936133822\times 10^{-8}
 & -2.5331337375936133822\times 10^{-8}\\
8 & -1.5279683494100548866\times 10^{-10}
 & -1.5279683494100548866\times 10^{-10}
 & -1.5279683494100548866\times 10^{-10}\\
10 & -8.5136386233028620694\times 10^{-13}
 & -8.5136386233028620694\times 10^{-13}
 & -8.5136386233028620694\times 10^{-13}\\
12 & -4.4933309172996436737\times 10^{-15}
 & -4.4933309172996436737\times 10^{-15}
 & -4.4933309172996436737\times 10^{-15}\\
14 & -2.2792824560768330713\times 10^{-17}
 & -2.279282456076833070\times 10^{-17}
 & -2.2792824560768330713\times 10^{-17}\\
16 & -1.1216308802833773975\times 10^{-19}
 & -1.1216308802833767\times 10^{-19}
 & -1.1216308802833773975\times 10^{-19}\\
18 & -5.3886761848431212239\times 10^{-22}
 & -5.388676184840\times 10^{-22}
 & -5.3886761848431212240\times 10^{-22}\\
20 & -2.5390072949282366326\times 10^{-24}
 & -2.5390072937\times 10^{-24}
 & -2.5390072949282366773\times 10^{-24}\\
22 & -1.1772061694462479128\times 10^{-26}
 & -1.1772057\times 10^{-26}
 & -1.1772061694462665630\times 10^{-26}\\
24 & -5.3846408912199178636\times 10^{-29}
 & -5.382\times 10^{-29}
 & -5.3846408912984788327\times 10^{-29}\\
\hline
\end{tabular}
\end{center}
\end{sidewaystable}

To calculate $\ftw$ and $\fdtw$, first evaluate the series for 
$\nfdtw^{\text{out}}$ \eqref{eq:fd2outh}, using $c_{3}$ from 
Table~\ref{tab:rotrlmc}.  Next, find $\nftw$ from \eqref{eq:f2rule}, and 
then get $\ftw$ and $\fdtw$ from \eqref{eq:nftwdef} 
and \eqref{eq:nfdtwdef}.  For dimensions, $\fdtw(r)=(2 M)^{3}\fdtw(x)$ 
and $\ftw(r)=(2 M)^{2}\ftw(x)$, and the same holds for $\nftw$ 
and $\nfdtw$.  

Equation \eqref{eq:d4eqnr} can be solved analytically for specific 
values of $l$ using the formula for the inhomogeneous solution 
of a fourth order differential equation found in 4.1.1 and 2.1.1 of 
\cite{odebook}.  A 
calculation for $l=2$ shows that $\bmtwa$ will diverge logarithmically 
as $r\to 2 M$ unless $c_{3}=-\frac{1}{2625}$, which is also the value 
in Table~\ref{tab:rotrlmc}.  This is similar to the effect of $\fdz$ 
on $\bmtwa$, as discussed following \eqref{eq:nfd0terms}.  Analytic 
solutions of this sort are constructed out of homogeneous solutions 
of \eqref{eq:d4eqnr}, which in turn are related to hypergeometric 
functions through the generalized Regge-Wheeler homogeneous solutions.  
For this reason, the series for $\nfdtw^{\text{out}}$ converges efficiently, 
provided $\frac{1}{x}\lesssim \frac{1}{2}$.

Regarding the number of constants, the four homogeneous 
solutions which are used in \eqref{eq:fd2outh} and \eqref{eq:fd2inh} 
lead to four constants of integration for the system 
\eqref{eq:ddf2eqn}-\eqref{eq:ddfd2eqn}.  Lengthy calculations show 
that two of the constants, $c_{2}$ and $c_{4}$, affect $\ftw$ and $\fdtw$, 
but not $\bmtwa$.  This is similar to the corresponding result for the 
system \eqref{eq:ddf0eqn}-\eqref{eq:ddfd0eqn}.  

The work in this subsection completes the 
zero frequency solutions for $l\ge 2$.  The results 
for $\fz$, $\fdz$, $\ftw$ and $\fdtw$ are substituted into the 
expressions from subsection~\ref{sec:zevparge}.  

\subsection{\label{sec:zevpareqo}Solutions for $l = 1$}

Most of the zero frequency solutions for $l=1$ have to be rederived, rather 
than using the solutions for $l\ge 2$.  
The latter generally do not reduce to the $l=1$ case in the way the 
non-zero frequency solutions do, as described in 
subsection~\ref{sec:nzevpareq}.  However, we can use the solutions 
for $\hz$ and $\bho$ in \eqref{eq:h0sol0} and \eqref{eq:bh1sol0}, 
with the substitution $\lambda\to 0$.  They solve the relevant 
field equations, \eqref{eq:ddbh1eqn} and \eqref{eq:ddh0eqn}, and the 
two first order equations, \eqref{eq:ndhzeqns0} and \eqref{eq:zdivheqn0}.  
The definition of $\po$ \eqref{eq:po0} and the associated differential 
equation \eqref{eq:po0eqn} also still apply.  

The remaining solutions are derived below.  
As before, the function $\bg$ is not present for $l=1$, so we need to find 
only $\bhz$, $\bhtw$, $\ho$ and $\bk$.  To do so, we start by deriving 
four first order differential equations for their derivatives.  After 
setting $\lambda=0$ in \eqref{eq:ndhgeqns0}, we rewrite that equation as
\begin{multline}
\label{eq:ndKeqns1}
\left(1-\frac{2 M}{r}\right) \dbk+\frac{3 (-2 M+r) }{r^2}\bhz
+\frac{2 (-2 M+r) }{r^3}\ho
\\+\frac{(2 M-r) }{r^2}\bhtw
=\frac{8 \pi  (-2 M+r)^2 }{M}Se_{11}+\frac{16\pi(-2 M+r)^2 }{M r}Se_{12}\;.
\end{multline}
Using \eqref{eq:ndKeqns1}, we eliminate $\dbk$ from \eqref{eq:ndh0eqns0} 
and \eqref{eq:ndh2eqns0} to obtain
\begin{multline}
\label{eq:ndH0eqns1}
\left(1-\frac{2 M}{r}\right) \dbhz+\frac{(-3 M+2 r) }{r^2}\bhz
+\frac{M }{r^2}\bhtw
\\=\frac{8\pi(-2 M+r)^2 }{M}Se_{11}
-16\pi\left(3-\frac{2 M}{r}-\frac{r}{M}\right)Se_{12}\;,
\end{multline}
\begin{multline}
\label{eq:ndH2eqns1}
\left(1-\frac{2 M}{r}\right)\dbhtw+\frac{(-7 M+4 r) }{r^2}\bhz
+\frac{(-3 M+2 r) }{r^2}\bhtw+\frac{(8 M-4 r) }{r^2}\bk
\\=\frac{8 \pi  (-2 M+r)^2  }{M}Se_{11}-16 \pi  \left(5-\frac{6 M}{r}
-\frac{r}{M}\right) Se_{12}\;.
\end{multline}
From the harmonic gauge condition \eqref{eq:divheqn2}, we have
\begin{equation}
\label{eq:ndh1eqns1}
\left(1-\frac{2 M}{r}\right) \dho+\frac{\bhz}{2}
-\frac{2 (M-r) }{r^2}\ho-\frac{\bhtw}{2}=0\;.
\end{equation}
These four equations, together with the two from \eqref{eq:ndhzeqns0} 
and \eqref{eq:zdivheqn0}, form a system of six first order differential 
equations for the six radial metric perturbation functions.  

We still can use the definition of $\pz$ from \eqref{eq:spin0func}
\begin{equation}
\label{eq:spin0func1}
\pz=r(-\bhz+\bhtw+2 \bk)\;,
\end{equation}
as well as its differential equation \eqref{eq:pzeqn} and its 
radial derivative
\begin{equation}
\label{eq:dspin0func1}
\dpz=-\bhz+\bhtw+2 \bk+r(-\dbhz+\dbhtw+2 \dbk)\;.
\end{equation}
We then manipulate the four equations 
\eqref{eq:ndKeqns1}-\eqref{eq:ndh1eqns1}, 
as well as \eqref{eq:spin0func1} and \eqref{eq:dspin0func1}, 
to express $\bhtw$, $\bhz$ and $\ho$ in terms of $\bk$ and $\pz$.  Doing so 
gives
\begin{equation}
\label{eq:trybh21}
\bhtw= \frac{1}{4} \left(\frac{3 }{r}\pz
-32 \pi  r  Se_{12}+2 r \dbk-\dpz\right)\;,
\end{equation}
\begin{equation}
\label{eq:trybh01}
\bhz= \frac{1}{4} \left(8 \bk-\frac{\pz}{r}
-32 \pi  r  Se_{12}+2 r \dbk-\dpz\right)\;,
\end{equation}
\begin{equation}
\label{eq:tryh01}
\ho=-3 r \bk+\frac{3}{4}\pz+\frac{4 \pi  r^{3} (-2 M+r)}{M}Se_{11}
-\frac{8\pi(M-r)r^{2}}{M}Se_{12}-r^{2}\dbk+\frac{r}{4}\dpz\;.
\end{equation}
Applying these results, we obtain an equation for $K$ from the field 
equation \eqref{eq:ddkeqn}:
\begin{multline}
\label{eq:ddkeqn1}
\left(1-\frac{2 M}{r}\right)^2 \ddbk
+\frac{\left(24 M^2-26 M r+7 r^2\right) }{r^3}\dbk
+\frac{4 (3 M-2 r) (2 M-r) }{r^4}\bk
\\=\frac{\left(4 M^2-8 M r+3 r^2\right) }{2 r^5}\pz
-8 \pi   Se_{00}-\frac{8\pi  (3 M-2 r) (-2 M+r)^2  }{M r^2}Se_{11}
\\-\frac{16 \pi  (M-2 r) (-2 M+r)^2}{M r^3}Se_{12}
+\frac{3 (-2 M+r)^2 }{2 r^4}\dpz\;.
\end{multline}
In this manner, we have reduced the problem to solving a single inhomogeneous 
second order differential equation, which we solve with the methods that 
we have previously used to solve similar equations.  The solution for 
$\bk$ and its derivative are in Appendix \ref{appc}, and they may be 
substituted into \eqref{eq:trybh21}-\eqref{eq:tryh01} to find the solutions 
for $\bhtw$, $\bhz$ and $\ho$.  Because of their complexity, these 
three additional solutions will not be written out in this thesis.  
For reasons given at the end of section~\ref{sec:ftmunu}, this particular mode 
(even parity, with $\omega=0$ and $l=1$) is not important for bound orbits, 
so it is not necessary to display $\bhtw$, $\bhz$ and $\ho$ in 
full.  

The solutions for $\bk$ refer to a second spin~$0$ generalized 
Regge-Wheeler function, $\pza$.  Its definition is
\begin{equation}
\pza=A(r)\left(-\bhz+\bhtw+2\bk\right)
+B(r)\left(-\dbhz+\dbhtw+2\dbk\right)+C(r)\bk+D(r)\dbk\;,
\end{equation}
where
\begin{multline}
A(r)=\frac{r}{960 M^6}\bigg\{\left(2 M^2-3 M r+r^2\right) \lnff
\bigg(3 \left(4 M^3+6 M^2 r-3 M r^2+9 r^3\right)
\\-4 M^2 r \lnr\bigg)-2 M \bigg(-8 M^4+46 M^3 r-88 M^2 r^2+93 M r^3
-27 r^4+4 M^2\\\quad\,\,\,\,\,\times \left(2 M^2-2 M r+r^2\right) \lnr\bigg)
-16 M^2 r \!\left(2 M^2-3 M r+r^2\right) \plog\bigg\}\;,
\end{multline}
\begin{multline}
B(r)=-\frac{(2 M-r) r^2}{1920 M^6}\bigg\{2 M \bigg(42 M^3-56 M^2 r
-21 M r^2+27 r^3-8 M^2 (M-r) \lnr\bigg)
\\+(M-r) \lnff\bigg(-3 r \left(-20 M^2+3 M r+9 r^2\right)
+8 M^2\\\times(M-r)\lnr\bigg)+32 M^2 (M-r)^2 \plog\bigg\}\;,
\end{multline}
\begin{equation}
C(r)=-\frac{r^3 \left(2 M \left(6 M^2-10 M r+3 r^2\right)
+3 r \left(2 M^2-3 M r+r^2\right)\lnff\right)}{32 M^6}\;,
\end{equation}
\begin{equation}
D(r)=-\frac{(2 M-r) r^4 \left(2 M (2 M-3 r)
+3 (M-r) r \lnff\right)}{64 M^6}\;.
\end{equation}
The differential equation for $\pza$ is
\begin{multline}
\mathcal{L}_{0}\pza=\frac{\pi  r}{60 M^6}A_{00}(r)Se_{00}
-\frac{\pi(-2 M+r)^2}{60 M^7 r}A_{11}(r)Se_{11}
+\frac{\pi  (-2 M+r)^2 }{2 M^7}A_{12}(r)Se_{12}
\\+\frac{\pi  (2 M-r)}{30 M^7 r^2} A_{22}(r) Ue_{22}
+\frac{\pi  (2 M-r) r^2}{120 M^6}A_{d00}(r)Se_{00}^{\prime}
\\-\frac{\pi  (2 M-r)^3}{120 M^7} A_{d11}(r) Se_{11}^{\prime}
+\frac{\pi  (-2 M+r)^2}{60 M^6 r} A_{d22}(r) Ue_{22}^{\prime}\;,
\end{multline}
where
\begin{equation}
\begin{split}
A_{00}(r)=&2 M \bigg(176 M^4-458 M^3 r+259 M^2 r^2-180 M r^3+81 r^4
-4 M^2 \left(8 M^2-18 M r\right.\\&\left.+7 r^2\right)\lnr\bigg)
+(2 M-r)\lnff \bigg(3 \left(4 M^4+42 M^3 r-75 M^2 r^2
\right.\\&\left.+33 M r^3-27 r^4\right)
+4 M^2 \left(4 M^2-11 M r+7 r^2\right)\lnr\bigg)
\\&+16 M^2 \left(8 M^3-26 M^2 r+25 M r^2-7 r^3\right) \plog\;,
\end{split}
\end{equation}
\begin{equation}
\begin{split}
A_{11}(r)=&-2 M \bigg(\!\!-92 M^5\!+346 M^4 r\!+239 M^3 r^2
\!-1656 M^2 r^3\!+1599 M r^4\!-450 r^5\!+4 M^3
\\& \times\!\left(4 M^2-14 M r\!+7 r^2\right)\! \lnr\bigg)\!
+\!\lnff\! \bigg(\!3 \left(8 M^6\!+40 M^5 r\!-146 M^4 r^2
\right.\\&\left.-177 M^3 r^3+735 M^2 r^4-603 M r^5+150 r^6\right)
+4 M^3 \left(4 M^3-18 M^2 r+21 M r^2\right.\\&\left.-7 r^3\right) \lnr\bigg)
+16 M^3 \left(4 M^3-18 M^2 r+21 M r^2-7 r^3\right) \plog\;,
\end{split}
\end{equation}
\begin{multline}
A_{12}(r)=2 M \left(4 M^3-28 M^2 r+49 M r^2-21 r^3\right)
\\+3 r \left(3 M^3-15 M^2 r+20 M r^2-7 r^3\right)\lnff\;,
\end{multline}
\begin{equation}
\begin{split}
A_{22}(r)=&2 M \bigg(50 M^5-248 M^4 r+116 M^3 r^2+63 M^2 r^3+48 M r^4
-45 r^5-4 M^3 \left(2 M^2\right.\\&\left.-10 M r+5 r^2\right) \lnr\bigg)
+\lnff \bigg(3 \left(8 M^6+20 M^5 r-103 M^4 r^2
\right.\\&\left.+46 M^3 r^3+15 M^2 r^4+21 M r^5-15 r^6\right)+4 M^3 
\left(2 M^3-12 M^2 r+15 M r^2\right.\\&\left.-5 r^3\right) \lnr\bigg)
+16 M^3 \left(2 M^3-12 M^2 r+15 M r^2-5 r^3\right) \plog\;,
\end{split}
\end{equation}
\begin{multline}
A_{d00}(r)=-2 M \left(-42 M^3+56 M^2 r-9 M r^2+18 r^3+8 M^2 (M-r)
\lnr\right)\\+(M-r)\lnff \bigg(3 r \left(20 M^2-3 M r+6 r^2\right)
+8 M^2\\\times (M-r) \lnr\bigg)+32 M^2 (M-r)^2 \plog\;,
\end{multline}
\begin{multline}
A_{d11}(r)=-2 M \bigg(\!-42 M^4+56 M^3 r+111 M^2 r^2-222 M r^3
+90 r^4+8 M^3 (M-r) \lnr\!\bigg)\\+(M-r)\lnff 
\bigg(3 r \left(20 M^3-3 M^2 r-54 M r^2+30 r^3\right)
\\+8 M^3 (M-r) \lnr\bigg)+32 M^3 (M-r)^2 \plog\;,
\end{multline}
\begin{multline}
A_{d22}(r)=2 M \left(42 M^3-56 M^2 r-21 M r^2+27 r^3-8 M^2 (M-r) \lnr\right)
\\+(M-r)\lnff \bigg(-3 r \left(-20 M^2+3 M r+9 r^2\right)
+8 M^2\\\times  (M-r)\lnr\bigg)+32 M^2 (M-r)^2 \plog\;.
\end{multline}
Like the non-zero frequency case for $l=1$, the zero frequency solution set 
depends on three generalized Regge-Wheeler functions, two with $s=0$ and 
one with $s=1$.

Gauge changes which preserve the harmonic gauge are given by
\begin{equation}
\label{eq:zm01}
\tbmz^{h}=\frac{(2 M-r) }{2 r}\dpo\;,
\end{equation}

\begin{equation}
\begin{split}
\label{eq:zm11}
\tbmo^{h}=&-\frac{2 \left(2 M \left(2 M^2-9 M r+6 r^2\right)
+3 r \left(2 M^2-5 M r+2 r^2\right)\lnff\right) }{r^2}\pza
\\&+\frac{1}{60 M r^2} \bigg\{M \left(4 M^2-17 M r-2 r^2\right)
\!-\!4 \bigg(\!M \left(M^2-6 M r+4 r^2\right)\!+r \left(2 M^2
\!-\!5 M r\right.\\&\left.+2 r^2\right)\lnff\bigg) \lnr
-4 r \left(2 M^2-5 M r+2 r^2\right) \plog\bigg\}\pz
\\&+\frac{1}{60 M r}\bigg\{M r (11 M+4 r)+4 \bigg(M \left(M^2-4 M r
+2 r^2\right)+r \left(2 M^2-3 M r+r^2\right)
\\&\lnff\bigg) \lnr+4 r \left(2 M^2-3 M r+r^2\right) \plog\bigg\} \dpz
\\&+\frac{2 \left(2 M \left(2 M^2-6 M r+3 r^2\right)+3 r \left(2 M^2
-3 M r+r^2\right)\lnff\right)}{r}\dpza\;,
\end{split}
\end{equation}

\begin{equation}
\begin{split}
\label{eq:zm21}
\tbmtw^{h}=&\frac{-4 M \left(7 M^2-15 M r+6 r^2\right)
+6 \left(2 M^3-7 M^2 r+7 M r^2-2 r^3\right)\lnff}{r}\pza
\\&+\frac{1}{120 M r} \bigg\{M \left(74 M^2-39 M r
-10 r^2\right)+8 \bigg(M \left(-5 M^2+10 M r-4 r^2\right)
\\&+\left(2 M^3-7 M^2 r+7 M r^2-2 r^3\right)\lnff\bigg) \lnr
+8 \left(2 M^3-7 M^2 r\right.\\&\left.+7 M r^2-2 r^3\right) \plog\bigg\}\pz
-\frac{(2 M-r) }{120 M}\bigg\{M (21 M+2 r)+8(M-r)\\&\times\bigg(\!-2 M+(M-r)
\lnff\bigg)\lnr\!+\!8 (M-r)^2\plog\bigg\}\dpz
\\&-6 \left(2 M^2-3 M r+r^2\right) \left(-2 M+(M-r)\lnff\right)\dpza\;.
\end{split}
\end{equation}
For this mode, gauge changes which preserve the harmonic gauge are made 
by adding homogeneous solutions of the generalized Regge-Wheeler equation 
for $s=0$ and $s=1$, just as is done for the corresponding non-zero 
frequency $l=1$ case.  The expressions for $\tbmo^{h}$ and $\tbmtw^{h}$ 
are complicated, but can be simplified somewhat by substituting 
explicit homogeneous spin~$0$ solutions into 
\eqref{eq:zm11}-\eqref{eq:zm21}.  

The $\widetilde{M}_{i}^{h}$ are solutions to the three differential 
equations \eqref{eq:ddm0eqn}-\eqref{eq:ddm2eqn}, which define a gauge 
change which preserves the harmonic gauge.  If we substitute $\omega=0$ 
and $\lambda=0$ into \eqref{eq:ddm0eqn}-\eqref{eq:ddm2eqn}, 
the resulting system is simple enough that we could 
have solved the equations directly, without writing the solutions 
in terms of generalized Regge-Wheeler functions.  This approach 
was taken by Ori \cite{ori04}, who derived expressions for the 
zero frequency, $l=1$ gauge change vectors and who did so without 
using the Regge-Wheeler formalism.  We can obtain Ori's published 
results by substituting explicit homogeneous 
spin~$0$ and spin~$1$ solutions into \eqref{eq:zm01}-\eqref{eq:zm21} 
and taking appropriate linear combinations of them.

\subsection{\label{sec:zevpareqz}Solutions for $l = 0$}

The zero frequency solutions for $l=0$ are written in terms of 
two $s=0$ generalized Regge-Wheeler functions, which have different 
source terms and participate in the metric perturbations in linearly 
independent ways.  There are only four radial perturbation functions, 
$\bho$, $\bhz$, $\bhtw$ and $\bk$.  In the perturbation \eqref{eq:ehmunu}, 
the spherical harmonic angular functions are constant, so this mode is 
spherically symmetric \cite{zerp70}.  For $l=0$, $\lambda=-1$.  As before, 
zero frequency solutions are time independent.

We begin by describing the rules for a change of gauge.  These 
rules are based in large part on the discussions in the appendices of 
\cite{dp04} and \cite{zerp70}, although these 
references use somewhat different notation.  Among other things, 
this thesis uses a covariant gauge change vector $\xi_{\mu}$, 
but the contravariant form $\xi^{\mu}$ is used in \cite{dp04}.  
Also, reference \cite{dp04} covers only circular orbits, for which 
the sole $l=0$ mode happens to be zero frequency.  The discussion 
below is more general.  To understand the gauge 
change rules, it is necessary use some time domain 
expressions, even though we will derive the solutions using 
Fourier transforms.  For zero frequency and $l=0$, it is possible 
to have a time dependent change of gauge, although the metric 
perturbation remains time independent.  Specifically, we may have a gauge 
change which is linear in time for this mode \cite{dp04}.

In the time domain, the gauge vector for $l=0$ is \cite{zerp70}
\begin{equation}
\label{eq:xi00}
\xi^{e,00}_{\mu}(t,r,\theta,\phi)
=\begin{pmatrix}\xi_{0}(t,r)Y_{00}(\theta,\phi),
\xi_{1}(t,r)Y_{00}(\theta,\phi),0,0\end{pmatrix}\;.
\end{equation}
The superscript ``$e$'' is short for ``even parity'', the notation ``$00$'' 
is ``$l=0$, $m=0$'', and $Y_{00}(\theta,\phi)=\frac{1}{\sqrt{4\pi}}$.  
Also in the time domain, the perturbation functions transform as
\begin{equation}
\label{eq:H0new00}
\bhz^{\text{new}}(t,r)=\bhz^{\text{old}}(t,r)+\frac{2 M }{r^2}\xi_{1}(t,r)
+\frac{2 r }{2 M-r}\frac{\partial\xi_{0}(t,r)}{\partial t}\;,
\end{equation}
\begin{equation}
\label{eq:H1new00}
\bho^{\text{new}}(t,r)=\bho^{\text{old}}(t,r)-\frac{2 M }{2 M r-r^2}\xi_{0}(t,r)
-\frac{\partial\xi_{0}(t,r)}{\partial r}
-\frac{\partial\xi_{1}(t,r)}{\partial t}\;,
\end{equation}
\begin{equation}
\label{eq:H2new00}
\bhtw^{\text{new}}(t,r)=\bhtw^{\text{old}}(t,r)-\frac{2 M }{r^2}\xi_{1}(t,r)
+\frac{(4 M-2 r) }{r}\frac{\partial\xi_{1}(t,r)}{\partial r}\;,
\end{equation}
\begin{equation}
\label{eq:Knew00}
\bk^{\text{new}}(t,r)=\bk^{\text{old}}(t,r)+\frac{2 (2 M-r) }{r^2}\xi_{1}(t,r)\;,
\end{equation}
which are time domain versions of \eqref{eq:bh0new}-\eqref{eq:bknew} 
for $l=0$ \cite{zerp70}.  From \eqref{eq:divchi}, the time domain 
equations for gauge changes which preserve the harmonic gauge are
\begin{multline}
\label{eq:tddm0eqn}
\frac{(-2 M+r)^2 }{r^2}\frac{\partial^{2}\xi_{0}(t,r)}{\partial r^{2}}
+\frac{2 (-2 M+r)^2 }{r^3}\frac{\partial\xi_{0}(t,r)}{\partial r}
\\-\frac{\partial^{2}\xi_{0}(t,r)}{\partial t^{2}}
+\frac{2 M (-2 M+r) }{r^3}\frac{\partial\xi_{1}(t,r)}{\partial t}=0\;,
\end{multline}
\begin{multline}
\label{eq:tddm1eqn}
\frac{(-2 M+r)^2 }{r^2}\frac{\partial^{2}\xi_{1}(t,r)}{\partial r^{2}}
+\frac{2 (-2 M+r) }{r^2}\frac{\partial\xi_{1}(t,r)}{\partial r}
-\frac{\partial^{2}\xi_{1}(t,r)}{\partial t^{2}}
\\-\frac{2 (-2 M+r)^2 }{r^4}\xi_{1}(t,r)
-\frac{2 M }{2 M r-r^2}\frac{\partial\xi_{0}(t,r)}{\partial t}=0\;,
\end{multline}
again assuming $l=0$.  The Fourier transforms 
of these time domain expressions may have both zero and non-zero 
frequency modes, so we need to separate out the zero frequency 
modes.  This is done below.

The metric perturbations have a time dependence of $e^{-\iom t}$.  This is 
because the stress energy tensor is decomposed that way, 
and the perturbations are related to the stress energy tensor 
through the field equations.  However, the gauge transformation 
vectors are not so restricted and may have an additional, 
different time dependence, as discussed in \cite{dp04} 
and \cite{zerp70}.  For $l=0$, we may set
\begin{equation}
\label{eq:m00}
\xi_{0}(t,r)=A_{0}(t,r)+\bmz(t,r)
=A_{0}(t,r)+\int_{\!-\infty}^{\infty} e^{-i\omega t} \bmz(\omega,r) d\omega\;,
\end{equation}
\begin{equation}
\label{eq:m10}
\xi_{1}(t,r)=A_{1}(t,r)+\bmo(t,r)
=A_{1}(t,r)+\int_{\!-\infty}^{\infty} e^{-i\omega t} \bmo(\omega,r) d\omega\;,
\end{equation}
where the quantities $A_{0}(t,r)$ and $A_{1}(t,r)$ have some time 
dependence other than $e^{-\iom t}$.  In the integrals, the zero 
frequency modes are $\bmz(\omega=0,r)$ and $\bmo(\omega=0,r)$.  
Equation \eqref{eq:Knew00} requires that $A_{1}(t,r)=0$, so 
that $\bk(t,r)$ will have a time dependence only of $e^{-\iom t}$.  
However, $A_{0}(t,r)$ need not be zero.  From \eqref{eq:H0new00}, 
$A_{0}(t,r)$ may be linear in time, without affecting the time 
dependence of $\bhz^{\text{new}}(t,r)$.  From \eqref{eq:H1new00}, $A_{0}(t,r)$ 
must be of the form
\begin{equation}
\label{eq:a0rt}
A_{0}(t,r)=C_{0}\ff t\;,
\end{equation}
where $C_{0}$ is a constant; otherwise, $\bho^{\text{new}}(t,r)$ would grow 
linearly with time.  An equivalent expression, but for a contravariant gauge 
vector $\xi^{\mu}$, is given in \cite{dp04}.  

Summarizing, the time domain gauge vector components for $l=0$ are
\begin{equation}
\label{eq:nm00}
\xi_{0}(t,r)=C_{0}\ff t
+\int_{\!-\infty}^{\infty} e^{-i\omega t} \bmz(\omega,r) d\omega\;,
\end{equation}
\begin{equation}
\label{eq:nm10}
\xi_{1}(t,r)=\int_{\!-\infty}^{\infty} e^{-i\omega t} \bmo(\omega,r) d\omega\;.
\end{equation}
A gauge transformation like this could not have been used 
for $l\ge 1$, because the first term of \eqref{eq:nm00} would 
cause the metric perturbation to grow linearly with 
time \eqref{eq:evhznew}.  The effect of this term is 
to rescale the coordinate time by a constant \cite{dp04}.  
From \eqref{eq:xnew}, we have $x^{\mu}_{\text{new}}=x^{\mu}_{\text{old}}+\xi^{\mu}$, so
\begin{equation}
t_{\text{new}}=t_{\text{old}}+\xi^{t}=t_{\text{old}}+g^{tt}\xi_{t}
=t_{\text{old}}-C_{0}t_{\text{old}}Y_{00}(\theta,\phi)
=\left(1-\frac{C_{0}}{\sqrt{4\pi}}\right)t_{\text{old}}\;.
\end{equation}

Substituting \eqref{eq:nm00} and \eqref{eq:nm10} into 
\eqref{eq:H0new00}-\eqref{eq:Knew00} and specializing to zero frequency, 
we obtain the gauge transformation rules for $l=0$, $\omega=0$:
\begin{equation}
\label{eq:nH0new00}
\bhz^{\text{new}}=\bhz^{\text{old}}+\frac{2 M }{r^2}\bmo-2 C_{0}\;,
\end{equation}
\begin{equation}
\label{eq:nH1new00}
\bho^{\text{new}}=\bho^{\text{old}}-\frac{2 M }{2 M r-r^2}\bmz
-\dbmz\;,
\end{equation}
\begin{equation}
\label{eq:nH2new00}
\bhtw^{\text{new}}=\bhtw^{\text{old}}-\frac{2 M }{r^2}\bmo
+\frac{(4 M-2 r) }{r}\dbmo\;,
\end{equation}
\begin{equation}
\label{eq:nKnew00}
\bk^{\text{new}}=\bk^{\text{old}}+\frac{2 (2 M-r) }{r^2}\bmo\;.
\end{equation}
Equivalent rules for circular orbits are in \cite{dp04}.  
Equations \eqref{eq:nH0new00}-\eqref{eq:nKnew00} are not time 
domain expressions.  Here, the metric perturbation functions 
are time independent, zero frequency modes of Fourier transforms, 
so that, for example, $\bhz^{\text{new}}=\bhz^{\text{new}}(\omega=0,r)$.  
Similarly, $\bmz=\bmz(\omega=0,r)$ and $\bmo=\bmo(\omega=0,r)$, which 
are from the Fourier integrals in equations \eqref{eq:nm00}-\eqref{eq:nm10}.  
The non-zero frequency mode contributions to the integrals are relevant only 
to gauge changes described in subsection~\ref{sec:nzevpareq}, in 
the discussion regarding non-zero frequency solutions for $l=0$.  

Equations \eqref{eq:nH0new00}-\eqref{eq:nKnew00} are general and are 
not limited to gauge changes which preserve the harmonic gauge. 
We still need to find expressions for $\bmz$ and $\bmo$ so that the 
harmonic gauge can be preserved.  To do so, substitute 
\eqref{eq:nm00}-\eqref{eq:nm10} into 
\eqref{eq:tddm0eqn}-\eqref{eq:tddm1eqn} and set $\omega=0$, which gives
\begin{equation}
\label{eq:ntddm0eqn}
\ddbmz+\frac{2}{r}\dbmz=0\;,
\end{equation}
\begin{equation}
\label{eq:ntddm1eqn}
\left(1-\frac{2 M}{r}\right)^2 \ddbmo
+\frac{2 (-2 M+r) }{r^2}\dbmo
-\frac{2 (-2 M+r)^2 }{r^4}\bmo=-\frac{2 M}{r^{2}}C_{0}\;.
\end{equation}
The gauge changes described by \eqref{eq:nH0new00}-\eqref{eq:nKnew00} 
will preserve the harmonic gauge if and only if $\bmz$ and $\bmo$ solve 
these two equations.  The solution to \eqref{eq:ntddm0eqn} is
\begin{equation}
\label{eq:m000sol}
\bmz=C_{01}-\frac{C_{02}}{r}\;,
\end{equation}
where $C_{01}$ and $C_{02}$ are constants.  The gauge function $\bmz$ affects 
only $\bho$.  From \eqref{eq:nH1new00},
\begin{equation}
\label{eq:cnH1new00}
\bho^{\text{new}}=\bho^{\text{old}}-\frac{2 M C_{01}}{2 M r-r^2}
+\frac{C_{02}}{2 M r-r^2}\;.
\end{equation}
Although \eqref{eq:ntddm0eqn} has two solutions, 
equation \eqref{eq:cnH1new00} shows they change 
$\bho$ in the same way, apart from a multiplicative constant 
of $-2M$.  In effect, there is only one undetermined constant for 
a harmonic gauge change to $\bho$.  The solution to \eqref{eq:ntddm1eqn} is
\begin{multline}
\label{eq:m100sol}
\bmo=\frac{C_{11}}{2 M r-r^2}+\frac{C_{12} r^2}{6 M-3 r}
\\+\frac{C_{0} \left(\left(-8 M^3+r^3\right) 
\lnff+M\left(-r (4 M+r)+8 M^2 \lnr\right)\right)}{3 (2 M-r) r}\;,
\end{multline}
which has three undetermined constants and which affects the metric 
perturbations through \eqref{eq:nH0new00} and 
\eqref{eq:nH2new00}-\eqref{eq:nKnew00}.  
The constant $C_{0}$ in $\bmo$ 
is the same as in the first term of $\xi_{0}$ \eqref{eq:nm00}.  Accordingly, 
an example of a gauge change which satisfies 
\eqref{eq:tddm0eqn}-\eqref{eq:tddm1eqn} 
and thereby preserves the harmonic gauge is
\begin{equation}
\label{eq:tm1gauge}
\begin{split}
\xi_{0}(t,r)=&\;C_{0}\ff t\;,
\\\xi_{1}(t,r)=&\;C_{0}\frac{ \left(\left(-8 M^3+r^3\right) 
\lnff+M\left(-r (4 M+r)+8 M^2 \lnr\right)\right)}{3 (2 M-r) r}\;,
\end{split}
\end{equation}
where $\xi_{0}$ and $\xi_{1}$ are components of the gauge 
vector $\xi^{e,00}_{\mu}$ \eqref{eq:xi00}.  Although \eqref{eq:tm1gauge} has 
time domain expressions, this particular gauge change will affect only 
the zero frequency mode for $l=0$.  

Detweiler and Poisson studied the $l=0$ multipole for circular orbits, 
including relevant gauge transformation rules \cite{dp04}.  They 
derived expressions equivalent to \eqref{eq:ntddm1eqn}, 
\eqref{eq:m100sol} and \eqref{eq:tm1gauge}, but in terms of a contravariant 
gauge vector $\xi^{\mu}$.

The equations for $\bho$ and $Se_{01}$ decouple from the others, so 
we will solve them first.  
Setting $\omega=0$ and $\lambda=-1$ in equations \eqref{eq:ddbh1eqn}, 
\eqref{eq:divheqn0} and \eqref{eq:divt0} gives, in order,
\begin{multline}
\label{eq:ddbh1eqn00}
\frac{(-2 M+r)^2 }{r^2}\ddbho
+\frac{2 \left(2 M^2-3 M r+r^2\right)}{r^3}\dbho
\\-\frac{2 \left(2 M^2-2 M r+r^2\right) }{r^4}\bho
=-\frac{16 \pi (-2 M+r) }{r}Se_{01}\;,
\end{multline}
\begin{equation}
\label{eq:divheqn000}
\left(1-\frac{2 M}{r}\right) \dbho
-\frac{2 (M-r) }{r^2}\bho=0\;,
\end{equation}
\begin{equation}
\label{eq:divt000}
\left(1-\frac{2 M}{r}\right) Se_{01}^{\prime}
-\frac{2 (M-r)}{r^2} Se_{01}=0\;.
\end{equation}
Equation \eqref{eq:divt000} has the solution
\begin{equation}
\label{eq:se01sol}
Se_{01}=\frac{C_{S}}{r(r-2 M)}\;,
\end{equation}
where $C_{S}$ is a constant.  Using \eqref{eq:divheqn000} and its derivative, 
we eliminate $\ddbho$ and then $\dbho$ from \eqref{eq:ddbh1eqn00}, 
which leads to 
\begin{equation}
\label{eq:nse0sol}
0=-\frac{16 \pi (-2 M+r) }{r}Se_{01}\;.
\end{equation}
It follows that $C_{S}=0$, so $Se_{01}=0$ for this mode.  Accordingly, $\bho$ 
is a homogeneous solution of \eqref{eq:ddbh1eqn00}, subject to 
\eqref{eq:divheqn000}.  If we use the derivative of \eqref{eq:divheqn000} to 
eliminate $\ddbho$ from the homogeneous form of \eqref{eq:ddbh1eqn00}, 
we get \eqref{eq:divheqn000} again, so any solution of \eqref{eq:divheqn000} 
is a homogeneous solution of \eqref{eq:ddbh1eqn00}.  
The only solution of \eqref{eq:divheqn000} is
\begin{equation}
\label{eq:bh1sol00}
\bho=\frac{C_{H}}{r(r-2 M)}\;,
\end{equation}
where $C_{H}$ is a constant that may be zero.  
This single solution resembles the gauge change rule \eqref{eq:cnH1new00}.  
Equation \eqref{eq:ddbh1eqn00} has a second homogeneous solution given by 
\begin{equation}
\frac{(3 M-r) r}{2 M- r}\;,
\end{equation}
but this is irrelevant because it 
is not also a solution of \eqref{eq:divheqn000}.  

We always can eliminate the solution \eqref{eq:bh1sol00} by a change of 
gauge which preserves the harmonic gauge.  Suppose $C_{H}\ne 0$.  
If we set $C_{01}=0$ and $C_{02}=C_{H}$ in $\bmz$ \eqref{eq:m000sol} and 
substitute into $\bho^{\text{new}}$ \eqref{eq:cnH1new00}, we find that
\begin{equation}
\bho^{\text{new}}=\bho^{\text{old}}+\frac{C_{02}}{2 M r-r^2}
=\frac{C_{H}}{r(r-2 M)}+\frac{C_{H}}{2 M r-r^2}=0\;.
\end{equation}
A null result also follows from $C_{01}=-C_{H}/2M$ and $C_{02}=0$.  
Because $\bho$ is entirely gauge dependent, we may choose 
$C_{H}=0$ and we have 
\begin{equation}
\label{eq:fbh1sol00}
\bho=0\;,\;\dbho=0\;.
\end{equation}
Moreover, we should set $\bho=0$.  The metric perturbation 
should depend on the motion of the smaller orbiting mass, but 
the field equation for $\bho$ \eqref{eq:ddbh1eqn00} does not have a 
non-zero source, because $Se_{01}=0$ for this mode.  

It is more complicated to solve for $\bhz$, $\bhtw$ and $\bk$.  Based 
on previous work, the remaining three field equations for 
$l=0$, $\omega=0$ are
\begin{multline}
\label{eq:ddbh0eqn00}
\frac{(-2 M+r)^2 }{r^2}\ddbhz
+\frac{2 (M-r) (2 M-r)}{r^3}\dbhz
-\frac{2 M^2 }{r^4}\bhz\\+\frac{2 M(3 M-2 r)}{r^4}\bhtw
+\frac{4 M (-2 M+r) }{r^4}\bk
=-8 \pi Se_{00}\\-\frac{8\pi (-2 M+r)^2 }{r^2}Se_{11}
-\frac{16 \pi (-2 M+r) }{r^3}Ue_{22}\;,
\end{multline}
\begin{multline}
\label{eq:ddbh2eqn00}
\frac{(-2 M+r)^2 }{r^2}\ddbhtw
+\frac{2 (M-r) (2 M-r)}{r^3}\dbhtw+\frac{2 M (3 M-2 r) }{r^4}\bhz
\\-\frac{2 \left(9 M^2-8 M r+2 r^2\right) }{r^4}\bhtw
+\frac{4 (2 M-r) (3 M-r) }{r^4}\bk\\=-8 \pi Se_{00}
-\frac{8 \pi (-2 M+r)^2}{r^2} Se_{11}
-\frac{16 \pi (2 M-r) }{r^3}Ue_{22}\;,
\end{multline}
\begin{multline}
\label{eq:ddbkeqn00}
\frac{(-2 M+r)^2 }{r^2}\ddbk
+\frac{2 (M-r) (2 M-r) }{r^3}\dbk
+\frac{2 M (-2 M+r) }{r^4}\bhz
\\+\frac{2 (2 M-r) (3 M-r) }{r^4}\bhtw
-\frac{2 (2 M-r) (4 M-r) }{r^4}\bk
\\=-8 \pi Se_{00}+\frac{8 \pi (-2 M+r)^2 }{r^2}Se_{11}\;.
\end{multline}
From \eqref{eq:divt1}, the remaining stress energy divergence equation is
\begin{equation}
\label{eq:divt100}
\left(1-\frac{2 M}{r}\right)Se_{11}^{\prime}
+\frac{M }{(-2 M+r)^2}Se_{00}-\frac{(M-2 r) }{r^2}Se_{11}
-\frac{2 }{r^3}Ue_{22}=0\;,
\end{equation}
and, from \eqref{eq:divheqn1}, the remaining harmonic gauge condition is
\begin{equation}
\label{eq:divheqn100}
\frac{\dbhz}{2}+\frac{\dbhtw}{2}-\dbk
-\frac{M }{2 M r-r^2}\bhz+\frac{(3 M-2 r) }{2 M r-r^2}\bhtw
-\frac{2 }{r}\bk=0\;.
\end{equation}
From \eqref{eq:ndh0eqns0} and \eqref{eq:ndh2eqns0}, two additional 
first order equations are
\begin{equation}
\label{eq:ndh0eqns00}
\left(1-\frac{2 M}{r}\right) \dbhz
+\left(-1+\frac{M}{r}\right)\dbk+\frac{\bhtw}{r}-\frac{\bk}{r}
=-8 \pi (-2 M+r) Se_{11}\;,
\end{equation}
\begin{multline}
\label{eq:ndh2eqns00}
\left(1-\frac{2 M}{r}\right) \dbhtw
+\left(-1+\frac{3 M}{r}\right) \dbk+\frac{2 M }{r^2}\bhz
+\frac{3 (-2 M+r) }{r^2}\bhtw\\+\frac{(8 M-3 r) }{r^2}\bk
=-8\pi (2 M-r) Se_{11}\;.
\end{multline}
Equation \eqref{eq:ndh0eqns00} is gauge invariant to linear order.  
There were additional equations for the modes with $l\ge 1$, 
but those equations are not applicable here, because the angular 
functions associated with them are zero for $l=0$.  

These equations are solved in terms of two spin~$0$ generalized 
Regge-Wheeler functions.  As before, one of those functions is
\begin{equation}
\label{eq:spin0func0}
\pz=r(-\bhz+\bhtw+2 \bk)\;,
\end{equation}
whose differential equation is still \eqref{eq:pzeqn} and whose derivative is
\begin{equation}
\label{eq:dspin0func0}
\dpz=-\bhz+\bhtw+2 \bk+r(-\dbhz+\dbhtw+2 \dbk)\;.
\end{equation}
To obtain an expression for $\bhz$, we apply 
\eqref{eq:dspin0func0}, \eqref{eq:ndh0eqns00} and 
\eqref{eq:spin0func0} to eliminate $\dbhtw$, $\dbhz$ and then 
$\bhtw$ from \eqref{eq:divheqn100}, which leads to
\begin{equation}
\label{eq:tempH0}
\bhz=-\frac{(-10 M+3 r)}{2 M-r}\bk-\frac{(4 M-r)}{4 M r-2 r^2}\pz
+8 \pi r^2 Se_{11}-\frac{r (-3 M+r)}{2 M-r}\dbk-\frac{\dpz}{2}\;.
\end{equation}
Using \eqref{eq:spin0func0} and \eqref{eq:tempH0}, we can eliminate 
$\bhtw$ and then $\bhz$ from \eqref{eq:ddbkeqn00}, which gives
\begin{equation}
\label{eq:nddbkeqn00}
\ddbk+\frac{4}{r}\dbk=-\frac{\pz}{r^3}-\frac{8 \pi r^2}{(-2 M+r)^2}Se_{00}
-8 \pi Se_{11}+\frac{\dpz}{r^2}\;.
\end{equation}
We solve this equation for $\bk$ in the same way that we have solved 
similar second order differential equations.  To obtain $\bhz$, we 
substitute $\bk$ and $\dbk$ into \eqref{eq:tempH0}.   Lastly, we 
substitute $\bk$ and $\bhz$ into $\pz$ \eqref{eq:spin0func0} and solve 
for $\bhtw$.  The solutions and their radial derivatives are listed in 
Appendix \ref{appd}.

The solutions refer to a second spin~$0$ generalized Regge-Wheeler 
function, $\pza$.  The differential equation for $\pza$ is
\begin{multline}
\label{eq:pzaeqn00}
\mathcal{L}_{0}\pza=S_{0a}=A_{00}(r)Se_{00}+A_{11}(r)Se_{11}+A_{22}(r)Ue_{22}
\\+A_{d00}(r)Se_{00}^{\prime}+A_{d11}(r)Se_{11}^{\prime}+A_{d22}(r)Ue_{22}^{\prime}\;,
\end{multline}
where
\begin{multline}
A_{00}(r)=\frac{\pi r}{3 M^3 (2 M-r)}\bigg\{-4 M 
\bigg(32 M^2-3 M r+3 r^2\\+12 M (2 M-r) \lnr\bigg)
+3 \lnff \bigg(48 M^3\\-8 M^2 r+8 M r^2-5 r^3+16 M (-2 M+r)^2
\lnr\bigg)\bigg\}\;,
\end{multline}
\begin{multline}
A_{11}(r)=-\frac{\pi (2 M-r)}{3 M^4 r}
\bigg\{-4 M \bigg(-76 M^3+3 M^2 r+3 M r^2+3 r^3
\\+12 M^2 (2 M-r) \lnr\bigg)+3 \lnff \bigg(-112 M^4+88 M^3 r
\\+4 M^2 r^2+5 M r^3-6 r^4+16 M^2 \left(2 M^2-3 M r+r^2\right)
\lnr\bigg)\bigg\}\;,
\end{multline}
\begin{multline}
A_{22}(r)=\frac{2 \pi (2 M-r)}{M^4 r^2}
\bigg\{2 M \left(8 M^2+4 M r+r^2-8 M^2 \lnr\right)+\lnff 
\\\times\left(-16 M^3+4 M^2 r+3 M r^2+2 r^3+8 M^2 (M-r)\lnr\right)
\bigg\}\;,
\end{multline}
\begin{equation}
A_{d00}(r)=\frac{\pi r^2 \left(-4 M^2+3 \lnff \left(8 M^2+r^2+4 M (2 M-r) 
\lnr\right)\right)}{3 M^3}\;,
\end{equation}
\begin{multline}
A_{d11}(r)=-\frac{\pi(-2 M+r)^2}{3 M^4} \bigg\{4 M^3+3 \lnff 
\\\times\bigg(-24 M^3+M r^2+r^3+4 M^2 (2 M-r)\lnr\bigg)\bigg\}\;,
\end{multline}
\begin{equation}
A_{d22}(r)=\frac{\pi (-2 M+r)^2 \lnff \left(-8 M^2-4 M r-r^2
+8 M^2 \lnr\right) }{M^4 r}\;.
\end{equation}
The definition of $\pza$ is
\begin{equation}
\begin{split}
\label{eq:spin0afunc00}
\pza=&\frac{r  \left(8 M^2+3 \lnff \left(-32 M^2
-4 M r-r^2+8 M^2 \lnr\right)\right)}{48 M^3}\bhz
-\frac{r}{48 M^4}\bigg\{8 M^3\\&
+3 \lnff\!\left(\!-48 M^3+4 M^2 r+5 M r^2
+2 r^3+8 M^2 (3 M\!-\!2 r) \lnr\right)\!\!\bigg\}\bhtw
\\&+\frac{r  \left(4 M^3+3 \lnff \left(-40 M^3+2 M r^2+r^3
+8 M^2 (2 M-r) \lnr\right)\right)}{24 M^4}\bk
\\&-\frac{(2 M-r) r^2 \lnff \left(-8 M^2-4 M r-r^2
+8 M^2 \lnr\right) }{16 M^4}\dbhtw
\\&+\frac{r^2 \left(8 M^3+\left(-96 M^3+3 r^3\right)
\lnff\right)}{48 M^4}\dbk\;.
\end{split}
\end{equation}
The derivation of these results resembles the method used 
previously for other modes.

The zero frequency generalized Regge-Wheeler equation for $s=0$, 
$l=0$ has two linearly independent homogeneous solutions
\begin{equation}
\label{eq:hpzio}
\pz^{\text{in}}=\frac{r}{2 M}\;,\;\pz^{\text{out}}=\frac{r\lnff}{2 M}\;.
\end{equation}
Inhomogeneous solutions are given by integrals over the 
source, using the formula \eqref{eq:retgrwsol}.  The asymptotic 
behavior of the inhomogeneous solutions $\pz$ and $\pza$ is 
\begin{equation}
\label{eq:asymp00out}
\pz=C_{0}^{\text{out}}\pz^{\text{out}}\;,\;\pza
=C_{0a}^{\text{out}}\pz^{\text{out}}\;,\;r\to\infty\;,
\end{equation}
\begin{equation}
\label{eq:asymp00in}
\pz=C_{0}^{\text{in}}\pz^{\text{in}}\;,\;\pza
=C_{0a}^{\text{in}}\pz^{\text{in}}\;,\;r\to 2 M\;,
\end{equation}
where $C^{\text{in}}$ and $C^{\text{out}}$ are constant ``ingoing'' and ``outgoing'' 
amplitudes.  To determine the asymptotic behavior of the metric 
perturbations, substitute \eqref{eq:asymp00out} or \eqref{eq:asymp00in} 
into the solutions in Appendix \ref{appd} and then use the even 
parity formula for $h_{\mu\nu}$ \eqref{eq:ehmunu}.  

In the limit $r\to\infty$, the metric perturbations 
for $l=0$, $\omega=0$ behave as
\begin{equation}
\label{eq:asymp00inf}
h_{tt}\sim h_{rr}\sim O\left(r^{-1}\right)\;,\;
h_{\theta\theta}\sim h_{\phi\phi}\sim O(r)\;.
\end{equation}
The perturbations go to zero relative to the background metric. 
In this sense the inhomogeneous solutions are asymptotically flat.  

The analysis of inhomogeneous solutions near the event horizon is 
more complicated.  As $r\to 2 M$, the perturbations for this mode are
\begin{equation}
\label{eq:asymp002M}
h_{tt}\sim O(X)\;,\;h_{rr}\sim O\left(X^{-1}\right)\;,\;
h_{\theta\theta}\sim h_{\phi\phi}\sim O(1)\;,
\end{equation}
where $X=\ff$.  The perturbations diverge, but no faster than the 
background metric.  However, the perturbations should be bounded in 
a system of coordinates where the background metric is finite, 
such as ingoing Eddington-Finkelstein coordinates \cite{dp04}, 
where the metric is  
\begin{equation}
\label{eq:ineddmet}
ds^{2}=-\ff dv^{2}+2 dr dv+r^2 d\theta^{2}+r^2 \sin^{2}\theta d\phi^{2}\;,
\end{equation}
and where $v=t+r_{*}$ \cite{mtw73}.  We transform to this coordinate 
system using the standard formula \cite{wein72}
\begin{equation}
\label{eq:ngmunu}
\tilde{g}_{\mu^{\prime}\nu^{\prime}}=\frac{\partial x^{\mu}}{\partial x^{\mu^{\prime}}}
\frac{\partial x^{\nu}}{\partial x^{\nu^{\prime}}}\tilde{g}_{\mu\nu}\;.
\end{equation}
Here, primes refer to the new coordinates, 
and $\tilde{g}_{\mu\nu}=g_{\mu\nu}+h_{\mu\nu}$ \eqref{eq:tilmet}.  
The metric perturbations for $l=0$, $\omega=0$ transform to
\begin{multline}
\label{eq:ghmat00}
h_{vv}=\ff\bhz\, Y_{00}\;,\;
h_{rv}=h_{vr}=\left(\bho-\bhz\right) Y_{00}\;,
\\h_{rr}=\frac{1}{\ff}\left(\bhz-2\bho+\bhtw\right) Y_{00}\;,\;
\\h_{\theta\theta}=\frac{h_{\phi\phi}}{\sin^{2}\theta}=r^{2}\bk\, Y_{00}\;,
\end{multline}
with the remaining perturbations being zero 
and $Y_{00}=\frac{1}{\sqrt{4\pi}}$.  
The components $h_{\theta\theta}$ and $h_{\phi\phi}$ are unchanged.  
In the Eddington coordinate system, the inhomogeneous solutions 
for $r\to 2M$ behave as 
\begin{equation}
\label{eq:easymp002M}
h_{vv}\sim O(X)\;,\;h_{rv}\sim O(1)\;,\;h_{rr}\sim O(X)\;,\;
h_{\theta\theta}\sim h_{\phi\phi}\sim O(1)\;,
\end{equation}
so they are bounded.  This analysis was done with $\bho=0$.  If that were 
not the case, we would find, using the harmonic gauge 
solution for $\bho$ \eqref{eq:bh1sol00}, that
\begin{equation}
\label{eq:easymp002MH1}
h_{rv}\sim O\left(X^{-1}\right)\;,\;
h_{rr}\sim O\left(X^{-2}\right)\;.
\end{equation}
The divergence suggests that we must set $\bho=0$ in the harmonic gauge 
for this mode.

The solutions in Appendix \ref{appd} are written in terms of spin~$0$ 
generalized Regge-Wheeler functions.  We also can write gauge 
changes which preserve the harmonic gauge in terms of generalized 
Regge-Wheeler functions.  For $\bmz$ \eqref{eq:m000sol}, we can show that
\begin{multline}
\label{eq:m000sol0}
\bmz=\frac{4 M (-M+r)+r (-2 M+r)\lnff}{r^2}\pz^{h}
\\+\frac{(2 M-r) \left(2 M+r \lnff\right)}{r}\big(\pz^{h}\big)^{\prime}\;,
\end{multline}
although this is not a unique way of rewriting $\bmz$.  Here, 
$\pz^{h}$ is a homogeneous solution of the generalized 
Regge-Wheeler equation with $s=0$, and is a linear combination of 
$\pz^{\text{in}}$ and $\pz^{\text{out}}$ from \eqref{eq:hpzio}.  If we choose
\begin{equation}
\pz^{h}=C_{01}\pz^{\text{in}}+\frac{C_{02}}{2 M}\pz^{\text{out}}\;,
\end{equation}
then $\bmz$ in \eqref{eq:m000sol0} simplifies to 
\begin{equation}
\bmz=C_{01}-\frac{C_{02}}{r}\;,
\end{equation}
which is the previous expression for $\bmz$ \eqref{eq:m000sol}.  
Whatever combination of functions is chosen, $\bmz$ 
will affect only $\bho$ and only in the manner specified by 
\eqref{eq:cnH1new00}, as explained there.

Similarly, the $C_{11}$ and $C_{12}$ terms of $\bmo$ \eqref{eq:m100sol} 
can be expressed in terms of spin~$0$ functions.  A comparison of 
the solution for $\bk$ in Appendix \ref{appd} and the gauge change 
formula for $\bk$ in \eqref{eq:nKnew00} suggests that
\begin{multline}
\label{eq:m100sol0}
\bmo=-\frac{\left(16 M^2+8 M r+3 r^2-8 M^2 \lnr\right)}{12 r^2}\pz^{h}
+\frac{1}{6 (2 M-r) r^2}\bigg\{-208 M^4+8 M^3 r
\\+6 M r^3+3 \left(64 M^4-32 M^3 r-2 M r^3+r^4\right)\lnff\bigg\}\pza^{h}
+\frac{1}{12 r}\bigg\{8 M^2+4 M r\\+r^2-8 M^2 \lnr\bigg\}
\big(\pz^{h}\big)^{\prime}
+\frac{\left(8 M^3+\left(-96 M^3+3 r^3\right)
\lnff\right)}{6 r}\big(\pza^{h}\big)^{\prime}\;,
\end{multline}
where $\pz^{h}$ and $\pza^{h}$ are homogeneous solutions of the 
generalized Regge-Wheeler equation with $s=0$.  The substitution 
$\pz^{h}=0$, $\pza^{h}=C_{12}\left(\frac{2}{3}\pz^{\text{in}}
-8 \pz^{\text{out}}\right)$ leads to
\begin{equation}
\bmo=\frac{C_{12} r^2}{6 M-3 r}\;,
\end{equation}
which is the $C_{12}$ term of $\bmo$ \eqref{eq:m100sol}.  The 
substitution $\pz^{h}=0$, $\pza^{h}=-\frac{3 C_{11}}{4 M^{3}}\pz^{\text{out}}$ 
gives
\begin{equation}
\bmo=\frac{C_{11}}{2 M r-r^2}\;,
\end{equation}
which is the $C_{11}$ term of $\bmo$.  Other combinations are 
also possible, including at least one which gives the $C_{0}$ term.  

Using spin~$0$ functions, the gauge change \eqref{eq:tm1gauge} can 
be restated as
\begin{multline}
\label{eq:tm1gauge0}
\xi_{0}(t,r)=C_{0}\bigg[\frac{(2 M-r) \left(-4 M+r+(2 M-r)\lnff
\right)}{r^2}\pz^{h}
\\-\frac{(-2 M+r)^2 \left(-1+\lnff\right)}{r}
\big(\pz^{h}\big)^{\prime}\bigg]t\;,
\end{multline}
\begin{equation}
\label{eq:tm1gauge1}
\begin{split}
\xi_{1}(&t,r)=C_{0}\bigg[\frac{\left(-4 M+r+(2 M-r)\lnff\right)
}{3 (2 M-r) r^2}\bigg\{\left(8 M^3-r^3\right)\lnff
\\&+M \left(r (4 M+r)-8 M^2\lnr\right)\bigg\}\pz^{h}
+\frac{\left(-1+\lnff\right)}{3 r}
\\&\times\bigg\{\left(-8 M^3+r^3\right)\lnff+M
\left(-r (4 M+r)+8 M^2\lnr\right)\bigg\}\big(\pz^{h}\big)^{\prime}\bigg]\;.
\end{split}
\end{equation}
If we set $\pz^{h}$ equal to either $\pz^{\text{in}}$ or $\pz^{\text{out}}$, then 
\eqref{eq:tm1gauge0} and \eqref{eq:tm1gauge1} simplify 
to \eqref{eq:tm1gauge}.  

The work above shows that, for $l=0$ and $\omega=0$, we can 
write gauge changes which preserve the harmonic gauge in terms of 
spin~$0$ generalized Regge-Wheeler functions.  This continues the 
pattern previously found for other modes, where such gauge changes 
can be written in terms of generalized Regge-Wheeler functions 
of $s=0$ or $s=1$.  However, it is simpler to use the expressions 
for $\bmz$ \eqref{eq:m000sol} and $\bmo$ \eqref{eq:m100sol} in 
calculations.

We can evaluate the solutions for $\bhz$, $\bhtw$ and $\bk$ 
analytically for the special case of a circular orbit of constant 
radius $R$, in the equatorial plane ($\thetap=\pi/2$).  From the 
discussion in Chapter~\ref{tmunuchap} of the stress energy tensor 
for circular orbits, we have
\begin{equation}
\label{eq:cirse00}
Se_{00}=\mz \enbar\,\frac{R-2 M}{2 \sqrt{\pi}R^{3}}
\,\delta(r-R)\delta(\omega)\;,
\end{equation}
\begin{equation}
\label{eq:cirse11}
Se_{11}=0\;,
\end{equation}
\begin{equation}
\label{eq:cirue22}
Ue_{22}=\mz \enbar\,\frac{M}{4 \sqrt{\pi}\left(R-2 M\right)}
\,\delta(r-R)\delta(\omega)\;.
\end{equation}
For circular orbits, only the zero frequency mode is needed when 
$l=0$ \eqref{eq:cirfft}.  
The specific energy $\enbar$ is given by \eqref{eq:cendef}.  The 
frequency delta function $\delta(\omega)$ is used to evaluate the 
inverse Fourier transform integrals in $h_{\mu\nu}$ \eqref{eq:hmunu}, 
so it will be omitted from subsequent equations in this discussion.  
The radial delta function indicates that mass distribution for 
the $l=0$ multipole is a thin spherical shell of constant radius $R$.  

The next step is to calculate $\pz$ and $\pza$ using the integral 
solution \eqref{eq:retgrwsol}.  That formula uses the homogeneous 
spin~$0$ solutions $\pz^{\text{in}}$ and $\pz^{\text{out}}$ \eqref{eq:hpzio}, the 
source terms $S_{0}$ and $S_{0a}$ from the differential equations 
for $\pz$ \eqref{eq:pzeqn} and $\pza$ \eqref{eq:pzaeqn00}, and the 
expressions above for $Se_{00}$ and $Ue_{22}$.  The derivatives 
$Se_{00}^{\prime}$ and $Ue_{22}^{\prime}$ in $S_{0a}$ are 
eliminated with integration by parts, and then the integrals are 
evaluated with the radial delta functions.  These calculations yield
\begin{multline}
\label{eq:cirpz}
\pz=-\frac{4 \enbar m_{0}\sqrt{\pi }r(3 M-R)}{M (2 M-R)}
\\\times\left(\lnbff\theta(R-r)+\lnff\theta(r-R)\right)
\end{multline}
and
\begin{multline}
\label{eq:cirpza}
\pza=-\frac{2 \enbar m_{0} \sqrt{\pi } r}{M (6 M-3 R)}\theta(R-r)
-\frac{\enbar m_{0} \sqrt{\pi } r \lnff}{4 M^3 (2 M-R)}
\\\times\left(8 M^2-4 M R+R^2+8 M (3 M-R) \lnbr\right)\theta(r-R)\;.
\end{multline}
We then substitute $\pz$ and $\pza$ into the solutions for $\bhz$, 
$\bhtw$ and $\bk$ and their derivatives. The metric perturbation 
functions in Appendix \ref{appd} contain terms with $Se_{00}$ and $Ue_{22}$, 
which have radial delta functions \eqref{eq:cirse00}, \eqref{eq:cirue22}.  
However, these terms are canceled by the delta functions from $\dpza$ 
that result from differentiating the theta functions in equation 
\eqref{eq:cirpza}, so the solutions and their derivatives are finite.  

For circular orbits, the solutions are
\begin{equation}
\label{eq:cirH0}
\bhz=\bhz^{\text{in}}\,\theta(R-r)+\bhz^{\text{out}}\,\theta(r-R)\;,
\end{equation}
\begin{equation}
\label{eq:cirH2}
\bhtw=\bhtw^{\text{in}}\,\theta(R-r)+\bhtw^{\text{out}}\,\theta(r-R)\;,
\end{equation}
\begin{equation}
\label{eq:cirK}
\bk=\bk^{\text{in}}\,\theta(R-r)+\bk^{\text{out}}\,\theta(r-R)\;,
\end{equation}
where, inside the orbit,
\begin{multline}
\label{eq:cirH0in}
\bhz^{\text{in}}=\frac{4 \enbar m_{0} \sqrt{\pi }}{3 r^3 (2 M-R)}
\bigg\{16 M^3+8 M^2 r+4 M r^2-3 r^3
\\+\left(4 M^2+2 M r+r^2\right) (3 M-R) \lnbff\bigg\}\;,
\end{multline}
\begin{multline}
\label{eq:cirH2in}
\bhtw^{\text{in}}=-\frac{4 \enbar m_{0} \sqrt{\pi }}{3 M r^3 (2 M-R)}
\bigg\{M \left(48 M^3-8 M^2 r-4 M r^2+r^3\right)
\\+\left(12 M^3-2 M^2 r-M r^2+r^3\right) (3 M-R)\lnbff\bigg\}\;,
\end{multline}
\begin{equation}
\label{eq:cirKin}
\bk^{\text{in}}=\frac{4 \enbar m_{0} \sqrt{\pi } 
\left(32 M^4-M r^3+\left(8 M^3-r^3\right) (3 M-R) 
\lnbff\right)}{3 M r^3 (2 M-R)}\;,
\end{equation}
and where, outside the orbit,
\begin{multline}
\label{eq:cirH0out}
\bhz^{\text{out}}=-\frac{4 \enbar m_{0} \sqrt{\pi }}{3 r^3 (-2 M+r) (2 M-R)}
\bigg\{32 M^4+12 M^3 r+3 M^2 r^2-9 M r^3-12 M^3 R
\\-4 M^2 r R-M r^2 R+3 r^3 R+M^2 R^2+\left(8 M^3-r^3\right) (3 M-R)
\\\times\lnff+8 M^3(3 M-R)\left(\lnbr-\lnr\right)\bigg\}\;,
\end{multline}
\begin{multline}
\label{eq:cirH2out}
\bhtw^{\text{out}}=\frac{4 \enbar m_{0} \sqrt{\pi }}{3 M r^3 (-2 M+r) (2 M-R)}
\bigg\{\left(24 M^4-16 M^3 r+3 M r^3-r^4\right) (3 M-R) 
\\\times\lnff+M \bigg(96 M^4-28 M^3 r-15 M^2 r^2+3 M r^3-36 M^3 R
\\\!+12 M^2 r R+5 M r^2 R-r^3 R+3 M^2 R^2-2 M r R^2
-8 M^2 (3 M-2 r)\\\times (3 M-R) \lnr
+8 M^2 (3 M-2 r) (3 M-R)\lnbr\bigg)\bigg\}\;,
\end{multline}
\begin{multline}
\label{eq:cirKout}
\bk^{\text{out}}=\frac{4 \enbar m_{0} \sqrt{\pi }}{3 M r^3 (2 M-R)}
\bigg\{\left(8 M^3-r^3\right) (3 M-R) \lnff
\\+M \bigg(32 M^3+12 M^2 r+3 M r^2-12 M^2 R-4 M r R-r^2 R+M R^2
\\+8 M^2 (-3 M+R)\lnr
+8 M^2 (3 M-R)\lnbr\bigg)\bigg\}\;.
\end{multline}
In \eqref{eq:cirH0}-\eqref{eq:cirK}, the coefficients of $\theta(R-r)$ 
and $\theta(r-R)$ are equal when $r=R$, so the metric perturbations 
are continuous.  The radial derivatives are discontinuous.  

For large $r$, the circular orbit perturbations behave as 
\begin{multline}
\label{eq:casymp00inf}
h_{tt}=\frac{2 m_{0}}{r}\enbar\frac{R-3 M}{R-2 M}+O\left(r^{-2}\right)\;,\;
h_{rr}=\frac{2 m_{0}}{r}\enbar\frac{R-3 M}{R-2 M}+O\left(r^{-2}\right)\;,
\\h_{\theta\theta}=\frac{h_{\phi\phi}}{\sin^{2}\theta}
=2 m_{0}r\enbar\frac{R-3 M}{R-2 M}+O\left(1\right)\;,
\end{multline}
which are calculated with \eqref{eq:cirH0out}-\eqref{eq:cirKout}.  
From the definition of $\enbar$ \eqref{eq:cendef},
\begin{equation}
\enbar\frac{R-3 M}{R-2 M}=\sqrt{1-\frac{3 M}{R}}\;.
\end{equation}
As $r\to\infty$, the perturbations go to zero relative to the background 
metric.  They are also isotropic.

To analyze behavior near the event horizon, we 
use $\bhz^{\text{in}}$, $\bhtw^{\text{in}}$ and $\bk^{\text{in}}$, which, 
by inspection, are finite and non-zero as $r\to 2 M$.  This implies that
\begin{equation}
h_{tt}\sim O(X)\;,\;h_{rr}\sim O\left(X^{-1}\right)\;,\;
h_{\theta\theta}\sim h_{\phi\phi}\sim O(1)\;,
\end{equation}
which diverges like the background metric.  In Eddington 
coordinates \eqref{eq:ghmat00}, we have
\begin{multline}
h_{vv}=-\frac{2 \enbar m_{0}\left(r-2 M\right)}{3 r^4(2 M-R)}
\bigg\{-16 M^3-8 M^2 r-4 M r^2+3 r^3
\\-\left(4 M^2+2 M r+r^2\right)(3 M-R)\lnbff\bigg\}\sim O(X)\;,
\end{multline}
\begin{multline}
h_{rv}=h_{vr}=-\frac{2 \enbar m_{0} }{3 r^3 (2 M-R)}\bigg\{16 M^3+8 M^2 r+4 M r^2
-3 r^3\\+\left(4 M^2+2 M r+r^2\right) (3 M-R) \lnbff\bigg\}\sim O(1)\;,
\end{multline}
\begin{equation}
h_{rr}=\frac{2 \enbar m_{0} (2 M-r) (2 M+r) 
\left(4 M+(3 M-R) \lnbff\right)}{3 M r^2 (2 M-R)}\sim O(X)\;,
\end{equation}
\begin{equation}
h_{\theta\theta}=\frac{h_{\phi\phi}}{\sin^{2}\theta}=r^{2}\bk^{\text{in}} Y_{00}\sim O(1)\;.
\end{equation}
These expressions, as well as their derivatives, are finite.  The 
solutions \eqref{eq:cirH0}-\eqref{eq:cirK} are both asymptotically flat for 
large $r$ and bounded near the event horizon.

We can use the solutions and their derivatives to calculate the bare 
force, which is given by \eqref{eq:fbare}.  Because the derivatives 
are discontinuous, the radial component of the bare force is also
discontinuous.  Calculating derivatives as $r\to R$ from inside 
the orbit gives
\begin{equation}
\label{eq:myfrin00}
f^{r}_{\text{in}}= m_{0}^2\enbar \frac{(2 M-R) 
\left(4 M^2+2 M R+R^2\right) }
{(3 M-R) R^5}\left(4 M+\left(3 M-R \right)\lnbff\right)\;,
\end{equation}
while the limit as $r\to R$ from outside the orbit is
\begin{equation}
\label{eq:myfrout00}
f^{r}_{\text{out}}=m_{0}^2\enbar\left(\frac{32 M^4-4 M R^3+R^4}
{(3 M-R) R^5}+\frac{(2 M-R) \left(4 M^2+2 M R+R^2\right)
}{R^5}\lnbff\right)\;.
\end{equation}
The difference is
\begin{equation}
\label{eq:myfrdiff}
f^{r}_{\text{out}}-f^{r}_{\text{in}}=\frac{m_{0}^2\enbar}{\left(3 M-R\right)R}\;.
\end{equation}
The other three components of the bare force, $f^{t}$, $f^{\theta}$ 
and $f^{\phi}$, are zero.  The last statement 
holds not just for circular orbits, but for arbitrary motion as well, 
when $l=0$, $\omega=0$.  Equations \eqref{eq:myfrin00} 
and \eqref{eq:myfrout00} apply only to circular orbits.  

We also can calculate the Newtonian limit of the spherically 
symmetric metric
\begin{equation}
\label{eq:met00}
ds^{2}=(g_{\mu\nu}+h_{\mu\nu}^{00})dx^{\mu}dx^{\nu}\;,
\end{equation}
where $h_{\mu\nu}^{00}$ is the perturbation for $l=0$, $\omega=0$.  
The Newtonian limit corresponds to a weak gravitational potential, 
small spatial velocities, and a metric of the form
\begin{equation}
\label{eq:newtmet}
ds^{2}=-\left(1+2\Phi\right) dt^{2}
+\left(1-2\Phi\right)\left(dx^{2}+dy^{2}+dz^{2}\right)\;,
\end{equation}
where $\Phi$ is the Newtonian gravitational potential \cite{schutz90}.  
We obtain the Newtonian limit by taking $M\to 0$, so that $\enbar\to 1$, 
which implies zero kinetic energy and no background potential 
energy.  The limit $M\to 0$ of \eqref{eq:met00} 
and \eqref{eq:cirH0}-\eqref{eq:cirK} is
\begin{equation}
\label{eq:mynewtmet}
ds^{2}=\begin{cases}
-\left(1-\frac{2 m_{0}}{R}\right) dt^{2}
+\left(1+\frac{2 m_{0}}{R}\right)\left(dr^{2}
+r^2 d\theta^{2}+r^2 \sin^{2}\theta d\phi^{2}\right)\;,\;r<R\;,
\\-\left(1-\frac{2 m_{0}}{r}\right) dt^{2}
+\left(1+\frac{2 m_{0}}{r}\right) \left(dr^{2}
+r^2 d\theta^{2}+r^2 \sin^{2}\theta d\phi^{2}\right)\;,\;r>R\;.
\end{cases}
\end{equation}
Comparing \eqref{eq:newtmet} and \eqref{eq:mynewtmet}, we have
\begin{equation}
\label{eq:newtpot}
\Phi=\begin{cases}
\displaystyle{-\frac{m_{0}}{R}}\;,\;r<R\;,
\\\displaystyle{-\frac{m_{0}}{r}}\;,\;r>R\;.
\end{cases}
\end{equation}
The metric \eqref{eq:mynewtmet} is equivalent to the Newtonian 
potential for a thin spherical shell of radius $R$, normalized to 
go to zero as $r\to\infty$.  This is reasonable, because the mass 
distribution for this multipole is a thin shell of constant 
radius \eqref{eq:cirse00}-\eqref{eq:cirue22}.  The limit does 
violate the assumption that $\frac{\mz}{M}\ll 1$.  However, this 
limit means that we are treating $\mz$ as a small perturbation of 
a flat background metric, which is permissible.  

Detweiler and Poisson have calculated the $l=0$ multipole for 
circular orbits \cite{dp04}.  Their methods and results are different from 
those given above.  Instead of solving the harmonic gauge equations 
directly, as we have done, they started in a different gauge, 
called the ``Zerilli gauge'', and then made a gauge transformation 
to the harmonic gauge.  They also did not write their solutions in 
terms of spin~$0$ generalized Regge-Wheeler functions.  Significantly, 
their metric perturbation functions are not equal to those listed 
in \eqref{eq:cirH0}-\eqref{eq:cirK}, leading to a different bare 
force.  However, their solutions do solve the field equations 
\eqref{eq:ddbh0eqn00}-\eqref{eq:ddbkeqn00}, which implies that 
their solutions and \eqref{eq:cirH0}-\eqref{eq:cirK} differ by only a 
homogeneous solution of the harmonic gauge field equations.

In the Zerilli gauge, Detweiler and Poisson found that
\begin{equation}
\label{eq:zerhtt}
h_{tt}^{Z}=2 m_{0} \enbar\left(\frac{1}{r}
-\frac{\ff}{R-2 M}\right)\,\theta(r-R)\;,
\end{equation}
\begin{equation}
\label{eq:zerhrr}
h_{rr}^{Z}=\frac{2 m_{0} \enbar\, r}{(r-2 M)^2}\,\theta(r-R)\;,
\end{equation}
\begin{equation}
\label{eq:zerhab}
\bho^{Z}=0\;,\;\bk^{Z}=0\;.
\end{equation}
The component $h_{tt}^{Z}$ goes to a constant as $r\to\infty$, rather 
than going to zero.  
Equations \eqref{eq:zerhtt}-\eqref{eq:zerhab} are not a solution of the 
harmonic gauge field equations \eqref{eq:hpertfeqn}, but do solve the 
general perturbed field equations \eqref{eq:pertfeqn}, which apply 
to any gauge.  Detweiler and Poisson explained their 
solution as follows:
\begin{quote}
It is easy to check that for $r>R$, $g_{\alpha\beta}+h_{\alpha\beta}^{Z}$ is 
another Schwarzschild metric with mass parameter $M+m_{[0]}\enbar$.  
The perturbation therefore describes the sudden shift in mass parameter 
that occurs at $r=R$.
\end{quote}
This reasoning merits some additional explanation.  Inside the 
orbit, the total metric $\tilde{g}_{\mu\nu}$ for this mode is equal 
to the background Schwarzschild 
metric, $g_{\mu\nu}$ \eqref{eq:schmet}.   The small mass $m_{0}$ affects 
only the exterior metric ($r>R$) and is incorporated only in the 
components $\tilde{g}_{rr}$ and $\tilde{g}_{tt}$.  For $r>R$, we have
\begin{equation}
\label{eq:tilgrr}
\tilde{g}_{rr}=\frac{1}{1-\frac{2\left(M+m_{0}\enbar\right)}{r}}
=g_{rr}+\frac{2 m_{0} \enbar\, r}{(r-2 M)^2}+O\left(m_{0}^{2}\right)
=g_{rr}+h_{rr}^{Z}+O\left(m_{0}^{2}\right)\;.
\end{equation}
To linear order in $\mz$, 
the component $\tilde{g}_{rr}$ is merely the Schwarzschild 
metric $g_{rr}$, with the substitution $M\to M+m_{0} \enbar$.  
Zerilli used similar reasoning to describe the radial infall of 
a small mass, also noting this follows from Birkhoff's 
theorem \cite{zerp70}.  The situation is somewhat different 
for $\tilde{g}_{tt}$.  Equation \eqref{eq:zerhtt} gives, for $r>R$, 
\begin{equation}
\label{eq:tilgtt}
\tilde{g}_{tt}=g_{tt}+h_{tt}^{Z}
=-1+\frac{2(M+m_{0}\enbar)}{r}
-\ff\frac{2 m_{0}\enbar}{R-2 M}\;.
\end{equation}
The first two terms of the right-hand equality are the Schwarzschild 
metric $g_{tt}$, with $M\to M+m_{0} \enbar$.  The last term is 
a constant multiple of $g_{tt}$.  In the Zerilli gauge, 
both $\tilde{g}_{rr}$ and $\tilde{g}_{tt}$ are Schwarzschild type solutions, 
modified by terms with $m_{0}$ outside the orbital radius.  
Birkhoff's theorem states the Schwarzschild metric is the unique 
spherically symmetric solution to the vacuum Einstein field 
equations, in the sense that that one may always make a coordinate 
transformation to bring the metric into the static Schwarzschild 
form \cite{mtw73}, \cite{wein72}.  The $l=0$ mode is spherically 
symmetric, and the Zerilli gauge is an extension of Birkhoff's 
theorem to linear perturbation theory for circular orbits.  A similar 
analysis for a small mass falling radially inward is in \cite{zerp70}.  
Although the perturbation stress energy tensor is non-zero at the 
location of the orbiting mass, the tensor is zero (a vacuum) elsewhere.

Detweiler and Poisson transformed from the Zerilli gauge to the 
harmonic gauge.  They did not publish their harmonic 
gauge solutions, but their results may be rederived using their analysis.  
More recently, their solutions were printed by others in \cite{bl05}.  
The Detweiler-Poisson radial perturbation 
functions $\bhz^{\text{DP}}$, $\bhtw^{\text{DP}}$ and $\bk^{\text{DP}}$ are not equal to 
$\bhz$, $\bhtw$ and $\bk$, as given by \eqref{eq:cirH0}-\eqref{eq:cirK}.  
The solutions differ by
\begin{equation}
\label{eq:cirH0diff}
\bhz\,Y_{00}-\bhz^{\text{DP}}
=\frac{2\mz  \enbar \left(4 M^3+2 M^2 r+M r^2-r^3\right)}{r^3 (2 M-R)}\;,
\end{equation}
\begin{equation}
\label{eq:cirH2diff}
\bhtw\,Y_{00}-\bhtw^{\text{DP}}
=-\frac{2\mz  \enbar M \left(12 M^2-2 M r-r^2\right)}{r^3 (2 M-R)}\;,
\end{equation}
\begin{equation}
\label{eq:cirKdiff}
\bk\,Y_{00}-\bk^{\text{DP}}=\frac{16\mz  \enbar M^3}{r^3 (2 M-R)}\;.
\end{equation}
For this comparison, it was necessary to multiply $\bhz$, $\bhtw$ 
and $\bk$ by the angular harmonic $Y_{00}(\theta,\phi)$, because 
Detweiler and Poisson absorbed this constant into their radial 
functions.  The differences \eqref{eq:cirH0diff}-\eqref{eq:cirKdiff} are 
a homogeneous solution of the field 
equations \eqref{eq:ddbh0eqn00}-\eqref{eq:ddbkeqn00}, 
which may be verified by substitution.  
They set $\bho=0$, ``on the grounds that the perturbation must 
be static.''  Detweiler and Poisson also 
showed that their solutions were bounded as $r\to 2M$, using 
ingoing Eddington coordinates.  For large $r$, their perturbed 
metric behaves as
\begin{multline}
\label{eq:dpcasymp00inf}
h_{tt}^{\text{DP}}=\frac{2\mz \enbar }{2 M-R}
-\frac{2\mz \enbar  R}{r (2 M-R)}+O\left(r^{-2}\right)\;,\;
h_{rr}^{\text{DP}}=\frac{2\mz\enbar}{r}+O\left(r^{-2}\right)\;,
\\h_{\theta\theta}^{\text{DP}}=\frac{h_{\phi\phi}^{\text{DP}}}{\sin^{2}\theta}
=2\mz \enbar\, r \frac{R-3 M}{R-2 M}+O\left(1\right)\;.
\end{multline}
The component $h_{tt}^{\text{DP}}$ is a constant as $r\to\infty$; 
the others go to zero relative to the background metric.  This 
behavior is different from the solutions \eqref{eq:cirH0}-\eqref{eq:cirK}, 
for which all components go to zero compared to the background.  

Using their harmonic gauge solutions, Detweiler and Poisson calculated 
the bare acceleration $a^{r}$, which, after multiplication by $\mz$, 
leads to
\begin{equation}
\label{eq:dpfrin00}
f^{r}_{\text{in,DP}}=m_{0}^2\enbar \frac{ (R-2 M) 
\left(R^2+2 M R+4 M^2\right) }{R^5}\left(\frac{M}{R-3 M}-\lnbff\right)\;,
\end{equation}
\begin{multline}
\label{eq:dpfrout00}
f^{r}_{\text{out,DP}}=-m_{0}^2\enbar \left(\frac{R^4-M R^3+8 M^4}
{R^5 (R-3 M)}\right.\\\left.+\frac{(R-2 M)\left(R^2+2 M R+4 M^2\right)}
{R^5}\lnbff\right)\;.
\end{multline}
These results differ from \eqref{eq:myfrin00}-\eqref{eq:myfrout00} by
\begin{equation}
\label{eq:mydpfrdiff}
f^{r}_{\text{in}}-f^{r}_{\text{in,DP}}=f^{r}_{\text{out}}-f^{r}_{\text{out,DP}}
=m_{0}^2\enbar\frac{3 M (R-2 M)
\left(4 M^2+2 M R+R^2\right)}{(R-3 M) R^5}\;.
\end{equation}
This discrepancy is due entirely to the fact that 
\eqref{eq:cirH0}-\eqref{eq:cirK} differ from the 
corresponding Detweiler-Poisson results by a homogeneous 
solution of the harmonic gauge field equations.  
The discontinuity in the Detweiler-Poisson bare force is
\begin{equation}
\label{eq:dpfrdiff}
f^{r}_{\text{out,DP}}-f^{r}_{\text{in,DP}}=\frac{m_{0}^2\enbar}{\left(3 M-R\right)R}\;.
\end{equation}
This agrees with \eqref{eq:myfrdiff} because the force discrepancy 
\eqref{eq:mydpfrdiff} is due to a homogeneous solution, which 
has continuous derivatives.  Also,
\begin{equation}
f^{r}_{\text{in}}-f^{r}_{\text{in,DP}}=f^{r}_{\text{out}}-f^{r}_{\text{out,DP}}
=\mz^{2}\frac{3 M}{R^{3}}+O(R^{-4})\;,
\end{equation}
for orbits of large radius $R$.

Taking the limit $M\to 0$ of the Detweiler-Poisson solutions 
($g_{\mu\nu}+h_{\mu\nu}^{\text{DP}}$) yields
\begin{equation}
\label{eq:dpnewtmet}
ds^{2}=\begin{cases}
-dt^{2}+\left(1+\frac{2 m_{0}}{R}\right)\left(dr^{2}
+r^2 d\theta^{2}+r^2 \sin^{2}\theta d\phi^{2}\right)\;,\;r<R\;,
\\-\left(1-\left(\frac{2 m_{0}}{r}-\frac{2 m_{0}}{R}\right)\right) dt^{2}
+\left(1+\frac{2 m_{0}}{r}\right) \left(dr^{2}
+r^2 d\theta^{2}+r^2 \sin^{2}\theta d\phi^{2}\right)\;,\;r>R\;.
\end{cases}
\end{equation}
This metric differs by a constant from the Newtonian 
formula \eqref{eq:newtmet}, because $h_{tt}^{\text{DP}}$ does not go to zero 
for large $r$.  In Newtonian physics, we may add a constant to 
the gravitational potential, without affecting the gravitational 
force.  Here, the different potentials result from different metric 
perturbations, which affect the relativistic bare force and 
cause the discrepancy given by \eqref{eq:mydpfrdiff}.

Detweiler and Poisson took the Newtonian 
limit in a different way.  They examined the bare acceleration in the 
limit of small $\frac{M}{R}$ and obtained
\begin{equation}
\label{eq:dpnewt}
a^{r}_{\text{in,DP}}\sim \frac{3 \mz M}{R^3}\;,\;
a^{r}_{\text{out,DP}}\sim -\frac{\mz}{R^2}+\frac{\mz M}{2 R^{3}}\;.
\end{equation}
They noted that, to leading order, this is consistent with 
a Newtonian gravitational field.  
Terms with $R^{-2}$ are Newtonian order, and terms with 
$R^{-3}$ are the first post-Newtonian order \cite{dp04}.  
After division by $\mz$, the equivalent 
expressions for \eqref{eq:myfrin00}-\eqref{eq:myfrout00} are
\begin{equation}
\label{eq:mynewt}
a^{r}_{\text{in}}\sim \frac{6 \mz M}{R^3}\;,\;
a^{r}_{\text{out}}\sim -\frac{\mz}{R^2}+\frac{7\mz M}{2 R^{3}}\;.
\end{equation}
Equations \eqref{eq:dpnewt} and \eqref{eq:mynewt} agree at Newtonian 
order ($R^{-2}$), but disagree at first post-Newtonian 
order ($R^{-3}$).  

The two solutions can be related by a change of gauge which preserves 
the harmonic gauge.  In the gauge vector $\bmo$ \eqref{eq:m100sol}, we set
\begin{equation}
\label{eq:todpconst}
C_{0}=0\;,\;C_{11}=-\mz\enbar\,\frac{16 M^{3}\sqrt{\pi}}{2 M-R}\;,\;
C_{12}=\mz\enbar\,\frac{6 \sqrt{\pi}}{2 M-R}\;,
\end{equation}
which gives
\begin{equation}
\label{eq:todpm1}
\bmo=-\mz\enbar\,\frac{2\sqrt{\pi }\left(4 M^2+2 M r+r^2\right)}{r(2 M-R)}\;.
\end{equation}
We then substitute \eqref{eq:todpm1} into the gauge change 
expressions \eqref{eq:nH0new00}, \eqref{eq:nH2new00} 
and \eqref{eq:nKnew00}, with the ``old'' metric perturbation functions 
being the solutions \eqref{eq:cirH0}-\eqref{eq:cirK}.  In the new gauge, 
we find that
\begin{equation}
\label{eq:gplusdp}
h_{\mu\nu}^{00}=\frac{2\mz\enbar}{2 M-R}\,g_{\mu\nu}+h_{\mu\nu}^{\text{DP}}\;,
\end{equation}
where $h_{\mu\nu}^{00}$ refers to $l=0$, $m=0$ multipole.  
The perturbation in the new gauge is the Detweiler-Poisson solution, 
plus a constant multiple of the background metric.  The $g_{\mu\nu}$ term 
does not affect the bare force \eqref{eq:fbare}, because the covariant 
derivative of the background metric is zero and because $\enbar$ and $R$ 
are constant at this order in perturbation theory.  In the new gauge, 
the only contribution to the bare force is from $h_{\mu\nu}^{\text{DP}}$, 
so the bare force is given by the Detweiler-Poisson expressions 
\eqref{eq:dpfrin00}-\eqref{eq:dpfrout00}.  

Moreover, the background 
metric term of \eqref{eq:gplusdp} may be absorbed in a rescaling 
of the spacetime interval $ds$, as follows.  Define $h_{\mu\nu}^{n}$ as 
the total perturbation (including \eqref{eq:gplusdp}), but summed 
over only the first $n$ multipoles, because the sum over all 
multipoles diverges at the location of the small mass $\mz$.  We have  
\begin{multline}
\label{eq:dssqrsc}
ds^{2}=\left(g_{\mu\nu}+h_{\mu\nu}^{n}\right)dx^{\mu}dx^{\nu}
=\left(g_{\mu\nu}+h_{\mu\nu}^{00}
+\sum_{l=1}^{n}\sum_{m=-l}^{l}h_{\mu\nu}^{lm}\right)dx^{\mu}dx^{\nu}
\\=\left(\left[1+\frac{2\mz\enbar}{2 M-R}\right]g_{\mu\nu}+h_{\mu\nu}^{\text{DP}}
+\sum_{l=1}^{n}\sum_{m=-l}^{l}h_{\mu\nu}^{lm}\right)dx^{\mu}dx^{\nu}
\\=\left(\left[1+\frac{2\mz\enbar}{2 M-R}\right]g_{\mu\nu}
+h_{\mu\nu}^{n,\text{DP}}\right)dx^{\mu}dx^{\nu}\;,
\end{multline}
where $h_{\mu\nu}^{n,\text{DP}}$ is the sum through $n$ using the Detweiler-Poisson 
solution for the $l=0$ multipole.  If we divide both sides 
of \eqref{eq:dssqrsc} by the factor in brackets, we obtain
\begin{equation}
\label{eq:tdssqrsc}
d\tilde{s}^{2}=\left(g_{\mu\nu}+h_{\mu\nu}^{n,\text{DP}}
+O\left(\mz^{2}\right)\right)dx^{\mu}dx^{\nu}\;,
\end{equation}
for which
\begin{equation}
\label{eq:dstil}
d\tilde{s}^{2}=\frac{ds^{2}}{1+\frac{2\mz\enbar}{2 M-R}}\;.
\end{equation}
The division does not affect $h_{\mu\nu}^{n,\text{DP}}$, because the perturbation 
is linear in $\mz$ and $h_{\mu\nu}^{n,\text{DP}}$ is already order $\mz$.  
The effect is to rescale the spacetime interval, as described 
in \eqref{eq:dstil}.  

Similarly, we can transform from the harmonic gauge solutions 
\eqref{eq:cirH0}-\eqref{eq:cirK} to the Zerilli gauge, given 
by \eqref{eq:zerhtt}-\eqref{eq:zerhab}.  We can not use $\bmo$ 
in \eqref{eq:m100sol} for this, because such a transformation 
does not preserve the harmonic gauge.  However, we can use the 
transformation formulae \eqref{eq:nH0new00}-\eqref{eq:nKnew00}.  
To do so, first solve \eqref{eq:nKnew00} for $\bmo$ and then 
set $\bk^{\text{old}}$ equal to \eqref{eq:cirK} and
\begin{equation}
\bk^{\text{new}}=\frac{4\mz\enbar\sqrt{\pi}}{2 M-R}\;.
\end{equation}
Substitute the resulting expression for $\bmo$ into $\bhz^{\text{new}}$
\eqref{eq:nH0new00} (with $C_{0}=0$) and $\bhtw^{\text{new}}$ \eqref{eq:nH2new00}.  
The new metric perturbation for $l=0$, $m=0$ is
\begin{equation}
\label{eq:gplusz}
h_{\mu\nu}^{00}=\frac{2\mz\enbar}{2 M-R}\,g_{\mu\nu}+h_{\mu\nu}^{Z}\;.
\end{equation}
The $g_{\mu\nu}$ term is the same as in \eqref{eq:gplusdp}, and it 
also may be absorbed by rescaling $ds^{2}$.  

Detweiler and Poisson argued their solution is the unique 
harmonic gauge solution that is bounded near the event horizon (in 
Eddington coordinates) and that does not diverge for large $r$.  
However, the solution derived in this thesis also meets these two 
boundary conditions, so their claim of uniqueness is incorrect.  
Their argument went as follows.  They started with their harmonic gauge 
solution and then tried finding a different solution by making a 
gauge change which preserved the harmonic gauge.  They showed 
that such a gauge change would either (1) introduce an unphysical  
divergence near the event horizon in Eddington coordinates, or (2) 
change the metric perturbation for large $r$ as
\begin{equation}
\label{eq:dpgasym}
\Delta h_{tt}\sim O\left(r^{-1}\right)\;,
\;\Delta h_{rr}\sim O\left(1\right)\;,
\;\Delta h_{\theta\theta}\sim \Delta h_{\phi\phi}\sim O(r^{2})\;.
\end{equation}
Equation \eqref{eq:dpgasym} does not explicitly appear in their paper, 
but may be inferred from their analysis.  They concluded that the 
perturbation change is ``ill behaved as $r\to\infty$'', 
presumably because $\Delta h_{\theta\theta}$ and $\Delta h_{\phi\phi}$ 
are order $r^{2}$.  

It is true that a gauge transformation between the Detweiler-Poisson 
solution and the solutions \eqref{eq:cirH0}-\eqref{eq:cirK} will 
introduce some components of order $r^{2}$.  This is shown by the 
discussion of \eqref{eq:gplusdp}, which involves a transformation to 
the Detweiler-Poisson solution.  However, the order $r^{2}$ components 
are not ``ill behaved'', because they are merely a constant multiple 
of the background metric components $g_{\theta\theta}$ and $g_{\phi\phi}$.  
Detweiler and Poisson did not consider this line of reasoning, so 
they overlooked the solution derived in this thesis.  

The different circular orbit solutions for $l=0$ produce different bare 
forces, so we need to determine which is correct.  Both are bounded 
as $r\to 2 M$, in the ingoing Eddington coordinate system.  However, 
their large $r$ 
behavior is different.  The solutions \eqref{eq:cirH0}-\eqref{eq:cirK} 
go to zero relative to the background metric as $r\to\infty$.  In 
contrast, the $tt$ component of the Detweiler-Poisson solution becomes 
constant as $r\to\infty$, although the other components go to zero.  
We normally would expect that the perturbation vanish asymptotically, 
consistent with Newtonian gravity.  Accordingly, 
equations \eqref{eq:cirH0}-\eqref{eq:cirK} should be used.  
Further discussion of this issue is given in section~\ref{sec:eqsf}.  
In any event, the difference between the bare forces is readily 
calculated using \eqref{eq:mydpfrdiff}.  

This completes the solution of the inhomogeneous even parity field 
equations \eqref{eq:ddbh0eqn}-\eqref{eq:ddgeqn}.  An interim summary of 
the odd and even parity results is in section~\ref{sec:intsum}, following 
a discussion of even parity homogeneous solutions in 
section~\ref{sec:hevpar}.  

\section{\label{sec:hevpar}Homogeneous Solutions}

This section analyzes the even parity homogeneous solutions and is 
patterned on the odd parity discussion in section \ref{sec:hoddpar}.  
We begin with the non-zero frequency solutions for $l\ge 2$.
The following form a system of eight first order differential equations:
\begin{equation}
\label{eq:d0dh0eqn}
\dhz-d_{0}=0
\end{equation}
\begin{multline}
\label{eq:hddh0eqn}
\frac{(-2 M+r)^2 }{r^2} d_{0}^{\,\prime}+\frac{\left(-8 M^2+4
 (2+\lambda) M r-r^2 \left(2+2 \lambda+(\iom)^2 r^2\right)\right) }{r^4}\hz
\\+\frac{2 \iom M(2 M-r) }{r^3}\ho
+\frac{2 (-2 M+r)^2 }{r^3}\bho=0\;,
\end{multline}
\begin{equation}
\label{eq:hdivheqn0}
\left(1-\frac{2 M}{r}\right) \dbho
-\frac{2 (1+\lambda) }{r^2}\hz+\frac{1}{2} \iom \bhz
-\frac{2 (M-r) }{r^2}\bho+\frac{1}{2} \iom \bhtw+\iom\bk=0\;,
\end{equation}
\begin{equation}
\label{eq:hdivheqn2}
\left(1-\frac{2 M}{r}\right) \dho-2 \lambda \bg
-\frac{\iom r }{2 M-r}\hz+\frac{\bhz}{2}-\frac{2 (M-r) }{r^2}\ho
-\frac{\bhtw}{2}=0\;,
\end{equation}
\begin{multline}
\label{eq:hndkeqns}
\left(1-\frac{2 M}{r}\right) \dbk
+\frac{2 (1+\lambda) M }{\iom r^4}\hz-\frac{(1+\lambda) (2 M-r) }{r^3}\ho
-\frac{(1+\lambda) (2 M-r) }{\iom r^3}\bho
\\+\frac{(2 M-r) }{r^2}\bhtw+\frac{(-3 M+r) }{r^2}\bk
+\frac{(1+\lambda) (2 M-r) }{\iom r^3}d_{0}=0\;,
\end{multline}
\begin{multline}
\label{eq:hndh2eqns}
\left(1-\frac{2 M}{r}\right)\dbhtw
-\frac{2 \lambda (1+\lambda)}{r}\bg
-\frac{2 (1+\lambda) \left(-3 M^2+M r
+(\iom)^2 r^4\right) }{\iom  (2 M-r) r^4}\hz
\\+\frac{(2 M+r+\lambda r) }{r^2}\bhz
+\frac{(1+\lambda) (3 M-r) }{r^3}\ho
-\frac{(1+\lambda) (3 M-r) }{\iom  r^3}\bho
\\+\frac{(-3 M+2 r) }{r^2}\bhtw
+\frac{\left(7 M^2-2 (5+\lambda) M r+r^2 \left(3+\lambda+(\iom)^2
r^2\right)\right)}{(2 M-r) r^2}\bk
\\+\frac{(1+\lambda) (3 M-r) }{\iom r^3}d_{0}=0\;,
\end{multline}

\begin{equation}
\begin{split}
\label{eq:hndh0eqns}
\bigg(1&-\frac{2 M}{r}\bigg) \dbhz
+\frac{2 \lambda (1+\lambda) }{r}\bg
+\frac{2 (1+\lambda) \left(M^2-M r+(\iom)^2 r^4\right) }{\iom (2 M-r)r^4}\hz
-\frac{(1+\lambda) }{r}\bhz\\&+\frac{(1+\lambda) (M-r) }{r^3}\ho
+\frac{\left(-(1+\lambda) M+r \left(1+\lambda
+2 (\iom)^2 r^2\right)\right) }{\iom r^3}\bho+\frac{M }{r^2}\bhtw
\\&+\frac{\left(-3 M^2+2 (2+\lambda) M r-r^2 \left(1+\lambda
+(\iom)^2 r^2\right)\right) }{(2 M-r) r^2}\bk
+\frac{(1+\lambda) (M-r) }{\iom r^3}d_{0}=0\;,
\end{split}
\end{equation}
\begin{equation}
\label{eq:hndgeqns}
\begin{split}
\bigg(1&-\frac{2 M}{r}\bigg)\lambda\dbg
+\frac{\lambda (1+\lambda) }{r}\bg
+\frac{\left(-(1+\lambda) M^2+2 (\iom)^2 M r^3
+(\iom)^2 \lambda r^4\right) }{\iom(2 M-r) r^4}\hz
\\&-\frac{(3 M+\lambda r) }{2 r^2}\bhz-\frac{\left(M-3 \lambda M
+2 \lambda r+(\iom)^2 r^3\right) }{2 r^3}\ho+\frac{\left(M+\lambda M
+(\iom)^2 r^3\right) }{2 \iom r^3}\bho
\\&+\frac{\left(3 M^2-M r+2 \lambda M r-\lambda r^2
-(\iom)^2 r^4\right) }{4 M r^2-2 r^3}\bk-\frac{\left(M+\lambda M+(\iom)^2
r^3\right) }{2 \iom r^3}d_{0}=0\;.
\end{split}
\end{equation}
Equation \eqref{eq:hddh0eqn} is from the field equation \eqref{eq:ddh0eqn}, 
written in terms of $d_{0}$ \eqref{eq:d0dh0eqn}.  Equations 
\eqref{eq:hdivheqn0} and \eqref{eq:hdivheqn2} are two of the harmonic 
gauge conditions.  The last four are homogeneous forms of the first 
order equations \eqref{eq:ndh0eqns}-\eqref{eq:ndh2eqns}.  This eight 
equation system can be derived from the field equations and harmonic 
gauge conditions.  In turn, the field equations and harmonic gauge 
conditions can be obtained from from the system of eight.  The 
different systems are equivalent, so the field equations 
have only eight linearly independent homogeneous solution vectors.  
Those vectors are formed out of the homogeneous solutions of the 
Zerilli and generalized Regge-Wheeler equations, namely, 
$\ptw^{\text{in}}$, $\po^{\text{in}}$, $\pz^{\text{in}}$ and $\pza^{\text{in}}$ 
and their outgoing counterparts.  This result is to be expected from 
the fact that the even parity inhomogeneous solutions can be written 
in terms of $\ptw$, $\po$, $\pz$ and $\pza$.  Although $\pz$ and $\pza$ 
both satisfy the same homogeneous differential equation, solution vectors 
formed from them are linearly independent because they participate 
in the metric perturbations in different ways.  The determinant of 
the $8\times 8$ matrix formed from the eight solution vectors is
\begin{equation}
\label{eq:evdet}
-\frac{\iom\lambda W_{0}^{2}W_{1}W_{2}}{r^5 (r-2 M)^2}\;,
\end{equation}
where $W_{s}$ is the Wronskian \eqref{eq:grwwr}.  The determinant is 
non-zero, which shows that the solution vectors are linearly 
independent.  However, the spin $0$ and spin $1$ solutions can be 
removed by a gauge transformation which preserves the harmonic 
gauge.  Such a transformation would be implemented using the gauge 
change vectors \eqref{eq:hm0sol}-\eqref{eq:hm2sol}.  Accordingly, 
the only physically significant homogeneous 
solutions for non-zero frequency and $l\ge 2$ are those 
constructed from $\ptw^{\text{in}}$ and $\ptw^{\text{out}}$, which 
are the gauge invariant solutions of the Zerilli equation.  

For $l=1$, the $\ptw$ functions are not present, so homogeneous solutions 
are constructed from spin $1$ and spin $0$ generalized Regge-Wheeler 
functions.  For $l=0$, only spin $0$ functions are used.  
Again, the spin $1$ and spin $0$ solutions can be removed by a gauge 
transformation which preserves the harmonic gauge.  For $l=0$, this 
result reflects Birkhoff's theorem, because the time dependent solution 
is removed by a coordinate transformation.

Turning to zero frequency solutions for $l\ge 2$, the field equations 
can be reduced to the following system of eight first order 
homogeneous differential equations:
\begin{equation}
\label{eq:dkdkeqn}
\dbk-d_{K}=0\;,
\end{equation}
\begin{multline}
\label{eq:ddkeqn0h}
\frac{(-2 M+r)^2 }{r^2}d_{K}^{\,\prime}+\frac{2 (M-r) (2 M-r)}{r^3}d_{K}
+\frac{2 M (-2 M+r) }{r^4}\bhz-\frac{4 (1+\lambda) (-2 M+r)^2 }{r^5}\ho
\\+\frac{2 (2 M-r) (3 M-r) }{r^4}\bhtw+\frac{\left(-16 M^2+4 (4+\lambda) M r
-r^2 \left(4+2 \lambda\right)\right) }{r^4}\bk=0\;,
\end{multline}
\begin{multline}
\label{eq:ndh0eqns0h}
\left(1-\frac{2 M}{r}\right) \dbhz
+\frac{2 \lambda (1+\lambda)}{r}\bg-\frac{(1+\lambda) }{r}\bhz
+\frac{2 (1+\lambda) (M-r) }{r^3}\ho\\+\frac{\bhtw}{r}+\frac{\lambda }{r}\bk
+\left(-1+\frac{M}{r}\right)d_{K}=0\;,
\end{multline}
\begin{multline}
\label{eq:ndh2eqns0h}
\left(1-\frac{2 M}{r}\right) \dbhtw
-\frac{2 \lambda (1+\lambda) }{r}\bg+\frac{(2 M+r+\lambda r)}{r^2}\bhz
+\frac{2 (1+\lambda) (3 M-r) }{r^3}\ho\\+\frac{3(-2 M+r) }{r^2}\bhtw
+\frac{(8 M-(4+\lambda) r) }{r^2}\bk
+\left(-1+\frac{3 M}{r}\right)d_{K}=0\;,
\end{multline}
\begin{multline}
\label{eq:ndhgeqns0h}
\lambda\left(1-\frac{2 M}{r}\right) \dbg
+\frac{\lambda (1+\lambda) }{r}\bg-\frac{(3 M+\lambda r) }{2 r^2}\bhz
-\frac{(M-\lambda M+\lambda r) }{r^3}\ho\\+\frac{M}{2 r^2}\bhtw
+\frac{\lambda }{2 r}\bk-\frac{M}{2 r}d_{K}=0\;,
\end{multline}
\begin{equation}
\label{eq:ndhzeqns0h}
\frac{(1+\lambda)(2 M-r)}{r}\dhz+\frac{2(1+\lambda)M}{r^2}\hz
-\frac{(1+\lambda) (2 M-r) }{r}\bho\\=0\;.
\end{equation}
\begin{equation}
\label{eq:divheqn0h}
\left(1-\frac{2 M}{r}\right) \dbho
-\frac{2 (1+\lambda) }{r^2}\hz-\frac{2 (M-r) }{r^2}\bho=0\;,
\end{equation}
\begin{equation}
\label{eq:divheqn2h}
\left(1-\frac{2 M}{r}\right) \dho-2 \lambda \bg
+\frac{\bhz}{2}-\frac{2 (M-r) }{r^2}\ho-\frac{\bhtw}{2}=0\;,
\end{equation}
Equation \eqref{eq:ddkeqn0h} is obtained from the field equation 
\eqref{eq:ddkeqn}, with $d_{K}$ replacing $\dbk$.  The next four 
equations are from the first order equations 
\eqref{eq:ndh0eqns0}-\eqref{eq:ndh2eqns0}.  
Equations \eqref{eq:divheqn0h}-\eqref{eq:divheqn2h} are two of the 
harmonic gauge conditions.  
Like the non-zero frequency case, the eight linearly independent solutions 
are combinations of the $\ptw$, $\po$, $\pz$ and $\pza$ homogeneous 
solutions.  Similarly, the spin $1$ and spin $0$ solutions can be removed 
by a gauge transformation which preserves the harmonic gauge, leaving 
only the gauge invariant $\ptw^{\text{in}}$ and $\ptw^{\text{out}}$ solutions.  
However, as explained in Chapter \ref{rweqnchap}, the $\ptw^{\text{in}}$ 
solution diverges at large $r$ and the $\ptw^{\text{out}}$ solution 
diverges logarithmically near the event horizon.  To prevent an 
unphysical divergence, we set the the spin $2$ solutions to zero by choice 
of constants.  The result is that there are no zero frequency 
homogeneous solutions for $l\ge 2$.  A similar conclusion was reached 
by Vishveshwara \cite{vish70}, although he did not work in the 
harmonic gauge.  Also, the spin $1$ and spin $0$ solutions have similar 
divergent behavior, so removing them by means of a gauge transformation 
is necessary and not elective.

The $l=1$ zero frequency homogeneous solutions are constructed from 
homogeneous $\po$, $\pz$ and $\pza$ homogeneous solutions, but these 
can be removed by a gauge transformation which preserves the 
harmonic gauge.  Moreover, such a transformation would seem to be 
required, because otherwise the solutions would be divergent for the 
reasons given above.  As mentioned earlier, the non-zero frequency 
homogeneous solutions for $l=1$ also can be removed by a gauge 
transformation which preserves the harmonic gauge.  Thus, the even parity 
$l=1$ homogeneous solutions are pure gauge.  This result agrees with 
previous work by Zerilli \cite{zerp70}.  Although Zerilli did not work 
in the harmonic gauge, he showed that his $l=1$ 
homogeneous solutions could be removed by a gauge transformation.  

The only case left is zero frequency, $l=0$.  As discussed in 
subsection \ref{sec:zevpareqz}, the homogeneous solution for $\bho$ 
is zero.  The field equations for $\bhz$, $\bhtw$ and $\bk$ can 
be reduced to a four equation first order system composed of 
\eqref{eq:dkdkeqn}-\eqref{eq:ndh2eqns0h}, modified by 
the substitution $\lambda\to -1$.  The possible homogeneous solutions are
\begin{multline}
\label{eq:sol100}
\bhz= \left(3-\frac{16 M^3}{r^3}
-\frac{8 M^2}{r^2}-\frac{4 M}{r}\right)C_{0a}^{\text{in}}
-\frac{\left(4 M^2+2 M r+r^2\right)}{6 r^3}C_{0}^{\text{in}}
\\+\frac{8  M^4}{3 r^3 (-2 M+r)}C_{0a}^{\text{out}}
+\frac{ 1}{6 (2 M-r) r^3}\bigg\{-8 M^3-4 M^2 r-M r^2
\\+3 r^3+\left(-8 M^3+r^3\right)\lnff+8 M^3 \lnr\bigg\}C_{0}^{\text{out}}\;,
\end{multline}
\begin{multline}
\label{eq:sol200}
\bhtw=\frac{8  M^3 (3 M-2 r)}{3 (2 M-r) r^3}C_{0a}^{\text{out}}
+\frac{ \left(48 M^3-8 M^2 r-4 M r^2+r^3\right)}{r^3}C_{0a}^{\text{in}}
\\+\frac{\left(12 M^3-2 M^2 r-M r^2+r^3\right)}{6 M r^3}C_{0}^{\text{in}}
+\frac{1}{6 M (2 M-r) r^3}
\\\times\bigg\{\left(24 M^4-16 M^3 r+3 M r^3-r^4\right)\lnff+M 
\\\times\left(24 M^3-4 M^2 r-5 M r^2+r^3-8 M^2(3 M-2 r) 
\lnr\right)\bigg\}C_{0}^{\text{out}}\;,
\end{multline}
\begin{multline}
\label{eq:sol300}
\bk= \left(1-\frac{32 M^3}{r^3}\right)C_{0a}^{\text{in}}
+\frac{ \left(1-\frac{8 M^3}{r^3}\right)}{6 M}C_{0}^{\text{in}}
-\frac{8  M^3}{3 r^3}C_{0a}^{\text{out}}
\\+\frac{\left(-8 M^3+r^3\right)\lnff
+M \left(-8 M^2-4 M r-r^2+8 M^2 \lnr\right)}{6 M r^3}C_{0}^{\text{out}}\;,
\end{multline}
\begin{multline}
\label{eq:sol400}
d_{K}^{}=\frac{96  M^3}{r^4}C_{0a}^{\text{in}}+\frac{4  M^2}{r^4}C_{0}^{\text{in}}
+\frac{8  M^3}{r^4}C_{0a}^{\text{out}}
\\+\frac{8 M^2+4 M r+r^2+8 M^2 \lnff-8 M^2 \lnr}{2 r^4}C_{0}^{\text{out}}\;.
\end{multline}
These solutions can be written in vector form as 
\begin{equation}
\label{eq:xvec00}
\bm{X}=C_{0}^{\text{in}}\bm{X_{1}}+C_{0}^{\text{out}}\bm{X_{2}}
+C_{0a}^{\text{in}}\bm{X_{3}}+C_{0a}^{\text{out}}\bm{X_{4}}\;,
\end{equation}
where the components of $\bm{X_{1}}$-$\bm{X_{4}}$ can be deduced from 
\eqref{eq:sol100}-\eqref{eq:sol400}.  A matrix formed from the column 
vectors $\bm{X_{1}}$-$\bm{X_{4}}$ has determinant
\begin{equation}
\label{eq:det00}
\frac{2 M^{2}}{(r-2 M)r^{4}}\;.
\end{equation}
This is non-zero, so $\bm{X_{1}}$-$\bm{X_{4}}$ are linearly independent.  
Accordingly, any homogeneous solution of the four equation system 
(and by extension, the equivalent field equations for $\bhz$, $\bhtw$ 
and $\bk$) can be written in the form \eqref{eq:xvec00}.  Suppose we 
make a gauge change to $\bm{X}$ \eqref{eq:xvec00} defined by
\begin{multline}
\label{eq:tobkgmult}
\xi_{0}(t,r)=\;C_{0}\ff t\;,\,
\xi_{1}(t,r)=\;\frac{C_{11}}{2 M r-r^2}+\frac{C_{12} r^2}{6 M-3 r}
\\+C_{0}\frac{\left(\left(-8 M^3+r^3\right) 
\lnff+M\left(-r (4 M+r)+8 M^2 \lnr\right)\right)}{3 (2 M-r) r}\;,
\end{multline}
where 
\begin{multline}
\label{eq:tobkgmultcc}
C_{0}=-\frac{C_{0}^{\text{out}}}{4 M}\;,\,
C_{11}=\frac{2 M^{2}}{3} \left(C_{0}^{\text{in}}+C_{0}^{\text{out}}
+24 C_{0a}^{\text{in}}  M+2 C_{0a}^{\text{out}}  M\right)\;,
\\C_{12}=-\frac{C_{0}^{\text{in}}+3 C_{0}^{\text{out}}+24 C_{0a}^{\text{in}} M}{4 M}\;.
\end{multline}
This gauge change will preserve the harmonic gauge and leave $\bho$ 
unchanged \eqref{eq:cnH1new00}, \eqref{eq:m100sol}-\eqref{eq:tm1gauge}.  
In the new gauge, the components of $\bm{X}$ become
\begin{equation}
\label{eq:tobkgmultnew}
\bhz^{\text{new}}=-C^{h}\;,\,\bhtw^{\text{new}}=\bk^{\text{new}}=C^{h}\;,\,
d_{K}=(\bk^{\text{new}})^{\prime}=0\;,
\end{equation}
where the constant $C^{h}$ is
\begin{equation}
\label{eq:tobkgmultcdef}
C^{h}=-3 C_{0a}^{\text{in}}-\frac{C_{0}^{\text{out}}}{2 M}\;.
\end{equation}
We then substitute the new gauge result \eqref{eq:tobkgmultnew} into the 
even parity metric perturbation $h^{e,lm}_{\mu\nu}$ \eqref{eq:ehmunu} 
to get
\begin{equation}
\label{eq:cgmunu}
h^{e,00}_{\mu\nu}=\frac{C^{h}}{\sqrt{4 \pi}}g_{\mu\nu}\;.
\end{equation}
The gauge transformation \eqref{eq:tobkgmult}-\eqref{eq:tobkgmultcc} 
changes an arbitrary homogeneous solution into a perturbation which 
is a constant multiple of the background metric.  We can not 
make a further gauge transformation to eliminate the homogeneous 
solution \eqref{eq:cgmunu}.  This is because the change required to 
remove $\bk$ \eqref{eq:nKnew00} will not also eliminate 
$\bhtw$ \eqref{eq:nH2new00}.  The fact we can transform 
a homogeneous perturbation to the Schwarzschild solution 
\eqref{eq:cgmunu} is consistent with Birkhoff's theorem.  

Working in a different gauge, Zerilli derived a homogenous solution 
for the $l=0$, zero frequency mode and showed it represented a change in 
the Schwarzschild mass $M$ \cite{zerp70}.  His solution does not solve 
the harmonic gauge field equations.  We could transform the 
homogeneous solution \eqref{eq:cgmunu} to Zerilli's gauge, which would 
relate the constant $C^{h}$ to the change in $M$ referred to by Zerilli.  
However, we will not be altering the value $M$, so we can set $C^{h}=0$.  
Also, a constant multiple of the background metric is, in effect, a 
trivial solution.

We summarize the even parity harmonic gauge homogeneous solutions as 
follows.  For non-zero frequency, the physically significant harmonic 
gauge solutions are constructed from the spin $2$ Zerilli functions, 
which are gauge invariant.  The other non-zero frequency homogeneous 
solutions can be removed by a gauge transformation which preserves the 
harmonic gauge.  The zero frequency solutions either 
can be removed by a gauge change or are divergent.  An exception is 
the $l=0$ solution, which can represent a change in the 
Schwarzschild mass $M$.

\section{\label{sec:intsum}Interim Summary of Odd and Even Parity 
Solutions}

Combining the odd and even parity results, we see that the harmonic gauge 
solutions can be expressed in terms of six functions which satisfy 
decoupled differential equations.  The odd parity solutions are 
written in terms of two generalized Regge-Wheeler functions, one with 
$s=2$ and one with $s=1$.  The even parity solutions contain the 
remaining four functions: three generalized Regge-Wheeler functions 
(two with $s=0$ and one with $s=1$) and the Zerilli function, which 
is related to the spin $2$ Regge-Wheeler function.  
The spin $2$ functions are gauge invariant and therefore physically 
meaningful.  The spin $1$ and spin $0$ functions are gauge dependent 
and do not appear in other gauges, such as the Regge-Wheeler gauge or the 
radiation gauge discussed in Chapter~\ref{radchap}.  Why do the 
harmonic gauge solutions break down this way?

One reason appears to be the form of the harmonic gauge field equations 
\eqref{eq:hpertfeqn}.  There are only two terms:  a wave operator 
term and a potential term due to the background spacetime 
curvature.  Similarly, the generalized Regge-Wheeler equation 
represents a wave interacting with a potential.  In contrast, the 
Regge-Wheeler field equations are the longer general equations 
\eqref{eq:pertfeqn}, which contain additional terms.  The harmonic 
gauge reduces the problem to the essentials -- the wave and the 
potential -- and its constituent elements replicate the pattern.

A second reason is that there is a class of gauge changes which 
preserve the harmonic gauge \eqref{eq:divchi}.  These gauge changes 
are made by adding homogeneous solutions of the generalized 
Regge-Wheeler equation for $s=1$ and $s=0$.  Because the spin $2$ 
functions are gauge invariant, they are not appropriate vehicles 
for implementing this gauge freedom.  There are also only two 
spin $2$ functions, but \eqref{eq:divchi} is a system of four equations.

The solutions derived in Chapters~\ref{oddpar} and~\ref{evpar} apply to 
arbitrary orbital motion, except where circular orbits are specifically 
discussed.
\chapter{\label{eqmochap}Equations of Motion}

Section \ref{sec:backgeo} discusses the geodesic equations for the 
background metric.  The equations are solved for relativistic 
elliptic and circular orbits.  Section \ref{sec:eqsf} explains the 
gravitational self-force equations, which give the first order 
perturbative corrections to the equations of motion.

\section{\label{sec:backgeo}Background Geodesic Equations}

The background geodesic equation is \cite{chandra92}
\begin{equation}
\label{eq:bckgdgeo}
{\frac{d^2 z}{d\tau^2}}^{\!\mu}
+\mathnormal{\Gamma}^{\mu}{}_{\!\alpha\beta}
{\frac{dz}{d\tau}}^{\!\alpha}{\frac{dz}{d\tau}}^{\!\beta}=0\;,
\end{equation}
which is the covariant derivative of the four-velocity.
For a timelike geodesic, the velocity normalization is
\begin{equation}
\label{eq:velnorm}
g_{\mu\nu}\,{\frac{dz}{d\tau}}^{\!\mu}{\frac{dz}{d\tau}}^{\!\nu}=-1\;.
\end{equation}
The parameter $\tau$ is the proper time.  
The components of \eqref{eq:bckgdgeo} are the equations of motion for 
a test mass $\mz$.  The background metric is spherically symmetric, so we 
can choose to have the orbital motion in the equatorial plane, for which 
$\theta=\pi/2$ and $\frac{d\theta}{d\tau}=0$.  This choice 
simplifies \eqref{eq:bckgdgeo} \cite{chandra92}.

As discussed in Schutz \cite{schutz90}, we can rewrite \eqref{eq:bckgdgeo} 
in terms of momenta, which leads to constants of the motion.  The 
contravariant four-momentum is $p^{\mu}=\mz \frac{dz^{\mu}}{d\tau}$.  
Replacing velocities with momenta in \eqref{eq:bckgdgeo} and lowering 
indices leads to
\begin{equation}
\label{eq:pdot}
\mz{\frac{d p}{d\tau}}\!{}^{\!\mu}
=\frac{1}{2}g^{}_{\alpha\beta,\mu}\,p^{\alpha}p^{\beta}\;.
\end{equation}
Because the background Schwarzschild metric does not depend on the 
coordinates $t$ and~$\phi$, equation \eqref{eq:pdot} implies that $p_{t}$ 
and $p_{\phi}$ are constants of the motion.  The constants are the 
energy $E$ and the z-component of angular momentum $L_{z}$, which are 
given by
\begin{equation}
\label{eq:enlzdef}
p_{t}=-E=\mz\,g_{tt}\frac{d t}{d\tau}=g_{tt}\,p^{t}\;,\;
p_{\phi}=L_{z}=\mz\,g_{\phi\phi}\frac{d\phi}{d\tau}=g_{\phi\phi}\,p^{\phi}\;.
\end{equation}
It is helpful to define the specific energy $\enbar=E/\mz$ and 
angular momentum $\elbar=L_{z}/\mz$.  Equation \eqref{eq:enlzdef} 
is rearranged to get
\begin{equation}
\label{eq:enbelb}
\frac{d t}{d\tau}=\frac{\enbar}{\ff} \;,\;
\frac{d\phi}{d\tau}=\frac{\elbar}{r^{2}}\;.
\end{equation}
Using \eqref{eq:enbelb}, we rewrite the normalization equation 
\eqref{eq:velnorm} as
\begin{equation}
\label{eq:drdtau}
\frac{d r}{d\tau}=\pm\sqrt{\enbar^{2}-V(r)}\;,
\end{equation}
where the effective potential $V$ is
\begin{equation}
\label{eq:vofrgeod}
V(r)=\ff\left(1+\frac{\elbar^{2}}{r^{2}}\right)\;.
\end{equation}
The sign of the square root depends on whether the radial coordinate 
is increasing or decreasing.  The first order equations \eqref{eq:enbelb} 
and \eqref{eq:drdtau} constitute the first integral of the second order 
geodesic equation \eqref{eq:bckgdgeo} and assume that the orbital motion 
is in the equatorial plane.  Equations \eqref{eq:enbelb}-\eqref{eq:drdtau} 
are the standard first order equations for timelike geodesics in the 
background Schwarzschild metric \cite{cut94}

Bound orbits have $E^{2}<1$ \cite{chandra92}.  We will solve the 
system \eqref{eq:enbelb}-\eqref{eq:drdtau} only for stable elliptic 
and circular orbits, which will be referred to collectively as 
bound orbits.  There are other types of bound orbits, such as various 
plunge orbits \cite{chandra92}, but we will not cover them here.  
The solutions below are not new and are taken mainly from the work 
of Darwin \cite{darwin59,darwin61}, Ashby \cite{ashby86} and 
Cutler \textit{et al.} \cite{cut94}.  

As in Newtonian mechanics, relativistic bound orbits are 
described in terms of the eccentricity $e$ and latus rectum~$p$ 
\cite{chandra92}, \cite{cut94}.  The semi-major axis $a$ also can 
be used.  These three are related by \cite{ashby86}
\begin{equation}
\label{eq:pdef}
p=a(1-e^{2})\;.
\end{equation}
Circular orbits have $e=0$.  Orbits with $0<e<1$ will be referred to 
as elliptic or eccentric orbits.  We will start with elliptic orbits.  
They move between a minimum radius $r_{\text{min}}$ (periastron) and 
maximum radius $r_{\text{max}}$ (apastron), which are
\begin{equation}
\label{eq:rminrmax}
r_{\text{min}}=a(1-e)=\frac{p}{1+e}\;,\;
r_{\text{max}}=a(1+e)=\frac{p}{1-e}\;.
\end{equation}
Some references define $p$ as a dimensionless quantity, in which 
case the numerators in \eqref{eq:rminrmax} would read $p M$ instead 
of $p$ \cite{cut94}.  At the turning points $r_{\text{min}}$ 
and $r_{\text{max}}$, the radial velocity \eqref{eq:drdtau} should be 
zero.  That will be the case if \cite{ashby86}
\begin{equation}
\label{eq:endef}
\enbar=\sqrt{\frac{1-4 M/p+4 M^{2}/a p}{D_{4}}}\;,\,
\elbar=\sqrt{\frac{M p}{D_{4}}}\;,
\end{equation}
where
\begin{equation}
\label{eq:d4def}
D_{4}=1-\frac{4 M}{p}+\frac{M}{a}\;.
\end{equation}
The radial velocity \eqref{eq:drdtau} is a cubic equation with three roots, 
two of which are $r_{\text{min}}$ and $r_{\text{max}}$.  The third root 
is \cite{darwin61}
\begin{equation}
\label{eq:r3def}
r_{3}=\frac{2 M p}{p-4 M}\;.
\end{equation}
For stability, we need $r_{3}<r_{\text{min}}$, because then the orbiting 
mass moves between $r_{\text{min}}$ and $r_{\text{max}}$ in a ``valley'' of 
the potential $V$ \cite{cut94}, \cite{darwin61}.  As discussed in 
these references, requiring $r_{3}<r_{\text{min}}$ leads to
\begin{equation}
\label{eq:pmin}
p>2 M(3+e)\;,
\end{equation}
which, after substitution into $r_{\text{min}}$ \eqref{eq:rminrmax}, gives
\begin{equation}
\label{eq:rminmin}
r_{\text{min}}>\frac{2 M(3+e)}{1+e}>4 M\;.
\end{equation}
Accordingly, the periastron of a stable elliptic orbit must be greater 
than $4 M$, which is approached only in the limit $e\to 1$ \cite{cut94}.

Applying the chain rule of differentiation, we combine \eqref{eq:enbelb} 
and \eqref{eq:drdtau} to obtain expressions for $\frac{d\phi}{d r}$ 
and $\frac{d t}{d r}$.  Integrating these with respect to $r$ gives
\begin{equation}
\label{eq:thatdef}
\hat{t}(r)=\enbar\int_{r_{\text{min}}}^{r}
\;\frac{d r^{\prime}}{\left(1-\frac{2 M}{r^{\prime}}\right)
\sqrt{\smash[b]{\enbar^{2}-V(r^{\prime})}}}\;,
\end{equation}
\begin{equation}
\label{eq:phatdef}
\hat{\phi}(r)=\elbar\int_{r_{\text{min}}}^{r}
\;\frac{d r^{\prime}}{{r^{\prime}}^{2}
\sqrt{\smash[b]{\enbar^{2}-V(r^{\prime})}}}\;.
\end{equation}
These integrals and their derivation are from \cite{cut94}.  The hats 
are used in \cite{cut94} to indicate that $t$ and $\phi$ are calculated 
along the orbit as $r$ increases from $r_{\text{min}}$ to $r_{\text{max}}$.  
Different formulae are needed for the return trip, during which $r$ 
decreases from $r_{\text{max}}$ to $r_{\text{min}}$.  As shown in 
\cite{cut94}, the coordinates $t$ and $\phi$ for a single elliptic 
orbit are given by
\begin{gather}
t=\hat{t}\;,\;\phi=\hat{\phi}\quad \quad 
r_{\text{min}}\;\text{to}\;\,r_{\text{max}}\;,\notag
\\t=P-\hat{t}\;,\;\phi=\Delta\phi-\hat{\phi}\quad \quad 
r_{\text{max}}\;\text{back to}\;\,r_{\text{min}}\;.
\label{eq:tpup}
\end{gather}
The radial period $P=2\,\hat{t}(r_{\text{max}})$ is the coordinate time 
for a single orbit, from periastron to the next periastron.  The 
periastron advance $\Delta\phi=2\,\hat{\phi}(r_{\text{max}})$ 
is the change in angular position from periastron to periastron.  
The derivation of \eqref{eq:tpup} in \cite{cut94} takes into account 
the two signs of the radial velocity \eqref{eq:drdtau}.  

Newtonian elliptic orbits are closed, with $\Delta\phi=2\pi$.  
Relativistic elliptic orbits are not closed and have $\Delta\phi>2\pi$, 
which goes to $2 \pi$ only in the weak-field Newtonian limit.  This 
point is discussed extensively by Cutler and his collaborators in 
\cite{cut94}.  They show that the radial coordinate $r$ has period $P$ 
for a single orbit, but the angular coordinate $\phi$ does not, because 
the orbits are not closed.  Instead, they prove that the quantity 
$\phi-\Omega_{\phi}\,t$ has period $P$.  
As a result, elliptic orbits can be described by two fundamental 
frequencies \cite{cut94}:
\begin{equation}
\label{eq:twofreq}
\Omega_{\phi}=\frac{\Delta\phi}{P}\;,\,\Omega_{r}=\frac{2\pi}{P}\;.
\end{equation}
These two orbital frequencies will reappear when we calculate 
the Fourier transform of the stress energy tensor for elliptic orbits in 
Chapter \ref{tmunuchap}.

The integrals $\hat{t}$ \eqref{eq:thatdef} and $\hat{\phi}$ 
\eqref{eq:phatdef} can be evaluated in several ways.  The main problem 
is that the denominators contain the radial velocity \eqref{eq:drdtau}, 
which is zero at the turning points.  This apparent singularity 
can be removed by a change of variable \cite{cut94}, \cite{darwin61}.  
One possibility is to replace $r$ with the eccentric anomaly $\psi$, 
defined by \cite{darwin61} as
\begin{equation}
\label{eq:rtopsi}
r=a(1-e\cos\psi)\;,\,dr=a e \sin\psi\;d\psi\;,\,
0\le\psi\le 2\pi\;.
\end{equation}
In terms of $\psi$, the radial velocity \eqref{eq:drdtau} is
\begin{equation}
\label{eq:potpsi}
\textstyle{\big\lvert\frac{dr}{d\tau}\big\rvert}
=\sqrt{\enbar^{2}-V}=\frac{a e}{2 M}\sin\psi
\left(\frac{\left(2 M\right)^{3}}{r^{3}}
\frac{\left(1-e\right)\left(\frac{p}{2 M}-3-e\right)
+2 e\left(\frac{p}{2 M}-2\right)\sin^{2}\frac{\psi}{2}}
{2\left(\frac{p}{2 M}\right)-3-e^{2}}\right)^{1/2}\;,
\end{equation}
on the interval $0\le\psi\le\pi$.  At the turning points, 
$\textstyle{\big\lvert\frac{dr}{d\tau}\big\rvert}=0$, because 
$\sin\psi=0$ there.  When \eqref{eq:rtopsi}-\eqref{eq:potpsi} are 
inserted into the integrals \eqref{eq:thatdef}-\eqref{eq:phatdef}, 
the numerator factor of $ae\sin\psi$ from $dr$ cancels the denominator 
factor of $ae\sin\psi$ from 
$\textstyle{\big\lvert\frac{dr}{d\tau}\big\rvert}$, 
preventing a singularity at the turning points.  The substitution 
also allows \eqref{eq:thatdef}-\eqref{eq:phatdef} to be used for 
circular orbits, for which $e=0$ and $\frac{dr}{d\tau}=0$.  
Another choice is
\begin{equation}
\label{eq:rtochi}
r=\frac{p}{1+e\cos\chi}\;,\,0\le\chi\le 2\pi\;,
\end{equation}
Darwin calls $\chi$ the ``relativistic anomaly'' \cite{darwin61}.  
Expressions for $\hat{t}$ and $\hat{\phi}$ in terms of $\chi$ are 
given in \cite{cut94}, based on Darwin's work \cite{darwin61}.  
The integrals also can be evaluated using elliptic integrals 
\cite{ashby86}.  For example, the periastron advance $\Delta\phi$ is 
\begin{equation}
\label{eq:delphiint}
\Delta\phi=\frac{4}{\sqrt{D_{6}}}K(m_{k})\;,
\end{equation}
where $K$ is the complete integral of the first kind \cite{hmf}, 
\cite{ashby86}, \cite{cut94}.  The definitions needed for 
\eqref{eq:delphiint} are
\begin{equation}
\label{eq:d6def}
D_{6}=1-2 M(3-e)/p\;,\;m_{k}=\frac{4 M e}{p D_{6}}\;.
\end{equation}
Elliptic integrals can be evaluated using the methods in \cite{hmf}, or 
with Carlson's elliptic integrals \cite{carl79}, \cite{numr}.  The 
elliptic integrals expressions in \cite{ashby86} give $t=0$ and $\phi=0$ 
at apastron.  Using the formulae from \cite{hmf} and \cite{carl79}, the 
expressions for $\hat{t}$ \eqref{eq:thatdef} and $\hat{\phi}$ 
\eqref{eq:phatdef} also can be written in terms of elliptic integrals 
which are zero at periastron.

The radial coordinate $r$ and angle $\phi$ are related by \cite{ashby86}
\begin{equation}
\label{eq:arofphi}
r=\frac{p}{1-e+2\,e\;\text{sn}^{2}
\left[\frac{1}{2}\sqrt{D_{6}}(\phi-\Delta\phi/2)\,;\,m_{k}\right]}\;,
\end{equation}
where Ashby's notation is modified to agree with this thesis.  
Here, sn is a Jacobian elliptic function.  Its angular argument is 
zero at apastron.  With \eqref{eq:delphiint}, this argument can be 
written as $u-K$, where $u=\sqrt{D_{6}}\,\phi/2$.  An alternative 
formulation is
\begin{equation}
\label{eq:profphi}
r=\frac{p}{1-e+2\,e\;\text{cd}^{2}
\left[\sqrt{D_{6}}\,\phi/2\,;\,m_{k}\right]}\;,
\end{equation}
which is obtained from \eqref{eq:arofphi} using the relation 
$\text{sn}(u-K)=-\text{cd}\,u$ (from 16.8.1 of \cite{hmf}).  
The function cd is the quotient of the elliptic functions cn and dn, 
and their angular argument is zero at periastron.  The Jacobian elliptic 
functions are calculated using the routine \textit{sncndn} from 
\textit{Numerical Recipes} \cite{numr}.

A circular orbit of constant radius $R$ has $e=0$, $r_{\text{max}}=r_{\text{min}}$ 
and $p=a=R$ \cite{chandra92}.  Stable circular orbits can be treated as 
a special case of elliptic orbits.  For zero eccentricity, the condition 
\eqref{eq:pmin} becomes $p> 6 M$.  This means that the innermost 
stable circular orbit (ISCO) has radius $R=6 M$ \cite{cut94}, 
\cite{drmwd04}.  It is possible to 
have unstable circular orbits of radius $3 M<R<6 M$ \cite{chandra92}, 
but we will not consider them further.  For circular orbits, the 
constants $\enbar$ and $\elbar$ \eqref{eq:endef} become
\begin{equation}
\label{eq:cendef}
\enbar=\frac{1-\frac{2 M}{R}}{\sqrt{1-\frac{3 M}{R}}}\;,\,
\elbar=\sqrt{\frac{M R}{1-\frac{3 M}{R}}}\;.
\end{equation}
The orbital angular frequency $\frac{d\phi}{dt}$ is the same as the 
Newtonian Kepler rule \cite{mtw73}
\begin{equation}
\label{eq:ciromeg}
\frac{d\phi}{dt}=\frac{d\phi}{d\tau}\frac{d\tau}{dt}
=\frac{\elbar}{R^{2}}\frac{\left(1-\frac{2 M}{R}\right)}{\enbar}
=\sqrt{\frac{M}{R^{3}}}\;,
\end{equation}
where we have used \eqref{eq:enbelb} and \eqref{eq:cendef}.  The 
derivative $\frac{d\phi}{dt}$ is constant for circular orbits, unlike 
elliptic orbits with non-zero eccentricity.  Applying the definition 
of $\Omega_{\phi}$ \eqref{eq:twofreq} to circular orbits gives
\begin{equation}
\label{eq:ciromeg2}
\Omega_{\phi}=\sqrt{\frac{M}{R^{3}}}\;,
\end{equation}
which is equal to $\frac{d\phi}{dt}$ \eqref{eq:ciromeg}.  
Because $\frac{d\phi}{dt}=\Omega_{\phi}$ is constant, we have \cite{pois93}
\begin{equation}
\label{eq:cirtp}
t=\Omega_{\phi}\,\phi\;,
\end{equation}
for circular orbits.

\section{\label{sec:eqsf}Equations for Gravitational Self-Force}

The formalism of the gravitational self-force was derived by Mino, 
Sasaki and Tanaka \cite{mst97} and by Quinn and Wald \cite{qwald97} 
in the harmonic gauge.  
Their derivations followed work on the electromagnetic and scalar 
self-forces by DeWitt and Brehme \cite{dewittbr60} and Hobbs 
\cite{hobbs68}.  The formalism and derivations are described 
in a lengthy review article by Poisson \cite{poisslrr}.  

\begin{figure}[htbp]
\begin{center}
\includegraphics*[width=4.98in]{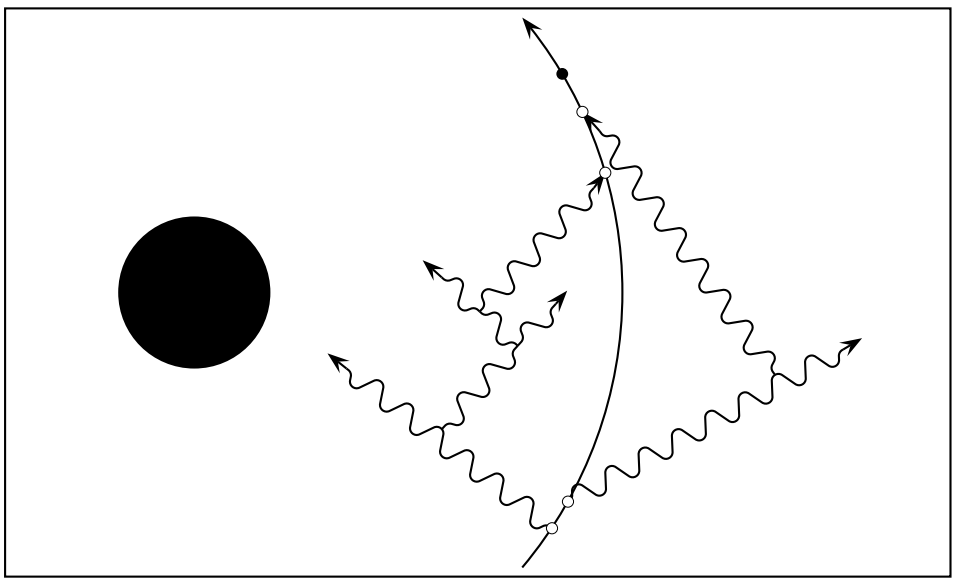}
\caption[Schematic of Self-Force for Circular Orbit]{\label{fig:sch}
Schematic of self-force for circular orbit.  A small mass $\mz$, 
which is represented by the small solid circle, orbits a much larger 
black hole.  Hollow circles represent previous positions of $\mz$.  The 
wavy lines represent four-dimensional gravitational waves which scatter 
off the background spacetime curvature.  Part radiates to infinity or 
into the central mass, and part returns to $\mz$, giving rise to the tail 
term of the self-force.} 
\end{center}
\end{figure}

Black hole perturbation theory treats the orbiting mass $\mz$ as 
a point mass.  This causes the perturbation to diverge at the 
location of $\mz$, which is where the force must be evaluated.  A 
further problem is that general relativity does not have point 
masses as such, but instead predicts black holes.  
As shown in the references above, the perturbation can be 
broken into two pieces, a direct part and a tail part:
\begin{equation}
\label{eq:dirtail}
h_{\mu\nu}=h_{\mu\nu}^{\text{dir}}+h_{\mu\nu}^{\text{tail}}\;.
\end{equation}
The direct part is the divergent part.  It is the relativistic analogue 
of the singularity 
in the Newtonian potential.  The tail part is an integral over the 
prior history of the orbiting mass.  A wave propagating in a curved 
spacetime will scatter off the background curvature, rather than 
propagating as a sharp pulse.  It develops a ``tail'' and may 
subsequently interact with the generating mass.  
The interaction between $\mz$, its field and the 
background spacetime curvature gives rise to the gravitational self-force.
A schematic of the tail term interaction is shown in Figure \ref{fig:sch}.  
Further discussion of the diagram is in section \ref{sec:nzrweqn}, 
in the explanation of iterative integral solutions.  

The tail term is not a homogeneous solution of the harmonic gauge field 
equations.  Detweiler and Whiting derived an alternative formulation 
of the self-force \cite{dw03}.  They also divide the perturbation into 
two pieces:
\begin{equation}
\label{eq:dwrs}
h_{\mu\nu}=h_{\mu\nu}^{S}+h_{\mu\nu}^{R}\;.
\end{equation}
The first term is the singular part.  The second is the regular part, 
which is a homogeneous solution of the field equations and which gives 
rise to the self-force.  It also is based on the prior history of the 
orbiting mass.  In this formulation, the small mass moves on a geodesic 
of the perturbed spacetime $g_{\mu\nu}+h_{\mu\nu}^{R}$.  
In Chapters~\ref{oddpar} and~\ref{evpar}, we derived homogeneous 
solutions to the harmonic gauge field equations.  The homogeneous 
perturbation $h_{\mu\nu}^{R}$ must be a linear combination of those 
homogeneous solutions, which suggests that the self-force is due 
to the non-zero frequency spin $2$ solutions.

The gravitational self-force gives the first order perturbative 
corrections to the background geodesic equations of motion, as discussed 
in the references above.  To see this, rewrite equation 
\eqref{eq:bckgdgeo} as 
\begin{equation}
\label{eq:aself}
\mz\left({\frac{d^2 z}{d\tau^2}}^{\!\mu}
+\mathnormal{\Gamma}^{\mu}{}_{\!\alpha\beta}
{\frac{dz}{d\tau}}^{\!\alpha}{\frac{dz}{d\tau}}^{\!\beta}\right)
=F^{\mu}_{\text{self}}\;,
\end{equation}
where
\begin{equation}
\label{eq:selfa}
F^{\mu}_{\text{self}}=-\mz\left(\delta^{\mu}_{\gamma}
+{\frac{dz}{d\tau}}^{\!\mu}{\frac{dz}{d\tau}}\!{}^{\!\gamma}\right)
\delta\!\mathnormal{\Gamma}^{\gamma}{}_{\!\alpha\beta}
{\frac{dz}{d\tau}}^{\!\alpha}{\frac{dz}{d\tau}}^{\!\beta}
\end{equation}
and
\begin{equation}
\label{eq:deltareg}
\delta\!\mathnormal{\Gamma}^{\gamma}{}_{\!\alpha\beta}
=\frac{1}{2}\,g^{\gamma\epsilon}\left(h^{\text{reg}}_{\epsilon\alpha;\beta}
+h^{\text{reg}}_{\epsilon\beta;\alpha}
-h^{\text{reg}}_{\alpha\beta;\epsilon}\right)\;.
\end{equation}
Here, ``reg'' refers to the regular part of the perturbation.  
In practice, it is difficult to calculate the regular part, because it 
involves an integral over prior history.  Instead, we regularize 
the bare force using ``mode-sum'' regularization.  The expression 
for the bare force is
\begin{equation}
\label{eq:fbare}
F^{\mu}_{\text{bare}}=-m_{0}\, \left(\delta^{\mu}_{\gamma}
+{\frac{dz}{d\tau}}^{\!\mu}{\frac{dz}{d\tau}}\!{}^{\!\gamma}\right)
\!\frac{1}{2}\,g^{\gamma\epsilon}\left(h^{}_{\epsilon\alpha;\beta}
+h^{}_{\epsilon\beta;\alpha}-h^{}_{\alpha\beta;\epsilon}\right)
{\frac{dz}{d\tau}}^{\!\alpha}{\frac{dz}{d\tau}}^{\!\beta}\;.
\end{equation}
This is decomposed into multipoles and referred to as $F^{\text{full}}$ 
below, in order to better track the notation of the references below.

The method of mode-sum regularization is described in \cite{bmnos02}.  
The derivations behind \cite{bmnos02} are in \cite{bo02s}, \cite{bo02g} 
and \cite{mns02}.  The summary below is taken mainly from these 
references.  The basic idea behind mode-sum regularization is this.  
Decompose the unregularized bare force $F^{\text{full}}_{\alpha}$ into 
a sum over individual modes of spherical harmonic index $l$, so that
\begin{equation}
\label{eq:flsum}
F^{\text{full}}_{\alpha}=\sum_{l=0}^{\infty}F^{\text{full}}_{\alpha l}\;.
\end{equation}
Each mode $F^{\text{full}}_{\alpha l}$ represents a sum over the index $m$ for 
that particular $l$.  Although the total force $F^{\text{full}}_{\alpha}$ 
diverges at the location of the orbiting mass, each separate mode 
$F^{\text{full}}_{\alpha l}$ is finite.  It is only the sum over $l$ that 
diverges.  The individual $l$-modes can be regularized by subtracting 
the divergent part.  Equivalently, the regularized self-force 
$F^{\text{self}}_{\alpha}$ is
\begin{equation}
\label{eq:abcd}
F^{\text{self}}_{\alpha}=\lim_{x\to z_{0}}\left[F^{\text{full}}_{\alpha}(x)
-F^{\text{dir}}_{\alpha}(x)\right]\;.
\end{equation}
where $F^{\text{dir}}_{\alpha}$ is the divergent part, $z_{0}$ is the 
position of the orbiting mass and $x$ is a field point in the 
neighborhood of $z_{0}$.  Expressed as a sum, the regularized self-force is 
\begin{equation}
\label{eq:fselfreg}
F^{\text{self}}_{\alpha}=\sum^{\infty}_{l=0}\left[\lim_{x\to z_{0}}F^{\text{full}}_{\alpha l}
-A_{\alpha}L-B_{\alpha}-C_{\alpha}/L\right]-D_{\alpha}\;,
\end{equation}
where $L=l+1/2$.  The quantities $A_{\alpha}$, $B_{\alpha}$, $C_{\alpha}$ 
and $D_{\alpha}$ are the so-called ``regularization parameters'' and 
are independent of $l$.  Each $l$-mode of the direct force has the form
\begin{equation}
\label{eq:fdirreg}
\lim_{x\to z_{0}}F^{\text{dir}}_{\alpha l}=A_{\alpha}L+B_{\alpha}+C_{\alpha}/L+O(L^{-2})\;.
\end{equation}
The first three terms on the right of \eqref{eq:fdirreg} are easily 
identified in \eqref{eq:fselfreg}.  The parameter $D_{\alpha}$ represents 
the sum over all $l$ of the $O(L^{-2})$ terms.  The $O(L^{-2})$ terms are 
not individually zero; however, their sum over all $l$ is.  Thus, 
$D_{\alpha}=0$.  Also, $C_{\alpha}=0$ for each individual mode.  The 
remaining parameters are
\begin{equation}
\label{eq:thsc}
A^{\text{sc}}_{\theta}=B^{\text{sc}}_{\theta}=0\;,
\end{equation}
\begin{equation}
\label{eq:asc}
A^{\text{sc}}_{\pm t}=\pm \frac{q^2}{r^2}\frac{u^{r}}{V}\;,
A^{\text{sc}}_{\pm r}=\mp \frac{q^2}{r^2}\frac{\enbar}{\ff V}\;,
A^{\text{sc}}_{\phi}=0\;,
\end{equation}
\begin{equation}
\label{eq:bsct}
B^{\text{sc}}_{t}=\frac{q^2}{r^2}
\frac{\enbar u^{r}\left[K(w)-2 E(w)\right]}{\pi V^{3/2}}\;,
\end{equation}
\begin{equation}
\label{eq:bscr}
B^{\text{sc}}_{r}=\frac{q^2}{r^2}
\frac{\big[(u^{r})^{2}-2 \enbar^{2}\big]K(w)
+\big[(u^{r})^{2}+\enbar^{2}\big]E(w)}{\pi\ff V^{3/2}}\;,
\end{equation}
\begin{equation}
\label{eq:bscphi}
B^{\text{sc}}_{\phi}=\frac{q^2}{r}
\frac{u^{r}\left[K(w)-E(w)\right]}
{\pi(\elbar/r)V^{1/2}}\;.
\end{equation}
In these expressions, $r$ is the radial coordinate of the orbiting mass, 
$\elbar$ is the specific angular momentum, 
$w=\elbar^{2}/(\elbar^{2}+r^{2})$, $V=1+\elbar^{2}/r^{2}$, and 
$u^{r}=\frac{dr}{d\tau}$.  The functions $K(w)$ and $E(w)$ are complete 
elliptic integrals of the first and second kinds \cite{hmf}.  
The different signs in $A^{\text{sc}}_{\pm t}$ and $A^{\text{sc}}_{\pm r}$ 
depend on the direction in which the limit $x\to z_{0}$ is taken.  
The upper sign means the limit is taken along the ingoing radial direction 
(from outside the orbit for a circular orbit); the lower, along the 
outgoing radial direction (from inside the orbit for a circular orbit).  
The parameters apply to any geodesic motion in the equatorial plane.  
For circular orbits, inspection of the parameters shows that only 
$F^{r}$ needs to be regularized, because $u^{r}=0$.

The superscript ``sc'' is short for ``scalar'', which requires some 
additional explanation.  The parameters above were first derived to 
regularize the self-force of a fictitious scalar charge $q$.  The 
scalar field equation is
\begin{equation}
\label{eq:scaleqn}
\Phi_{;\alpha}{}^{;\alpha}=-4\pi\rho\;,
\end{equation}
where $\Phi$ is the scalar field and where the source is
\begin{equation}
\label{eq:scalsour}
\rho=q \int_{\!-\infty}^{\infty}\frac{\delta^{4}(x-z(\tau))}{\sqrt{-g}}\,d\tau\;.
\end{equation}
The scalar bare force is
\begin{equation}
\label{eq:scalfor}
F_{\alpha}^{\text{sc}}=q\,\Phi_{,\alpha}\;.
\end{equation}
The scalar field is specified by the single differential 
equation \eqref{eq:scaleqn}, which is simpler to solve than the ten 
coupled gravitational field equations of perturbation theory.  The 
scalar field is analogous, because it also has a wave equation with 
a delta function source.  It was simpler to derive the scalar 
regularization parameters first \cite{bo02s}, \cite{bo02g} and 
\cite{mns02}.  

However, we do not wish to calculate the scalar self-force, because 
it does not describe an actual, physical field.  Instead, we wish to find  
the gravitational self-force due to the metric perturbation $h_{\mu\nu}$.  
The gravitational parameters (``gr'') are related to the scalar 
parameters by \cite{bmnos02}
\begin{equation}
\label{eq:sctogr}
A^{\text{gr}}_{\alpha}=A^{\text{sc}}_{\alpha}\,,\,
B^{\text{gr}}_{\alpha}=\left(\delta^{\lambda}_{\alpha}+u_{\alpha}u^{\lambda}\right)
B^{\text{sc}}_{\lambda}\,,\,C^{\text{gr}}_{\alpha}=D^{\text{gr}}_{\alpha}=0\;,
\end{equation}
where $u^{\lambda}$ are the four-velocity components.  We also replace 
$q$ in \eqref{eq:thsc}-\eqref{eq:bscphi} with $\mz$.  The bare force 
is calculated using the unregularized metric perturbation $h_{\mu\nu}$.

The regularization parameters above were calculated in the harmonic 
gauge.  The gravitational self-force is gauge dependent, as shown by 
Barack and Ori \cite{bo01}.  The gauge dependence reflects the principle of 
equivalence \cite{sago03},~n.~18.  The regularization may be different 
in another gauge.  For example, the regularization in the 
Regge-Wheeler gauge is the same for radial infall, but different and 
impractical for circular orbits \cite{bl05}.  

The expression for the direct force \eqref{eq:fdirreg} does not 
include regularization parameters for the $O(L^{-2})$ terms.  This 
is because the sum of these terms over all $l$ is zero, even though 
individually these higher order terms are non-zero.  When summed 
over $l$, the difference $F^{\text{full}}_{\alpha l}-F^{\text{dir}}_{\alpha l}$ 
converges slowly, as $O(L^{-2})$.  The speed of convergence may be 
accelerated if the higher order terms are included.  A procedure 
for doing so is described by Detweiler and his collaborators, who 
used it to calculate the scalar self-force for circular orbits 
\cite{dmw03}.  Their method is described below.  To the direct force, 
add terms of the form
\begin{equation}
\label{eq:hotlterms}
E^{k}_{\alpha}\mathcal{A}^{k+1/2}_{l}\;,\,k=1,2,\dots\;,
\end{equation}
where
\begin{equation}
\label{eq:hotltermsa}
\mathcal{A}^{k+1/2}_{l}=\frac{(2 l+1)\mathcal{P}_{k+1/2}}
{(2 l-2 k-1)(2 l-2 k+1)\cdots(2 l+2 k+1)(2 l+2 k+3)}
\end{equation}
and
\begin{equation}
\label{eq:hotltermsp}
\mathcal{P}_{k+1/2}=(-1)^{k+1}2^{k+3/2}
\left[(2 k+1)!!\right]^{2}\;.
\end{equation}
For a given $k$, $E^{k}_{\alpha}\mathcal{A}^{k+1/2}_{l}$ is 
$O(L^{-2 k})$.  The parameters $E^{k}_{\alpha}$ are independent of $l$, so
\begin{equation}
\label{eq:teles}
\sum_{l=0}^{\infty}E^{k}_{\alpha}\mathcal{A}^{k+1/2}_{l}=0\;,
\end{equation}
which follows from the form of $\mathcal{A}^{k+1/2}_{l}$.  
Detweiler and his collaborators derived an analytical 
scalar force expression for $E^{1}_{r}$.  They determined higher order 
parameters by numerical fit to the force modes for larger $l$, after 
subtraction of the analytic regularization parameters.  

The regularization parameters were derived by expanding the direct 
force in scalar spherical harmonics.  However, the bare force is 
calculated numerically from the perturbation $h_{\mu\nu}$, which is 
expressed in terms of tensor harmonics.  Because the 
regularization subtraction is implemented $l-$mode by $l-$mode, it is 
necessary to convert the bare force modes from tensor harmonics to 
spherical harmonics.  Detailed formulae for doing so were not published 
until 2007 \cite{basa07}, where different notation is used for the 
metric perturbation.  However, the numerical calculations of self-force 
in Chapter~\ref{numchap} do not use the expressions in \cite{basa07}.  
Instead, the numerical results use angular expressions derived from 
brief hints in articles published several years earlier 
\cite{bmnos02}, \cite{bo02g}.  An example of the angular expressions is
\begin{multline}
\label{eq:dyrule}
\sin\theta\frac{\partial Y_{lm}(\theta,\phi)}{\partial\theta}
=-(l+1)\sqrt{\frac{(l-m) (l+m)}{(2 l-1) (1+2 l)}}\; Y_{l-1\, m}(\theta,\phi)
\\+l\sqrt{\frac{(1+l-m) (1+l+m)}{(1+2 l) (3+2 l)}}\;Y_{l+1\, m}(\theta,\phi)\;,
\end{multline}
which appears in the odd parity metric perturbation \eqref{eq:ohmunu}.  
It is derived from the definition of spherical harmonics and the 
recursion relations for associated Legendre polynomials \cite{arfken85}.  
The bare force contains more complicated angular expressions, which 
are evaluated with the help of triple spherical harmonic integrals 
\cite{arfken85}, \cite{zare88}.  The angular conversion formulae are 
lengthy and will not be set forth here.

Chapter~\ref{numchap} contains numerical calculations of the 
gravitational self-force for circular orbits using mode-sum regularization.  
The bare perturbation is calculated using the harmonic gauge solutions 
derived in Chapters~\ref{oddpar} and~\ref{evpar}.  Convergence of the 
regularization series is accelerated using the method of Detweiler 
and his collaborators \cite{dmw03}.  Earlier this year, Barack and 
Sago published calculations of the self-force for circular orbits 
\cite{basa07}.  They calculated the harmonic gauge metric perturbation 
in the time domain, using a method designed mainly by Barack and 
Lousto \cite{bl05}.  They solved the field equations directly, instead 
of doing the analytic calculations described in 
Chapters~\ref{oddpar} and~\ref{evpar}.  A comparison is made to 
their numerical results in Chapter~\ref{numchap}.

Barack and Sago also give expressions relating the self-force to 
parameters of the orbital motion for circular orbits of radius $R$.  
Their analysis and results are summarized below.  The self-force has 
two aspects: the dissipative, or radiation reaction part, and the 
conservative part.  The dissipative part is the rate of energy and 
angular momentum loss to the gravitational waves.  These rates are given by
\begin{equation}
\label{eq:edotldot}
\frac{d\enbar}{d\tau}=-\mz^{-1}F_{t}\;,\,
\frac{d\elbar}{d\tau}=\mz^{-1}F_{\phi}\;.
\end{equation}
Using the definitions of $\frac{dt}{d\tau}$ \eqref{eq:enbelb} and $\enbar$ 
and the relation $F^{t}=g^{tt}F_{t}$, we can rewrite the first equation 
in \eqref{eq:edotldot} as
\begin{equation}
\label{eq:edotsf4}
\dot{E}^{\text{sf}}\equiv\frac{d E}{d t}=\left(1-\frac{2 M}{R}\right)^{2}
\frac{F^{t}}{\enbar}\;,
\end{equation}
which will be used for numerical calculations in Chapter \ref{numchap}.  
The conservative part gives non-radiative corrections to the orbital 
parameters and is attributable to the radial component of the self-force.  
The orbital frequency is changed by
\begin{equation}
\label{eq:omegcorr}
\Omega=\Omega_{0}\left[1-\left(\frac{R(R-3 M)}{2 M \mz}\right)F_{r}\right]\;,\,
\Omega_{0}=\sqrt{\frac{M}{R^{3}}}=\Omega_{\phi}\;.
\end{equation}
The change in orbital frequency reflects the fact that both bodies are 
moving around the center of mass, instead of test mass motion where 
the central mass is fixed \cite{dp04}.  There are also non-radiative 
corrections to the energy and orbital angular momentum~\cite{basa07}.

Because the gravitational self-force is gauge dependent, it is necessary 
to identify gauge invariant quantities, which represent physical observables 
\cite{det05}.  As discussed by Barack and Sago \cite{basa07}, three gauge 
invariant quantities are $\frac{d\enbar}{d\tau}$ \eqref{eq:edotldot}, 
$\frac{d\elbar}{d\tau}$ \eqref{eq:edotldot} and $\Omega$ 
\eqref{eq:omegcorr}.  There is also a gauge invariant relation between 
the energy and angular momentum \cite{basa07}, \cite{det05}.  
Finally, the corrections to the orbital motion must be incorporated 
into the gravitational waveforms.  In order to do this in a gauge 
invariant manner, it is necessary to extend perturbation theory to 
second order in the mass ratio $\mz/M$, which is beyond the scope of this 
thesis \cite{poisslrr}, \cite{rose05}-\cite{rose06}.  

In subsection \ref{sec:zevpareqz}, we derived solutions for the zero 
frequency, $l=0$ multipole.  The solutions derived there differ 
from the Detweiler-Poisson result for circular orbits~\cite{dp04}.  
The different solutions are related by the gauge transformation 
\eqref{eq:todpm1}, which preserves the harmonic gauge and which 
represents a change $\xi^{r}$ in the radial 
coordinate.  After substituting $\bmo$ \eqref{eq:todpm1} into 
$\xi_{\mu}$ \eqref{eq:chimu}, we find that
\begin{equation}
\label{eq:xirtodp}
\xi^{r}=g^{rr}\xi_{r}=\ff\bmo Y_{00}(\theta,\phi)
=\mz\enbar\,\frac{\left(r-2 M\right)\left(4 M^{2}+2 M r+r^{2}\right)}
{r^{2}\left(R-2 M\right)}\;.
\end{equation}
In terms of the coordinate change relation 
$x^{\mu}_{\text{new}}=x^{\mu}_{\text{old}}+\xi^{\mu}$ 
\eqref{eq:xnew}, we have
\begin{equation}
\label{eq:rnewdp}
r_{\text{new}}=r_{\text{old}}+\xi^{r}\;.
\end{equation}
In this expression, $r_{\text{new}}$ is the radial coordinate for the 
gauge used by Detweiler and Poisson and $r_{\text{old}}$ is the 
radial coordinate for the gauge used to derive the solutions in this 
thesis.  At the orbital radius $R$, equation \eqref{eq:xirtodp} 
simplifies to
\begin{equation}
\label{eq:xirtodpr}
\xi^{r}=\mz\enbar\,\frac{\left(4 M^{2}+2 M R+R^{2}\right)}
{R^{2}}\;.
\end{equation}
From \eqref{eq:mydpfrdiff}, this gauge change alters the self-force by
\begin{equation}
\label{eq:mydpfrdiff4}
F_{r}^{\text{new}}=F_{r}^{\text{old}}-m_{0}^2\enbar\,\frac{3 M 
\left(4 M^2+2 M R+R^2\right)}{(R-3 M) R^4}\;.
\end{equation}
Here, $F_{r}=g_{rr}F^{r}$, and ``new'' and ``old'' have the same meanings 
as in equation \eqref{eq:rnewdp}.  Although this gauge change affects 
the self-force, it will not change the value of the orbital frequency.  
To show this, it is helpful to rewrite \eqref{eq:omegcorr} as
\begin{equation}
\label{eq:omegcorr2}
\Omega^{2}=\frac{M}{R^{3}}-\frac{R-3 M}{R^{2}}\frac{F_{r}}{\mz}\;.
\end{equation}
The coordinate change $\xi^{r}$ affects the first term.  The force change 
affects the second term, but with the opposite sign.  The two changes cancel 
to order $\mz$, leaving $\Omega$ unchanged to that order.  
The zero frequency, $l=0$ multipole does not contribute to $F^{t}$ 
and $F^{\phi}$, so equation \eqref{eq:edotldot} implies that 
$\frac{d\enbar}{d\tau}$ and $\frac{d\elbar}{d\tau}$ also are unaffected 
by the change of gauge.  At present, it is not possible to calculate 
explicitly the effect on the waveforms, but they should be invariant 
\cite{poisslrr}.  Based on the discussion above and in Barack and 
Sago \cite{basa07}, the difference in solutions should not affect the 
gauge invariant physical observables.			
\chapter{\label{tmunuchap}Calculation of the Stress Energy 
Tensor for a Point Mass}

In this chapter, we derive the components of the stress energy tensor 
for an orbiting point mass $m_{0}$.  The stress energy tensor contains 
information about the position and velocity of the orbiting mass.  Its 
components are the source terms for the field equations.  The main result 
of this chapter is the calculation of the radial coefficients (such as 
$Se^{lm}_{00}(\omega,r)$) of the angular functions in equations 
\eqref{eq:otmunu} and \eqref{eq:etmunu}.

The standard stress energy tensor for a point mass is 
\begin{equation}
T^{\mu\nu}=m_{0}\int_{\!-\infty}^{\infty}
\frac{\delta^{4}(x-z(\tau))}{\sqrt{-g}}\,{\frac{dz}{d\tau}}^{\!\mu}
{\frac{dz}{d\tau}}^{\!\nu}\,d\tau\,\,.
\end{equation}
This expression is used for the stress energy tensor because the divergence 
equation $T^{\mu\nu}\!{}_{;\nu}=0$ gives the geodesic equation of motion with 
respect to the background spacetime \cite{det05}, \cite{poisslrr}, \cite{zerp70}.  
The vector $x$ represents a field point having coordinates 
$(t,r,\theta,\phi)$.  The spacetime coordinates of the orbiting mass 
are $z^{\mu}(\tau)$, where $\tau$ is the proper time.  The components 
of the vector $z$ are $(t^{\prime},r^{\prime},\thetap,\phip)$.  
The determinant of the background metric tensor is $g$, so that, 
for the Schwarzschild metric, $\sqrt{-g}=r^{2}\sin\theta$.

Following Zerilli \cite{zerp70}, we simplify the stress energy tensor as 
follows.  First, change the variable of integration 
from $\tau$ to $t^{\prime}$ using
\begin{equation}
\int_{\!-\infty}^{\infty}d\tau \to 
\int_{\!-\infty}^{\infty}dt^{\prime}\frac{d\tau}{dt^{\prime}}
=\int_{\!-\infty}^{\infty}\frac{dt^{\prime}}{\gamma}\;,
\end{equation}
where $\gamma=\frac{dt^{\prime}}{d\tau}$.  Using the chain rule to 
rewrite the velocities as 
${\frac{dz}{d\tau}}^{\!\nu}=\gamma {\frac{dz}{dt^{\prime}}}^{\!\nu}$, we have
\begin{equation}
T^{\mu\nu}=m_{0}\int_{\!-\infty}^{\infty}
\frac{\delta^{4}(x-z(t^{\prime}))}{\sqrt{-g}}\,{\frac{dz}{dt^{\prime}}}^{\!\mu}
{\frac{dz}{dt^{\prime}}}^{\!\nu}\gamma\, dt^{\prime}\,\,.
\end{equation}
We then integrate with the delta function $\delta(t-t^{\prime})$ to get
\begin{equation}
\label{eq:delttmunu}
T^{\mu\nu}=m_{0}\gamma\frac{\delta(r-r^{\prime}(t))
\delta^{2}(\Omega-\Omega^{\prime}(t))}{r^{2}}\dot{z}^{\mu}\dot{z}^{\nu}\;,
\end{equation}
where $\deltaom=\frac{\delta(\theta-\theta^{\prime})\delta(\phi-\phi^{\prime})}
{\sin \theta}$ and where we have 
defined $\dot{z}^{\nu}={\frac{dz}{dt}}^{\!\nu}$.  
The perturbed field equations \eqref{eq:hpertfeqn} use the covariant 
form of the stress energy tensor, so we need to lower indices with
\begin{equation}
\label{eq:covdelttmunu}
T_{\mu\nu}=g_{\mu\rho}g_{\nu\sigma}T^{\rho\sigma}\;,
\end{equation}
where $T^{\rho\sigma}$ is from \eqref{eq:delttmunu}.

Section~\ref{sec:ttmunu} explains the multipole decomposition of the covariant 
stress energy tensor \eqref{eq:covdelttmunu}.  
Section~\ref{sec:ftmunu} shows how to compute the Fourier transform 
of the stress energy tensor for circular and elliptic orbits.  

\section{\label{sec:ttmunu}Multipole Decomposition}

The angular delta function, $\deltaom$, contains the $\theta$ and $\phi$ 
dependence of the stress energy tensor, as given by equations 
\eqref{eq:delttmunu} and \eqref{eq:covdelttmunu}.  The multipole 
decomposition consists of expanding the delta function in terms of 
spin-weighted spherical harmonics, which are described below.  
The derivation in this section 
is done in the time domain, rather than using Fourier transforms.  
The Fourier transform of the stress energy tensor depends on the 
orbital motion, but the results derived in this section~\ref{sec:ttmunu} 
will be applicable to arbitrary orbital motion.

In the time domain, the multipole decomposition of the covariant 
stress energy tensor is 
\begin{equation}
\label{eq:timtmunu}
T_{\mu\nu}(t,r,\theta,\phi)=\sum_{l=0}^{\infty}\sum_{m=-l}^{l}
\left(T^{o,lm}_{\mu\nu}(t,r,\theta,\phi)
+T^{e,lm}_{\mu\nu}(t,r,\theta,\phi)\right)\;.
\end{equation}
Here, $T^{o,lm}_{\mu\nu}(t,r,\theta,\phi)$ and $T^{e,lm}_{\mu\nu}(t,r,\theta,\phi)$ 
are given by \eqref{eq:otmunu} and \eqref{eq:etmunu}, respectively, with 
the substitution $\omega\to t$.  The remainder of this section shows how to 
calculate the time-radial coefficients (such as $Se^{lm}_{00}(t,r)$) of the 
angular functions.  In doing so, we will convert the tensor 
harmonics, which are the angular functions used in \eqref{eq:timtmunu} 
(and \eqref{eq:otmunu}-\eqref{eq:etmunu}), to spin-weighted spherical 
harmonics.  We also will temporarily use a tetrad basis from the 
Newman-Penrose formalism, as discussed below.  The following 
derivation is different from the usual method.  
Zerilli and others used the orthogonality of the tensor 
harmonics to derive the coefficients, which requires evaluating 
integrals of inner products of the tensor 
harmonics \cite{ashby}, \cite{sago03}, \cite{zerp70}.  The method 
below is algebraic, does not require integration, and shows how 
the angular functions are derived from the delta function $\deltaom$.

The spin-weighted spherical harmonics are described in 
\cite{gj67}, \cite{gn67}, and \cite{np66}.  Relevant points from these 
references are summarized below.  The notation for the 
spin-weighted spherical harmonics is ${}_{s}Y_{lm}(\theta,\phi)$, 
where $s$ is the spin weight.  The familiar spherical harmonics 
have spin weight~$0$, that is, $Y_{lm}(\theta,\phi)={}_{0}Y_{lm}(\theta,\phi)$.  
We will consider only integral values of $s$, although the 
harmonics may be extended to half-integral spin weights.  Harmonics 
of different spin weight are related by raising and lowering 
operators.  The raising, or ``edth'' operator $\eth$, is defined as
\begin{equation}
\label{eq:edthop}
\eth\,{}_{s}Y_{lm}(\theta,\phi)=-(\sin\theta)^{s}\left[
\frac{\partial}{\partial\theta}+i\csc\theta\frac{\partial}{\partial\phi}
\right](\sin\theta)^{-s}{}_{s}Y_{lm}(\theta,\phi)\;.
\end{equation}
It increases spin weight by one, so that
\begin{equation}
\label{eq:ethup}
\eth\,{}_{s}Y_{lm}(\theta,\phi)
=\sqrt{(l-s)(l+s+1)}\;{}_{s+1}Y_{lm}(\theta,\phi)\;.
\end{equation}
The lowering operator $\overline{\eth}$ is
\begin{equation}
\label{eq:edthbar}
\overline{\eth}\,{}_{s}Y_{lm}(\theta,\phi)=-(\sin\theta)^{-s}\left[
\frac{\partial}{\partial\theta}-i\csc\theta\frac{\partial}{\partial\phi}
\right](\sin\theta)^{s}{}_{s}Y_{lm}(\theta,\phi)\;,
\end{equation}
which lowers spin weight by one as
\begin{equation}
\label{eq:ethdown}
\overline{\eth}\,{}_{s}Y_{lm}(\theta,\phi)=-\sqrt{(l+s)(l-s+1)}
\;{}_{s-1}Y_{lm}(\theta,\phi)\;.
\end{equation}
Using $\eth$ and $\overline{\eth}$, we can construct 
spin-weighted spherical harmonics of non-zero $s$ from the 
spherical harmonics $Y_{lm}(\theta,\phi)$.  
Equations \eqref{eq:edthop} and \eqref{eq:edthbar} imply that 
\begin{equation}
\label{eq:lgts}
{}_{s}Y_{lm}(\theta,\phi)=0\;,\text{ for }\lvert s\rvert>l\;.
\end{equation}
The spin-weighted spherical harmonics satisfy a second order 
differential equation, 
\begin{equation}
\label{eq:yseqn}
\overline{\eth}\eth\,{}_{s}Y_{lm}(\theta,\phi)=-(l-s)(l+s+1)
\;{}_{s}Y_{lm}(\theta,\phi)\;.
\end{equation}
The function $\eth\,{}_{s}Y_{lm}(\theta,\phi)$ has spin weight $s+1$, so 
$\overline{\eth}$ in \eqref{eq:yseqn} is calculated by applying 
\eqref{eq:edthbar} with the replacement $s\to s+1$, which leads to 
the operator expression
\begin{equation}
\label{eq:ddop}
\overline{\eth}\eth=\frac{\partial^{2}}{\partial\theta^{2}}
+\cot\theta\frac{\partial}{\partial\theta}-\frac{m^{2}}{\sin^{2}\theta}
-\frac{2 m s\cos\theta}{\sin^{2}\theta}-s^{2}\cot^{2}\theta+s\;.
\end{equation}
The harmonics form a complete set for angular functions of 
spin weight~$s$ on the unit sphere.  The completeness relation is 
\begin{equation}
\label{eq:compl}
\deltaom=\sum^{\infty}_{l\ge \lvert s\rvert}\,\sum^{l}_{m=-l}
{}_{s}\overline{Y}_{lm}(\thetap,\phip)\;{}_{s}Y_{lm}(\theta,\phi)\;.
\end{equation}
The overbar signifies complex conjugation.  
Harmonics of the same spin weight are orthonormal in the sense that
\begin{equation}
\label{eq:ysorthog}
\int {}_{s}\overline{Y}_{l^{\prime}m^{\prime}}(\theta,\phi)\;
{}_{s}Y_{lm}(\theta,\phi)\;d\Omega
=\delta_{l l^{\prime}}\,\delta_{m m^{\prime}}\;,
\end{equation}
where
\begin{equation}
\int d\Omega
=\int_{0}^{2\pi}d\phi \int_{0}^{\pi} \sin \theta\, d\theta\;.
\end{equation}
The spin-weighted spherical harmonics may be defined so that
\begin{equation}
\label{eq:ysconj}
{}_{s}\overline{Y}_{lm}(\theta,\phi)
=(-1)^{m+s}\,{}_{-s}Y_{l\,-m}(\theta,\phi)\;,
\end{equation}
which is given in \cite{gj67} and \cite{np66} and misprinted 
in \cite{gn67}.  Equation \eqref{eq:ysconj} can be used to 
evaluate harmonics for negative $s$ and $m$.  

Following Arfken \cite{arfken85}, we define the 
spherical harmonics as
\begin{equation}
\label{eq:sphdef}
Y_{lm}(\theta,\phi)
=(-1)^{m}\sqrt{\frac{2 l+1}{4 \pi}\frac{(l-m)!}{(l+m)!}}
\,P_{lm}(\cos\theta)\;.
\end{equation}
Here, the associated Legendre functions $P_{lm}(\cos\theta)$ are 
\begin{equation}
\label{eq:pldef}
P_{lm}(x)=\frac{1}{2^{l}\,l\,!}\left(1-x^{2}\right)^{m/2}
\frac{d^{l+m}}{dx^{l+m}}\left(x^{2}-1\right)^{l}\,,\,-l\le m\le l\;,
\end{equation}
and the factor of $(-1)^{m}$ is the so-called Condon-Shortley phase.  
The spherical harmonic differential equation is
\begin{equation}
\label{eq:ddylm}
\frac{\partial^{2}Y_{lm}(\theta,\phi)}{\partial\theta^{2}}
+\cot\theta\frac{\partial Y_{lm}(\theta,\phi)}{\partial\theta}
+\frac{1}{\sin^{2}\theta}\frac{\partial^{2}
Y_{lm}(\theta,\phi)}{\partial\phi^{2}}
=-l(l+1)Y_{lm}(\theta,\phi)\;.
\end{equation}
The definition \eqref{eq:sphdef} implies that
\begin{equation}
\label{eq:yconj}
\overline{Y}_{lm}(\theta,\phi)=(-1)^{m}\,Y_{l\,-m}(\theta,\phi)\;.
\end{equation}
With the definition \eqref{eq:sphdef} and the operators 
$\eth$ and $\overline{\eth}$, we can calculate spin-weighted 
spherical harmonics for non-zero $s$.  For our purposes, 
we will need harmonics of $s=\pm 2$, $\pm 1$, and $0$.  

In the stress energy tensor expressions \eqref{eq:timtmunu} 
and \eqref{eq:tmunu}-\eqref{eq:etmunu}, the angular functions 
are written in terms of tensor harmonics.  The tensor harmonics can 
be related to, and therefore written in terms of, the spin-weighted 
spherical harmonics \cite{th80}.  Using the operators 
$\eth$ \eqref{eq:edthop} and $\overline{\eth}$ \eqref{eq:edthbar} 
and the definitions of $W_{lm}(\theta,\phi)$ \eqref{eq:wlm} 
and $X_{lm}(\theta,\phi)$ \eqref{eq:xlm}, we can show that
\begin{equation}
\label{eq:yptwo}
{}_{2}Y_{lm}(\theta,\phi)= \frac{W_{lm}(\theta,\phi)
+i X_{lm}(\theta,\phi)}{\sqrt{l(l+1)(l-1)(l+2)}}\;,
\end{equation}

\begin{equation}
\label{eq:ymtwo}
{}_{-2}Y_{lm}(\theta,\phi)= \frac{W_{lm}(\theta,\phi)
-i X_{lm}(\theta,\phi)}{\sqrt{l(l+1)(l-1)(l+2)}}\;,
\end{equation}

\begin{equation}\label{eq:ypone}
{}_{1}Y_{lm}(\theta,\phi)
=-\frac{\frac{\partial Y_{lm}(\theta,\phi)}{\partial\theta}
+i\csc\theta \frac{\partial Y_{lm}(\theta,\phi)}{\partial\phi}}
{\sqrt{l(l+1)}}\;,
\end{equation}

\begin{equation}\label{eq:ymone}
{}_{-1}Y_{lm}(\theta,\phi)
= \frac{\frac{\partial Y_{lm}(\theta,\phi)}{\partial\theta}
-i\csc\theta \frac{\partial Y_{lm}(\theta,\phi)}{\partial\phi}}
{\sqrt{l(l+1)}}\;.
\end{equation}
One can verify that the ${}_{s}Y_{lm}(\theta,\phi)$ given above and 
their conjugates satisfy \eqref{eq:ysconj}, provided that the 
spherical harmonics $Y_{lm}(\theta,\phi)$ meet \eqref{eq:yconj}.  
Equations \eqref{eq:yptwo}-\eqref{eq:ymone} can be inverted to give
\begin{equation}
\label{eq:wy}
W_{lm}(\theta,\phi)= \frac{1}{2} \sqrt{l(l+1)(l-1)(l+2)}
({}_{-2}Y_{lm}(\theta,\phi)+{}_{2}Y_{lm}(\theta,\phi))\;,
\end{equation}
\begin{equation}
\label{eq:xy}
X_{lm}(\theta,\phi)= \frac{1}{2} i \sqrt{l(l+1)(l-1)(l+2)}
({}_{-2}Y_{lm}(\theta,\phi)-{}_{2}Y_{lm}(\theta,\phi))\;,
\end{equation}

\begin{equation}
\label{eq:dyphi}
\frac{\partial Y_{lm}(\theta,\phi)}{\partial\phi}
= \frac{1}{2} i \sqrt{l(l+1)} \sin\theta 
({}_{-1}Y_{lm}(\theta,\phi)+{}_{1}Y_{lm}(\theta,\phi))\;,
\end{equation}

\begin{equation}
\label{eq:dyth}
\frac{\partial Y_{lm}(\theta,\phi)}{\partial\theta}
=\frac{1}{2} \sqrt{l(l+1)} 
({}_{-1}Y_{lm}(\theta,\phi)-{}_{1}Y_{lm}(\theta,\phi))\;.
\end{equation}
Equations \eqref{eq:wy}-\eqref{eq:xy} and the orthogonality 
integral \eqref{eq:ysorthog} may be used to evaluate
\begin{multline}
\label{eq:wxint}
\int\left[\,\overline{W}_{l^{\prime}m^{\prime}}(\theta,\phi)W_{lm}(\theta,\phi)
+\overline{X}_{l^{\prime}m^{\prime}}(\theta,\phi)X_{lm}(\theta,\phi)\right] d\Omega
\\=l(l+1)(l-1)(l+2)\delta_{l l^{\prime}}\,\delta_{m m^{\prime}}
=4\lambda(1+\lambda)\delta_{l l^{\prime}}\,\delta_{m m^{\prime}}\;,
\end{multline}
which will be used in Chapter~\ref{radchap}.  Alternatively, the 
integral \eqref{eq:wxint} may be evaluated by writing 
$W_{lm}(\theta,\phi)$ and $X_{lm}(\theta,\phi)$ in terms of associated 
Legendre functions using \eqref{eq:sphdef}-\eqref{eq:pldef} \cite{ashby}.  
However, it is much simpler to use the spin-weighted spherical 
harmonics.  

We could substitute \eqref{eq:wy}-\eqref{eq:dyth} into 
$T_{\mu\nu}$ \eqref{eq:timtmunu}, which would replace the tensor harmonics 
with spin-weighted spherical harmonics.  The resulting expressions 
would be more complicated than the original ones.  This is because 
the spin-weighted spherical harmonics are more useful in a different 
coordinate system, which uses a tetrad basis from the Newman-Penrose 
formalism \cite{chandra92}, \cite{np62}.  
These two references use a metric signature of $+$$-$$-$$-$.  
Because a signature of $-$$+$$+$$+$ is used in 
this thesis, there will be some differences; however, 
they will be minor, since we will use the Newman-Penrose formalism 
only to a limited extent.  

The discussion of the tetrad basis below is taken mainly from 
Chandrasekhar \cite{chandra92}, but other references are also noted.
The Newman-Penrose tetrad basis consists of four null vectors, 
which are
\begin{equation}
\label{eq:npvec}
\bm{e}_{(1)}=\bm{l}\;,\;\bm{e}_{(2)}=\bm{n}\;,\;
\bm{e}_{(3)}=\bm{m}\;,\;\bm{e}_{(4)}=\bm{\overline{m}}\;.
\end{equation}
They are referred to as null vectors, because their norms are zero.  
Indices in the tetrad frame will be enclosed in parentheses.  
In Schwarzschild spacetime, the four vectors may be chosen so that 
their components are
\begin{equation}
\label{eq:nplvec}
l^{\mu}=\left(\frac{r}{r-2 M},1,0,0\right)\;,
\end{equation}
\begin{equation}
\label{eq:npnvec}
n^{\mu}=\frac{1}{2}\left(1,-1+\frac{2M}{r},0,0\right)\;,
\end{equation}
\begin{equation}
\label{eq:npmvec}
m^{\mu}=\frac{1}{\sqrt{2} r}\left(0,0,1,i\csc\theta\right)\;,
\end{equation}
\begin{equation}
\label{eq:npmbvec}
\overline{m}^{\mu}=\frac{1}{\sqrt{2} r}\left(0,0,1,-i\csc\theta\right)\;.
\end{equation}
This basis is often called the Kinnersly tetrad \cite{chrz75}, \cite{teukpr74}.  
As defined above, the vector $\bm{l}$ is tangent 
to outgoing radial null geodesics, and $\bm{n}$ is tangent to ingoing 
radial null geodesics \cite{chandra92} (pp. 124, 134), \cite{poisrt04} 
(pp. 52, 193, in a different  notation).  The symmetric scalar inner 
product of two basis vectors is
\begin{equation}
\bm{e}_{(a)}\cdot \bm{e}_{(b)}=g_{\mu\nu}e_{(a)}^{\mu}e_{(b)}^{\nu}\;,
\end{equation}
where we will take $g_{\mu\nu}$ from the Schwarzschild 
metric \eqref{eq:schmet}.  The basis is normalized as
\begin{equation}
\label{eq:ldotnmdotmb}
\bm{l}\cdot \bm{n}=-1\;,\;\bm{m}\cdot \bm{\overline{m}}=1\;.
\end{equation}
Other inner products are
\begin{equation}
\label{eq:miscdot}
\bm{l}\cdot \bm{l}=\bm{n}\cdot \bm{n}=\bm{m}\cdot \bm{m}=
\bm{\overline{m}}\cdot \bm{\overline{m}}=\bm{l}\cdot \bm{m}=
\bm{l}\cdot \bm{\overline{m}}=\bm{n}\cdot \bm{m}=
\bm{n}\cdot \bm{\overline{m}}=0\;,
\end{equation}
where the first four inner products are zero because the tetrad 
is a null basis and the last four are zero because of orthogonality.  
The metric tensor in the tetrad basis is 
$\eta_{(a)(b)}$, where
\begin{equation}
\label{eq:tetmet}
\eta_{(a)(b)}
=g_{\mu\nu}e_{(a)}^{\mu}e_{(b)}^{\nu}=
\begin{pmatrix}
 0 & -1 & 0 & 0 \\
 -1 & 0 & 0 & 0 \\
 0 & 0 & 0 & 1 \\
 0 & 0 & 1 & 0
\end{pmatrix} \,.
\end{equation}
Equation \eqref{eq:tetmet} is the matrix form of \eqref{eq:ldotnmdotmb} 
and \eqref{eq:miscdot}.  
Because of different metric signatures, the signs of 
\eqref{eq:ldotnmdotmb} and \eqref{eq:tetmet} are opposite those 
given in \cite{chandra92}, \cite{np62}.  

The null basis also can be used to explain the parameter $s$ of the 
spin-weighted spherical harmonics.  As noted above, the vector $\bm{l}$ 
is tangent to outgoing null geodesics and is orthogonal to $\bm{m}$.  
The real and imaginary parts of $\bm{m}$ are spacelike vectors 
(orthogonal to each other) which may be rotated in their plane 
about $\bm{l}$.  A quantity $\eta$ has spin weight~$s$ 
if $\eta\to e^{i s\psi}\eta$ under a rotation of the real and 
imaginary parts of $\bm{m}$ through the angle $\psi$ \cite{np66}.  
In other words, $s$ describes how $\eta$ transforms under a 
rotation about the direction of propagation along a null geodesic.  
In this sense, spin weight is equivalent to helicity, as defined in 
equation \eqref{eq:heldef} of the plane wave example.  
The factor $e^{i s\psi}$ does not appear in 
the spin-weighted spherical harmonic expressions above, because 
there the third angle $\psi$ is set equal to zero.  In general 
relativity, transverse gravitational waves have spin 
weight~$\pm 2$ \cite{ell731}-\cite{ell732}, \cite{th80}.  
Electromagnetic waves have spin weight~$\pm 1$ \cite{th80}.  

Components of the stress energy tensor in the tetrad frame are 
obtained by projecting the tensor onto the basis vectors 
using \cite{chandra92}
\begin{equation}
T_{(a)(b)}=T_{\mu\nu}e_{(a)}^{\mu}e_{(b)}^{\nu}\;.
\end{equation}
The notation for the tetrad frame multipole expansion will be
\begin{equation}
T_{(a)(b)}=\sum_{l=0}^{\infty}\sum_{m=-l}^{l}T^{lm}_{(a)(b)}
=\sum_{l=0}^{\infty}\sum_{m=-l}^{l}
T^{lm}_{\mu\nu}e_{(a)}^{\mu}e_{(b)}^{\nu}\;.
\end{equation}
We have two representations of the stress energy tensor.  The first 
is the tensor harmonic multipole expansion \eqref{eq:timtmunu}, 
whose time-radial coefficients (such as $Se^{lm}_{00}(t,r)$) need 
to be determined.  The second representation is the delta function 
tensor \eqref{eq:delttmunu}, whose angular delta function needs to 
be expanded in multipoles using the completeness 
relation \eqref{eq:compl}.  

We project the first representation \eqref{eq:timtmunu} onto the tetrad 
basis and replace the tensor harmonic angular functions with spin-weighted 
spherical harmonics, using \eqref{eq:wy}-\eqref{eq:dyth}.  
Subject to the restriction $\lvert s\rvert>l$ \eqref{eq:lgts}, 
the resulting multipole components are  
\begin{equation}
\label{eq:t11first}
T^{lm}_{(1)(1)}=\left(\frac{r^2 Se^{lm}_{00}(t,r)}{(-2 M+r)^2}
-\frac{2 r Se^{lm}_{01}(t,r)}{2 M-r}+Se^{lm}_{11}(t,r)\right) Y_{lm}(\theta,\phi)\;,
\end{equation}
\begin{equation}
T^{lm}_{(1)(2)}=\frac{\left(-r^2 Se^{lm}_{00}(t,r)
+(-2 M+r)^2 Se^{lm}_{11}(t,r)\right)}{2 (2 M-r) r}Y_{lm}(\theta,\phi)\;,
\end{equation}

\begin{multline}
T^{lm}_{(1)(3)}=\frac{\sqrt{l(l+1)}}{\sqrt{2} (2 M-r) r}
\left[r Se^{lm}_{02}(t,r)+(-2 M+r) Se^{lm}_{12}(t,r)
\right.\\\left.-i (r So^{lm}_{02}(t,r)+(-2 M+r) So^{lm}_{12}(t,r))\right]
{}_{1}Y_{lm}(\theta,\phi)\;,
\end{multline}

\begin{multline}
T^{lm}_{(1)(4)}=-\frac{\sqrt{l(l+1)}}{\sqrt{2} (2 M-r) r}
\left[r Se^{lm}_{02}(t,r)+(-2 M+r) Se^{lm}_{12}(t,r)
\right.\\\left.+i (r So^{lm}_{02}(t,r)+(-2 M+r) So^{lm}_{12}(t,r))\right]
{}_{-1}Y_{lm}(\theta,\phi)\;,
\end{multline}

\begin{equation}
T^{lm}_{(2)(2)}=\frac{\left(r^2 Se^{lm}_{00}(t,r)+(2 M-r) (2 r Se^{lm}_{01}(t,r)
+(2 M-r) Se^{lm}_{11}(t,r))\right) }{4 r^2}Y_{lm}(\theta,\phi)\;,
\end{equation}

\begin{multline}
T^{lm}_{(2)(3)}=\frac{\sqrt{l(l+1)}}{2 \sqrt{2} r^2}
\left[-r Se^{lm}_{02}(t,r)+(-2 M+r) Se^{lm}_{12}(t,r)
\right.\\\left.+i (r So^{lm}_{02}(t,r)+(2 M-r) So^{lm}_{12}(t,r))\right]
{}_{1}Y_{lm}(\theta,\phi)\;,
\end{multline}

\begin{multline}
T^{lm}_{(2)(4)}=\frac{\sqrt{l(l+1)}}{2 \sqrt{2} r^2}
\left[r Se^{lm}_{02}(t,r)+(2 M-r) Se^{lm}_{12}(t,r)
\right.\\\left.+i (r So^{lm}_{02}(t,r)+(2 M-r) So^{lm}_{12}(t,r))\right]
{}_{-1}Y_{lm}(\theta,\phi)\;,
\end{multline}

\begin{equation}
T^{lm}_{(3)(3)}=\frac{\sqrt{l(l+1)(l-1)(l+2)} 
(Se^{lm}_{22}(t,r)+i So^{lm}_{22}(t,r))}{r^2}{}_{2}Y_{lm}(\theta,\phi)\;,
\end{equation}

\begin{equation}
T^{lm}_{(3)(4)}=\frac{Ue^{lm}_{22}(t,r)}{r^2}Y_{lm}(\theta,\phi)\;,
\end{equation}
\begin{equation}\label{eq:t44first}
T^{lm}_{(4)(4)}=\frac{\sqrt{l(l+1)(l-1)(l+2)} 
(Se^{lm}_{22}(t,r)-i So^{lm}_{22}(t,r))}{r^2}{}_{-2}Y_{lm}(\theta,\phi)\;.
\end{equation}
The remaining components are determined by symmetry of the indices 
for $T_{(a)(b)}$.  In the tetrad frame, each component has a single spin 
weight.  This is different from the original representation, where some 
of the components would have more than one spin weight after we replaced 
the tensor harmonics with spin-weighted harmonics.

The next step is to project the second representation of the stress 
energy tensor \eqref{eq:delttmunu} (with indices lowered 
\eqref{eq:covdelttmunu}) onto the tetrad basis.  After doing so, 
we expand the angular delta function $\deltaom$ in terms of spin-weighted 
spherical harmonics using the completeness relation \eqref{eq:compl}.  
The harmonic for each component is chosen to match the spin weight 
given in \eqref{eq:t11first}-\eqref{eq:t44first}.  This procedure leads to
\begin{equation}\label{eq:t11sec}
T^{lm}_{(1)(1)}=m_{0}\gamma\frac{(2 M-r+r \rdot)^2}{r^2 (-2 M+r)^2}
\deltar\overline{Y}_{lm}(\thetap,\phip) Y_{lm}(\theta,\phi)\;,
\end{equation}

\begin{equation}
T^{lm}_{(1)(2)}=-m_{0}\gamma\frac{(-4 M^2+4 M r-r^2+r^2 (\rdot)^{2})}
{2 r^3 (-2 M+r)}\deltar\overline{Y}_{lm}(\thetap,\phip) Y_{lm}(\theta,\phi)\;,
\end{equation}

\begin{equation}
T^{lm}_{(1)(3)}=-m_{0}\gamma\frac{(2 M-r+r \rdot) 
(\thdot+i\sin\thetap \phdot)}{\sqrt{2} (2 M-r) r}\deltar
{}_{1}\overline{Y}_{lm}(\thetap,\phip){}_{1}Y_{lm}(\theta,\phi)\;,
\end{equation}

\begin{equation}
T^{lm}_{(1)(4)}=-m_{0}\gamma\frac{(2 M-r+r \rdot) (\thdot-i\sin\thetap \phdot)}
{\sqrt{2} (2 M-r) r}\deltar
{}_{-1}\overline{Y}_{lm}(\thetap,\phip){}_{-1}Y_{lm}(\theta,\phi)\;,
\end{equation}

\begin{equation}
T^{lm}_{(2)(2)}=m_{0}\gamma\frac{(-2 M+r+r \rdot)^2}{4 r^4}
\deltar\overline{Y}_{lm}(\thetap,\phip) Y_{lm}(\theta,\phi)\;,
\end{equation}

\begin{equation}
T^{lm}_{(2)(3)}=m_{0}\gamma\frac{ (2 M-r-r \rdot) 
(\thdot+i\sin\thetap \phdot)}{2 \sqrt{2} r^2}\deltar
{}_{1}\overline{Y}_{lm}(\thetap,\phip){}_{1}Y_{lm}(\theta,\phi)\;,
\end{equation}

\begin{equation}
T^{lm}_{(2)(4)}=m_{0}\gamma\frac{ (2 M-r-r 
\rdot) (\thdot-i\sin\thetap \phdot)}{2 \sqrt{2} r^2}\deltar
{}_{-1}\overline{Y}_{lm}(\thetap,\phip){}_{-1}Y_{lm}(\theta,\phi)\;,
\end{equation}
\begin{equation}
T^{lm}_{(3)(3)}= m_{0}\gamma \frac{1}{2}
(\thdot+i \sin\thetap\phdot)^2\deltar
{}_{2}\overline{Y}_{lm}(\thetap,\phip){}_{2}Y_{lm}(\theta,\phi)\;,
\end{equation}

\begin{equation}
T^{lm}_{(3)(4)}= m_{0}\gamma \frac{1}{2}   
\left((\thdot)^{2}+\sin^{2}\thetap(\phdot)^{2}\right)
\deltar\overline{Y}_{lm}(\thetap,\phip) Y_{lm}(\theta,\phi)\;,
\end{equation}

\begin{equation}\label{eq:t44sec}
T^{lm}_{(4)(4)}= m_{0}\gamma\frac{1}{2}(\thdot-i \sin\thetap\phdot)^2\deltar
{}_{-2}\overline{Y}_{lm}(\thetap,\phip){}_{-2}Y_{lm}(\theta,\phi)\;.
\end{equation}
Again, those not listed are found by symmetry, and components 
are zero when $\lvert s\rvert>l$.  In some ways, this step resembles 
the derivation of the source term for the Teukolsky equation, as 
described in \cite{cut94}, \cite{pois93}.

We equate corresponding components of \eqref{eq:t11first}-\eqref{eq:t44first} 
and \eqref{eq:t11sec}-\eqref{eq:t44sec} 
to form a system of ten equations.  Solving this system, we obtain the 
following time-radial coefficients of the stress energy tensor:

\begin{equation}
\label{eq:firstrct}
Se^{lm}_{00}(t,r)=m_{0}\gamma\frac{(2 M-r)^2}{r^4}
\deltar\overline{Y}_{lm}(\thetap,\phip)\;,
\end{equation}

\begin{equation}
Se^{lm}_{01}(t,r)=-m_{0}\gamma\frac{\rdot}{r^2}
\deltar\overline{Y}_{lm}(\thetap,\phip)\;,
\end{equation}

\begin{equation}
Se^{lm}_{11}(t,r)=m_{0}\gamma\frac{(\rdot)^{2}}{(2 M-r)^2}
\deltar\overline{Y}_{lm}(\thetap,\phip)\;,
\end{equation}

\begin{equation}
Ue^{lm}_{22}(t,r)=m_{0}\gamma\frac{1}{2} r^2
\left((\thdot)^{2}+\sin^{2}\thetap(\phdot)^{2}\right)
\deltar\overline{Y}_{lm}(\thetap,\phip)\;,
\end{equation}

\begin{multline}
So^{lm}_{02}(t,r)=m_{0}\gamma\frac{(2 M-r)}{l(l+1) r}\deltar
\\\times\left(\csc\thetap\frac{\partial\overline{Y}_{lm}(\thetap,\phip)}
{\partial\phi}\thdot-\sin\thetap \frac{\partial
\overline{Y}_{lm}(\thetap,\phip)}{\partial\theta}\phdot \right)\;,
\end{multline}
\begin{multline}
So^{lm}_{12}(t,r)=m_{0}\gamma\frac{r \rdot}{l(l+1) (-2 M+r)}\deltar 
\\\times\left(\csc\thetap \frac{\partial\overline{Y}_{lm}(\thetap,\phip)}
{\partial\phi}\thdot-\sin\thetap \frac{\partial
\overline{Y}_{lm}(\thetap,\phip)}{\partial\theta}\phdot \right)\;,
\end{multline}

\begin{equation}
Se^{lm}_{02}(t,r)=m_{0}\gamma\frac{(2 M-r)}{l(l+1) r}\deltar  
\left(\frac{\partial\overline{Y}_{lm}(\thetap,\phip)}{\partial\phi}\phdot
+\frac{\partial\overline{Y}_{lm}(\thetap,\phip)}{\partial\theta}\thdot\right)\;,
\end{equation}

\begin{equation}
Se^{lm}_{12}(t,r)=-m_{0}\gamma\frac{r \rdot}{l(l+1) (2 M-r)}\deltar
\left(\frac{\partial\overline{Y}_{lm}(\thetap,\phip)}{\partial\phi}\phdot
+\frac{\partial\overline{Y}_{lm}(\thetap,\phip)}{\partial\theta}\thdot\right)\;,
\end{equation}

\begin{multline}
Se^{lm}_{22}(t,r)=m_{0}\gamma\frac{r^2}{l(l+1)(l-1)(l+2)}\deltar
\\\times\left[\overline{X}_{lm}(\thetap,\phip)\sin\thetap \thdot\phdot
+\frac{1}{2}\overline{W}_{lm}(\thetap,\phip) 
\left((\thdot)^{2}-\sin^{2}\thetap (\phdot)^{2}\right)\right]\;,
\end{multline}

\begin{multline}
\label{eq:lastrct}
So^{lm}_{22}(t,r)=m_{0}\gamma\frac{r^2}{l(l+1)(l-1)(l+2)}\deltar
\\\times\left[\overline{W}_{lm}(\thetap,\phip)\sin\thetap \thdot\phdot
+\frac{1}{2}\overline{X}_{lm}(\thetap,\phip) 
\left(\sin^{2}\thetap(\phdot)^{2}-(\thdot)^{2}\right)\right]\;.
\end{multline}
Some of these are zero for certain values of $l$ \cite{zerp70}.  The 
angular functions in $So^{lm}_{22}$ and $Se^{lm}_{22}$ are zero for $l<2$, 
and $So^{lm}_{02}$, $So^{lm}_{12}$, $Se^{lm}_{02}$ and $Se^{lm}_{12}$ are zero 
for $l=0$.  Taking into account differences in notation, the components 
above agree with those derived by Zerilli \cite{zerp70}, 
as corrected by others \cite{ashby}, \cite{sago03}.

Equations \eqref{eq:firstrct}-\eqref{eq:lastrct} simplify when the 
orbital motion is in the equatorial plane, for which $\thetap=\pi/2$ 
and $\thdot=0$ \cite{ashby}. 
For example, the definition of $W_{lm}(\theta,\phi)$ \eqref{eq:wy} and the 
spherical harmonic differential equation \eqref{eq:ddylm} give \cite{ashby}
\begin{equation}
\overline{W}_{lm}\big(\textstyle{\frac{\pi}{2}},\phip\big)
=\left(2 m^{2}-l(l+1)\right)
\overline{Y}_{lm}\big(\textstyle{\frac{\pi}{2}},\phip\big)\;.
\end{equation}
With these substitutions, the odd parity source terms simplify to
\begin{equation}
\label{eq:firstrco}
So^{lm}_{02}(t,r)=-m_{0}\gamma\frac{(2 M-r)\phdot}{l(l+1) r}\deltar\dylmpi\;,
\end{equation}
\begin{equation}
So^{lm}_{12}(t,r)=-m_{0}\gamma\frac{r \rdot\phdot}{l(l+1) (-2 M+r)}
\deltar\dylmpi\;,
\end{equation}
\begin{equation}
So^{lm}_{22}(t,r)=m_{0}\gamma\frac{r^2(\phdot)^{2}}{2 l(l+1)(l-1)(l+2)}\deltar
(-2im)\dylmpi\;.
\end{equation}
The even parity source terms reduce to 
\begin{equation}
Se^{lm}_{00}(t,r)=m_{0}\gamma\frac{(2 M-r)^2}{r^4}
\deltar\ylmpi\;,
\end{equation}
\begin{equation}
Se^{lm}_{01}(t,r)=-m_{0}\gamma\frac{\rdot}{r^2}
\deltar\ylmpi\;,
\end{equation}
\begin{equation}
Se^{lm}_{11}(t,r)=m_{0}\gamma\frac{(\rdot)^{2}}{(2 M-r)^2}
\deltar\ylmpi\;,
\end{equation}
\begin{equation}
Ue^{lm}_{22}(t,r)=m_{0}\gamma\frac{1}{2} r^2(\phdot)^{2}
\deltar\ylmpi\;,
\end{equation}
\begin{equation}
Se^{lm}_{02}(t,r)=m_{0}\gamma\frac{(2 M-r)\phdot}{l(l+1) r}\deltar  
(-im)\ylmpi\;,
\end{equation}
\begin{equation}
Se^{lm}_{12}(t,r)=-m_{0}\gamma\frac{r \rdot\phdot}{l(l+1) (2 M-r)}\deltar
(-im)\ylmpi\;,
\end{equation}
\begin{equation}
\label{eq:lastrco}
Se^{lm}_{22}(t,r)=m_{0}\gamma\frac{r^2(\phdot)^{2}}{2l(l+1)(l-1)(l+2)}\deltar
\left(l(l+1)-2m^{2}\right)\ylmpi\;,
\end{equation}
most of which were also calculated by \cite{ashby}.  
All even parity source terms have an angular factor of 
$\overline{Y}_{lm}\big(\textstyle{\frac{\pi}{2}},0\big)$, while 
all the odd parity components have a factor of 
$\frac{\partial\overline{Y}_{lm}\left(\frac{\pi}{2},0\right)}
{\partial\theta}$.  From the definition of spherical 
harmonics \eqref{eq:sphdef} and the discussion 
of parity in \cite{arfken85}, the even parity angular factor 
is non-zero only if the sum $l+m$ is even, and the odd factor 
is non-zero only if $l+m$ is odd.  This means that we 
need to solve only the even parity field equations for even $l+m$ 
and only the odd equations for odd $l+m$.  Similar reasoning applies to 
source terms of the Teukolsky equation \cite{pois93}.  The spherical 
harmonics may be calculated numerically using routines from \cite{numr}.

\section{\label{sec:ftmunu}Fourier Transforms}

In this section, we calculate the Fourier transforms of the time-radial 
coefficients in (\ref{eq:firstrco})-(\ref{eq:lastrco}).  For convenience, 
the coefficients can be written in the following form
\begin{equation}
\label{eq:slmtr}
S^{lm}(t,r)=f^{lm}(r)(\rdot)^{n}\delta(r-r^{\prime}(t)) e^{-i m \phip(t)}
\;,\;\;n=0,1,2\;.
\end{equation}
where, as before, $\rdot=\frac{d r^{\prime}}{dt}$.  The factor $f^{lm}(r)$ is 
different for each coefficient.  The Fourier transform 
$S^{lm}(\omega,r)$ is defined as  
\begin{equation}
\label{eq:sft}
S^{lm}(\omega,r)=\frac{1}{2\pi}\int_{\!-\infty}^{\infty} e^{i\omega t}S^{lm}(t,r)\,dt\;,
\end{equation}
and the inverse transform is
\begin{equation}
\label{eq:sinvft}
S^{lm}(t,r)=\int_{\!-\infty}^{\infty}e^{-i\omega t}S^{lm}(\omega,r)\,d\omega\;.
\end{equation}
Evaluation of the transform integral in (\ref{eq:sft}) depends on the orbital 
motion.  Two cases are calculated below:  circular orbits and 
elliptic orbits.  

The derivation for circular orbits is based on Poisson's, as 
described in \cite{pois93}.  He calculated the circular orbit source 
term for the Teukolsky equation, which is different from, but related to, 
the Regge-Wheeler equation.  His method can be adapted to the 
Fourier transform of the stress energy tensor components and the 
source terms for the generalized Regge-Wheeler equations.  
For circular orbits, the orbital radius is constant and $\rdot=0$, so the 
radial factors can be moved outside the transform integral.  Since $\rdot=0$, 
we replace $(\rdot)^{n}$ in \eqref{eq:slmtr} with the Kronecker delta $\delta_{n0}$.  
Further, the azimuthal angle is related to the time by $\phip(t)=\Omega_{\phi}t$ 
\eqref{eq:cirtp}, where $\Omega_{\phi}$ is the orbital angular frequency.  
The integral \eqref{eq:sft} can be rewritten as
\begin{equation}
S^{lm}(\omega,r)=f^{lm}(r)\delta_{n0}\deltar \left[\frac{1}{2\pi}\int_{\!-\infty}^{\infty} 
e^{i\omega t}e^{-i m\Omega_{\phi}t}\,dt\right]\;,\;\;n=0,1,2\;.
\end{equation}
The quantity in brackets is the integral representation of a delta 
function \cite{arfken85}.  The Fourier transform for circular orbits is simply 
\begin{equation}
\label{eq:cirfft}
S^{lm}(\omega,r)=f^{lm}(r)\delta_{n0}\deltar\delta(\omega-m\Omega_{\phi})\;,\;\;n=0,1,2\;.
\end{equation}
Because of the second delta function factor, the frequency for each mode is an 
integral multiple of the orbital angular frequency.  The leading 
radiation multipole is the quadrupole moment, so
the dominant gravitational wave frequency for circular orbits is twice 
the orbital frequency \cite{pois93}.

Elliptic orbits are more complicated.  The derivation below is adapted 
from the work of Cutler and others in \cite{cut94}, which also was for the 
Teukolsky equation.  It is desirable to express the Fourier 
integral (\ref{eq:sft}) as a sum over discrete frequencies, in order to 
simplify calculations.  If $g(t)$ is a periodic function with period $P$, 
then \cite{mandw}
\begin{equation}
\label{eq:gfour}
g(t)=\sum^{\infty}_{k=-\infty}a_{k}e^{-i\frac{2\pi}{P}k t}\;,
\end{equation}
where
\begin{equation}
\label{eq:angfour}
a_{k}=\frac{1}{P}\int_{0}^{P}e^{i\frac{2\pi}{P}k t^{\prime}}g(t^{\prime})dt^{\prime}\;.
\end{equation}
The function $g(t)$ is periodic if $g(t+P)=g(t)$.

Because elliptic orbits in general relativity are not closed, $S^{lm}(t,r)$ 
is not periodic.  As explained in \cite{cut94}, an elliptic orbit has 
a radial period $P$, meaning the orbiting mass returns to the same 
radial coordinate $r$ after a time $P$ has elapsed. However, the angular 
position $\phip(t)$ is 
different:  $\phip(t+P)=\phip(t)+\Delta\phi$, where $\Delta\phi>2\pi$.  
In particular,
\begin{equation}
\begin{split}
S^{lm}(t+P,r)&=f^{lm}(r)(\rdot)^{n}\deltar e^{-i m(\phip+\Delta\phi) }
\\&=S^{lm}(t,r) e^{-i m\Delta\phi}\;.
\end{split}
\end{equation}
Because $m\Delta\phi$ is not an integral multiple of $2\pi$, the factor 
$e^{-i m\Delta\phi}$ is not unity.  The function $S^{lm}(t,r)$ is not 
periodic, because $S^{lm}(t+P,r)\neq S^{lm}(t,r)$.

To circumvent this obstacle, we find a new quantity which is periodic.  
The procedure for doing so is described in \cite{cut94}; however, that 
reference uses the Teukolsky equation, so our results and notation will 
be different.  Define 
\begin{equation}
\label{eq:slmtildef}
\widetilde{S}^{lm}(t,r)=S^{lm}(t,r)e^{i m\Omega_{\phi}t}\;,
\end{equation}
where $\Omega_{\phi}=\frac{\Delta\phi}{P}$.  
The function $\widetilde{S}^{lm}(t,r)$ is periodic with a period $P$, because
\begin{equation}
\begin{split}
\widetilde{S}^{lm}(t+P,r)&=f^{lm}(r)(\rdot)^{n}\deltar e^{-i m (\phip+\Delta\phi)}
e^{i m\Omega_{\phi}(t+P)}
\\&=\left\{f^{lm}(r)(\rdot)^{n}\deltar e^{-i m \phip} e^{i m\Omega_{\phi}t}\right\}
e^{-i m \Delta\phi}e^{i m \Omega_{\phi}P}
\\&=\widetilde{S}^{lm}(t,r)\;.
\end{split}
\end{equation}
In the last step, the relation $\Omega_{\phi}P=\Delta\phi$ has been used.  
Using (\ref{eq:gfour}) and (\ref{eq:angfour}), $\widetilde{S}^{lm}(t,r)$ can be 
expressed as a Fourier series with discrete frequencies 
$k\Omega_{r}=k\frac{2\pi}{P}$:
\begin{equation}
\label{eq:sfourser}
\widetilde{S}^{lm}(t,r)=\sum^{\infty}_{k=-\infty}\widetilde{S}^{lmk}(\omega,r)
e^{-i k \Omega_{r} t}\;,
\end{equation}
where
\begin{equation}
\label{eq:slmk}
\widetilde{S}^{lmk}(\omega,r)=\frac{1}{P}\int_{0}^{P}e^{i k \Omega_{r} t^{\prime}}
\widetilde{S}^{lm}(t^{\prime},r)dt^{\prime}\;.
\end{equation}
Solving (\ref{eq:slmtildef}) for $S^{lm}(t,r)$ and substituting the result 
into (\ref{eq:sft}) leads to
\begin{equation}
S^{lm}(\omega,r)=\frac{1}{2\pi}\int_{\!-\infty}^{\infty} e^{i(\omega-m\Omega_{\phi}) t}
\widetilde{S}^{lm}(t,r)\,dt\;.
\end{equation}
Inserting (\ref{eq:sfourser}) and rearranging terms gives
\begin{equation}
S^{lm}(\omega,r)=\sum^{\infty}_{k=-\infty}\widetilde{S}^{lmk}(\omega,r)
\left[\frac{1}{2\pi}\int_{\!-\infty}^{\infty} 
e^{i(\omega-m\Omega_{\phi}-k\Omega_{r}) t}\,dt\right]\;.
\end{equation}
The integral is a delta function \cite{arfken85}, so 
\begin{equation}
\begin{split}
\label{eq:nsfourser}
S^{lm}(\omega,r)&=\sum^{\infty}_{k=-\infty}\widetilde{S}^{lmk}(\omega,r)
\delta(\omega-(m\Omega_{\phi}+k\Omega_{r}))\\
&=\sum^{\infty}_{k=-\infty}\widetilde{S}^{lmk}(\omega,r)
\delta(\omega-\omega_{mk})\;.
\end{split}
\end{equation}
The delta function implies that the frequency spectrum is discrete, 
with 
\begin{equation}
\label{eq:omegamk}
\omega=\omega_{mk}= m\Omega_{\phi}+k\Omega_{r}\,,\;
\Omega_{\phi}=\frac{\Delta\phi}{P}\,,\;\Omega_{r}=\frac{2\pi}{P}\;.
\end{equation}
Each discrete angular frequency is a linear combination of the two 
fundamental orbital frequencies, $\Omega_{\phi}$ and $\Omega_{r}$ \cite{cut94}.

The next step is to express $\widetilde{S}^{lmk}(\omega,r)$ in terms of 
$S^{lm}(t,r)$ and substitute the result into (\ref{eq:nsfourser}).
Starting with (\ref{eq:slmk}) and substituting in succession 
(\ref{eq:slmtildef}) and (\ref{eq:slmtr}) produces
\begin{equation}
\begin{split}
\label{eq:nslmk}
\widetilde{S}^{lmk}(\omega,r)&=\frac{1}{P}\int_{0}^{P}e^{i k \Omega_{r} t^{\prime}}
\widetilde{S}^{lm}(t^{\prime},r)dt^{\prime}
\\&=\frac{1}{P}\int_{0}^{P}e^{i (k \Omega_{r}+m\Omega_{\phi})t^{\prime}}
S^{lm}(t^{\prime},r)dt^{\prime}
\\&=\frac{\Omega_{r}}{2\pi}\int_{0}^{P}e^{i(\omega_{mk}t^{\prime}- m\phip)}
f^{lm}(r)(\rdot)^{n}\deltar dt^{\prime}\;.
\end{split}
\end{equation}
In the final line, the definition $\Omega_{r}=\frac{2\pi}{P}$ has been 
used.  Combining (\ref{eq:nsfourser}) and (\ref{eq:nslmk}) gives the 
Fourier transform of $S^{lm}(t,r)$ as
\begin{equation}
\label{eq:ftint}
S^{lm}(\omega,r)=\sum^{\infty}_{k=-\infty}\delta(\omega-\omega_{mk})
\frac{\Omega_{r}}{2\pi}\int_{0}^{P}e^{i(\omega_{mk}t^{\prime}- m\phip)}
f^{lm}(r)(\rdot)^{n}\deltar dt^{\prime}\;,\;\;n=0,1,2\;.
\end{equation}

The integral in \eqref{eq:ftint} is evaluated by changing the variable of 
integration from $t^{\prime}$ to $r^{\prime}$ and using the radial delta 
function.  The analysis below follows the steps taken in \cite{cut94} for 
the source term of the Teukolsky equation.  As explained in \cite{cut94}, 
a single orbit is divided into two parts.  During the first part, the 
orbiting mass moves from periastron ($r^{\prime}=r_{\text{min}}$, $t^{\prime}=0$ and 
$\phip=0$) to apastron ($r^{\prime}=r_{\text{max}}$, $t^{\prime}=\frac{P}{2}$ and 
$\phip=\frac{\Delta\phi}{2}$).   During the second part, the mass moves from 
apastron back to periastron ($t^{\prime}=P$ and $\phip=\Delta\phi$).  
The limits of integration must take into account this division, so the 
integral is split.  Further, $\rdot> 0$ when $r^{\prime}$ is increasing
($0<t^{\prime}<\frac{P}{2}$), but $\rdot< 0$ for $\frac{P}{2}<t^{\prime}<P$.  The 
integral in (\ref{eq:ftint}) becomes
\begin{equation}
\begin{split}
\int_{0}^{P}&e^{i(\omega_{mk}t^{\prime}- m\phip)}f(r)(\rdot)^{n}\deltar dt^{\prime}\\&=
\int_{0}^{\frac{P}{2}}e^{i(\omega_{mk}t^{\prime}- m\phip)}f(r)(\rdot)^{n}\deltar dt^{\prime}
+\int_{\frac{P}{2}}^{P}e^{i(\omega_{mk}t^{\prime}- m\phip)}f(r)(\rdot)^{n}\deltar dt^{\prime}
\\&=\!\!\int_{r_{\text{min}}}^{r_{\text{max}}}\!\!e^{i(\omega_{mk}t^{\prime}- m\phip)}f(r)(\rdot)^{n}
\frac{\deltar}{\rdot} dr^{\prime}
\!+\!\int_{r_{\text{max}}}^{r_{\text{min}}}\!\!e^{i(\omega_{mk}t^{\prime}- m\phip)}
f(r)(\rdot)^{n} \frac{\deltar}{\rdot} dr^{\prime}
\\&=\!\!\int_{r_{\text{min}}}^{r_{\text{max}}}\!\!e^{i(\omega_{mk}t^{\prime}- m\phip)}f(r)(\rdot)^{n}
\frac{\deltar}{\ardot} dr^{\prime}\!+\!\int_{r_{\text{min}}}^{r_{\text{max}}}
\!\!e^{i(\omega_{mk}t^{\prime}- m\phip)}
f(r)(\rdot)^{n} \frac{\deltar}{\ardot} dr^{\prime}\!.
\end{split}
\end{equation}
Reversing the limits of integration in the second integral on the last line 
gives a minus sign, which is negated by $\ardot=-\rdot$ in the denominator.   
In the first integral, $t^{\prime}=\hat{t}$ and $\phip=\hat{\phi}$.  
In the second integral, $t^{\prime}=P-\hat{t}$ and $\phip=
\Delta\phi-\hat{\phi}$, so that
\begin{equation}
e^{i(\omega_{mk}t^{\prime}- m\phip)}=e^{i(\omega_{mk}P- m\Delta\phi)}
e^{-i(\omega_{mk}\hat{t}- m\hat{\phi})}=e^{i k 2\pi}e^{-i(\omega_{mk}\hat{t}- m\hat{\phi})}
=e^{-i(\omega_{mk}\hat{t}- m\hat{\phi})}\;.
\end{equation}
Also in the second integral, $(\rdot)^{n}=(-1)^{n}\ardot^{n}$.  
We make these substitutions and then use the delta functions to evaluate 
the integrals.  The integration gives
\begin{multline}
\label{eq:intslmk}
\int_{0}^{P}e^{i(\omega_{mk}t^{\prime}- m\phip)}f^{lm}(r)(\rdot)^{n}\deltar dt^{\prime}
=f^{lm}(r)\theta(r-r_{\text{min}})\theta(r_{\text{max}}-r)
\\\times\left\{e^{i(\omega_{mk}\hat{t}(r)- m\hat{\phi}(r))}\frac{\ardot^{n}}{\ardot}
+e^{-i(\omega_{mk}\hat{t}(r)- m\hat{\phi}(r))}\frac{(-1)^{n}\ardot^{n}}{\ardot}\right\}\;,
\end{multline}
where $n=0,1,2$.  We define $\theta(x)=1$, $x>0$, and $\theta(x)=0$, $x<0$.  
The theta functions replace the limits of integration and 
restrict $r$ to the radial range of orbital motion, 
because the product $\theta(r-r_{\text{min}})\theta(r_{\text{max}}-r)$ 
implies $r_{\text{min}} \le r\le  r_{\text{max}}$.  
Equation \eqref{eq:intslmk} is the evaluation of the integral in the 
expression for $S^{lm}(\omega,r)$ \eqref{eq:ftint}.

Further simplification of (\ref{eq:intslmk}) depends on the value of $n$.  
The exponentials can be expressed as trigonometric functions using the 
identities $\cos z=\frac{e^{i z}+e^{-iz}}{2}$ and 
$\sin z=\frac{e^{i z}-e^{-iz}}{2i}$ \cite{hmf}.  For $n=0$, the factor in 
curly brackets simplifies to 
$\frac{2}{\ardot}\cos(\omega_{mk}\hat{t}-m\hat{\phi})$.  
The corresponding results for $n=1$ and $n=2$ are 
$2 i\sin(\omega_{mk}\hat{t}- m\hat{\phi})$ and 
$2\ardot\cos(\omega_{mk}\hat{t}- m\hat{\phi})$, respectively.  

To summarize, the Fourier transform of $S^{lm}(t,r)$ for elliptic orbits is 
given by
\begin{equation}
\label{eq:slmft}
\begin{split}
S^{lm}(\omega,r)=&\frac{1}{2\pi}\int_{\!-\infty}^{\infty} e^{i\omega t}S^{lm}(t,r)\,dt
=\frac{1}{2\pi}\int_{\!-\infty}^{\infty} e^{i\omega t}f^{lm}(r)
(\rdot)^{n}\delta(r-r^{\prime}(t)) e^{-i m \phip(t)}dt
\\=&\sum^{\infty}_{k=-\infty}\delta(\omega-\omega_{mk})
\theta(r-r_{\text{min}})\theta(r_{\text{max}}-r)\frac{\Omega_{r}}{2\pi}f^{lm}(r)
\\&\quad\quad\times
\begin{cases}
\!\frac{2}{\ardot}\cos(\omega_{mk}\hat{t}- m\hat{\phi})&\quad n=0\;,
\\2 i\sin(\omega_{mk}\hat{t}- m\hat{\phi})&\quad n=1\;,
\\2\ardot\cos(\omega_{mk}\hat{t}- m\hat{\phi})&\quad n=2\;.
\end{cases}
\end{split}
\raisetag{27pt}
\end{equation}
The functions $f^{lm}(r)$ are found by inspecting the time-radial 
coefficients listed in \eqref{eq:firstrco}-\eqref{eq:lastrco}.  
In these expressions, we will use the chain rule to substitute $\axdot$ 
for $\gamma\ardot$ and $\dphids$ for $\gamma\phdot$, 
where, as before, $\gamma=\frac{dt^{\prime}}{d\tau}$.

After substituting the various $f^{lm}(r)$ into \eqref{eq:slmft}, we 
find that the radial coefficients of the stress energy tensor are
\begin{equation}
\label{eq:firstrc}
\pi So^{lm}_{02}(\omega,r)=m_{0}\sum_{k=-\infty}^{\infty}
\frac{\Omega_{r}\gamma(r-2 M)}
{2(\lambda+1)r\axdot}\dphids\costp\ndylmpi\;,
\end{equation}

\begin{equation}
\pi So^{lm}_{12}(\omega,r)=m_{0}\sum_{k=-\infty}^{\infty}\frac{-i\Omega_{r}r}
{2(\lambda+1)(r-2 M)}\dphids\sintp\ndylmpi\;,
\end{equation}

\begin{equation}
\pi So^{lm}_{22}(\omega,r)=m_{0}\sum_{k=-\infty}^{\infty}\frac{-i m\Omega_{r}r^{2}}
{4\lambda(\lambda+1)\axdot}\left(\dphids\right)^{2}\costp\ndylmpi\;,
\end{equation}

\begin{equation}
\pi Se^{lm}_{00}(\omega,r)=m_{0}\sum_{k=-\infty}^{\infty}
\frac{\Omega_{r}\gamma^{2}(r-2 M)^{2}}{r^{4}\axdot}\costp\nylmpi\;,
\end{equation}

\begin{equation}
\pi Se^{lm}_{01}(\omega,r)=m_{0}\sum_{k=-\infty}^{\infty}\frac{-i\Omega_{r}\gamma}
{r^{2}}\sintp\nylmpi\;,
\end{equation}

\begin{equation}
\pi Se^{lm}_{02}(\omega,r)=m_{0}\sum_{k=-\infty}^{\infty}
\frac{i m\Omega_{r}\gamma(r-2 M)}{2(\lambda+1)r\axdot}\dphids
\costp\nylmpi\;,
\end{equation}

\begin{equation}
\pi Se^{lm}_{11}(\omega,r)=m_{0}\sum_{k=-\infty}^{\infty}\frac{\Omega_{r}\axdot}
{(r-2 M)^{2}}\costp\nylmpi\;,
\end{equation}

\begin{equation}
\pi Se^{lm}_{12}(\omega,r)=m_{0}\sum_{k=-\infty}^{\infty}
\frac{m\Omega_{r}r}{2(\lambda+1)(r-2 M)}\dphids\sintp\nylmpi\;,
\end{equation}
\begin{equation}
\pi Se^{lm}_{22}(\omega,r)=m_{0}
\sum_{k=-\infty}^{\infty}\frac{\Omega_{r}(\lambda+1-m^{2}) r^{2}}
{4\lambda(\lambda+1)\axdot}\left(\dphids\right)^{2}\costp\nylmpi\;,
\end{equation}

\begin{equation}
\label{eq:lastrc}
\pi Ue^{lm}_{22}(\omega,r)=m_{0}\sum_{k=-\infty}^{\infty}\frac{\Omega_{r}r^{2}}
{2\axdot}\left(\dphids\right)^{2}\costp\nylmpi\;.
\end{equation}
Each of these should be multiplied by
\begin{equation}
\label{eq:thedelta}
\theta(r-r_{\text{min}})\theta(r_{\text{max}}-r)\delta(\omega-\omega_{mk})\;.
\end{equation}
Equations \eqref{eq:firstrc}-\eqref{eq:lastrc} have been derived 
for elliptic orbits, but may also be used for circular orbits.  
Expressions for circular orbits are obtained by restricting the range 
of $k$ to $k=0$, so that $\omega_{mk}=m \Omega_{\phi}$ \cite{cut94}.  
In turn, this implies $\omega_{mk}\hat{t}-m\hat{\phi}=0$ because 
$\phi=\Omega_{\phi} t$ \eqref{eq:cirtp} for circular orbits.  Four of the 
radial coefficients -- $So_{12}$, $Se_{01}$, $Se_{11}$ and $Se_{12}$ -- 
represent components of the stress energy tensor $T_{\mu\nu}$ 
\eqref{eq:covdelttmunu} that have factors of the radial velocity, which is zero 
for circular orbits.  These four are zero, either because $\axdot=0$ ($Se_{11}$) 
or because $\sintp=0$ for circular orbits ($So_{12}$, $Se_{01}$, $Se_{12}$).

Some of the radial coefficients have a factor of $\axdot$ in the 
denominator.  These factors will be zero for circular orbits and zero 
at the turning points $r_{\text{min}}$ and $r_{\text{max}}$ of elliptic 
orbits \cite{cut94}.  The treatment of 
these singularities is discussed in section~\ref{sec:irweqn}, 
following equation~\eqref{eq:cov}.

In deriving the Fourier transforms for elliptic orbits, we have 
followed the analogous treatment of the Teukolsky source terms 
in \cite{cut94}.  A different derivation was given by Tanaka and 
others in \cite{tsstn93}.  They solved the Regge-Wheeler equation for 
both even and odd parity modes, 
but with a source derived from the Teukolsky equation.  The Fourier 
transform in \cite{tsstn93} has multiple radial integrals and is 
more complicated than that derived above, although it also relies 
on the two fundamental frequencies.  

For bound orbits, the frequency is zero only if 
$\omega_{mk}=k\Omega_{r}+m\Omega_{\phi}=0$.  From the definitions 
$\Omega_{r}=\frac{2\pi}{P}$ and $\Omega_{\phi}=\frac{\Delta\phi}{P}$, 
that equality will be satisfied only when:  (1) $k=m=0$, or (2) $\Delta\phi$ 
is a rational fraction of $2 \pi$.  The latter condition generally 
will not be met because of the definition of $\Delta\phi$ \cite{cut94}.  
Accordingly, bound orbit zero frequency modes have $k=m=0$.  
Section~\ref{sec:zevpareqo} describes the zero frequency even parity 
solutions for $l=1$ and notes that this particular mode is not 
important for bound orbits.  The Fourier transforms show why this 
is so.  From the discussion at the end of section~\ref{sec:ttmunu}, 
the even parity $l=1$ modes are non-zero only when $m=\pm 1$, 
because the even parity angular functions are zero unless $l+m$ is even.  
The requirement $k=m=0$ is not met.  However, this reasoning does not 
preclude even parity $l=1$ zero frequency modes for orbital motion which 
is not circular or elliptic.  

The calculation of the stress energy tensor is now complete, for bound orbits.  
The radial factors in equations (\ref{eq:firstrc})-(\ref{eq:lastrc}) may be 
substituted into the multipole expansion of the stress energy tensor 
in equations \eqref{eq:otmunu} and \eqref{eq:etmunu}.  The Fourier 
decomposition reveals the frequency spectrum of the gravitational 
radiation.  For circular orbits, the characteristic frequencies are 
integral multiples of the orbital angular 
frequency $\Omega_{\phi}$ \cite{pois93}.  
For elliptic orbits, the frequencies are linear combinations of two 
fundamental frequencies:  the orbital angular frequency $\Omega_{\phi}$ 
and the radial angular frequency $\Omega_{r}$ \cite{cut94}.		
\chapter{\label{rweqnchap}Solution of Generalized Regge-Wheeler 
Equations}

From equation \eqref{eq:grweqn}, the generalized Regge-Wheeler equation is
\begin{equation}
\label{eq:nzfgrw}
\frac{d^2 \psi_s(r_{*})}{d r_{*}^2}+\omega^2 \psi_s(r_{*})
-\ff\left(\frac{l(l+1)}{r^2}+(1-s^{2})\frac{2M}{r^3}\right)\psi_s(r_{*})
=S_{slm}(\omega,r_{*})\;,
\end{equation}
where $r_{*}=r+2 M \ln\!\left[r/(2 M)-1\right]$, $l(l+1)=2(\lambda+1)$ 
and $s=0,1,2$ \cite{hs00}, \cite{leav86}, \cite{leopois97}.  
Inspection of the odd and even parity harmonic gauge 
solutions shows that we need to solve \eqref{eq:nzfgrw} only for 
$l\ge s$.  Section~\ref{sec:nzrweqn} discusses 
non-zero frequency homogeneous solutions to \eqref{eq:nzfgrw} and 
concludes with a discussion of the Zerilli equation.  
Section~\ref{sec:zrweqn} does the same for zero frequency.  Finally, 
section~\ref{sec:irweqn} explains the construction of inhomogeneous 
solutions.

\section{\label{sec:nzrweqn}Non-Zero Frequency Homogeneous 
Solutions}

Non-zero frequency homogeneous solutions for the case $s=2$ are 
discussed by Chandrasekhar \cite{chandra92}.  His work is also 
applicable to the cases $s=0$ and $s=1$.  Chandrasekhar's notation is 
different from that below.  For example, his solutions have a time 
dependence of $e^{i\sigma t}$, instead of the $e^{-i\omega t}$ factor used 
in this thesis.  The discussion in the next four paragraphs is 
taken mainly from his book \cite{chandra92}.

Chandrasekhar points out that the homogeneous Regge-Wheeler equation 
resembles the one-dimensional, time-independent Schrodinger equation, 
with $\omega^{2}$ taking the place of the energy eigenvalue.  
Both equations represent a wave interacting with a potential, so 
similar solution methods can be used for each.  In equation 
\eqref{eq:nzfgrw}, the potential is gravitational and results from 
the background spacetime curvature due to the central mass $M$.  The 
coordinate $r_{*}$ has range $-\infty<r_{*}<\infty$.  The potential goes 
to zero for large $r$ and near the event horizon at $2 M$, so 
asymptotically equation \eqref{eq:nzfgrw} becomes
\begin{equation}
\label{eq:asympgrweqn}
\frac{d^2 \psi_s(r_{*})}{d r_{*}^2}+\omega^2 \psi_s(r_{*})=0\;,\,
r_{*}\to\pm\infty\;,
\end{equation}
with solutions $e^{\pm\iom r_{*}}$.  Because the second order differential 
equation has only two linearly independent homogeneous solutions, 
the asymptotic forms must be linear combinations of the exponentials, 
chosen to represent the scattered waves.  Accordingly, one homogeneous 
solution is
\begin{equation}
\label{eq:inexp}
\psi^{\text{in}}_{s}\sim e^{-\iom r_{*}},r\to 2 M\;;\;
\psi^{\text{in}}_{s}\sim B^{\text{in}}e^{-\iom r_{*}}
+B^{\text{out}}e^{\iom r_{*}},r\to\infty\;.  
\end{equation}
This is an incoming wave at large $r$ of amplitude $B^{\text{in}}$, 
a reflected wave of amplitude $B^{\text{out}}$ and a transmitted, ingoing 
wave of unit amplitude near the event horizon.  
A second homogeneous solution is
\begin{equation}
\label{eq:outexp}
\psi^{\text{out}}_{s}\sim e^{\iom r_{*}},r\to \infty\;;\;
\psi^{\text{out}}_{s}\sim A^{\text{in}}e^{-\iom r_{*}}
+A^{\text{out}}e^{\iom r_{*}},r\to\ 2 M\;.  
\end{equation}
This is an outgoing wave that starts near the event horizon with 
amplitude $A^{\text{out}}$.  Part is reflected back, with amplitude 
$A^{\text{in}}$, and part goes outwards to infinity, with unit amplitude.  
More generally, we can write, for all $r_{*}$,
\begin{equation}
\label{eq:inexpout}
\psi^{\text{in}}_{s}=B^{\text{in}}\overline{\psi}^{\text{out}}_{s}
+B^{\text{out}}\psi^{\text{out}}_{s}\;,\;
\psi^{\text{out}}_{s}=A^{\text{in}}\psi^{\text{in}}_{s}
+A^{\text{out}}\overline{\psi}^{\text{in}}_{s}\;,
\end{equation}

The Wronskian of two linearly independent homogeneous solutions is 
constant because there is no first derivative term \cite{arfken85}.  
To calculate the Wronskian $W_s$ of $\psi^{\text{out}}_s$ and 
$\psi^{\text{in}}_s$, it is convenient to use the asymptotic solutions 
\eqref{eq:inexp} and \eqref{eq:outexp} for large $r$, which gives
\begin{equation}
\label{eq:grwwr}
W_s\equiv\frac{d\psi^{\text{out}}_s(r_{*}) }{d r_{*}}\psi^{\text{in}}_s(r_{*})-
\frac{d\psi^{\text{in}}_s(r_{*}) }{d r_{*}}\psi^{\text{out}}_s(r_{*})
=2 i \omega B_s^{\text{in}}(\omega)\;.
\end{equation}
Similarly, the constant ${B}_s^{\text{out}}$ is obtained from a 
different Wronskian
\begin{equation}
\label{eq:bouteqn}
\frac{d\psi^{\text{out}}_s(r_{*}) }{d r_{*}}\overline{\psi}^{\text{in}}_s(r_{*})-
\frac{d\overline{\psi}^{\text{in}}_s(r_{*}) }{d r_{*}}\psi^{\text{out}}_s(r_{*})
=2 i \omega \overline{B}_s^{\text{out}}(\omega)\;.
\end{equation}
Substituting the solutions near the event horizon instead into 
\eqref{eq:grwwr}-\eqref{eq:bouteqn} and comparing the results to the 
large $r$ case leads to
\begin{equation}
\label{eq:atob}
A^{\text{in}}=-\overline{B}^{\text{out}}\;,\,A^{\text{out}}=B^{\text{in}}\;.
\end{equation}

Given an incident wave of unit magnitude, the reflection 
coefficient $\mathbb{R}$ and transmission coefficient $\mathbb{T}$ 
are related by
\begin{equation}
\label{eq:rpteq1}
\mathbb{R}+\mathbb{T}=1\;,
\end{equation}
which represents flux conservation.  The coefficients $\mathbb{R}$ 
and $\mathbb{T}$ are the squared complex magnitudes of the reflected 
and transmitted wave amplitudes.  Equation \eqref{eq:rpteq1} follows 
from the constancy of the Wronskian.  It is derived by calculating 
the Wronskian for a homogeneous solution and its conjugate at 
$r_{*}\to\infty$ and $r_{*}\to -\infty$ and requiring that the Wronskians 
for the two limits be equal.  If we divide $\psi^{\text{in}}$ 
\eqref{eq:inexp} by $B^{\text{in}}$, then 
\begin{equation}
\label{eq:psiinrt}
\mathbb{R}=\frac{\lvert B^{\text{out}}\rvert^{2}}{\lvert B^{\text{in}}\rvert^{2}}\;,\,
\mathbb{T}=\frac{1}{\lvert B^{\text{in}}\rvert^{2}}\;,
\end{equation}
which implies \cite{ashby}
\begin{equation}
\label{eq:bin2bout2}
\lvert B^{\text{in}}\rvert^{2}-\lvert B^{\text{out}}\rvert^{2}=1\;.
\end{equation}
This relation may be derived from $\psi^{\text{out}}$ \eqref{eq:outexp} 
as well.  Chandrasekhar also shows $\mathbb{R}$ and $\mathbb{T}$ are 
the same for the Regge-Wheeler equation ($s=2$) and the Zerilli equation, 
for incident waves of unit magnitude.

Chandrasekhar has different notation for the constants in his 
discussion. The notation above is typical of that used 
elsewhere \cite{ashby}, \cite{pois93}, \cite{tsstn93}.

Additionally, Chandrasekhar derives a solution in the form of an integral 
equation, which can be solved by iteration to give an infinite series 
\cite{chandra92}.  In quantum mechanics, successive iterations form a 
Born series, which represents multiple scattering 
interactions \cite{griffiths}, \cite{sakurai}.  
The integral solution suggests that the waves may scatter off the 
background spacetime curvature multiple times, as shown in 
Figure~\ref{fig:sch}.  Solution by iteration can be used to study 
scattering of late time tails \cite{ching95}.

The homogeneous solutions are calculated numerically.  Usually, 
this is done by starting with series solutions for $\psi^{\text{out}}$ 
at large $r$ and $\psi^{\text{in}}$ near the event horizon \cite{ashby}, 
\cite{cufps93}, \cite{cut94}, \cite{dmw03}.  
In terms of the dimensionless variables $x=r/(2 M)$ and $\Omega=2 M \omega$, 
the homogeneous generalized Regge-Wheeler equation is
\begin{multline}
\label{eq:nzfgrwx}
\frac{(x-1)^2}{x^2}\frac{d^{2}\psi}{dx^{2}}
+\frac{(x-1)}{x^3}\frac{d\psi}{dx}+\Omega^2\psi
\\+\frac{\left(1+s^2 (x-1)+\left(-1+l+l^2\right) x-l (1+l) x^2
\right)}{x^4}\psi=0\;.
\end{multline}
The outgoing series solution is
\begin{equation}
\label{eq:psioutx}
\psi^{\text{out}}(x)=e^{i\Omega x_{*}}\sum_{n=0}^{\infty}\frac{a_{n}}{x^{n}}\;,\,
x_{*}=x+\ln[x-1]\;.
\end{equation}
The recursion relation for the series coefficients is
\begin{equation}
\label{eq:anout}
a_{n}=-\frac{\left(l+l^2+n-n^2\right)}{2 i\Omega n}a_{n-1}
-\frac{\left(1-2 n+n^2-s^2\right)}{2 i\Omega n}a_{n-2}\;,
\end{equation}
where $a_{0}=1$ and $a_{-1}=0$.  
For $\psi^{\text{in}}$, we change the independent variable in the differential 
equation \eqref{eq:nzfgrwx} to $X=1-1/x$ and obtain
\begin{multline}
\label{eq:nzfgrwbigx}
(X-1)^4 X^2 \frac{d^{2}\psi}{dX^{2}}
+(X-1)^3 X (3 X-1)\frac{d\psi}{dX}
\\+\left[\Omega^2-\left(l+l^2+(-1+s^2)
(X-1)\right) (X-1)^2 X\right]\psi=0\;.
\end{multline}
The ingoing series solution is
\begin{equation}
\label{eq:nzfpin}
\psi^{\text{in}}(X)=e^{-i\Omega X_{*}} \sum^{\infty}_{n=0}a_{n}X^{n}\;,\,
X_{*}=\frac{1}{1-X}+\ln\left[\frac{X}{1-X}\right]\;,
\end{equation}
and the recursion relation is
\begin{equation}
\label{eq:anin}
a_{n}=-\frac{\left(1+l+l^2-2 n+2 n^2-s^2\right)}{(2 i\Omega-n)n}a_{n-1}
+\frac{\left(1-2 n+n^2-s^2\right)}{(2 i\Omega-n) n}a_{n-2}\;,
\end{equation}
where $a_{0}=1$ and $a_{-1}=0$.  The series above agree with those derived 
by others for particular spins \cite{ashby} ($s=2$), 
\cite{dmw03} ($s=0$, outgoing), \cite{cut94} (first three terms of $s=2$).

The series for $\psi^{\text{in}}$ \eqref{eq:nzfpin} converges very slowly, 
unless evaluated near the event horizon.  The series for 
$\psi^{\text{out}}$ \eqref{eq:psioutx} converges only for large $r$ and only 
for a finite number of terms.  If expanded to a large number of terms, 
it starts to diverge.  In this sense, it is an asymptotic series 
\cite{ashby}, \cite{mandw}.  The orbits of interest are in an intermediate 
region, so we need to use a differential equation solver to go outwards 
from the $\psi^{\text{in}}$ series evaluation point and inward from the 
$\psi^{\text{out}}$ series.  The Bulirsch-Stoer method, which is described in 
\textit{Numerical Recipes} \cite{numr}, is often used for this 
purpose \cite{cufps93}, \cite{cut94}.

Numerical calculations in this thesis were done with a different method, 
involving iterated power series.  We can expand $\psi^{\text{in}}$ and 
$\psi^{\text{out}}$ as power series about a non-singular point $x_{0}$:
\begin{equation}
\label{eq:psitay}
\psi(x)=\sum_{n=0}^{\infty}a_{n}(x-x_{0})^n\;,
\end{equation}
where $\psi(x)$ is either $\psi^{\text{in}}$ or $\psi^{\text{out}}$.
The recursion relation is
\begin{equation}
\begin{split}
\label{eq:antay}
a_{n}=&\frac{1}{n(n-1)(x_{0}-1)^2 x_{0}^2}\Big\{
-\left[(n-1)(x_{0}-1)x_{0}(5-8 x_{0}+n (-2+4 x_{0}))\right]a_{n-1}
\\&+\left[-9-s^2 (x_{0}-1)+41 x_{0}-l(l+1) x_{0}-36 x_{0}^2+l(l+1) x_{0}^2
-\Omega^2 x_{0}^4\right.\\&\left.+n^2 \left(-1+6 x_{0}-6 x_{0}^2\right)
+n \left(6-32 x_{0}+30 x_{0}^2\right)\right]a_{n-2}
-\left[-28+l(l+1)\right.\\&\left.+s^2+n (15-28 x_{0})+48 x_{0}-2 l(l+1) x_{0}
+4 \Omega^2 x_{0}^3+n^2 (-2+4 x_{0})\right]a_{n-3}
\\&+\left[-20+l(l+1)+9 n-n^2-6 \Omega^2 x_{0}^2\right]a_{n-4}
-4 \Omega^2  x_{0}a_{n-5}-\Omega^2 a_{n-6}\Big\}\;,
\end{split}
\end{equation}
where the initial values are
\begin{equation}
\label{eq:antaymisc}
a_{0}=\psi(x_{0})\;,\,a_{1}=\left.\frac{d\psi}{dx}\right|_{x=x_{0}},\;
a_{n}=0 \;\,\text{for}\;\, n<0\;.  
\end{equation}
The expansion around a non-singular point is a Taylor series, 
because power series are unique \cite{arfken85}.  
The series converges slowly if the difference $x-x_{0}$ is too large, so 
the series is applied by successive iterations.  The starting values 
$a_{0}$ and $a_{1}$ for the first iteration are taken from the 
$\psi^{\text{in}}$ \eqref{eq:nzfpin} and 
$\psi^{\text{out}}$ \eqref{eq:psioutx} series.  The next iteration 
uses the results of the first iteration and so on.

Homogeneous solutions of the Zerilli equation \eqref{eq:zereqn} can 
be obtained from solutions of the Regge-Wheeler equation by applying 
differential operators \cite{ashby}, \cite{chandra92}.  The relations 
are 
\begin{equation}
\label{eq:oddevout}
\psi^{\text{out}}_{2,Z}=\frac{1}{\lambda+\lambda^2+3\iom M}\!
\left[\!\left(\lambda+\lambda^2+\frac{9 M^2(r-2 M)}
{r^2(3 M+\lambda r)}\right)\psi^{\text{out}}_{2,\text{RW}}
+3 M \left(1-\frac{2 M}{r}\right)
\frac{d\psi^{\text{out}}_{2,\text{RW}}}{dr}\right]
\end{equation}
and
\begin{equation}
\label{eq:oddevin}
\psi^{\text{in}}_{2,Z}=\frac{1}{\lambda+\lambda^2-3\iom M}\!
\left[\!\left(\lambda+\lambda^2+\frac{9 M^2(r-2 M)}
{r^2(3 M+\lambda r)}\right)\psi^{\text{in}}_{2,\text{RW}}
+3 M \left(1-\frac{2 M}{r}\right)
\frac{d\psi^{\text{in}}_{2,\text{RW}}}{dr}\right]\!,
\end{equation}
where ``Z'' refers to a homogeneous solution of the Zerilli equation 
and ``RW'' means a homogeneous solution of the generalized 
Regge-Wheeler equation with $s=2$.  The differential operators are 
normalized so that $\psi^{\text{out}}_{2,Z}\to e^{\iom r_{*}}$ 
as $r\to\infty$ and $\psi^{\text{in}}_{2,Z}\to e^{-\iom r_{*}}$ 
as $r\to 2 M$, and to this extent the operators given differ from those 
in the two references above.

\section{\label{sec:zrweqn}Zero Frequency Homogeneous Solutions}

For zero frequency, the homogeneous generalized Regge-Wheeler equation is
\begin{equation}
\label{eq:zfgrw}
\frac{d^2 \psi_s(r_{*})}{d r_{*}^2}
-\ff\left(\frac{l(l+1)}{r^2}+(1-s^{2})\frac{2M}{r^3}\right)\psi_s(r_{*})=0\;.
\end{equation}
Solutions of (\ref{eq:zfgrw}) are related to hypergeometric 
functions \cite{cmp78} (cases $s=1,2$), \cite{leav86} (case $s=2$).  
Cf.~\cite{chrz75} (Teukolsky equation), 
\cite{rw57} (odd parity field equations), 
\cite{zerp70} (even parity field equations).  

Using the notation of \cite{hmf}, the hypergeometric series 
is defined as
\begin{equation}
\label{eq:hypser}
{}_{2}F_{1}(a,b;c;x)=\sum^{\infty}_{n=0}
\frac{(a)_{n}(b)_{n}}{(c)_{n}}\frac{x^{n}}{n!}\;.
\end{equation}
The quantity $(a)_{n}$ is Pochhammer's symbol, given by
\begin{equation}
(a)_{n}=a(a+1)(a+2)\cdots(a+n-1)=\frac{\Gamma(a+n)}{\Gamma(a)}\;.
\end{equation}
Here, $\Gamma(a)$ is the gamma function.
The hypergeometric functions satisfy a second order differential 
equation
\begin{equation}
\label{eq:ddhyp}
x(1-x)\frac{d^2 y}{dx^2}+[c-(a+b+1)x]\frac{d y}{dx}-ab\,\,y=0\;,
\end{equation}
where $y(x)={}_{2}F_{1}(a,b;c;x)$.

The hypergeometric series in (\ref{eq:hypser}) converges within the unit 
circle $|x|=1$ and, in some cases, on the unit circle \cite{hmf}.  Because 
$2 M<r<\infty$, we change variables in (\ref{eq:zfgrw}) from $r$ to 
$z=\frac{2 M}{r}$, with $0\le z\le 1$.  Using the chain rule of 
differentiation, $\frac{d\psi(r)}{dr}=-\frac{z^{2}}{2M}\frac{d\psi(z)}{dz}$.  
In terms of $z$, the Regge-Wheeler equation (\ref{eq:zfgrw}) is
\begin{equation}
\label{eq:zrweqn}
(z-1) z^2 \frac{d^2\psi(z)}{dz^2}
+z (3 z-2) \frac{d\psi(z)}{dz}
+\left(l+l^2+z-s^2 z\right)\psi(z)=0\;.
\end{equation}
To solve (\ref{eq:zrweqn}), we substitute $\psi(z)=g(z)y(z)$ and solve for 
$g(z)$ so that the resulting differential equation for $y(z)$ is in the 
form of 
(\ref{eq:ddhyp}).  If $g(z)=z^{-l-1}$, we find
\begin{equation}
\label{eq:ddhypin}
z(1-z)\frac{d^2 y}{dz^2}+(2 l (z-1)-z)\frac{d y}{dz}
-\left(l^2-s^2\right)y=0\;,
\end{equation}
which is a hypergeometric equation with $a=-l-s$, $b=-l+s$ and 
$c=-2 l$. Accordingly, one solution to (\ref{eq:zrweqn}) is
\begin{equation}
\label{eq:psinz}
\psi^{\text{in}}(z)=z^{-l-1}\,{}_{2}F_{1}(-l-s,-l+s;-2 l;z)\;.
\end{equation}
Because $b=-l+s\le 0$, the hypergeometric series 
terminates and is a polynomial of degree $z^{l-s}$ \cite{hmf}.  
This solution is finite as $r\to 2M$ and diverges like $r^{l+1}$ 
as $r\to\infty$.  Since it is bounded near the horizon, it is labeled 
$\psi^{\text{in}}$.  Similarly, setting $g(z)=z^{l}$ gives a different 
hypergeometric equation
\begin{equation}
z(1-z)\frac{d^2 y}{dz^2}+(2+2 l (1-z)-3 z)\frac{d y}{dz}
-(1+2 l+l^2-s^2)y=0
\end{equation}
and a second solution to (\ref{eq:zrweqn}),
\begin{equation}
\label{eq:psoutz}
\psi^{\text{out}}(z)=z^{l}\,{}_{2}F_{1}(1+l-s,1+l+s;2+2 l;z)\;.
\end{equation}  
The second solution is designated $\psi^{\text{out}}$ because it is bounded 
as $r\to\infty$, where it behaves as $r^{-l}$.  It is an infinite series.  Equation 
\eqref{eq:psoutz} agrees with the 
$s=1$ and $s=2$ solutions given in \cite{cmp78} and \cite{leav86}.  
The Wronskian of the two solutions, as defined in (\ref{eq:grwwr}), is
\begin{equation}
\label{eq:zwron}
W_{s}=-\frac{1+2l}{2M}\;.
\end{equation}
Because the Wronskian is non-zero, $\psi^{\text{in}}$ and $\psi^{\text{out}}$ 
are linearly independent.

The solutions also can be expressed in terms of the variable 
$X=1-\frac{2 M}{r}=1-z$, where $0\le X \le 1$.  This form is more suitable 
for $r$ near $2 M$.  Using equation (15.3.10) of \cite{hmf} to change 
variables in the hypergeometric function, $\psi^{\text{out} }$ becomes
\begin{multline}
\label{eq:psioutofx}
\psi^{\text{out}}(X)=-\frac{\Gamma(2+2 l)}{\Gamma(1+l-s) 
\Gamma(1+l+s)}(1-X)^{l}\Big\{{}_{2}F_{1}(1+l-s,1+l+s;1;X)\ln[X]
\\\shoveright{+\sum^{\infty}_{n=0}\Big[\frac{(1+l-s)_{n}(1+l+s)_{n}}{(n!)^{2}}
\big(\psi_{d}(1+l-s+n)}
\\+\psi_{d}(1+l+s+n)-2\psi_{d}(1+n)\big)X^{n}\Big]\Big\}\;.
\end{multline}
The symbol $\psi_{d}$ refers to  the digamma function, which is also 
called  $\psi$ function in \cite{hmf}.  For integral $n$, 
\begin{equation}
\psi_{d}(n)=-\gamma+\sum^{n-1}_{k=1}\frac{1}{k}\;,
\end{equation}
where $\gamma=0.5772156649015329\ldots$ is Euler's constant.  
Equation (\ref{eq:psioutofx}) shows
that $\psi^{\text{out}}$ diverges logarithmically as $r\to 2M$, because 
$\ln[X]=\ln\left[1-\frac{2 M}{r}\right]$.  

To convert $\psi^{\text{in}}$ to a function of $X$, we change variables in 
(\ref{eq:ddhypin}) and obtain 
\begin{equation}
X(1-X)\frac{d^2 y}{dX^2}+(1+(2 l-1) X)\frac{d y}{dX}
-\left(l^2-s^2\right)y=0\;,
\end{equation}
which has a solution 
\begin{equation}
\label{eq:yhypin}
y(X)={}_{2}F_{1}(-l-s,-l+s;1;X)
\end{equation}
that is a polynomial of degree $X^{l-s}$.  The hypergeometric function 
in (\ref{eq:yhypin}) is not equal to the hypergeometric function in 
(\ref{eq:psinz}).  From equation (15.1.20) of \cite{hmf},
\begin{equation}
\label{eq:hypinf}
{}_{2}F_{1}(a,b;c;1)
=\frac{\Gamma(c)\Gamma(c-a-b)}{\Gamma(c-a)\Gamma(c-b)}\;, 
\end{equation}
provided $c\neq 0,-1,-2,\ldots$ and $\Re(c-a-b)>0$.  Applying 
(\ref{eq:hypinf}) to (\ref{eq:yhypin}) gives
\begin{equation}
\label{eq:hypfac}
{}_{2}F_{1}(-l-s,-l+s;1;X)\to
\frac{\Gamma(1)\Gamma(1+2 l)}{\Gamma(1+l+s)\Gamma(1+l-s)}
\end{equation}
as  $r\to \infty$ and $X\to 1$.  However, as $r\to\infty$, 
${}_{2}F_{1}(-l-s,-l+s;-2 l;z)\to 1$.  
For equality, we need to multiply (\ref{eq:hypinf}) by the inverse 
of (\ref{eq:hypfac}).  This leads to 
\begin{equation}
\label{eq:psinzx}
\psi^{\text{in}}(X)=\frac{\Gamma(1+l-s) \Gamma(1+l+s)}{\Gamma(1+2 l)}
\left[(1-X)^{-l-1}\,{}_{2}F_{1}(-l-s,-l+s;1;X)\right]\;,
\end{equation}
which is equal to $\psi^{\text{in}}(z)$ from (\ref{eq:psinz}).  
The part in brackets can be expanded as a series in $X$. The resulting 
series is equal to the series for non-zero frequency 
$\psi^{\text{in}}$ from (\ref{eq:nzfpin}), in the limit $\omega\to 0$.  

Hypergeometric functions can be calculated numerically using the program 
\textit{hypser} from \textit{Numerical Recipes} \cite{numr}, which 
calculates hypergeometric series in the form
\begin{multline}
\label{eq:numrhyp}
{}_{2}F_{1}(a,b;c;x)=1+\frac{a b}{c}\frac{x}{1!}
+\frac{a(a+1)b(b+1)}{c(c+1)}\frac{x^{2}}{2!}+\cdots
\\+\frac{a(a+1)\ldots(a+n-1)b(b+1)\ldots(b+n-1)}
{c(c+1)\ldots(c+n-1)}\frac{x^{n}}{n!}+\cdots\cdot
\end{multline}
Calculation of $\psi^{\text{out}}$ involves summing an infinite series.  
The series (\ref{eq:numrhyp}) converges quickly for $|x|\le\frac{1}{2}$
\cite{numr}.  From the definition of $z$, that inequality corresponds 
to $r\ge 4 M$.  Numerical calculations in this thesis will have $r\ge 4 M$.  
Accordingly, \textit{hypser} can be used efficiently to calculate 
$\psi^{\text{out}}$, as given by (\ref{eq:psoutz}).  For $\psi^{\text{in}}$, the 
hypergeometric functions are finite series, so speed of convergence 
is not an issue.  The hypergeometric function in (\ref{eq:psinz}) is 
an alternating series, which causes a loss of significant figures 
for larger $l$ and increasing $z$.  
On the other hand, all terms of the hypergeometric series in 
(\ref{eq:psinzx}) are positive, so (\ref{eq:psinzx}) is better suited 
than (\ref{eq:psinz}) for calculating $\psi^{\text{in}}$ numerically using 
finite precision arithmetic.

Zero frequency homogeneous solutions for the Zerilli equation can 
be obtained using the operators in \eqref{eq:oddevout} 
and \eqref{eq:oddevin}, with the substitution $\omega=0$.  The 
Zerilli solution Wronskian is given by \eqref{eq:zwron}.

\section{\label{sec:irweqn}Inhomogeneous Solutions}

The generalized Regge-Wheeler and Zerilli equations are second order 
differential equations with source terms derived from the stress 
energy tensor for a point mass.  These equations can be written in the 
form \eqref{eq:grweqnop}
\begin{equation}
\label{eq:grweqnop6}
\mathcal{L}_{s}\psi_{s}=S_{s}\;.
\end{equation}
Inhomogeneous solutions 
are obtained from Green's functions, which are constructed from 
homogeneous solutions by the usual methods described in 
\cite{jack75}, \cite{mandw}.  Following~\cite{mandw}, the particular 
solution to \eqref{eq:grweqnop6} is
\begin{equation}
\label{eq:grweqnsol}
\psi_s(\omega,r_{*})=\int_{\!-\infty}^{\infty}
G_s(\omega,r_{*},r^{\prime}_*) S_{s}(\omega,r^{\prime}_*)\,dr^{\prime}_*\;,
\end{equation}
where the Green's function $G_s$ satisfies
\begin{equation}
\label{eq:grweqngf}
\mathcal{L}_{s} G_s=\delta(r_{*}-r^{\prime}_*)\;.
\end{equation}
If we substitute \eqref{eq:grweqnsol} into the left side of 
\eqref{eq:grweqnop6} and apply \eqref{eq:grweqngf} to the integral, 
we get $S_{s}$ on the right.  A homogeneous solution may be added 
to the particular solution \eqref{eq:grweqnsol}, subject to the 
boundary conditions of the problem \cite{mandw}.  As before 
\eqref{eq:rstar}, 
\begin{equation}
\label{eq:diffrstar}
dr_{*}^{\prime}=\frac{dr^{\prime}}{1-\frac{2 M}{r^{\prime}}}\;.
\end{equation}
An unprimed $r$ represents a field point, 
while $r^{\prime}$ is the radial coordinate of the orbiting mass.  

For inhomogeneous solutions, we will follow standard practice and 
use the retarded Green's function
\begin{equation}
\label{eq:retgf}
G^{\text{ret}}_s(\omega,r_{*},r^{\prime}_*)
=\frac{\psi^{\text{out}}_s(r_{*})
\psi^{\text{in}}_s(r^{\prime}_*)}{W_s}\,\theta(r_{*}-r^{\prime}_*)
+\frac{\psi^{\text{in}}_s(r_{*})\psi^{\text{out}}_s(r^{\prime}_*)}{W_s}
\,\theta(r^{\prime}_*-r_{*})\;.
\end{equation}
Substituting $G^{\text{ret}}_s$ into \eqref{eq:grweqnsol} leads to the 
retarded solution
\begin{equation}
\label{eq:retgrwsol}
\psi^{\text{ret}}_s(r_{*})
=\frac{\psi^{\text{out}}_s(r_{*})}{W_s}\,\int_{\!-\infty}^{r_{*}}
\psi^{\text{in}}_s(r^{\prime}_*) S_{s}(r^{\prime}_*)\,d r^{\prime}_*
+\frac{\psi^{\text{in}}_s(r_{*})}{W_s}\,\int^{\infty}_{r_{*}}
\psi^{\text{out}}_s(r^{\prime}_*) S_{s}(r^{\prime}_*)\,d r^{\prime}_*\;.
\end{equation}
The Wronskian $W_s$ is given by \eqref{eq:grwwr} for non-zero frequency 
and \eqref{eq:zwron} for zero frequency.  The reader may verify by 
substitution that \eqref{eq:retgf} is a solution of \eqref{eq:grweqngf} 
and that \eqref{eq:retgrwsol} is a solution of \eqref{eq:grweqnop6}.  
For non-zero frequency, the boundary conditions are that radiation 
does not come from outside the black hole system or from inside the 
event horizon of the large mass.  In other words, the radiation is 
caused by the orbital motion of the small mass. Equation 
\eqref{eq:retgrwsol} is called the causal, or retarded, solution 
because it represents outgoing radiation as $r\to\infty$ ($r>r^{\prime}$) 
and ingoing radiation as $r\to 2 M$ ($r<r^{\prime}$).  This reasoning 
and the retarded solutions above are not new and can be found elsewhere 
in various places \cite{ashby}, \cite{cut94}, 
\cite{drfhs05}, \cite{leav86}, \cite{pois93}, \cite{tsstn93}.  

For zero frequency, we still use the solution \eqref{eq:retgrwsol}, 
but the justification is somewhat different.  As discussed in section 
\ref{sec:zrweqn}, the homogeneous solution $\psi^{\text{out}}$ is bounded 
for large $r$, but diverges logarithmically near the event horizon.  
In contrast, the solution $\psi^{\text{in}}$ diverges as $r\to\infty$, 
but is bounded as $r\to 2 M$.  The boundary conditions are that the 
zero frequency solutions be bounded for both large and small $r$.  
Cf.~\cite{vish70} and~\cite{zerp70}, which apply this requirement to the 
metric perturbation.  The form of $\psi^{\text{ret}}$ is necessary for 
non-divergent behavior in each case.  Even though the zero frequency 
solutions are time independent, we will still use the superscript 
``ret'' for simplicity, although ``bounded'' is a better description.

Using the homogeneous solutions derived in sections~\ref{sec:nzrweqn} 
and~\ref{sec:zrweqn}, we can construct asymptotic solutions from 
\eqref{eq:retgrwsol}, as is also done in the references above.  We will 
start with non-zero frequency.  For large $r$, we have $r>r^{\prime}$, 
so that
\begin{equation}
\label{eq:psiinf}
\psi^{\text{ret}}_s(\omega,r)=A^{\infty}_{slm\omega}e^{i\omega r_{*}}
+O(r^{-1})\;,\; r\to\infty\;,
\end{equation}
where the amplitude constant $A^{\infty}_{slm\omega}$ is
\begin{equation}
\label{eq:ainf}
A^{\infty}_{slm\omega}=\frac{1}{W_s}\,\int_{\!-\infty}^{\infty}
\psi^{\text{in}}_s(r^{\prime}_*) S_{s}(r^{\prime}_*)\,d r^{\prime}_*\;.
\end{equation}
Near the event horizon, $r<r^{\prime}$ and
\begin{equation}
\label{eq:psihor}
\psi^{\text{ret}}_s(\omega,r)=A^{2 M}_{slm\omega}e^{-i\omega r_{*}}+O(X)\;,
\; r\to 2 M\;,
\end{equation}
where
\begin{equation}
\label{eq:ahor}
A^{2 M}_{slm\omega}=\frac{1}{W_s}\,\int^{\infty}_{\!-\infty}
\psi^{\text{out}}_s(r^{\prime}_*) S_{s}(r^{\prime}_*)\,d r^{\prime}_*\;,
\;X=\ff\;.
\end{equation}
As the source $S_{s}$ is proportional to $\frac{\mz}{M}$, so are the 
amplitudes.  Some of the source terms contain radial derivatives of the 
stress energy tensor.  For example, the even parity $S_{2}$ has a 
derivative of $Se_{02}$ \eqref{eq:zereqns1}.  We integrate by parts to 
remove these derivatives, which leads to some integrand terms 
having $\psi_{s}^{\prime}$ instead of $\psi_{s}$.  This is done elsewhere 
for solutions of the Teukolsky equation \cite{cut94}.  When integrating 
by parts, we assume that surface terms vanish; if they did not, we 
could add a homogeneous solution to remove them.

The integral limits are generic and should be replaced 
by limits restricted to the motion of the source.  For elliptic orbits, 
the range of orbital motion is 
$r_{\text{min}}\le r^{\prime} \le r_{\text{max}}$, as discussed in 
Chapters~\ref{eqmochap} and~\ref{tmunuchap}.  In such case, the retarded 
solution is
\begin{multline}
\label{eq:retgrwsole}
\psi^{\text{ret}}_s(r)
=\frac{\psi^{\text{out}}_s(r)}{W_s}\,\int_{r_{\text{min}}}^{r}
\psi^{\text{in}}_s(r^{\prime}) S_{s}(r^{\prime})
\left(1-\tfrac{2 M}{r^{\prime}}\right)^{-1}\,d r^{\prime}
\\+\frac{\psi^{\text{in}}_s(r)}{W_s}\,\int^{r_{\text{max}}}_{r}
\psi^{\text{out}}_s(r^{\prime}) S_{s}(r^{\prime})
\left(1-\tfrac{2 M}{r^{\prime}}\right)^{-1}\,d r^{\prime}\;,
\end{multline}
where \eqref{eq:omegamk}
\begin{equation}
\omega=\omega_{mk}= m\Omega_{\phi}+k\Omega_{r}\;.
\end{equation}
The amplitudes are
\begin{equation}
\label{eq:ainfe}
A^{\infty}_{slm\omega}=\frac{1}{W_s}\,\int_{r_{\text{min}}}^{r_{\text{max}}}
\psi^{\text{in}}_s(r^{\prime}) S_{s}(r^{\prime})
\left(1-\tfrac{2 M}{r^{\prime}}\right)^{-1}\,d r^{\prime}\;,
\,r>r_{\text{max}}\;,
\end{equation}
\begin{equation}
\label{eq:ahore}
A^{2 M}_{slm\omega}=\frac{1}{W_s}\,\int^{r_{\text{max}}}_{r_{\text{min}}}
\psi^{\text{out}}_s(r^{\prime}) S_{s}(r^{\prime})
\left(1-\tfrac{2 M}{r^{\prime}}\right)^{-1}\,d r^{\prime}\;.
\,r<r_{\text{min}}\;,
\end{equation}
As a result,
\begin{equation}
\label{eq:psiretgtle}
\psi^{\text{ret}}_{s}(r)=\psi^{\text{out}}_{s}(r)A^{\infty}_{slm\omega}\;,\,r>r_{\text{max}}\;,\,
\psi^{\text{ret}}_{s}(r)=\psi^{\text{in}}_{s}(r)A^{2 M}_{slm\omega}\;,\,r<r_{\text{min}}\;.
\end{equation}
Similar integrals have been used elsewhere for elliptic orbits 
\cite{cut94}, \cite{tsstn93}.  

As discussed in Chapter \ref{tmunuchap} in the first full paragraph 
following equation \eqref{eq:thedelta}, some of the stress energy tensor 
coefficients \eqref{eq:firstrc}-\eqref{eq:lastrc} have denominator 
factors of~$\axdot$, which will be zero at the turning points of 
eccentric orbits and zero for circular orbits \cite{cut94}.  Because 
the source terms $S_{s}$ are constructed from the 
coefficients, the integrals \eqref{eq:ainfe}-\eqref{eq:ahore} appear 
to be singular at $r^{\prime}=r_{\text{min}}$ and $r^{\prime}=r_{\text{max}}$.  
To avoid this problem, it is necessary to change the variable of 
integration, as explained elsewhere~\cite{cut94}.  We will change to 
the eccentric anomaly $\psi$, defined by \eqref{eq:rtopsi} as 
\cite{darwin61}
\begin{equation}
\label{eq:cov}
r^{\prime}=a(1-e\cos\psi)\;,\,dr^{\prime}=a e \sin\psi\;d\psi\;,\,
0\le\psi\le 2\pi\;.
\end{equation}
After the change of variable, the retarded solution is
\begin{multline}
\label{eq:retgrwsolepsi}
\psi^{\text{ret}}_s(r)
=\frac{\psi^{\text{out}}_s(r)}{W_s}\,\int_{0}^{\psi}
\psi^{\text{in}}_s(r^{\prime}) S_{s}(r^{\prime})
\left(1-\tfrac{2 M}{r^{\prime}}\right)^{-1}
\,a e \sin\psi\,d\psi
\\+\frac{\psi^{\text{in}}_s(r)}{W_s}\,\int^{\pi}_{\psi}
\psi^{\text{out}}_s(r^{\prime}) S_{s}(r^{\prime})
\left(1-\tfrac{2 M}{r^{\prime}}\right)^{-1}
\,a e \sin\psi\,d\psi\;.
\end{multline}
The amplitude integrals \eqref{eq:ainfe}-\eqref{eq:ahore} become
\begin{equation}
\label{eq:ainfepsi}
A^{\infty}_{slm\omega}=\frac{1}{W_s}\,\int_{0}^{\pi}
\psi^{\text{in}}_s(r^{\prime}_*) S_{s}(r^{\prime}_*)
\,a e \sin\psi\,d\psi\;,
\,r>r_{\text{max}}\;,
\end{equation}
\begin{equation}
\label{eq:ahorepsi}
A^{2 M}_{slm\omega}=\frac{1}{W_s}\,\int^{\pi}_{0}
\psi^{\text{out}}_s(r^{\prime}_*) S_{s}(r^{\prime}_*)
\,a e \sin\psi\,d\psi\;,
\,r<r_{\text{min}}\;.
\end{equation}
The limits of integration $0$ and $\pi$ correspond to $r_{\text{min}}$ 
and $r_{\text{max}}$, respectively.  In terms of~$\psi$, the velocity $\axdot$ 
is~\eqref{eq:potpsi}
\begin{equation}
\label{eq:potpsi6}
\textstyle{\axdot}=\frac{a e}{2 M}\sin\psi
\left(\frac{\left(2 M\right)^{3}}{r^{3}}
\frac{\left(1-e\right)\left(\frac{p}{2 M}-3-e\right)
+2 e\left(\frac{p}{2 M}-2\right)\sin^{2}\frac{\psi}{2}}
{2\left(\frac{p}{2 M}\right)-3-e^{2}}\right)^{1/2}
,\,0\le\psi\le\pi\;,
\end{equation}
which is zero at the turning points and for circular orbits because 
$a e\sin\psi$ is zero then.  When $\axdot$ is in the denominator, this 
factor of $a e \sin\psi$ is canceled by the numerator factor of 
$a e\sin\psi$ that comes from $d r^{\prime}$, removing the singularity 
from the integrand.  A different variable change was 
used by Cutler and his collaborators \cite{cut94}, but for the same 
reasons.  They made the substitution
\begin{equation}
\label{eq:cutchi}
r^{\prime}=\frac{p}{1+e \cos\chi}\;,\,
dr^{\prime}=\frac{p e \sin\chi}{\left(1+e \cos\chi\right)^{2}}\,d\chi
\;,\,0\le\chi\le 2\pi\;.
\end{equation}
Their version of $\axdot$ also has a factor of $\sin\chi$, which is likewise 
canceled in the integral.  
% The parameter $\psi$ is preferable to $\chi$ for larger $e$, because 
% then the denominator factor $(1+e \cos\chi)^{-2}$ causes the integrand 
% to peak near apastron.  Cf.~\cite{darwin}. 

An additional issue is that the stress energy tensor coefficients also 
appear in the harmonic gauge solutions.  Some of these coefficients have 
a factor of $\axdot$ in the denominator.  The solutions also have terms 
with derivatives of the functions $\psi_{s}$.  It turns out that when we 
differentiate the integrals in $\psi^{\text{ret}}_s$ \eqref{eq:retgrwsol}, 
we get additional terms which cancel out all of the coefficients, except 
$So_{12}$, $Se_{01}$, $Se_{11}$ and $Se_{12}$.  These four coefficients do 
not have $\axdot$ factors in the denominator, which can be verified 
from the list \eqref{eq:firstrc}-\eqref{eq:lastrc}.  Accordingly, the 
stress energy tensor coefficients do not cause a singularity in the 
solutions.  

For circular orbits, $r_{\text{min}}=r_{\text{max}}$.  Nevertheless, the retarded 
solution \eqref{eq:retgrwsolepsi} can still be used, provided 
$S_{s}$ is constructed from the stress energy tensor 
radial coefficients listed in \eqref{eq:firstrc}-\eqref{eq:lastrc}.  
Alternatively, we could use the radial integrals in the retarded solution 
form \eqref{eq:retgrwsol}, together with the circular orbit stress energy 
tensor \eqref{eq:cirfft}, and evaluate the integrals using the radial delta 
function in \eqref{eq:cirfft}.  In effect, the latter approach was 
followed by Poisson to calculate circular orbit solutions of the Teukolsky 
equation \cite{pois93}.  For calculations in this thesis, we will use 
the formula \eqref{eq:retgrwsolepsi}, because it also can be used for 
elliptic orbits.  In Chapter~\ref{radchap}, we will use 
\eqref{eq:ainfepsi}-\eqref{eq:ahorepsi} (with $s=2$) as gravitational 
wave amplitudes for bound orbits, both circular and elliptic.

For zero frequency, we use the same integrals as for non-zero 
frequency, except that $\psi^{\text{in}}_s$ and $\psi^{\text{out}}_s$ are taken 
from section \ref{sec:zrweqn}.  As $r\to\infty$, the behavior of 
$\psi^{\text{ret}}_s$ is $O(r^{-l})$.  Near the event horizon, 
$\psi^{\text{ret}}_s$ goes to a constant as $r\to 2 M$.  

The inhomogeneous solutions above have source terms constructed from the 
stress energy tensor, which is non-zero only at the location of the 
orbiting mass and which is a known function.  A different type of 
inhomogeneous equation is 
\begin{equation}
\label{eq:ddfmtwa6}
\mathcal{L}_{0}\fmtwa=\ff\pz\;,
\end{equation}
which is \eqref{eq:ddfmtwa}.  For $\pz$, we substitute the retarded solution 
$\psi^{\text{ret}}_{0}$ \eqref{eq:retgrwsol}.  This means $\pz$ extends over 
the range $2 M<r<\infty$ and is an integral solution itself, so it is 
more difficult to use a Green's function here.  Instead, we use the 
``shooting method'', which is described in \textit{Numerical Recipes} 
\cite{numr}.  We first find two series solutions, one at large $r$ and 
one near the event horizon.  We then match the two solutions and their 
derivatives at an intermediate point.  

In terms of the dimensionless variables $x=\frac{r}{2 M}$ and 
$\Omega=2 M \omega$, equation \eqref{eq:ddfmtwa6} is
\begin{multline}
\label{eq:newm2aeqn0x}
\frac{(-1+x)^2}{x^2}\frac{d^{2}\fmtwa}{dx^{2}}
+\frac{(-1+x)}{x^3}\frac{d\fmtwa}{dx}
\\+\frac{\left(1+x+2\lambda x-2 (1+\lambda) x^2
-(i\Omega)^2 x^4\right)}{x^4}\fmtwa
=\left(1-\frac{1}{x}\right)\pz\;.
\end{multline}
The series solution for large $r$ is
\begin{equation}
\label{eq:m2aout}
\fmtwa^{\text{out}}(x)=A_{0lm\omega}^{\infty}\bigg\{e^{i\Omega x_{*}}
\sum_{n=-1}^{\infty}\frac{b_{n}}{x^{n}}+c_{\text{out}}\,\pz^{\text{out}}(x)\bigg\}\;,
\end{equation}
where
\begin{equation}
\label{eq:bnout}
b_{n}=-\frac{1}{2 i\Omega n}\Big\{\left(2+2\lambda+n-n^2\right)b_{n-1}
+(-1+n)^2 b_{n-2}+a_{n+1}\Big\} \text{ for }n\ge1\;,
\end{equation}
\begin{equation}
\label{eq:bnouts}
b_{n}=0 \text{ for }n<-1,\,b_{-1}=\frac{1}{2 i\Omega},\,b_{0}=0\;.
\end{equation}
The coefficient $a_{n+1}$ follows the recursion relation 
\eqref{eq:anout}.  Equation \eqref{eq:m2aout} is derived with the 
assumption that the amplitude $A_{0lm\omega}^{\infty}$ is a constant, which will 
be the case as long as $r$ is greater than the maximum source position.  
For bound orbits, this means $r>r_{\text{max}}$, and $A_{0lm\omega}^{\infty}$ is 
taken from \eqref{eq:ainfepsi}.  
Normally, we also need $r\gg r_{\text{max}}$, in order that $r$ be large 
enough for the series to converge.  The constant $c_{\text{out}}$ is 
discussed below.

Changing variables from $x$ to $X=\ff$ in \eqref{eq:newm2aeqn0x} gives
\begin{multline}
\label{eq:newm2aeqn0bx}
(-1+X)^4 X^2\frac{d^{2}\fmtwa}{dX^{2}}
+(-1+X)^3 X (-1+3 X)\frac{d\fmtwa}{dX}
\\+\left(-(i\Omega)^2-(3+2\lambda-X) (-1+X)^2 X\right)\fmtwa
=X\pz\;.
\end{multline}
The series solution near the event horizon is
\begin{equation}
\label{eq:m2ain}
\fmtwa^{\text{in}}(X)=A_{0lm\omega}^{2 M}\bigg\{e^{-i\Omega X_{*}} 
\sum^{\infty}_{n=1}b_{n}X^{n}+c_{\text{in}}\,\pz^{\text{in}}(X)\bigg\}\;,
\end{equation}
where $c_{\text{in}}$ and $A_{0lm\omega}^{2 M}$ are constants.  For bound 
orbits, $A_{0lm\omega}^{2 M}$ is given by \eqref{eq:ahorepsi}, 
and $X$ is chosen so that $r<r_{\text{min}}$.  The recursion relation 
for $b_{n}, n\ge 1$, is
\begin{multline}
\label{eq:bnin}
b_{n}=\frac{1}{(2 i\Omega-n)n}\Big\{
-\left(5+2 \lambda-4 i\Omega(-1+n)-6 n+4 n^2\right)b_{n-1}
\\+\left(19+4 \lambda-2 i\Omega(-2+n)-18 n+6 n^2\right)b_{n-2}
\\-\left(23+2 \lambda-18 n+4 n^2\right)b_{n-3}+(-3+n)^2 b_{n-4}
-a_{n-1}\Big\}\;,
\end{multline}
where $b_{n}=0$ for $n<1$ and $a_{n-1}$ is from \eqref{eq:anin}.  

Starting with $\fmtwa^{\text{out}}$ and its derivative as initial values, 
we use a numerical differential equation solver to integrate equation 
\eqref{eq:newm2aeqn0x} inwards.  Similarly, we integrate outwards 
from $\fmtwa^{\text{in}}$ until the two solutions meet.  
We solve for the constants $c_{\text{in}}$ and $c_{\text{out}}$ by requiring 
that $\fmtwa^{\text{in}}$ and $\fmtwa^{\text{out}}$, as well as their 
derivatives, match at some intermediate point.  For circular orbits, 
the matching is done at the orbital radius.  The Bulirsch-Stoer method, 
as implemented in \textit{Numerical Recipes} \cite{numr}, is a suitable 
differential equation solver for this purpose.

It is numerically easier to solve \eqref{eq:ddfmtwa6} than the 
alternative differential equations for $\fz$ and $\fdz$, which are  
\eqref{eq:nzf0eqn}-\eqref{eq:nzfd0eqn}. For large $r$, inhomogeneous 
series solutions to \eqref{eq:nzf0eqn}-\eqref{eq:nzfd0eqn} are
\begin{multline}
\label{eq:nf0sol}
\fz^{\infty}=\frac{M}{2 (\iom)^2 r}
-\frac{(1+\lambda) M}{2 (\iom)^4 r^3}
+\frac{(11+16 \lambda) M^2}{4 (\iom)^4 r^4}
\\+\frac{M \left(3 \lambda^2+(\iom)^2 M^2
+\lambda \left(3-12 (\iom)^2 M^2\right)\right)}{2 (\iom)^6 r^5}
+O(r^{-6})\;,
\end{multline}
\begin{multline}
\label{eq:nfd0sol}
\fdz^{\infty}=\frac{r}{2 (\iom)^2}-\frac{2 M^2}{(\iom)^2 r}
-\frac{(1+\lambda)M}{2 (\iom)^4 r^2}+\frac{(13+18 \lambda) M^2}
{6 (\iom)^4 r^3}+\frac{M}{12 (\iom)^6 r^4}
\left[9\lambda+9 \lambda^2-4 (\iom)^2\right.\\\left.
\times(1+12 \lambda) M^2\right]
+\frac{M^2 \left(133+25 \lambda-183 \lambda^2
-120 (\iom)^2 M^2\right)}{30 (\iom)^6 r^5}+O(r^{-6})\;.
\end{multline}
We also can derive derive inhomogeneous series solutions in powers of $X$ 
near the event horizon.  The large $r$ and near horizon series can be 
matched using the shooting method and the homogeneous solutions
\begin{equation}
\label{eq:f0h}
\fz^{h}=c_{4}-\tfrac{1}{2}\left(1-\tfrac{2 M}{r}\right)
\left(c_{1}\psi^{\text{in}}_{0}\psi^{\text{in}}_{0}
+c_{2}\psi^{\text{out}}_{0}\psi^{\text{in}}_{0}
+c_{3}\psi^{\text{out}}_{0}\psi^{\text{out}}_{0}\right)_{\!,r}\;,
\end{equation}
\begin{equation}
\label{eq:fd0h}
\fdz^{h}=\left(1-\tfrac{2 M}{r}\right)
\left(c_{1}\psi^{\text{in}}_{0}\psi^{\text{in}}_{0}
+c_{2}\psi^{\text{out}}_{0}\psi^{\text{in}}_{0}
+c_{3}\psi^{\text{out}}_{0}\psi^{\text{out}}_{0}\right)\;.
\end{equation}
The numerical problem is that the homogeneous solutions are quadratic 
in the generalized Regge-Wheeler functions $\psi^{\text{in}}_{0}$ 
and $\psi^{\text{out}}_{0}$.  This results in cancellations and loss of 
significant figures when matching solutions at some intermediate point, 
particularly as the spherical harmonic index $l$ increases.  In 
contrast, the homogeneous solutions to \eqref{eq:ddfmtwa6} are 
linear in $\psi^{\text{in}}_{0}$ and $\psi^{\text{out}}_{0}$, resulting in much 
less cancellation.

\chapter{\label{radchap}Radiation}

Using the results of previous chapters, we can calculate the 
gravitational radiation emitted as the small mass orbits 
the central black hole.  
Section~\ref{sec:wavs} contains the derivation of waveforms in a radiation 
gauge, which is suitable for observers at large distances from the source.  
Section~\ref{sec:wavenerg} shows how to calculate the energy and 
angular momentum carried away from the orbiting mass by the 
gravitational waves.  The main results of this chapter are not new.  
They have been derived by others using different methods in the references 
discussed below.  What is new is that we will obtain the results from the 
harmonic gauge solutions derived in Chapters~\ref{oddpar} and~\ref{evpar}.

\section{\label{sec:wavs}Waveforms}

Gravitational waves have two polarization tensors, designated $\hp$ 
and $\hc$.  Following convention, we define
\begin{equation}
\label{eq:hphc}
\hp=\frac{1}{2}\left(h_{\hat{\theta}\hat{\theta}}
-h_{\hat{\phi}\hat{\phi}}\right)\;,\;\hc=h_{\hat{\theta}\hat{\phi}}\;,
\end{equation}
where the hats indicate that the components are written in an 
orthonormal basis \cite{bcp07}, \cite{nagar05}.  
To derive expressions for $\hp$ and $\hc$, we find a 
gauge transformation from the harmonic gauge to a suitable radiation gauge 
and then project the resulting polarization tensors onto an orthonormal 
basis.  

First, we transform from the harmonic gauge to a radiation gauge.  For 
large $r$, the radiation gauge will be a transverse-traceless gauge.  
Chrzanowski specified \cite{chrz75} an outgoing radiation gauge by 
imposing the conditions
\begin{equation}
\label{eq:outrg}
h_{\mu\nu}n^{\nu}=0\;,\;h=g^{\mu\nu}h_{\mu\nu}=0\;,
\end{equation}
and an ingoing radiation gauge by requiring
\begin{equation}
\label{eq:inrg}
h_{\mu\nu}l^{\nu}=0\;,\;h=0\;.
\end{equation}
Here, $l^{\nu}$ and $n^{\nu}$ are components of the Newman-Penrose basis 
vectors $\bm{l}$ and $\bm{n}$ \eqref{eq:npvec}.  
We will use \eqref{eq:outrg} to find outgoing waveforms for 
large $r$, with the components of $\bm{n}$ given by \eqref{eq:npnvec}.  

Combining the gauge transformation formula \eqref{eq:hnew} and the 
first equation of \eqref{eq:outrg}, we need to 
find a gauge transformation vector $\xi_{\mu}$ such that
\begin{equation}
\label{eq:noutrg}
h^{\text{RA}}_{\mu\nu}n^{\nu}=\left(h_{\mu\nu}^{\text{HA}}
-\xi_{\mu;\nu}-\xi_{\nu;\mu}\right)n^{\nu}=0\;.
\end{equation}
The superscripts ``RA'' and ``HA'' refer to the radiation and 
harmonic gauges, respectively.  Because we are using separation of 
variables, we must solve \eqref{eq:noutrg} separately 
for each Fourier mode specified by a combination of $lm\omega$.  Only 
large $r$ behavior is needed for waveforms, so the usual way to change 
to a radiation gauge is to expand the metric perturbations and gauge 
transformation vectors in series of decreasing (mainly inverse) powers 
of $r$.  This was done by Zerilli \cite{zerp70} and Ashby \cite{ashby} 
to transform from the Regge-Wheeler gauge to a radiation gauge, although 
they did not use Chrzanowski's conditions.  The radiative modes are 
non-zero frequency modes for $l\ge2$ \cite{zerp70}.  In the harmonic 
gauge solutions from Chapters~\ref{oddpar} and~\ref{evpar}, we set 
the stress energy tensor coefficients (such as $Se_{00}$) equal to zero 
and substitute the outgoing radiation solutions derived in 
Chapter~\ref{rweqnchap} for the generalized Regge-Wheeler, Zerilli and 
related functions.  Doing so gives asymptotic series for the radial coefficients 
of the metric perturbation in the harmonic gauge.  We then write out the 
components of equation \eqref{eq:noutrg} in series form and solve 
term-by-term for the series coefficients of the gauge transformation 
vectors.  This yields an odd parity series for the radial gauge 
transformation function $Z$ \eqref{eq:ochimu} and even parity series for 
$\bmz$, $\bmo$ and $\bmtw$ \eqref{eq:echimu}.  We substitute the various 
series into the gauge transformation formulae 
\eqref{eq:hznew}-\eqref{eq:htwnew} and \eqref{eq:bh0new}-\eqref{eq:gnew} 
and obtain asymptotic series for the radial coefficients of the 
metric perturbation in the new, radiation gauge.  
The leading order radiation gauge behavior is described below, following 
an explanation of orthonormal bases.

The summary of orthonormal bases below is taken mainly from Hartle 
\cite{hartle03}.  Components in an orthonormal basis $e_{\hat{\mu}}^{\alpha}$ 
are signified by ``hats''.  For example,
\begin{equation}
\label{eq:coortohat}
h_{\hat{\mu}\hat{\nu}}=e_{\hat{\mu}}^{\alpha}e_{\hat{\nu}}^{\beta}h_{\alpha\beta}\;.
\end{equation}
Here, $h_{\alpha\beta}$ is the perturbation \eqref{eq:hmunu}, written in the 
non-orthonormal coordinate basis we use normally use.  ``Orthonormal'' 
means
\begin{equation}
\label{eq:orthdef}
g_{\alpha\beta}e_{\hat{\mu}}^{\alpha}e_{\hat{\nu}}^{\beta}
=\eta_{\hat{\mu}\hat{\nu}}=\text{diag}(-1,1,1,1)\;.
\end{equation}
The ``hat'' indices are raised and lowered with $\eta_{\hat{\mu}\hat{\nu}}$, 
rather than the background metric $g_{\alpha\beta}$.  
For a diagonal background metric, one possible orthonormal basis is
\begin{equation}
\label{eq:onbasis}
e_{\hat{t}}^{\alpha}=[(-g_{tt})^{-1/2},0,0,0]\;,
e_{\hat{r}}^{\alpha}=[0,(g_{rr})^{-1/2},0,0]\;,
\end{equation}
and so on for $e_{\hat{\theta}}^{\alpha}$ and $e_{\hat{\phi}}^{\alpha}$.  We will use 
this basis because it is also used in the conventions for $\hp$ and 
$\hc$ \eqref{eq:hphc}.  
As discussed by Hartle, an orthonormal basis defines a laboratory 
frame where physical measurements are made.  
The vector $e_{\hat{t}}^{\alpha}$ above is equal to the four-velocity 
$u^{\alpha}={\frac{dz}{d\tau}}^{\!\alpha}$ of a reference frame at rest with 
respect to the origin of the black hole system.  To show this, solve the 
velocity normalization condition $g_{\alpha\beta}u^{\alpha}u^{\beta}=-1$ for 
$u^{\alpha}=u^{t}\delta^{\alpha}_{t}$.  We want to express the components of 
the waveforms in such a frame, so we will use the basis 
\eqref{eq:onbasis}.  If another 
reference frame is desired, a subsequent coordinate transformation may 
be made.  
A different formulation of orthonormal bases is given by Price and 
Thorne \cite{prth69} as
\begin{equation}
\label{eq:ptortho}
h_{\hat{\mu}\hat{\nu}}=\lvert g^{\mu\mu}\rvert^{1/2} 
\lvert g^{\nu\nu}\rvert^{1/2} h_{\mu\nu}\;,
\end{equation}
which leads to the same result here.  
They refer to $h_{\hat{\mu}\hat{\nu}}$ as the ``physical components'' of the 
perturbation.  In \eqref{eq:ptortho}, we do not sum over repeated 
indices, contrary to our usual practice.

The radiation gauge perturbation is traceless, which is different from 
the harmonic gauge.  The gauge transformation yields 
infinite series of inverse powers of $r$, so we calculate 
only the first few terms.  The series for the trace is zero to at least 
$O(r^{-5})$, and probably to higher inverse orders as well.  This result 
was obtained by applying only the first of Chrzanowski's conditions 
\eqref{eq:outrg}, namely, $h_{\mu\nu}n^{\nu}=0$.  We could also 
require $h=0$ explicitly, like \eqref{eq:outrg}.  
However, when transforming from the harmonic gauge to the outgoing 
radiation gauge, the traceless result follows from the first condition 
alone, at least asymptotically.

Because the radiation gauge is traceless, equation \eqref{eq:hphc} 
simplifies to
\begin{equation}
\label{eq:shphc}
\hp=h_{\hat{\theta}\hat{\theta}}=-h_{\hat{\phi}\hat{\phi}}\;,\;
\hc=h_{\hat{\theta}\hat{\phi}}=h_{\hat{\phi}\hat{\theta}}\;.
\end{equation}
The gravitational waveforms are given by equations \eqref{eq:hthth} 
and \eqref{eq:hthphi} below.  The plus polarization is
\begin{equation}
\label{eq:hthth}
\hp=h_{\hat{\theta}\hat{\theta}}=\sum_{l=2}^{\infty}\sum_{m=-l}^{l}
\int_{\!-\infty}^{\infty} e^{-\iom t}\;
h_{\hat{\theta}\hat{\theta}}^{lm}(\omega,r,\theta,\phi)\;d\omega\;,
\end{equation}
where, for odd parity,
\begin{equation}
\label{eq:hththodd}
h_{\hat{\theta}\hat{\theta}}^{lm}(\omega,r,\theta,\phi)=
-\frac{e^{\iom r_{*}}}{\iom r} A^{\infty}_{2lm\omega}X_{lm}(\theta,\phi)+O(r^{-2})\;,
\end{equation}
and, for even parity,
\begin{equation}
\label{eq:hththev}
h_{\hat{\theta}\hat{\theta}}^{lm}(\omega,r,\theta,\phi)=
\frac{e^{\iom r_{*}}}{2 r} A^{\infty}_{2lm\omega}W_{lm}(\theta,\phi)+O(r^{-2})\;.
\end{equation}
The cross polarization is
\begin{equation}
\label{eq:hthphi}
\hc=h_{\hat{\theta}\hat{\phi}}=\sum_{l=2}^{\infty}\sum_{m=-l}^{l}
\int_{\!-\infty}^{\infty} e^{-\iom t}\;
h_{\hat{\theta}\hat{\phi}}^{lm}(\omega,r,\theta,\phi)\;d\omega\;,
\end{equation}
where, for odd parity,
\begin{equation}
\label{eq:hthphiodd}
h_{\hat{\theta}\hat{\phi}}^{lm}(\omega,r,\theta,\phi)=
\frac{e^{\iom r_{*}}}{\iom r} A^{\infty}_{2lm\omega}W_{lm}(\theta,\phi)+O(r^{-2})\;,
\end{equation}
and, for even parity,
\begin{equation}
\label{eq:hthphiev}
h_{\hat{\theta}\hat{\phi}}^{lm}(\omega,r,\theta,\phi)=
\frac{e^{\iom r_{*}}}{2 r} A^{\infty}_{2lm\omega}X_{lm}(\theta,\phi)+O(r^{-2})\;.
\end{equation}
The angles $\theta$ and $\phi$ are the observer's angular coordinates, 
and $r$ is the distance to the observer.  For plotting waveforms, 
the exponentials may be rewritten as $e^{-\iom u}$, where the retarded 
time $u=t-r_{*}$ \cite{gkenn02}.  The even parity modes are due to the 
radial function $\bg$, not $\bk$, and the odd are attributable to $\htw$.  
The outgoing amplitude constants $A^{\infty}_{2lm\omega}$ are the even 
and odd parity retarded 
solution source integrals \eqref{eq:ainf}, with $s=2$.  For bound orbits, 
we use \eqref{eq:ainfepsi}.  As explained in Chapter~\ref{tmunuchap}, bound 
orbits have a discrete frequency spectrum, symbolized by a frequency delta 
function.  We use the delta function to evaluate the frequency integrals 
in \eqref{eq:hthth} and \eqref{eq:hthphi}.  For elliptic orbits, the frequency 
integrals become sums, so that
\begin{equation}
\label{eq:inttosum}
\int_{\!-\infty}^{\infty} e^{-\iom t}f(\omega)\delta(\omega-\omega_{mk})d\omega\to
\sum_{k=-\infty}^{\infty}e^{-i\omega_{mk} t}f(\omega_{mk})\;,
\end{equation}
where $\omega_{mk}= m\Omega_{\phi}+k\Omega_{r}$ \eqref{eq:omegamk}.  
For circular orbits, the index $k$ is restricted to zero.

Combining the odd and even results, the other metric perturbations 
in the radiation gauge behave as
\begin{multline}
\label{eq:othpert}
h_{\hat{t}\hat{t}}\sim O(r^{-3})\;,
h_{\hat{t}\hat{r}}\sim O(r^{-3})\;,
h_{\hat{t}\hat{\theta}}\sim O(r^{-2})\;,
h_{\hat{t}\hat{\phi}}\sim O(r^{-2})\;,
\\h_{\hat{r}\hat{r}}\sim O(r^{-3})\;,
h_{\hat{r}\hat{\theta}}\sim O(r^{-2})\;,
h_{\hat{r}\hat{\phi}}\sim O(r^{-2})\;,
\end{multline}
with those not listed determined by symmetry.  Projecting the 
divergence $h_{\alpha\beta}{}^{;\beta}$ onto the orthonormal basis, we have
\begin{equation}
\label{eq:hraddiv}
h_{\hat{t}\nu}{}^{;\nu}\sim O(r^{-4})\;,
h_{\hat{r}\nu}{}^{;\nu}\sim O(r^{-4})\;,
h_{\hat{\theta}\nu}{}^{;\nu}\sim O(r^{-3})\;,
h_{\hat{\phi}\nu}{}^{;\nu}\sim O(r^{-3})\;.
\end{equation}
For a traceless gauge, 
$\overline{h}_{\alpha\beta}{}^{;\beta}=h_{\alpha\beta}{}^{;\beta}$.  In the radiation 
gauge, $\overline{h}_{\alpha\beta}{}^{;\beta}$ is only asymptotically zero, 
not identically zero.  This is different from the harmonic gauge.  

Misner, Thorne and Wheeler discuss the ``transverse-traceless'' 
(TT) gauge and define it by the eight constraints \cite{mtw73}
\begin{equation}
\label{eq:htt}
h_{\mu 0}=h_{\mu\nu}u^{\nu}=0\;,h_{i j}{}^{,j}=0\;,h=h_{k}{}^{k}=0\;.
\end{equation}
Here, the index 0 is the time coordinate, the indices $i$, $j$ 
and $k$ represent spatial coordinates, and $\mu$ can be any of the four.  
Counting up the components, the three equations in \eqref{eq:htt} contain 
eight conditions.  A partial rather than covariant derivative is used, 
because the discussion in \cite{mtw73} concerns perturbations of a flat 
background metric $\eta_{\mu\nu}$.  The first equality in \eqref{eq:htt} 
means that the the wave has only spatial components.  The first and second 
equations imply that the wave is transverse to its direction of propagation, 
like a plane wave.  The last condition is that the trace is zero.  
The perturbation $h_{\mu\nu}$ is symmetric, so it has at most ten 
independent components.  The significance of the transverse-traceless 
gauge is that the eight constraints reduce the number of free components 
from ten to two.  The remaining two represent the two physically significant 
degrees of freedom, or polarizations, of the gravitational waves 
\cite{mtw73}, \cite{wein72}.  

We can show that our radiation gauge is a transverse-traceless 
gauge.  Condensing the radiation gauge results, we have
\begin{equation}
\label{eq:httrad}
h_{\hat{\mu}\hat{0}}=h_{\hat{\mu}\nu}u^{\nu}\sim O(r^{-2})\;,
h_{\hat{\mu}\nu}{}^{;\nu}\sim O(r^{-3})\;,
h_{\mu}{}^{\mu}=0\;.
\end{equation}
Asymptotically, these results are equivalent to the eight constraints in 
\eqref{eq:htt} which define the transverse-traceless gauge.  
Imposing Chrzanowski's condition $h_{\mu\nu}n^{\nu}=0$ \eqref{eq:outrg} has 
lead to a transformation from the harmonic gauge to a radiation gauge 
which is transverse-traceless for large $r$.  
The polarizations depend only on the asymptotic amplitudes 
of the even and odd parity spin~$2$ ($\ptw$) functions, which are gauge 
invariant.  The other generalized Regge-Wheeler functions, which have 
$s=0$ or $s=1$ and which are gauge dependent, do not contribute to the 
radiation.

Another way of deriving the waveforms is through the Newman-Penrose 
formalism \cite{chandra92}, \cite{np62}.  This method does not require 
gauge transformation calculations.  The discussion of the formalism 
below is taken largely from Chandrasekhar \cite{chandra92}.  
The Newman-Penrose formalism defines five complex Weyl scalars, which 
are constructed by projecting the ten independent components of the 
Weyl tensor onto a null tetrad basis.  The five scalars are
\begin{equation}
\label{eq:weyl0}
\Psi_{0}=-C_{\alpha\beta\gamma\delta}l^{\alpha}m^{\beta}l^{\gamma}m^{\delta}\;,
\end{equation}
\begin{equation}
\label{eq:weyl1}
\Psi_{1}=-C_{\alpha\beta\gamma\delta}l^{\alpha}n^{\beta}l^{\gamma}m^{\delta}\;,
\end{equation}
\begin{equation}
\label{eq:weyl2}
\Psi_{2}=-C_{\alpha\beta\gamma\delta}l^{\alpha}m^{\beta}\overline{m}^{\gamma}n^{\delta}\;,
\end{equation}
\begin{equation}
\label{eq:weyl3}
\Psi_{3}=-C_{\alpha\beta\gamma\delta}l^{\alpha}n^{\beta}\overline{m}^{\gamma}n^{\delta}\;,
\end{equation}
\begin{equation}
\label{eq:weyl4}
\Psi_{4}=-C_{\alpha\beta\gamma\delta}
n^{\alpha}\overline{m}^{\beta}n^{\gamma}\overline{m}^{\delta}\;.
\end{equation}
The Weyl tensor is represented by $C_{\alpha\beta\gamma\delta}$ and, in a 
vacuum, is equal to the covariant Riemann curvature tensor.  For the 
unperturbed Schwarzschild metric, only $\Psi_{2}$ is non-zero.  

We will use the unperturbed tetrad basis in 
\eqref{eq:nplvec}-\eqref{eq:npmbvec}.  
With our metric signature $-$$+$$+$$+$ and the 
basis \eqref{eq:nplvec}-\eqref{eq:npmbvec}, 
equation \eqref{eq:weyl2} gives $M/r^{3}$ for the background $\Psi_{2}$.  
Combining the above results, we may rewrite the vacuum Weyl scalars as
\begin{equation}
\label{eq:pweyl0}
\Psi_{0}=-\delta\!R_{\alpha\beta\gamma\delta}l^{\alpha}m^{\beta}l^{\gamma}m^{\delta}\;,
\end{equation}
\begin{equation}
\label{eq:pweyl1}
\Psi_{1}=-\delta\!R_{\alpha\beta\gamma\delta}l^{\alpha}n^{\beta}l^{\gamma}m^{\delta}\;,
\end{equation}
\begin{equation}
\label{eq:pweyl2}
\Psi_{2}=\frac{M}{r^{3}}
-\delta\!R_{\alpha\beta\gamma\delta}l^{\alpha}m^{\beta}\overline{m}^{\gamma}n^{\delta}\;,
\end{equation}
\begin{equation}
\label{eq:pweyl3}
\Psi_{3}=-\delta\!R_{\alpha\beta\gamma\delta}l^{\alpha}n^{\beta}
\overline{m}^{\gamma}n^{\delta}\;,
\end{equation}
\begin{equation}
\label{eq:pweyl4}
\Psi_{4}=-\delta\!R_{\alpha\beta\gamma\delta}
n^{\alpha}\overline{m}^{\beta}n^{\gamma}\overline{m}^{\delta}\;,
\end{equation}
where $\delta\!R_{\alpha\beta\gamma\delta}$ is the perturbed covariant 
Riemann curvature tensor.  

As discussed by Chandrasekhar \cite{chandra92}, both $\Psi_{4}$ 
and $\Psi_{0}$ are invariant under infinitesimal 
tetrad rotations and coordinate transformations, so these two are 
physically significant Weyl scalars of gravitational perturbations, for $l\ge2$.  
In contrast, $\Psi_{1}$, $\Psi_{3}$ and at least the even parity perturbation 
of $\Psi_{2}$ are not invariant.  To simplify matters, one procedure is to choose 
the tetrad and coordinates so that 
$\Psi_{1}$ and $\Psi_{3}$ are zero and $\Psi_{2}$ is equal to its 
unperturbed background value.  This is done, for example, in section~82 
of \cite{chandra92}, but that treatment of the odd parity $\Psi_{2}$ is disputed 
by Hamilton \cite{hampc}.  
However, we have not tried to choose such a tetrad and coordinates 
for equations \eqref{eq:pweyl0}-\eqref{eq:pweyl4}.  

Because $\Psi_{4}$ and $\Psi_{0}$ are gauge invariant, we use them 
to describe gravitational radiation.  Outgoing radiation for large $r$ 
is obtained from $\Psi_{4}$; ingoing radiation as $r\to 2M$, from 
$\Psi_{0}$ \cite{chandra92}, \cite{teuk73}.  For an individual frequency 
mode of a Fourier transform at large $r$, Teukolsky found that
\begin{equation}
\label{eq:psi4radf}
\Psi_{4}=-\left(\delta\!R_{\hat{t}\hat{\theta} \hat{t}\hat{\theta}}
-i\,\delta\!R_{\hat{t}\hat{\theta} \hat{t}\hat{\phi}}\right)
=-\frac{\omega^{2}}{2}\left(h_{\hat{\theta}\hat{\theta}}
-i h_{\hat{\theta}\hat{\phi}}\right)\;.
\end{equation}
Teukolsky's derivation assumes $h_{\hat{\theta}\hat{\theta}}$ and 
$h_{\hat{\theta}\hat{\phi}}$ are in a transverse-traceless gauge.  However, 
$\Psi_{4}$ is gauge invariant, so we do not have have to go through 
the mechanics of a gauge transformation.  
Equivalently, we can write a time domain expression
\begin{equation}
\label{eq:psi4radt}
\Psi_{4}=\frac{1}{2}\frac{\partial^{2}}{\partial t^{2}}\left(\hp-i\hc\right)\;,
\end{equation}
which is the form used in equations (3.54) of \cite{drhs06} and 
(2$\cdot$30) of \cite{progsupp97}.  Equations \eqref{eq:psi4radf} and 
\eqref{eq:psi4radt} are related using the definitions of $\hp$ 
and $\hc$ \eqref{eq:shphc}.  Also, the results \eqref{eq:psi4radf} and 
\eqref{eq:psi4radt} presuppose the definition of $\Psi_{4}$ 
\eqref{eq:weyl4} and tetrad basis \eqref{eq:nplvec}-\eqref{eq:npmbvec}.  
Other conventions may result in slightly different expressions, although 
presumably not different waveforms \cite{bcp07}, \cite{bucent07}.

We could calculate $\Psi_{0}$ and $\Psi_{4}$ by solving the Teukolsky 
equation, which is a linear partial differential equation that 
describes gravitational, electromagnetic and 
scalar perturbations \cite{teuk73}.  For the Schwarzschild metric, 
the Teukolsky equation is
\begin{multline}
\label{eq:teukpde}
\frac{r^{4}}{\Delta}\frac{\partial^{2}\psi}{\partial t^{2}}
-\frac{1}{\sin^{2}\theta}\frac{\partial^{2}\psi}{\partial \phi^{2}}
-\Delta^{-s}\frac{\partial}{\partial r}
\left(\Delta^{s+1}\frac{\partial\psi}{\partial r}\right)
-\frac{1}{\sin\theta}\frac{\partial}{\partial\theta}
\left(\sin\theta\frac{\partial\psi}{\partial\theta}\right)
\\-\frac{2 i s \cos\theta}{\sin^{2}\theta}\frac{\partial\psi}{\partial\phi}
-2 s\left[\frac{M r^{2}}{\Delta}-r\right]\frac{\partial\psi}{\partial t}
+\left(s^{2}\cot^{2}\theta-s\right)\psi=4\pi r^{2} T\;,
\end{multline}
where $\Delta=r^{2}-2 M r$ and $T$ is the source 
constructed from the stress energy tensor $T_{\mu\nu}$.  The spin weight 
$s$ is $\pm 2$ for gravitational perturbations, $\pm 1$ for electromagnetic 
perturbations and $0$ for scalar perturbations.  The definition of the 
function $\psi$ depends on the spin weight.  For $s=2$, $\psi=\Psi_{0}$; 
for $s=-2$, $\psi=r^{4}\Psi_{4}$.

Instead of solving the Teukolsky equation, we will use the 
harmonic gauge solutions derived in this thesis and the definitions 
of $\Psi_{0}$ \eqref{eq:pweyl0} and $\Psi_{4}$ \eqref{eq:pweyl4}.  
To do so, we first calculate the perturbed Riemann curvature tensor 
$\delta\!R_{\alpha\beta\gamma\delta}$ to linear order in the mass ratio $\mz/M$, 
using the results of exercise 35.11 of \cite{mtw73}.  The total Riemann 
curvature tensor $\widetilde{R}^{\alpha}{}_{\!\beta\gamma\delta}$ can be split 
into background and perturbed parts, so that
\begin{equation}
\label{eq:riem7}
\widetilde{R}^{\alpha}{}_{\!\beta\gamma\delta}=R^{\alpha}{}_{\!\beta\gamma\delta}
+\delta\!R^{\alpha}{}_{\!\beta\gamma\delta}
+O\left(\tfrac{\mz}{M}\right)^{2}\;.
\end{equation}
Here, $R^{\alpha}{}_{\!\beta\gamma\delta}$ is the curvature tensor computed 
with the background metric, and $\delta\!R^{\alpha}{}_{\!\beta\gamma\delta}$ 
is the first order perturbation of the curvature tensor.  From \cite{mtw73}, 
we have
\begin{equation}
\label{eq:a1a}
\delta\!R^{\alpha}{}_{\!\beta\gamma\delta}
=\delta\!\mathnormal{\Gamma}^{\alpha}{}_{\!\beta\delta;\gamma}
-\delta\!\mathnormal{\Gamma}^{\alpha}{}_{\!\beta\gamma;\delta}\;,
\end{equation}
where
\begin{equation}
\label{eq:deltareg7}
\delta\!\mathnormal{\Gamma}^{\gamma}{}_{\!\alpha\beta}
=\frac{1}{2}\,g^{\gamma\epsilon}\left(h_{\epsilon\alpha;\beta}
+h_{\epsilon\beta;\alpha}-h_{\alpha\beta;\epsilon}\right)
\end{equation}
is the first order perturbation of the Christoffel symbol of the 
second kind.  The covariant Riemann curvature tensor is
\begin{equation}
\label{eq:a2}
\widetilde{R}_{\alpha\beta\gamma\delta}=\tilde{g}_{\alpha\epsilon}
\widetilde{R}^{\epsilon}{}_{\!\beta\gamma\delta}
=g_{\alpha\epsilon}R^{\epsilon}{}_{\!\beta\gamma\delta}
+\left\{h_{\alpha\epsilon}R^{\epsilon}{}_{\!\beta\gamma\delta}
+g_{\alpha\epsilon}\delta\!R^{\epsilon}{}_{\!\beta\gamma\delta}\right\}
+O\left(\tfrac{\mz}{M}\right)^{2}\;,
\end{equation}
where, as before \eqref{eq:tilmet},
\begin{equation}
\label{eq:tilmet7}
\tilde{g}_{\mu\nu}=g_{\mu\nu}+h_{\mu\nu}\;.
\end{equation}
In equation \eqref{eq:a2}, the part in curly brackets is
$\delta\!R_{\alpha\beta\gamma\delta}$, so \eqref{eq:a2} may be rewritten as
\begin{equation}
\label{eq:a3}
\widetilde{R}_{\alpha\beta\gamma\delta}=R_{\alpha\beta\gamma\delta}
+\delta\!R_{\alpha\beta\gamma\delta}
+O\left(\tfrac{\mz}{M}\right)^{2}\;.
\end{equation}

The next step is to project $\delta\!R_{\alpha\beta\gamma\delta}$ onto the 
tetrad basis, using \eqref{eq:pweyl0} and \eqref{eq:pweyl4}.  This 
procedure gives
\begin{equation}
\label{eq:psi4sep}
\Psi_{4}=\sum_{l=2}^{\infty}\sum_{m=-l}^{l}{}_{-2}Y_{lm}(\theta,\phi)
\int_{\!-\infty}^{\infty} e^{-\iom t}\Psi_{4}^{lm}(\omega,r)d\omega\;.
\end{equation}
Equation \eqref{eq:lgts} implies ${}_{-2}Y_{lm}(\theta,\phi)=0$ for $l<2$.  
For odd parity, the radial coefficient $\Psi_{4}^{lm}(\omega,r)$ is
\begin{multline}
\label{eq:oddpsi4}
\Psi_{4}^{lm}(\omega,r)=\frac{i \sqrt{\lambda (1+\lambda)}}{4 \lambda r^4}
\Big[\iom r \left(\lambda (-2 M+r)+\iom r \left(-M+\iom r^2\right)\right)\ho
\\+2 (\lambda-\iom r) \left(-M+\iom r^2\right)\hz
-2\iom \lambda \left(-3 M+r+\iom r^2\right)\htw
\\+r \left(\lambda (-2 M+r)+\iom r 
\left(-M+\iom r^2\right)\right)\dhz\Big]\;.
\end{multline}
For even parity, we have
\begin{equation}
\label{eq:evpsi4}
\Psi_{4}^{lm}(\omega,r)=\sqrt{\lambda (1+\lambda)}
\left[\alpha \bg+\beta \hz+\gamma \bhz+\delta \ho
+\epsilon \bhtw+\zeta \bk+\kappa \dbk\right]\;,
\end{equation}
where
\begin{equation}
\alpha=\frac{\left(M (1+\lambda-3 \iom r)+\iom r^2 
(-\lambda+\iom r)\right)}{2 r^3}\;,
\end{equation}
\begin{equation}
\beta=-\frac{\iom\left(-3 M^2+\lambda r^2 (1+\iom r)+3 M r 
(-\lambda+\iom r)\right)}{2 \lambda (2 M-r) r^3}\;,
\end{equation}
\begin{equation}
\gamma=\frac{\left(-3 M^2+\lambda r^2 (1+\iom r)+3 M r 
(-\lambda+\iom r)\right)}{4 \lambda r^4}\;,
\end{equation}
\begin{equation}
\delta=-\frac{\left((1+\lambda) M^2+\iom M r^2 (-1+\lambda+\iom r)
-\iom r^3 \left(\lambda+(\iom)^2 r^2\right)\right)}{2 \lambda r^5}\;,
\end{equation}
\begin{equation}
\epsilon=\frac{\left((1+\lambda) M^2+\iom M r^2 (-1
+\lambda+\iom r)-\iom r^3 \left(\lambda+(\iom)^2 r^2\right)\right)}
{4 \lambda (1+\lambda) r^4}\;,
\end{equation}
\begin{equation}
\begin{split}
\zeta=\frac{1}{4 \lambda (1+\lambda) (2 M-r) r^3}
\left[\lambda^2 (2 M-r) \left(M-\iom r^2\right)
+3 (\iom)^2 M r^2 \left(-M+\iom r^2\right)
\right.\\\left.+\lambda\left((\iom)^3 r^5+M^2 (2-6 \iom r)
-M r \left(1-3 \iom r+(\iom)^2 r^2\right)\right)\right]\;,
\end{split}
\end{equation}
\begin{equation}
\kappa=-\frac{\left((1+\lambda) M^2+\iom M r^2 (-1+\lambda
+\iom r)-\iom r^3 \left(\lambda+(\iom)^2 r^2\right)\right)}
{4 \lambda (1+\lambda) r^3}\;.
\end{equation}
Similarly, we find that $\Psi_{0}$ is
\begin{equation}
\label{eq:psi0sep}
\Psi_{0}=\sum_{l=2}^{\infty}\sum_{m=-l}^{l}{}_{2}Y_{lm}(\theta,\phi)
\int_{\!-\infty}^{\infty} e^{-\iom t}\Psi_{0}^{lm}(\omega,r)d\omega\;.
\end{equation}
Equation \eqref{eq:lgts} implies ${}_{2}Y_{lm}(\theta,\phi)=0$ for $l<2$.  
For odd parity, we obtain
\begin{multline}
\label{eq:oddpsi0}
\Psi_{0}^{lm}(\omega,r)=\frac{i\sqrt{\lambda(1+\lambda)}}{\lambda r^2(-2 M+r)^2}
\Big[\iom r \left(\lambda (-2 M+r)+\iom r \left(M+\iom r^2\right)\right)\ho
\\-2(\lambda+\iom r) \left(M+\iom r^2\right)\hz
+2 \iom \lambda (3 M+r (-1+\iom r))\htw
\\+r \left(\lambda (-2 M+r)+\iom r \left(M+\iom r^2\right)\right)\dhz\Big]\;.
\end{multline}
The even parity radial coefficient is
\begin{equation}
\label{eq:evpsi0}
\Psi_{0}^{lm}(\omega,r)=\sqrt{\lambda(1+\lambda)}
\left[\alpha \bg+\beta \hz+\gamma \bhz
+\delta \ho+\epsilon \bhtw+\zeta \bk+\kappa \dbk\right]\;,
\end{equation}
where
\begin{equation}
\alpha=\frac{2 \left(\iom r^2 (\lambda+\iom r)+M (1+\lambda
+3 \iom r)\right)}{r (-2 M+r)^2}\;,
\end{equation}
\begin{equation}
\beta=\frac{2 \iom \left(3 M^2+\lambda r^2 (-1+\iom r)+3 M r 
(\lambda+\iom r)\right)}{\lambda (2 M-r)^3 r}\;,
\end{equation}
\begin{equation}
\gamma=\frac{\left(-3 M^2+\lambda r^2 (1-\iom r)
-3 M r (\lambda+\iom r)\right)}{\lambda r^2 (-2 M+r)^2}\;,
\end{equation}
\begin{equation}
\delta=-\frac{2 \left((1+\lambda) M^2+\iom M r^2 (1-\lambda+\iom r)
+\iom r^3 \left(\lambda+(\iom)^2 r^2\right)\right)}{\lambda r^3(-2 M+r)^2}\;,
\end{equation}
\begin{equation}
\epsilon=\frac{\left((1+\lambda) M^2+\iom M r^2 (1-\lambda+\iom r)
+\iom r^3 \left(\lambda+(\iom)^2 r^2\right)\right)}
{\lambda(1+\lambda)r^2 (-2 M+r)^2}\;,
\end{equation}
\begin{equation}
\begin{split}
\zeta=\frac{1}{\lambda(1+\lambda) (2 M-r)^3 r}
\left[\lambda^2 (2 M-r) \left(M+\iom r^2\right)
-3 (\iom)^2 M r^2 \left(M+\iom r^2\right)
\right.\\\left.+\lambda \left(-(\iom)^3 r^5+M^2 (2+6 \iom r)
-M r \left(1+3 \iom r+(\iom)^2 r^2\right)\right)\right]\;,
\end{split}
\end{equation}
\begin{equation}
\kappa=-\frac{\left((1+\lambda) M^2+\iom M r^2 (1-\lambda
+\iom r)+\iom r^3 \left(\lambda+(\iom)^2 
r^2\right)\right)}{\lambda (1+\lambda) r (-2 M+r)^2}\;.
\end{equation}
These are vacuum expressions.  We can show that combinations of radial 
functions in $\Psi_{0}^{lm}(\omega,r)$ 
and $\Psi_{4}^{lm}(\omega,r)$ are gauge invariant, in the same sense 
that the odd parity $\ptw$ \eqref{eq:spin2func} and even parity 
$\ptw$ \eqref{eq:evspin2func} are gauge invariant.  Nevertheless, these 
expressions are not unique, because the first order differential 
identities \eqref{eq:dhzeqn} and \eqref{eq:ndh0eqns}-\eqref{eq:ndgeqns} 
are also gauge invariant.  Using the homogeneous forms of those 
identities, we may rewrite $\Psi_{0}^{lm}(\omega,r)$ and 
$\Psi_{4}^{lm}(\omega,r)$ in terms of different combinations of the 
radial perturbation functions, just as we can rewrite the definitions 
of $\ptw$ in the alternative forms $\ptw^{\text{JT}}$ 
\eqref{eq:jtspin2func} and $\ptw^{\text{Mon}}$ \eqref{eq:evspin2funcmon}.  
However, doing so will not change the value of $\Psi_{0}^{lm}(\omega,r)$ and 
$\Psi_{4}^{lm}(\omega,r)$, because the first order differential equations 
are identities.

The next step is to substitute the non-zero frequency harmonic gauge 
solutions derived in Chapters~\ref{oddpar} and~\ref{evpar} into 
the expressions for $\Psi_{0}^{lm}(\omega,r)$ and $\Psi_{4}^{lm}(\omega,r)$.  
For odd parity, this gives
\begin{multline}
\label{eq:oddpsi4rlm}
\Psi_{4}^{lm}(\omega,r)=\frac{i \sqrt{\lambda (1+\lambda)}}{2 \iom r^5}
\bigg\{r(2 M-r) \left[-3 M+r+\iom r^2\right]\dptw
\\-\left[6 M^2-M r (5+2 \lambda+3 \iom r)+r^2 
\left(1+\lambda+\iom r+(\iom)^2 r^2\right)\right]\ptw\bigg\}\;,
\end{multline}
\begin{multline}
\label{eq:oddpsi0rlm}
\Psi_{0}^{lm}(\omega,r)=-\frac{2 i \sqrt{\lambda (1+\lambda)} }
{\iom (2 M-r)^2 r^3}\bigg\{r(2 M-r)[3 M+r (-1+\iom r)]\dptw
\\+\left[6 M^2+M r (-5-2 \lambda+3 \iom r)+r^2 \left(1+\lambda
-\iom r+(\iom)^2 r^2\right)\right]\ptw\bigg\}\;.
\end{multline}
For even parity, we get
\begin{equation}
\label{eq:evpsi4rlm}
\begin{split}
\Psi_{4}^{lm}(\omega,r&)=\frac{ \sqrt{\lambda (1+\lambda)}}
{4 r^5 (3 M+\lambda r)^{2}}\bigg\{r(3 M+\lambda r)(2 M-r)
\left[3 M^2+3 M r (\lambda-\iom r)
\right.\\&\left.-\lambda r^2 (1
+\iom r)\right]\dptw+\left[-18 M^4-9 M^3 r (-1+2 \lambda+\iom r)
\right.\\&\left.+\lambda^2 r^4 \left(1+\lambda+\iom r
+(\iom)^2 r^2\right)+3 M^2 r^2\left(-2 \lambda^2+3 (\iom)^2 r^2
\right.\right.\\&\left.\left.+\lambda(3-4 \iom r)\right)
+\lambda M r^3\left(\lambda-2 \lambda^2
-3 \iom \lambda r+3 \iom r (1+2 \iom r)\right)\right]\ptw\bigg\}\;,
\end{split}
\end{equation}
\begin{equation}
\label{eq:evpsi0rlm}
\begin{split}
\Psi_{0}^{lm}(\omega,r&)=\frac{\sqrt{\lambda(1+\lambda)}}
{r^3 (2 M-r)^{2} (3 M+\lambda r)^{2}}\bigg\{(2 M-r) r (3 M+\lambda r)
\left[3 M^2+\lambda r^2 \right.\\&\left.\times(-1+\iom r)+3 M r 
(\lambda+\iom r)\right]\dptw+\left[-18 M^4+9 M^3 r (1-2 \lambda+\iom r)
\right.\\&\left.+\lambda^2 r^4 \left(1+\lambda-\iom r+(\iom)^2 r^2\right)
+\lambda M r^3 \left(\lambda-2 \lambda^2+3 \iom \lambda r
\right.\right.\\&\left.\left.+3 \iom r (-1+2 \iom r)\right)
+3 M^2 r^2 \left(-2 \lambda^2+3 (\iom)^2 r^2
+\lambda (3+4 \iom r)\right)\right]\ptw\bigg\}\;.
\end{split}
\end{equation}
Again, these are vacuum expressions.  We can verify by substitution that 
they are homogeneous solutions of the Teukolsky equation 
\eqref{eq:teukpde}.   The odd parity $\ptw$ is 
defined in \eqref{eq:spin2func} and is a homogeneous solution of 
the Regge-Wheeler equation \eqref{eq:ptweqn}, \eqref{eq:ptwsource}.  
If we had used $\ptw^{\text{JT}}$ instead, we would have slightly 
different odd parity forms, which may be obtained by applying 
the vacuum form of \eqref{eq:jtspin2funcps}:  $\ptw=-\iom\ptw^{\text{JT}}$.  
The even parity $\ptw$ is defined in \eqref{eq:evspin2func} and is a 
homogeneous solution of the Zerilli equation 
\eqref{eq:zereqn}-\eqref{eq:zereqns1}.  Both $\Psi_{0}^{lm}(\omega,r)$ 
and $\Psi_{4}^{lm}(\omega,r)$ depend on the even and odd parity 
$\ptw$ functions.  This is not an accident.  As explained above, the 
perturbations of $\Psi_{0}$ and $\Psi_{4}$ are gauge invariant, 
physically meaningful Weyl scalars.  The $\ptw$ functions are also gauge 
invariant.  

For large $r$, we approximate $\Psi_{4}^{lm}(\omega,r)$ and 
$\Psi_{0}^{lm}(\omega,r)$ using the outgoing radiation 
series expansions from section~\ref{sec:nzrweqn}.  Asymptotically, we get
\begin{equation}
\label{eq:apsi4odd}
\Psi_{4}^{lm}(\omega,r)=-\frac{\omega^{2}}{2}\frac{e^{\iom r_{*}}}{\iom r}
\left[-A^{\infty}_{2lm\omega}2 i\sqrt{\lambda(1+\lambda)}\right]+O(r^{-2})
\end{equation}
for odd parity and 
\begin{equation}
\label{eq:apsi4ev}
\Psi_{4}^{lm}(\omega,r)=-\frac{\omega^{2}}{2}\frac{e^{\iom r_{*}}}{r}
A^{\infty}_{2lm\omega}\sqrt{\lambda(1+\lambda)}+O(r^{-2})
\end{equation}
for even parity.  Because we have substituted the Regge-Wheeler and Zerilli 
solutions into $\Psi_{4}^{lm}(\omega,r)$, the amplitudes $A^{\infty}_{2lm\omega}$ 
are the same as those used in $\hp$ \eqref{eq:hthth} and 
$\hc$ \eqref{eq:hthphi}.  For both parities, outgoing solutions give 
$\Psi_{0}^{lm}(\omega,r)\sim O(r^{-5})$ as $r\to\infty$.  The 
$O(r^{-1})$ and $O(r^{-5})$ behaviors are those that would be obtained if 
we had solved Teukolsky's equation directly \cite{teuk73}.  
We now can apply \eqref{eq:psi4radf} to derive expressions for 
$h_{\hat{\theta}\hat{\theta}}$ and $h_{\hat{\theta}\hat{\phi}}$ from $\Psi_{4}$.  To 
do so, we replace the spin-weighted spherical harmonic 
${}_{-2}Y_{lm}(\theta,\phi)$ used in \eqref{eq:psi4sep} with the tensor 
harmonics $W_{lm}(\theta,\phi)$ and $X_{lm}(\theta,\phi)$ \eqref{eq:ymtwo}.  
This gives
\begin{equation}
\label{eq:psi4sepxw}
\Psi_{4}=\sum_{l=2}^{\infty}\sum_{m=-l}^{l}\int_{\!-\infty}^{\infty} 
e^{-\iom t}\Psi_{4}^{lm}(\omega,r,\theta,\phi)d\omega\;,
\end{equation}
where, for odd parity,
\begin{equation}
\Psi_{4}^{lm}(\omega,r,\theta,\phi)=-\frac{\omega^2}{2}
\left\{\frac{e^{\iom r_{*}} }{\iom r}A^{\infty}_{2lm\omega}
\left[- X_{lm}(\theta,\phi)-i W_{lm}(\theta,\phi)\right]\right\}+O(r^{-2})
\end{equation}
and, for even parity,
\begin{equation}
\Psi_{4}^{lm}(\omega,r,\theta,\phi)=-\frac{\omega^2}{2}
\left\{\frac{e^{\iom r_{*}}}{2 r}A^{\infty}_{2lm\omega}
\left[W_{lm}(\theta,\phi)-i X_{lm}(\theta,\phi)\right]\right\}+O(r^{-2})\;.
\end{equation}
The terms in curly brackets are equal to 
$h_{\hat{\theta}\hat{\theta}} -i h_{\hat{\theta}\hat{\phi}}$, 
in agreement with the previous results \eqref{eq:hthth} and 
\eqref{eq:hthphi}.  

We will use $\Psi_{0}$ near the event horizon to calculate the energy 
and angular momentum fluxes in section~\ref{sec:wavenerg}.  We substitute 
the ingoing radiation series expansions from section~\ref{sec:nzrweqn} 
into \eqref{eq:oddpsi0rlm} and \eqref{eq:evpsi0rlm}.  As $r\to 2 M$, we 
find
\begin{equation}
\label{eq:psi02modd}
\Psi_{0}^{lm}(\omega,r)=-\frac{e^{-\iom r_{*}}}{(2 M)^{2}X^2} 
2 i A^{2 M}_{2lm\omega}(1+4\iom M)\sqrt{\lambda(1+\lambda)}+O(X^{-1})
\end{equation}
for odd parity and 
\begin{equation}
\label{eq:psi02mev}
\Psi_{0}^{lm}(\omega,r)=\frac{e^{-\iom r_{*}}}{(2 M)^{2}X^2} 
\iom A^{2 M}_{2lm\omega}(1+4\iom M)\sqrt{\lambda(1+\lambda)}+O(X^{-1})
\end{equation}
for even parity.  As before, $X=\ff$.  The ingoing amplitude constants 
$A^{2 M}_{2lm\omega}$ are the odd and even parity retarded solution source 
integrals \eqref{eq:ahor} and, in the case of bound orbits, 
\eqref{eq:ahorepsi}.  The expressions for $\Psi_{0}$ diverge quadratically 
near the event horizon.  As discussed elsewhere \cite{teukpr74}, 
the problem is that the tetrad \eqref{eq:nplvec}-\eqref{eq:npmbvec}, 
which we have been using, is singular as $r\to 2 M$.  A slightly different 
basis, the Hawking-Hartle (HH) basis, is not.  For the Schwarzschild 
background metric, the two bases and forms of $\Psi_{0}$ are related 
by \cite{teukpr74}
\begin{equation}
\label{eq:hhbasis}
\bm{l}^{\text{HH}}=\frac{r-2 M}{2 r}\bm{l}\;,
\bm{n}^{\text{HH}}=\frac{2 r}{r-2 M}\bm{n}\;,
\bm{m}^{\text{HH}}=\bm{m}\;,
\Psi_{0}^{\text{HH}}=\left(\frac{r-2 M}{2 r}\right)^{2}\Psi_{0}\;.
\end{equation}
Explicitly, we can write
\begin{equation}
\label{eq:psi0hh}
\Psi_{0}^{\text{HH}}=\sum_{l=2}^{\infty}\sum_{m=-l}^{l}{}_{2}Y_{lm}(\theta,\phi)
\int_{\!-\infty}^{\infty} e^{-\iom t}\Psi_{0}^{\text{HH},lm}(\omega,r)d\omega\;,
\end{equation}
where
\begin{equation}
\label{eq:psi0hhodd}
\Psi_{0}^{\text{HH},lm}(\omega,r)=-\frac{e^{-\iom r_{*}}}{8 M^{2}} 
i A^{2 M}_{2lm\omega}(1+4\iom M)\sqrt{\lambda(1+\lambda)}+O(X)
\end{equation}
for odd parity and 
\begin{equation}
\label{eq:psi0hhev}
\Psi_{0}^{\text{HH},lm}(\omega,r)=\frac{e^{-\iom r_{*}}}{16 M^{2}} 
\iom A^{2 M}_{2lm\omega}(1+4\iom M)\sqrt{\lambda(1+\lambda)}+O(X)
\end{equation}
for even parity as $r\to 2 M$.  

We also can calculate the perturbations of $\Psi_{1}$, $\Psi_{2}$ and 
$\Psi_{3}$, using equations \eqref{eq:pweyl1}-\eqref{eq:pweyl3}.  
As mentioned above, it is possible to choose a gauge where most of these 
quantities are zero for $l\ge 2$, but the harmonic gauge is apparently not such a 
gauge.  Only key points about these functions are mentioned below.  
The perturbation of $\Psi_{2}$ is proportional to $Y_{lm}(\theta,\phi)$ and
\begin{equation}
\label{eq:psisang}
\Psi_{1}\propto{}_{1}Y_{lm}(\theta,\phi)\;,
\Psi_{3}\propto{}_{-1}Y_{lm}(\theta,\phi)\;.
\end{equation}
For odd parity, $\Psi_{1}$ and $\Psi_{3}$ depend on the odd parity 
generalized Regge-Wheeler functions $\ptw$ and $\po$, while the 
perturbation of $\Psi_{2}$ depends only on $\ptw$.  For even parity, 
all three perturbations contain the even parity functions 
$\ptw$, $\po$, $\pz$ and $\pza$.  We also can solve for the metric 
perturbation radial factors in terms of the radial factors of the 
Weyl scalars, but the results are more complicated than the solutions 
derived in Chapters~\ref{oddpar} and~\ref{evpar} and will not be set 
forth here.  These calculations were done with the unperturbed tetrad 
\eqref{eq:nplvec}-\eqref{eq:npmbvec}.  A different approach might simplify 
the expressions.

\section{\label{sec:wavenerg}Energy and Angular Momentum Flux}

In this section, we calculate the average energy and angular momentum carried 
by gravitational waves, both outwards to large distances and inwards through 
the event horizon.  The main results will be for bound orbits.

Expressions for the flux outwards are obtained from the Isaacson tensor.  
Using perturbation theory, Isaacson derived a stress energy tensor 
for gravitational waves \cite{isaac167}, \cite{isaac267}.  
For an arbitrary gauge, the Isaacson gravitational wave (``GW'') 
tensor is \cite{mtw73}
\begin{equation}
\label{eq:tmngw}
T_{\mu\nu}^{\scriptscriptstyle{\text{(GW)}}}
=\frac{1}{32\pi}\left\langle 
\overline{h}_{\alpha\beta;\mu}\overline{h}^{\alpha\beta}\!{}_{;\nu}
-\frac{1}{2}\overline{h}_{;\mu}\overline{h}_{;\nu}
-2\overline{h}^{\alpha\beta}\!{}_{;\beta}
\overline{h}_{\alpha\scriptscriptstyle{(}\scriptstyle{\mu;\nu}
\scriptscriptstyle{)}}\right\rangle\;.
\end{equation}
In the harmonic gauge, $\overline{h}^{\alpha\beta}\!{}_{;\beta}=0$ 
\eqref{eq:divheqn}.  Substituting the definition of 
$\overline{h}_{\mu\nu}$ \eqref{eq:hbar}, we can 
rewrite \eqref{eq:tmngw} in terms of $h_{\mu\nu}$ as
\begin{equation}
\label{eq:tmngwh}
T_{\mu\nu}^{\scriptscriptstyle{\text{(GW)}}}
=\frac{1}{32\pi}\left\langle 
h_{\alpha\beta;\mu}h^{\alpha\beta}\!{}_{;\nu}
-\frac{1}{2}h_{;\mu}h_{;\nu}\right\rangle\;.
\end{equation}
This step simplifies calculations, because the perturbation is 
given in terms of $h_{\mu\nu}$.  

The brackets in \eqref{eq:tmngw} and \eqref{eq:tmngwh} represent 
averaging over several wavelengths.  The Isaacson tensor is 
valid only in the shortwave, or high frequency, approximation limit, 
meaning that the wavelength is much smaller than the radius of 
the background gravitational curvature. 
In this limit, the averaged tensor is gauge invariant.  
The high frequency prerequisite can always be met 
at a large distance from an isolated source, such as the black hole 
system studied here \cite{isaac167}, \cite{mtw73}.  

We can use the Isaacson tensor to calculate the energy flux of the 
waves as $r\to\infty$.  Based on \cite{th80}, 
the average power radiated outwards in the radial 
direction through a large sphere of radius $r$ is
\begin{equation}
\label{eq:thdedt}
\left\langle\frac{d E^{\infty}}{dt}\right\rangle
=-\int T_{t}{}^{r} r^{2} d\Omega\;,\,r\to\infty\;,       
\end{equation}
where
\begin{equation}
\label{eq:t0r}
T_{t}{}^{r}=g^{rr}T_{t r}^{\scriptscriptstyle{\text{(GW)}}}\;.
\end{equation}
We will apply \eqref{eq:thdedt} to circular and elliptic orbits only.  

For these orbits, averaging over several wavelengths 
is equivalent to a time integral over an orbital period $P$, so that
\begin{equation}
\label{eq:aveint}
\bigg\langle f(t)\bigg\rangle=\frac{1}{P}\int_{0}^{P}f(t) dt\;.
\end{equation}
Because we are using Fourier transforms, the time dependence of 
\eqref{eq:tmngwh} is only in exponential factors $e^{-\iom t}$.  
This form of averaging has been used elsewhere, such as \cite{gkenn02}.  
One could argue that several wavelengths would be equivalent to several 
periods.  However, we calculate the first order perturbation ($O(\mz/M)$)
by assuming that the orbiting mass travels on a geodesic of the 
background spacetime.  To this order, successive orbits are repeating, 
in the sense that each orbit gives the same integral per period.  

We will evaluate $T_{\mu\nu}^{\scriptscriptstyle{\text{(GW)}}}$ in the harmonic 
gauge, using \eqref{eq:tmngwh}.  We start by expanding the derivatives of the 
harmonic gauge metric perturbations in series of inverse powers of $r$, 
as we did in section~\ref{sec:wavs}.  Frequency integrals become 
sums \eqref{eq:inttosum}.  Because $T_{\mu\nu}^{\scriptscriptstyle{\text{(GW)}}}$ 
is quadratic in the perturbation, we must multiply two multipole 
expansions, one with indices 
$l$, $m$ and $k$ and the other with indices $l^{\prime}$, 
$m^{\prime}$ and $k^{\prime}$.  The angular and time integrals kill cross terms.  
By orthogonality, the integral over all angles \eqref{eq:thdedt} is 
non-zero only if $l^{\prime}=l$, $m^{\prime}=-m$.  The angular integral also 
generates a factor of $l(l+1)(l-1)(l+2)$ \eqref{eq:wxint}.  
Recalling that $\omega=\omega_{mk}=m \Omega_{\phi}+k \Omega_{r}$ and 
$\Omega_{r}=\frac{2\pi}{P}$ \eqref{eq:omegamk}, we 
evaluate the time average integral \eqref{eq:aveint} to get
\begin{multline}
\label{eq:naveint}
\frac{1}{P}\int_{0}^{P}e^{-i(\omega_{m^{\prime}k^{\prime}}+\omega_{mk})t} dt
=\frac{1}{P}\int_{0}^{P}e^{-i(\omega_{-m k^{\prime}}+\omega_{mk})t} dt
\\=\frac{1}{P}\int_{0}^{P}e^{-i(k^{\prime}+k)\frac{2\pi}{P}t} dt=\delta_{-k^{\prime}k}\;.
\end{multline}
We have $m^{\prime}=-m$, $k^{\prime}=-k$ and 
$\omega_{m^{\prime}k^{\prime}}=-\omega_{mk}$, so the primed index multipole 
expansion is the complex conjugate of the other.  The methods described 
in this paragraph are based on those used elsewhere for the 
Regge-Wheeler gauge \cite{ashby}, \cite{zerp70}, and for bound orbit 
solutions of the Teukolsky equation \cite{cut94}, \cite{gkenn02}.  

At the end of our computations, we find that the outgoing energy flux is
\begin{equation}
\label{eq:edotout}
\dot{E}^{\infty}=\left\langle\frac{dE^{\infty}}{dt}\right\rangle=\frac{1}{16\pi}
\sum^{\infty}_{l=2}\,\sum^{l}_{m=-l}\,\sum^{\infty}_{k=-\infty}f^{}_{lmk}
\lvert A^{\infty}_{2lm\omega}\rvert^{2} l(l+1)(l-1)(l+2)\;.
\end{equation}
The amplitude $A^{\infty}_{2lm\omega}$ is the source integral 
\eqref{eq:ainfepsi}, with $s=2$.  Vertical bars denote the magnitude of 
a complex quantity.  Here and elsewhere in this section, we define
\begin{equation}
\label{eq:flmk}
f^{}_{lmk}=\begin{cases}
\quad 1\;,\;\text{odd parity modes}\;,
\\\left(\frac{\omega_{mk}}{2}\right)^{2}\;,\;\text{even parity modes}\;,
\end{cases}
\end{equation}
where $\omega_{mk}=m \Omega_{\phi}+k \Omega_{r}$ \eqref{eq:omegamk}.  
As discussed at the end of section~\ref{sec:ttmunu}, 
the even parity modes are non-zero only for $l+m$ 
even and the odd parity modes are non-zero only for $l+m$ odd, 
provided the orbit is in the equatorial plane.  

The choice of gauge affects the manner in which $\dot{E}^{\infty}$ 
is calculated, but not the end result \eqref{eq:edotout}.  In the harmonic gauge, 
the trace is non-zero and we find that
\begin{equation}
\label{eq:harmedotout}
\dot{E}^{\infty}=\frac{1}{16\pi}
\sum^{\infty}_{l=2}\,\sum^{l}_{m=-l}\,\sum^{\infty}_{k=-\infty}\!\!f^{}_{lmk}
\bigg\{\!\Big[\lvert A^{\infty}_{2lm\omega}\rvert^{2} l(l+1)(l-1)(l+2)
+\lvert A^{\infty}_{0lm\omega}\rvert^{2}\Big]
-\lvert A^{\infty}_{0lm\omega}\rvert^{2}\bigg\},
\end{equation}
where $A^{\infty}_{0lm\omega}$ is the outgoing amplitude \eqref{eq:ainfepsi} 
of the even parity function $\pz$.  
Inside the curly brackets, the first two terms (in the square brackets) 
come from the $h_{\alpha\beta;\mu}h^{\alpha\beta}\!{}_{;\nu}$ term of 
$T_{\mu\nu}^{\scriptscriptstyle{\text{(GW)}}}$ \eqref{eq:tmngwh} and the last term 
comes from the $\frac{1}{2}h_{;\mu}h_{;\nu}$ term of 
$T_{\mu\nu}^{\scriptscriptstyle{\text{(GW)}}}$.  The two spin~$0$ terms cancel, 
leaving $\dot{E}^{\infty}$ \eqref{eq:edotout}.  This harmonic gauge calculation is 
instructive, but more complicated than necessary.  Away from the source, 
we may remove the trace, as well as the other spin~$0$  and spin~$1$ pieces, by 
means of a gauge transformation which preserves 
the harmonic gauge, as shown in Chapter~\ref{evpar}.  Such a gauge 
transformation would not affect the spin~$2$ pieces of the perturbation, so 
we would still obtain \eqref{eq:edotout}.  
If we use the radiation gauge instead, we return to the original expression 
for $T_{\mu\nu}^{\scriptscriptstyle{\text{(GW)}}}$, which is \eqref{eq:tmngw}.  
Because the radiation gauge is asymptotically transverse-traceless 
\eqref{eq:httrad}, only the first term of \eqref{eq:tmngw} 
(which reduces to $h_{\alpha\beta;\mu}h^{\alpha\beta}\!{}_{;\nu}$) will contribute, 
and it gives \eqref{eq:edotout} also.  
The different gauges yield the same end result for $\dot{E}^{\infty}$ \eqref{eq:edotout}, 
because the Isaacson tensor is gauge invariant.  

Alternatively, we can use the Newman-Penrose formalism.  Teukolsky showed
\begin{equation}
\label{eq:dedtout}
\frac{d^{2}E^{\infty}}{dt\,d\Omega}
=\lim_{r\to\infty}\frac{r^{2}\omega^{2}}
{16\pi}\left[\big(h_{\hat{\theta}\hat{\theta}}\big)^{2}
+\big(h_{\hat{\theta}\hat{\phi}}\big)^{2}\right]
=\lim_{r\to\infty}\frac{r^{2}}{4\pi\omega^{2}}\lvert \Psi_{4}\rvert^{2}\;,
\end{equation}
for a single frequency mode \cite{teuk73}.  Substituting the asymptotic expansions 
of $\Psi_{4}$ \eqref{eq:apsi4odd}-\eqref{eq:apsi4ev} into \eqref{eq:dedtout} 
leads to the previous expression for $\dot{E}^{\infty}$ \eqref{eq:edotout}.

The angular momentum and energy fluxes per frequency mode are related by
\begin{equation}
\label{eq:enlz}
dL_{z}=\frac{m}{\omega}\,dE\;,
\end{equation}
which is based in part on the energy and angular momentum relations of 
quantum mechanics \cite{bek73}, \cite{cut94}, \cite{gkenn02}, \cite{teukpr74}.  
Applying \eqref{eq:enlz} to  $\dot{E}^{\infty}$ \eqref{eq:edotout}, we get
\begin{equation}
\label{eq:lzdotout}
\dot{L}_{z}^{\infty}=\left\langle\frac{dL^{\infty}_{z}}{dt}\right\rangle
=\frac{1}{16\pi}\sum^{\infty}_{l=2}\,\sum^{l}_{m=-l}\,\sum^{\infty}_{k=-\infty}
\frac{m}{\omega_{mk}}f^{}_{lmk}\lvert A^{\infty}_{2lm\omega}\rvert^{2} 
l(l+1)(l-1)(l+2)
\end{equation}
for the angular momentum flux outward.  As defined above, $\dot{E}^{\infty}$ 
and $\dot{L}_{z}^{\infty}$ are only averages, not instantaneous rates of 
change.  Adjusting for differences in notation, the expressions for 
$\dot{E}^{\infty}$ and $\dot{L}_{z}^{\infty}$ 
agree with those derived elsewhere \cite{ashby}, \cite{martel04}, \cite{nagar05}.  
A time domain harmonic gauge expression for $\dot{E}^{\infty}$ is derived in  
\cite{bl05}, in a different manner.

Teukolsky and Press \cite{teukpr74} derived a Kerr metric expression for 
the energy flux inward though the event horizon of the central black 
hole.  Specialized to the Schwarzschild metric, their result is
\begin{equation}
\label{eq:dedtin}
\frac{d^{2}E^{2 M}}{dt\,d\Omega}=\frac{M^{2}}{\pi}
\lvert \sigma^{\text{HH}}\rvert^{2}\;,
\end{equation}
where
\begin{equation}
\label{eq:dedtindef}
\sigma^{\text{HH}}=-\frac{\Psi_{0}^{\text{HH}}}{i\omega+2 \epsilon}
=-\frac{4 M\Psi_{0}^{\text{HH}}}{1+4 i\omega M}\;,
\;\epsilon=\frac{1}{8 M}\;.
\end{equation}
They did not use the Isaacson tensor, but instead derived \eqref{eq:dedtin} 
from the Hawking-Hartle formula for the increase in event horizon area 
due to the ingoing radiation energy and angular momentum 
flux \cite{hawkhart72}.  
We substitute $\Psi_{0}^{\text{HH}}$ from \eqref{eq:psi0hh}-\eqref{eq:psi0hhev} 
into \eqref{eq:dedtin}, integrate over all angles and average over 
time as in \eqref{eq:aveint}, all of which gives
\begin{equation}
\label{eq:edotin}
\dot{E}^{2 M}=\left\langle\frac{dE^{2 M}}{dt}\right\rangle=\frac{1}{16\pi}
\sum^{\infty}_{l=2}\,\sum^{l}_{m=-l}\,\sum^{\infty}_{k=-\infty}f^{}_{lmk}
\lvert A^{2 M}_{2lm\omega}\rvert^{2} l(l+1)(l-1)(l+2)\
\end{equation}
for circular and elliptic orbits.  
The ingoing amplitude $A^{2 M}_{2lm\omega}$ is the source integral 
\eqref{eq:ahorepsi}, with $s=2$.  A more complicated expression for the 
flux in terms of $\Psi_{4}$ can be derived 
from \eqref{eq:dedtin} \cite{teukpr74}.

The expression for $\dot{E}^{2 M}$ has the same form as 
$\dot{E}^{\infty}$ \eqref{eq:edotout}, except that the amplitudes are 
different.  The similarity is due to flux conservation, as 
embodied in the identity \eqref{eq:bin2bout2} \cite{ashby}, \cite{davrt72}, 
\cite{zerp70}.  

For the angular momentum flux through the event horizon, we can 
use \eqref{eq:enlz} again \cite{gkenn02}, \cite{teukpr74}.  Doing so gives
\begin{equation}
\label{eq:lzdotin}
\dot{L}_{z}^{2 M}=\left\langle\frac{dL^{2 M}_{z}}{dt}\right\rangle=\frac{1}{16\pi}
\sum^{\infty}_{l=2}\,\sum^{l}_{m=-l}\,\sum^{\infty}_{k=-\infty}\frac{m}{\omega_{mk}}
f^{}_{lmk}\lvert A^{2 M}_{2lm\omega}\rvert^{2} l(l+1)(l-1)(l+2)\;.
\end{equation}
This is the same form as $\dot{L}_{z}^{\infty}$ \eqref{eq:lzdotout}.  The 
expressions for $\dot{E}^{2 M}$ and $\dot{L}_{z}^{2 M}$ agree with those 
derived elsewhere using the Regge-Wheeler gauge \cite{ashby}.  

The expressions above are the rates of energy and angular momentum 
transport by the waves.  The time averaged rates of energy 
$\left(\left\langle\frac{dE}{dt}\right\rangle\right)$ and angular 
momentum $\left(\left\langle\frac{dL_{z}}{dt}\right\rangle\right)$ lost 
by the orbiting mass are the opposite \cite{gkenn02}, so that
\begin{equation}
\label{eq:edotm0}
\dot{E}=\left\langle\frac{dE}{dt}\right\rangle=
-\left(\left\langle\frac{dE^{\infty}}{dt}\right\rangle
+\left\langle\frac{dE^{2 M}}{dt}\right\rangle\right)\;,
\end{equation}
\begin{equation}
\label{eq:lzdotm0}
\dot{L}_{z}=\left\langle\frac{dL_{z}}{dt}\right\rangle=
-\left(\left\langle\frac{dL^{\infty}_{z}}{dt}\right\rangle
+\left\langle\frac{dL^{2 M}_{z}}{dt}\right\rangle\right)\;.
\end{equation}
As before, $\dot{E}$ and $\dot{L}_{z}$ are only averaged quantities.  

The main results of this chapter are the waveforms $\hp$ \eqref{eq:hthth} 
and $\hc$ \eqref{eq:hthphi} and the bound orbit expressions for 
$\dot{E}^{\infty}$ \eqref{eq:edotout}, $\dot{L}_{z}^{\infty}$ \eqref{eq:lzdotout}, 
$\dot{E}^{2 M}$ \eqref{eq:edotin} and $\dot{L}_{z}^{2 M}$ \eqref{eq:lzdotin}.  
Numerical calculations of some of these quantities for selected orbits are in 
the following chapter.
\chapter{\label{numchap}Numerical Results}

\begin{figure}[htbp]
\begin{center}
\includegraphics[width=5.0in]{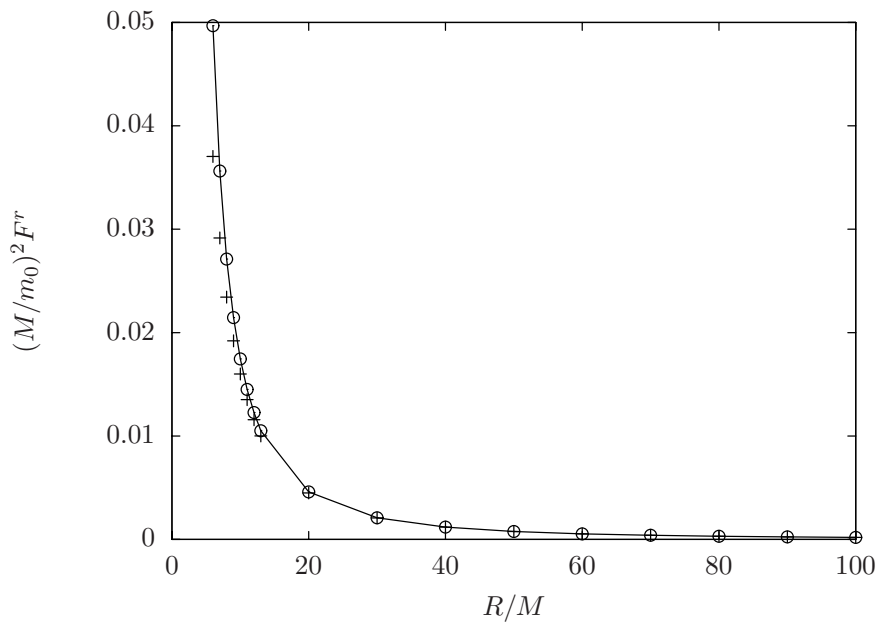}
\caption[Plot of Radial Self-Force for Circular Orbits]{\label{fig:frplot}
Plot of radial self-force for circular orbits.  
Circles are data points from Table~\ref{tab:frtable}.  
Pluses are the approximation $F^{r}\sim \frac{2 m_{0}^{2}}{R^2}
\left(1-\frac{2 M}{R}\right)$.}
\end{center}
\end{figure}

This chapter discusses numerical calculations.  The main result is 
Table \ref{tab:frtable}, which gives the radial component of the self-force
for a variety of circular orbits.  The data for $R\le 100 M$ are plotted 
in Figure \ref{fig:frplot}.  The data points terminate at $R=6 M$, which 
is the innermost stable circular orbit.  

The leading order behavior is $\frac{2 m_{0}^{2}}{R^2}$.  This is characterized 
as the Newtonian self-force by Detweiler and Poisson \cite{dp04}.  It gives 
the shift in orbital angular frequency that occurs because both bodies are 
now moving around the center of mass.  To leading order in $R$, the perturbed 
orbital angular frequency is
\begin{equation}
\label{eq:omegsh}
\Omega^{2}=\frac{M-2 \mz}{R^3}\;,
\end{equation}
which also can be obtained from \eqref{eq:omegcorr}.  Following Detweiler and 
Poisson, we interpret $R$ as the radial coordinate with respect to the center 
of mass.  In terms of the total separation $s$ between $M$ and $\mz$, we have 
\cite{dp04}
\begin{equation}
\label{eq:omegshs}
\Omega^{2}=\frac{M+\mz}{s^3}\;,
\end{equation}
the usual Keplerian form of the frequency.  

\newcolumntype{q}{D{.}{.}{14}}
\begin{table}[htbp]
\caption[Radial Component of Self-Force for Circular Orbits]
{\label{tab:frtable}
Below is a table of the radial component 
of the self-force, $F^{r}$, for circular orbits of radius R.  
For large R, $F^{r}\sim \frac{2 m_{0}^{2}}{R^2}
\left(1-\frac{2 M}{R}\right)$.}
\begin{center}
\begin{tabular}{|r|q||r|q|}
\hline
\multicolumn{1}{|c|}{$ R/M $}& \multicolumn{1}{c||}{$ (M/m_{0})^{2}F^{r}$}&
\multicolumn{1}{c|}{$ R/M $}& \multicolumn{1}{c|}{$ (M/m_{0})^{2}F^{r}$}\\
\hline
\rule[2mm]{0mm}{2mm}
   6     &  4.9685669\times 10^{-2} &   110    &  1.6237973\times 10^{-4}\\
   7     &  3.5624667\times 10^{-2} &   120    &  1.3664172\times 10^{-4}\\
   8     &  2.7112763\times 10^{-2} &   130    &  1.1657168\times 10^{-4}\\
   9     &  2.1452689\times 10^{-2} &   140    &  1.0061965\times 10^{-4}\\
  10     &  1.7454613\times 10^{-2} &   150    &  8.7731466\times 10^{-5}\\
  11     &  1.4507231\times 10^{-2} &   200    &  4.9508804\times 10^{-5}\\
  12     &  1.2263358\times 10^{-2} &   300    &  2.2075818\times 10^{-5}\\
  13     &  1.0511248\times 10^{-2} &   400    &  1.2438053\times 10^{-5}\\
  20     &  4.5872951\times 10^{-3} &   500    &  7.9682266\times 10^{-6}\\
  30     &  2.0912401\times 10^{-3} &   600    &  5.5371464\times 10^{-6}\\
  40     &  1.1929325\times 10^{-3} &   700    &  4.0700299\times 10^{-6}\\
  50     &  7.7022778\times 10^{-4} &   800    &  3.1172221\times 10^{-6}\\
  60     &  5.3811284\times 10^{-4} &   900    &  2.4636705\times 10^{-6}\\
  70     &  3.9708290\times 10^{-4} &  1000    &  1.9960142\times 10^{-6}\\
  80     &  3.0502873\times 10^{-4} &  10000   &  1.9996001\times 10^{-8}\\
  90     &  2.4163987\times 10^{-4} &  100000  &  1.9999600\times 10^{-10}\\
  100    &  1.9614005\times 10^{-4} &  1000000 &  1.9999960\times 10^{-12}\\
\hline
\end{tabular}
\end{center}
\end{table}

\begin{figure}[htbp]
\begin{center}
\includegraphics{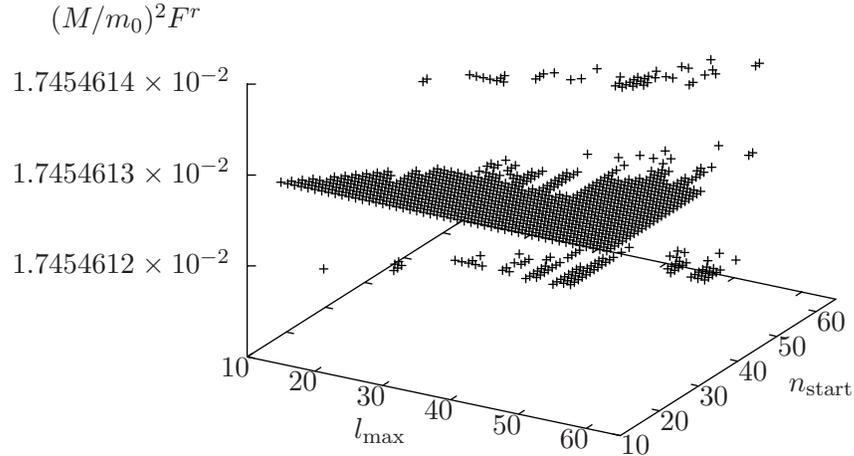}
\caption[Sample Plot of Least Squares Fit for $R=10 M$]{\label{fig:leastsq}
Sample plot of least squares fit for $R=10 M$. The self-force $F^{r}$ on 
the left is the value in Table \ref{tab:frtable}.  A numerical fit is 
calculated for $l$-modes ranging from $n_{\text{start}}$ to 
$l_{\text{max}}$.  The regularized self-force for that fit is the 
mode sum taken from $l=0$ to $l=l_{\text{max}}$. To a high degree of 
precision, the calculated self-force is independent of the particular 
$l$-modes fitted, for a broad range of numerical fits.}
\end{center}
\end{figure}

As discussed in Chapter \ref{eqmochap}, we can accelerate the convergence 
of the self-force regularization using a numerical fit to find the higher 
order regularization parameters.  Figure \ref{fig:leastsq} gives an example 
of this for $R=10 M$, using the LAPACK least squares routine DGELSS 
\cite{laug}.  The plot shows that the calculated self-force is consistent 
for a broad range of numerical fits.  Similar plots could have been 
prepared for the other radii in Table \ref{tab:frtable}, although the 
highest value of $l_{\text{max}}$ was decreased as the orbital radius 
increased.  The higher order terms go through $k=4$ in \eqref{eq:hotlterms}.

For circular orbits, the radial component of the self-force is conservative 
and not dissipative.  The temporal component $F^{t}$ is dissipative and should 
be offset by the energy flux of the gravitational waves \cite{basa07}.  The wave 
energy flux to infinity is $\dot{E}^{\infty}$ \eqref{eq:edotout} and the energy 
flux down the horizon is $\dot{E}^{2 M}$ \eqref{eq:edotin}.  
From equation \eqref{eq:edotsf4}, the rate of energy loss due to the 
gravitational self-force is
\begin{equation}
\label{eq:edotsf}
\dot{E}^{\text{sf}}=\left(1-\frac{2 M}{R}\right)^{2}\frac{F^{t}}{\enbar}\;,
\end{equation}
where ``sf'' indicates that this is calculated using the self-force component $F^{t}$.  
Table~\ref{tab:edotcomp} shows that 
$\dot{E}^{\infty}+\dot{E}^{2 M}+\dot{E}^{\text{sf}}=0$, to good precision.  
An example is $R=10 M$. The energy flux to infinity is 
$(M/\mz)^{2}\dot{E}^{\infty}$=$6.15037255\times 10^{-5}$ and the 
energy flux down the horizon is 
$(M/\mz)^{2}\dot{E}^{2 M}$=$1.25912942\times 10^{-8}$.  The sum is 
$(M/\mz)^{2}\dot{E}$=$6.15163168\times 10^{-5}$.  The self-force 
gives an energy loss:  $(M/\mz)^{2}\dot{E}^{\text{sf}}$=$-6.15163168\times 10^{-5}$, 
which is opposite the wave energy flux.  The offset occurs even though 
$\dot{E}^{\infty}$ and $\dot{E}^{2 M}$ are quadratic in the perturbation, while 
$\dot{E}^{\text{sf}}$ is linear in the perturbation.  

\newcolumntype{w}{D{.}{.}{15}}
\begin{sidewaystable}
\caption[Comparison of Radiation Reaction Self-Force with Gravitational Wave 
Energy Flux]
{\label{tab:edotcomp} Comparison of radiation reaction self-force with gravitational 
wave energy flux.  The columns $\dot{E}^{\infty}$ and $\dot{E}^{2 M}$ give the rates 
at which gravitational waves carry energy to infinity and down the event horizon, 
respectively.  Their sum is $\dot{E}=\dot{E}^{\infty}+\dot{E}^{2 M}$.  The column 
$\dot{E}^{\text{sf}}$ gives the rate of energy loss obtained from the temporal 
component of the gravitational self-force.  The sum of the last two columns is zero.}
\begin{center}
\begin{tabular}{|r|w|w|w|w|}
\hline
\multicolumn{1}{|c|}{$\!R/M\!$} 
& \multicolumn{1}{c|}{\rule[2mm]{0mm}{2.5mm}$(M/\mz)^{2}\dot{E}^{\infty}$}
& \multicolumn{1}{c|}{$(M/\mz)^{2}\dot{E}^{2 M}$}
& \multicolumn{1}{c|}{$(M/\mz)^{2}\dot{E}$} 
& \multicolumn{1}{c|}{$(M/\mz)^{2}\dot{E}^{\text{sf}}$}\\
\hline
\rule[2mm]{0mm}{2mm}
6&9.37270411\times 10^{-4}&3.06894559\times 10^{-6}&9.40339356\times 10^{-4}&-9.40339356\times 10^{-4}\\
7&3.99633989\times 10^{-4}&5.29300869\times 10^{-7}&4.00163290\times 10^{-4}&-4.00163290\times 10^{-4}\\
8&1.95979479\times 10^{-4}&1.25069497\times 10^{-7}&1.96104549\times 10^{-4}&-1.96104549\times 10^{-4}\\
9&1.05896576\times 10^{-4}&3.66762344\times 10^{-8}&1.05933252\times 10^{-4}&-1.05933252\times 10^{-4}\\
10&6.15037255\times 10^{-5}&1.25912942\times 10^{-8}&6.15163168\times 10^{-5}&-6.15163168\times 10^{-5}\\
11&3.77867502\times 10^{-5}&4.87560894\times 10^{-9}&3.77916258\times 10^{-5}&-3.77916258\times 10^{-5}\\
12&2.42896246\times 10^{-5}&2.07631371\times 10^{-9}&2.42917009\times 10^{-5}&-2.42917009\times 10^{-5}\\
13&1.62065198\times 10^{-5}&9.55161446\times 10^{-10}&1.62074749\times 10^{-5}&-1.62074749\times 10^{-5}\\
20&1.87145474\times 10^{-6}&1.61665964\times 10^{-11}&1.87147091\times 10^{-6}&-1.87147091\times 10^{-6}\\
30&2.48647170\times 10^{-7}&3.80318286\times 10^{-13}&2.48647550\times 10^{-7}&-2.48647550\times 10^{-7}\\
40&5.95015183\times 10^{-8}&2.73219859\times 10^{-14}&5.95015456\times 10^{-8}&-5.95015456\times 10^{-8}\\
50&1.96245750\times 10^{-8}&3.57741633\times 10^{-15}&1.96245786\times 10^{-8}&-1.96245786\times 10^{-8}\\
60&7.92644417\times 10^{-9}&6.82440618\times 10^{-16}&7.92644485\times 10^{-9}&-7.92644485\times 10^{-9}\\
70&3.68188111\times 10^{-9}&1.68566659\times 10^{-16}&3.68188127\times 10^{-9}&-3.68188127\times 10^{-9}\\
80&1.89453586\times 10^{-9}&5.02733130\times 10^{-17}&1.89453591\times 10^{-9}&-1.89453591\times 10^{-9}\\
90&1.05411228\times 10^{-9}&1.73092826\times 10^{-17}&1.05411230\times 10^{-9}&-1.05411230\times 10^{-9}\\
100&6.23820341\times 10^{-10}&6.67326986\times 10^{-18}&6.23820347\times 10^{-10}&-6.23820347\times 10^{-10}\\
120&2.51576768\times 10^{-10}&1.28399905\times 10^{-18}&2.51576769\times 10^{-10}&-2.51576769\times 10^{-10}\\
150&8.27445791\times 10^{-11}&1.71112004\times 10^{-19}&8.27445793\times 10^{-11}&-8.27445793\times 10^{-11}\\
\hline
\end{tabular}
\end{center}
\end{sidewaystable}

Fujita and Tagoshi made precise numerical calculations 
of the outgoing energy flux carried by the gravitational waves 
for circular orbits using a different numerical method \cite{fujitag04}.  
Table \ref{tab:en10table} shows good agreement between their results 
and calculations done using the methods described in this thesis, for $R=10 M$.  
\newcolumntype{v}{D{.}{.}{21}}
\begin{table}[htbp]
\caption[Circular Orbit Energy Flux Comparison for $R=10 M$]
{\label{tab:en10table}Circular orbit energy flux comparison for $R=10 M$.  
The column on the right is taken from Table VIII 
of \cite{fujitag04}, rounded to fifteen digits.  The column on the left 
was calculated using the methods described in this thesis.}
\begin{center}
\begin{tabular}{|r|r|v|v|}
\hline
\multicolumn{1}{|c|}{$l $}& \multicolumn{1}{c|}{$ m $}&
\multicolumn{1}{c|}{\rule[2mm]{0mm}{2.5mm}$(M/\mz)^{2}\dot{E}^{\infty}$ Thesis}
& \multicolumn{1}{c|}{$(M/\mz)^{2}\dot{E}^{\infty}$ Fujita and Tagoshi}\\
\hline
\rule[2mm]{0mm}{2mm}
2 & 1 & 1.93160935115669 \times 10^{-7}  & 1.93160935115669 \times 10^{-7}\\
2 & 2 & 5.36879547910210 \times 10^{-5}  & 5.36879547910214 \times 10^{-5}\\
3 & 1 & 5.71489891261480 \times 10^{-10} & 5.71489891261478 \times 10^{-10}\\
3 & 2 & 4.79591646159026 \times 10^{-8}  & 4.79591646159025 \times 10^{-8}\\
3 & 3 & 6.42608275624719 \times 10^{-6}  & 6.42608275624724 \times 10^{-6}\\
4 & 1 & 1.45758564229714 \times 10^{-13} & 1.45758564229713 \times 10^{-13}\\
4 & 2 & 5.26224530895924 \times 10^{-10} & 5.26224530895930 \times 10^{-10}\\
4 & 3 & 8.77875752521502 \times 10^{-9}  & 8.77875752521507 \times 10^{-9}\\
4 & 4 & 9.53960039485201 \times 10^{-7}  & 9.53960039485188 \times 10^{-7}\\
5 & 1 & 2.36763718744954 \times 10^{-16} & 2.36763718744955 \times 10^{-16}\\
5 & 2 & 3.81935323719895 \times 10^{-13} & 3.81935323719893 \times 10^{-13}\\
5 & 3 & 1.82910132522830 \times 10^{-10} & 1.82910132522831 \times 10^{-10}\\
5 & 4 & 1.49211627485282 \times 10^{-9}  & 1.49211627485280 \times 10^{-9}\\
5 & 5 & 1.52415476457987 \times 10^{-7}  & 1.52415476457990 \times 10^{-7}\\
6 & 1 & 3.59779535991180 \times 10^{-20} & 3.59779535991173 \times 10^{-20}\\
6 & 2 & 1.97636895352003 \times 10^{-15} & 1.97636895352005 \times 10^{-15}\\
6 & 3 & 2.12388274763689 \times 10^{-13} & 2.12388274763686 \times 10^{-13}\\
6 & 4 & 4.66333988474111 \times 10^{-11} & 4.66333988474121 \times 10^{-11}\\
6 & 5 & 2.47463869472717 \times 10^{-10} & 2.47463869472724 \times 10^{-10}\\
6 & 6 & 2.51821315681017 \times 10^{-8}  & 2.51821315681016 \times 10^{-8}\\
7 & 1 & 3.29136294915892 \times 10^{-23} & 3.29136294915887 \times 10^{-23}\\
7 & 2 & 9.08415089084877 \times 10^{-19} & 9.08415089084875 \times 10^{-19}\\
7 & 3 & 2.03736275096858 \times 10^{-15} & 2.03736275096860 \times 10^{-15}\\
7 & 4 & 6.99409365020717 \times 10^{-14} & 6.99409365020741 \times 10^{-14}\\
7 & 5 & 1.03409891279350 \times 10^{-11} & 1.03409891279349 \times 10^{-11}\\
7 & 6 & 4.06799480917117 \times 10^{-11} & 4.06799480917109 \times 10^{-11}\\
7 & 7 & 4.23452267128467 \times 10^{-9}  & 4.23452267128478 \times 10^{-9}\\
\hline
\end{tabular}
\end{center}
\end{table}

\begin{figure}[htbp]
\begin{center}
\includegraphics[width=5.0in]{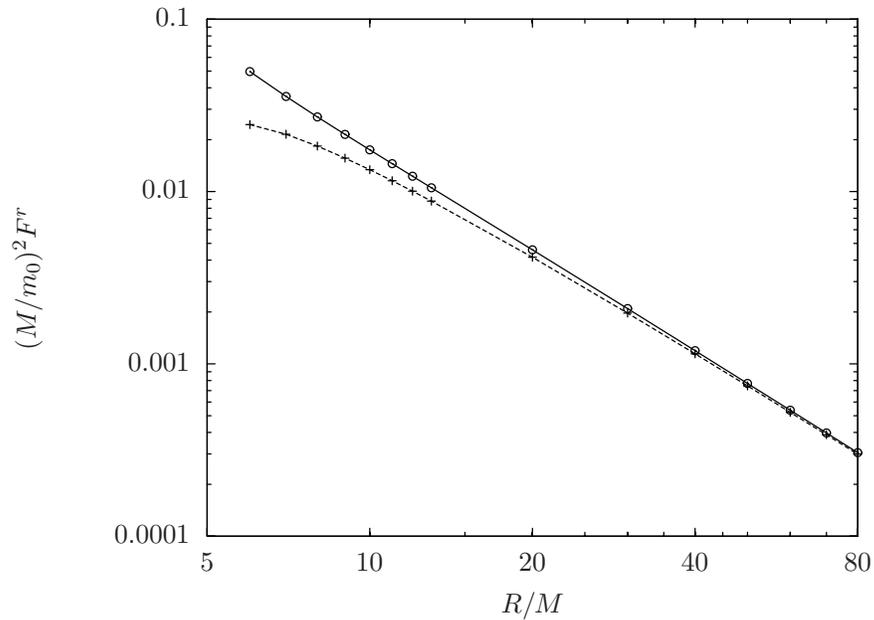}
\caption[Comparison Plot of Radial Self-Force for Circular Orbits]{\label{fig:frplot2}
Comparison plot of radial self-force for circular orbits.  
Circles on the solid line are data points from Table~\ref{tab:frtable}.  
Pluses on the dashed line are data points from Table~\ref{tab:frtabledp}.}
\end{center}
\end{figure}

As discussed in Chapter \ref{evpar}, Detweiler and Poisson derived a different 
solution for the even parity $l=0$ multipole.  Table \ref{tab:frtabledp} 
converts the self-force values in Table \ref{tab:frtable} to their equivalents, 
using equation \eqref{eq:mydpfrdiff}.  Figure \ref{fig:frplot2} is a log-log 
plot comparing the different self-forces for $6 M\le R\le 80 M$.  
The self-force with the Detweiler-Poisson $l=0$ solution curves noticeably.

\begin{table}[htbp]
\caption[Radial Component of Self-Force for Circular Orbits Using 
Detweiler and Poisson Formula]
{\label{tab:frtabledp}
Below is a table of the radial component of the self-force for circular 
orbits, based on the Detweiler-Poisson formula \cite{dp04} for 
the $l=0$ mode of the bare force.  For large R, 
$F^{r}\sim \frac{2 m_{0}^{2}}{R^2}\left(1-\frac{7 M}{2 R}\right)$.  
Only seven digits are given for $R/M=13$ and $R/M=140$, because one 
significant figure is lost due to subtraction in 
going from Table~\ref{tab:frtable} to this table.}

\begin{center}
\begin{tabular}{|r|q||r|q|}
\hline
\multicolumn{1}{|c|}{$ R/M $}& \multicolumn{1}{c||}{$ (M/m_{0})^{2}F^{r}$}&
\multicolumn{1}{c|}{$ R/M $}& \multicolumn{1}{c|}{$ (M/m_{0})^{2}F^{r}$}\\
\hline
\rule[2mm]{0mm}{2mm}
   6     &  2.4466497\times 10^{-2} &   110    &  1.6007306\times 10^{-4}\\
   7     &  2.1499068\times 10^{-2} &   120    &  1.3486847\times 10^{-4}\\
   8     &  1.8357824\times 10^{-2} &   130    &  1.1517927\times 10^{-4}\\
   9     &  1.5637098\times 10^{-2} &   140    &  9.950639\times 10^{-5}\\
  10     &  1.3389470\times 10^{-2} &   150    &  8.6827447\times 10^{-5}\\
  11     &  1.1551745\times 10^{-2} &   200    &  4.9129042\times 10^{-5}\\
  12     &  1.0046239\times 10^{-2} &   300    &  2.1963771\times 10^{-5}\\
  13     &  8.804886\times 10^{-3} &   400    &  1.2390883\times 10^{-5}\\
  20     &  4.1570550\times 10^{-3} &   500    &  7.9441058\times 10^{-6}\\
  30     &  1.9698169\times 10^{-3} &   600    &  5.5231993\times 10^{-6}\\
  40     &  1.1428832\times 10^{-3} &   700    &  4.0612522\times 10^{-6}\\
  50     &  7.4494860\times 10^{-4} &   800    &  3.1113443\times 10^{-6}\\
  60     &  5.2361368\times 10^{-4} &   900    &  2.4595438\times 10^{-6}\\
  70     &  3.8800965\times 10^{-4} &  1000    &  1.9930067\times 10^{-6}\\
  80     &  2.9897883\times 10^{-4} &  10000   &  1.9993000\times 10^{-8}\\
  90     &  2.3740623\times 10^{-4} &  100000  &  1.9999300\times 10^{-10}\\
  100    &  1.9306263\times 10^{-4} &  1000000 &  1.9999930\times 10^{-12}\\
\hline
\end{tabular}
\end{center}
\end{table}

Table \ref{tab:compfrtable} compares the radial self-force in 
Table \ref{tab:frtable} to the results of Barack and Sago, who 
did their calculations in the time domain by solving the field 
equations directly.  They used the Detweiler-Poisson solution.  Once 
this difference is taken into account, there is good agreement.

\newcolumntype{x}{D{.}{.}{11}}
\newcolumntype{y}{D{.}{}{0}}
\newcolumntype{z}{D{.}{.}{13}}
\begin{table}[htbp]
\caption[Radial Component of Self-Force Comparison]
{\label{tab:compfrtable} Column (a) is the radial self-force from 
Table \ref{tab:frtable}. Column (b) is the radial self-force 
from Table IV of \cite{basa07}. Column (c) is the estimated fractional 
error therein, also from \cite{basa07}. Column (d) is the radial 
self-force from Table \ref{tab:frtabledp}.}
\begin{center}
\begin{tabular}{|r|z|x|y|z|}
\hline
\multicolumn{1}{|c|}{$ R/M $\rule[2mm]{0mm}{2mm}}& 
\multicolumn{1}{c|}{$ (M/m_{0})^{2}F^{r}$ (a)}&
\multicolumn{1}{c|}{$ (M/m_{0})^{2}F^{r}$ (b)}&
\multicolumn{1}{c|}{Error (c)}& 
\multicolumn{1}{c|}{$ (M/m_{0})^{2}F^{r}$ (d)}\\
\hline
\rule[2mm]{0mm}{2mm}
  6 & 4.9685669\times 10^{-2} & 2.44661 \times 10^{-2} & 9 \times 10^{-4} & 2.4466497\times 10^{-2}\\
  7 & 3.5624667\times 10^{-2} & 2.14989 \times 10^{-2} & 6 \times 10^{-4} & 2.1499068\times 10^{-2}\\
  8 & 2.7112763\times 10^{-2} & 1.83577 \times 10^{-2} & 5 \times 10^{-4} & 1.8357824\times 10^{-2}\\
  9 & 2.1452689\times 10^{-2} & 1.56369 \times 10^{-2} & 4 \times 10^{-4} & 1.5637098\times 10^{-2}\\
 10 & 1.7454613\times 10^{-2} & 1.33895 \times 10^{-2} & 8 \times 10^{-5} & 1.3389470\times 10^{-2}\\
 11 & 1.4507231\times 10^{-2} & 1.15518 \times 10^{-2} & 6 \times 10^{-5} & 1.1551745\times 10^{-2}\\
 12 & 1.2263358\times 10^{-2} & 1.00463 \times 10^{-2} & 5 \times 10^{-5} & 1.0046239\times 10^{-2}\\
 13 & 1.0511248\times 10^{-2} & 8.80489 \times 10^{-3} & 4 \times 10^{-5} & 8.804886\;\,\times 10^{-3}\\
 20 & 4.5872951\times 10^{-3} & 4.15706 \times 10^{-3} & 1 \times 10^{-5} & 4.1570550\times 10^{-3}\\
 30 & 2.0912401\times 10^{-3} & 1.96982 \times 10^{-3} & 5 \times 10^{-6} & 1.9698169\times 10^{-3}\\
 40 & 1.1929325\times 10^{-3} & 1.14288 \times 10^{-3} & 2 \times 10^{-6} & 1.1428832\times 10^{-3}\\
 50 & 7.7022778\times 10^{-4} & 7.44949 \times 10^{-4} & 1 \times 10^{-6} & 7.4494860\times 10^{-4}\\
 60 & 5.3811284\times 10^{-4} & 5.23613 \times 10^{-4} & 2 \times 10^{-5} & 5.2361368\times 10^{-4}\\
 70 & 3.9708290\times 10^{-4} & 3.88010 \times 10^{-4} & 1 \times 10^{-5} & 3.8800965\times 10^{-4}\\
 80 & 3.0502873\times 10^{-4} & 2.98979 \times 10^{-4} & 8 \times 10^{-6} & 2.9897883\times 10^{-4}\\
 90 & 2.4163987\times 10^{-4} & 2.37406 \times 10^{-4} & 7 \times 10^{-6} & 2.3740623\times 10^{-4}\\
100 & 1.9614005\times 10^{-4} & 1.93063 \times 10^{-4} & 5 \times 10^{-6} & 1.9306263\times 10^{-4}\\
120 & 1.3664172\times 10^{-4} & 1.34868 \times 10^{-4} & 4 \times 10^{-6} & 1.3486847\times 10^{-4}\\
150 & 8.7731466\times 10^{-5} & 8.68274 \times 10^{-5} & 2 \times 10^{-6} & 8.6827447\times 10^{-5}\\
\hline
\end{tabular}
\end{center}
\end{table}

The numerical results of this chapter show that the harmonic gauge solutions 
derived in this thesis can be used to accurately calculate the gravitational 
self-force for circular orbits.  An effort was made to calculate the self-force for 
elliptic orbits.  It is possible to calculate efficiently the waveforms and energy 
flux at infinity and near the event horizon.  From the wave energy flux, the radiation 
reaction in an average sense can be calculated for elliptic orbits, but this approach 
does not give the self-force -- including the conservative part -- as such.  To 
calculate the self-force, we need to evaluate the solutions along the orbit itself.  
Unfortunately, solutions along the orbit have larger oscillations than at infinity and 
the event horizon.  As a result, the sum over the frequency index $k$ converges very 
slowly along an elliptic orbit, except at the turning points of the 
orbit.  For the solutions to be useful for elliptic orbits, it seems 
necessary to regularize the retarded Green's function in the frequency 
domain.  This is left for future work, although numerical calculations 
show that the imaginary part of the Green's function is small and 
finite.  For circular orbits, the self-force components 
$F^{t}$ and $F^{\phi}$ do not require regularization.  Numerical calculations 
show they are due entirely to the imaginary part of the spin $2$ retarded 
Green's functions.  Note that the imaginary part is a homogeneous solution 
of the Green's function equation.  Detweiler and Whiting showed that the 
self-force can be calculated from homogeneous solutions \cite{dw03}.  
However, the imaginary part does not give the conservative component, 
$F^{r}$, which must be due to the real part of the Green's function.  In 
a related area, Gralla and collaborators calculated the dissipative part of 
the scalar self-force for circular orbits using the imaginary part of the scalar 
Green's function, which represents one-half the difference between the retarded and 
advanced solutions  \cite{grallafw05}.  We must leave these issues and their application to 
elliptic orbits for future work.
\chapter{\label{concchap}Conclusion}

The main research result of this thesis consists of the harmonic gauge 
solutions derived in Chapters \ref{oddpar} and \ref{evpar}, using 
separation of variables and Fourier transforms.  The solutions 
are written in terms of six functions of various spin weights, which 
satisfy decoupled ordinary differential equations.  For odd parity, 
the solutions are given in terms of two generalized Regge-Wheeler 
functions, one with $s=2$ and one with $s=1$.  The even parity solutions 
contain the remaining four functions: three generalized Regge-Wheeler 
functions (two with $s=0$ and one with $s=1$) and the Zerilli function, 
which is related to the spin $2$ Regge-Wheeler function.  
The spin $2$ functions are gauge invariant and therefore physically 
meaningful.  Gauge changes which preserve the harmonic gauge are 
implemented by adding homogeneous spin 1 and spin 0 solutions.

Chapter \ref{eqmochap} discusses the background equations of motion and 
shows how the harmonic gauge solutions can be applied to calculate the 
gravitational self-force, which gives the first order perturbative corrections to the 
equations of motion for a small mass orbiting a much larger black hole.  
Chapter \ref{tmunuchap} provides Fourier transforms for 
the stress energy tensor.  Chapter \ref{rweqnchap} shows how to solve 
the generalized Regge-Wheeler and Zerilli equations.  Chapter \ref{radchap} 
explains how to obtain expressions for gravitational waveforms and energy 
flux from the solutions derived in Chapters \ref{oddpar} and \ref{evpar}.

The harmonic gauge solutions yield accurate calculations of the gravitational 
self-force for circular orbits, as demonstrated in Chapter \ref{numchap}.  
However, there are open issues.  The $l=0$ solution 
for circular orbits conflicts with the published Detweiler-Poisson solution 
for that multipole.  As discussed at the end of Chapter \ref{eqmochap}, 
the discrepancy does not appear to affect gauge invariant observables, but 
nevertheless should be resolved.  Another issue is how to calculate the 
gravitational self-force for elliptic orbits, and the problems here are 
briefly explained at the end of Chapter \ref{numchap}.  
Even if the harmonic gauge solutions derived in this thesis are not 
practical for additional numerical work, they still would be useful for 
analytic approximations, which in turn could further our understanding of 
the gravitational self-force.

\bibliographystyle{plain}

\nocite{*}

\bibliography{refs}

\begin{thebibliography}{100}

\bibitem{hmf}
M.~Abramowitz and I.~Stegun, editors.
\newblock {\em Handbook of Mathematical Functions}.
\newblock Dover Publications, New York, 1972.

\bibitem{laug}
E.~Anderson, Z.~Bai, C.~Bischof, S.~Blackford, J.~Demmel, J.~Dongarra,
  J.~Du~Croz, A.~Greenbaum, S.~Hammarling, A.~McKenney, and D.~Sorensen.
\newblock {\em {LAPACK} Users' Guide}.
\newblock Society for Industrial and Applied Mathematics, Philadelphia, third
  edition, 1999.

\bibitem{arfken85}
G.~Arfken.
\newblock {\em Mathematical Methods for Physicists}.
\newblock Academic Press, New York, third edition, 1985.

\bibitem{ashby}
N.~Ashby.
\newblock Personal communication, as described at the end of the {I}ntroduction
  to this thesis.

\bibitem{ashby86}
N.~Ashby.
\newblock Planetary perturbation equations based on relativistic {K}eplerian
  motion.
\newblock In J.~Kovalevsky and V.~Brumberg, editors, {\em Relativity in
  Celestial Mechanics and Astrometry}, page~41, Leningrad, 1985. IAU Symposium
  No. 114, Reidel, Netherlands, 1986.

\bibitem{bl02}
L.~Barack and C.~Lousto.
\newblock Computing the gravitational self-force on a compact object plunging
  into a {S}chwarzschild black hole.
\newblock {\em Phys. Rev. D}, 66:061502(R), 2002.

\bibitem{bl05}
L.~Barack and C.~Lousto.
\newblock Perturbations of {S}chwarzschild black holes in the {L}orenz gauge:
  Formulation and numerical implementation.
\newblock {\em Phys. Rev. D}, 72:104026, 2005.

\bibitem{bmnos02}
L.~Barack, Y.~Mino, H.~Nakano, A.~Ori, and M.~Sasaki.
\newblock Calculating the gravitational self-force in {S}chwarzschild
  spacetime.
\newblock {\em Phys. Rev. Lett.}, 88:091101, 2002.

\bibitem{bo00}
L.~Barack and A.~Ori.
\newblock Mode sum regularization approach for the self-force in black hole
  spacetime.
\newblock {\em Phys. Rev. D}, 61:061502(R), 2000.

\bibitem{bo01}
L.~Barack and A.~Ori.
\newblock Gravitational self-force and gauge transformations.
\newblock {\em Phys. Rev. D}, 64:124003, 2001.

\bibitem{bo02s}
L.~Barack and A.~Ori.
\newblock Regularization parameters for the self-force in {S}chwarzschild
  spacetime: {S}calar case.
\newblock {\em Phys. Rev. D}, 66:084022, 2002.

\bibitem{bo02g}
L.~Barack and A.~Ori.
\newblock Regularization parameters for the self-force in {S}chwarzschild
  spacetime. {II}. {G}ravitational case.
\newblock {\em Phys. Rev. D}, 67:024029, 2003.

\bibitem{basa07}
L.~Barack and N.~Sago.
\newblock Gravitational self-force on a particle in circular orbit around a
  {S}chwarzschild black hole.
\newblock {\em Phys. Rev. D}, 75:064021, 2007.

\bibitem{bek73}
J.~Bekenstein.
\newblock Extraction of energy and charge from a black hole.
\newblock {\em Phys. Rev. D}, 7:949, 1973.

\bibitem{biww}
L.~Blanchet, B.~Iyer, C.~Will, and A.~Wiseman.
\newblock Gravitational waveforms from inspiralling compact binaries to
  second-post-{N}ewtonian order.
\newblock {\em Class. Quantum Grav.}, 13:575, 1996.

\bibitem{brscal73}
R.~Breuer, P.~Chrzanowski, H.~Hughes, and C.~Misner.
\newblock Geodesic synchroton radiation.
\newblock {\em Phys. Rev. D}, 8:4309, 1973.

\bibitem{brownetc07}
D.~Brown, S.~Fairhurst, B.~Krishnan, R.~Mercer, R.~Kopparapu, L.~Santamaria,
  and J.~Whelan.
\newblock Data formats for numerical relativity waves, September 2007.
\newblock gr-qc/0709.0093v1.

\bibitem{bcp07}
A.~Buonanno, G.~Cook, and F.~Pretorius.
\newblock Inspiral, merger, and ring-down of equal-mass black-hole binaries.
\newblock {\em Phys. Rev. D}, 75:124018, 2007.

\bibitem{bucent07}
A.~Buonanno, Y.~Pan, J.~Baker, J.~Centrella, B.~Kelly, S.~McWilliams, and
  J.~van Meter.
\newblock Toward faithful templates for non-spinning binary black holes using
  the effective-one-body approach, June 2007.
\newblock gr-qc/0706.3732v1.

\bibitem{burko00}
L.~Burko.
\newblock Self-force on a particle in orbit around a black hole.
\newblock {\em Phys. Rev. Lett.}, 84:4529, 2000.

\bibitem{carl79}
B.~Carlson.
\newblock Computing elliptic integrals by duplication.
\newblock {\em Numer. Math.}, 33:1, 1979.

\bibitem{chandra92}
S.~Chandrasekhar.
\newblock {\em The Mathematical Theory of Black Holes}.
\newblock Oxford University Press, New York, 1992.

\bibitem{ching95}
E.~Ching, P.~Leung, W.~Suen, and K.~Young.
\newblock Wave propagation in gravitational systems: late time behavior.
\newblock {\em Phys. Rev. D}, 52:2118, 1995.

\bibitem{chrz75}
P.~Chrzanowski.
\newblock Vector potential and metric perturbations of a rotating black hole.
\newblock {\em Phys. Rev. D}, 11:2042, 1975.

\bibitem{chrzmi74}
P.~Chrzanowski and C.~Misner.
\newblock Geodesic synchrotron radiation in the {K}err geometry by the method
  of asymptotically factorized {G}reen's functions.
\newblock {\em Phys. Rev. D}, 10:1701, 1974.

\bibitem{cmp78}
C.~Cunningham, R.~Price, and V.~Moncrief.
\newblock Radiation from collapsing relativistic stars. {I}. {L}inearized
  odd-parity radiation.
\newblock {\em Astrophys. J.}, 224:643, 1978.

\bibitem{cufps93}
C.~Cutler, L.~Finn, E.~Poisson, and G.~Sussman.
\newblock Gravitational radiation from a particle in circular orbit around a
  black hole. {II}. {N}umerical results for the nonrotating case.
\newblock {\em Phys. Rev. D}, 47:1511, 1993.

\bibitem{cut94}
C.~Cutler, D.~Kennefick, and E.~Poisson.
\newblock Gravitational radiation reaction for bound motion around a
  {S}chwarzschild black hole.
\newblock {\em Phys. Rev. D}, 50:3816, 1994.

\bibitem{darwin59}
C.~Darwin.
\newblock The gravity field of a particle.
\newblock {\em Proc. R. Soc. A}, 249:180, 1959.

\bibitem{darwin61}
C.~Darwin.
\newblock The gravity field of a particle. {II}.
\newblock {\em Proc. R. Soc. A}, 263:39, 1961.

\bibitem{davrt72}
M.~Davis, R.~Ruffini, and J.~Tiomno.
\newblock Pulses of gravitational radiation of a particle falling radially into
  a {S}chwarzschild black hole.
\newblock {\em Phys. Rev. D}, 5:2932, 1972.

\bibitem{det05}
S.~Detweiler.
\newblock Perspective on gravitational self-force analyses.
\newblock {\em Class. Quantum Grav.}, 22:S681, 2005.

\bibitem{dmw03}
S.~Detweiler, E.~Messaritaki, and B.~Whiting.
\newblock Self-force of a scalar field for circular orbits about a
  {S}chwarzschild black hole.
\newblock {\em Phys. Rev. D}, 67:104016, 2003.

\bibitem{dp04}
S.~Detweiler and E.~Poisson.
\newblock Low multipole contributions to the gravitational self-force.
\newblock {\em Phys. Rev. D}, 69:084019, 2004.

\bibitem{dw03}
S.~Detweiler and B.~Whiting.
\newblock Self-force via a {G}reen's function decomposition.
\newblock {\em Phys. Rev. D}, 67:024025, 2003.

\bibitem{dewittbr60}
B.~DeWitt and R.~Brehme.
\newblock Radiation damping in a gravitational field.
\newblock {\em Ann. of Phys. (N.Y.)}, 9:220, 1960.

\bibitem{drmwd04}
L.~Diaz-Rivera, E.~Messaritaki, B.~Whiting, and S.~Detweiler.
\newblock Scalar field self-force effects on orbits about a {S}chwarzschild
  black hole.
\newblock {\em Phys. Rev. D}, 70:124018, 2004.

\bibitem{drfhs05}
S.~Drasco, \'E. Flanagan, and S.~Hughes.
\newblock Computing inspirals in {K}err in the adiabatic regime: {I}. {T}he
  scalar case.
\newblock {\em Class. Quantum Grav.}, 22:S801, 2005.

\bibitem{drhs06}
S.~Drasco and S.~Hughes.
\newblock Gravitational wave snapshots of generic extreme mass ratio inspirals.
\newblock {\em Phys. Rev. D}, 73:024027, 2006.

\bibitem{ell731}
D.~Eardley, D.~Lee, and A.~Lightman.
\newblock Gravitational-wave observations as a tool for testing relativistic
  gravity.
\newblock {\em Phys. Rev. D}, 8:3308, 1973.

\bibitem{ell732}
D.~Eardley, D.~Lee, A.~Lightman, R.~Wagoner, and C.~Will.
\newblock Gravitational-wave observations as a tool for testing relativistic
  gravity.
\newblock {\em Phys. Rev. Lett.}, 30:884, 1973.

\bibitem{fujitag04}
R.~Fujita and H.~Tagoshi.
\newblock New numerical methods to evaluate homogeneous solutions of the
  {T}eukolsky equation.
\newblock {\em Prog. Theor. Phys.}, 112:415, 2004.

\bibitem{galtsov82}
D.~Gal'tsov.
\newblock Radiation reaction in the {K}err gravitational field.
\newblock {\em J. Phys. A}, 15:3737, 1982.

\bibitem{gs79}
U.~Gerlach and U.~Sengupta.
\newblock Gauge-invariant perturbations on most general spherically symmetric
  space-times.
\newblock {\em Phys. Rev. D}, 19:2268, 1979.

\bibitem{gkenn02}
K.~Glampedakis and D.~Kennefick.
\newblock Zoom and whirl: Eccentric equatorial orbits around spinning black
  holes and their evolution under gravitational radiation reaction.
\newblock {\em Phys. Rev. D}, 66:044002, 2002.

\bibitem{gj67}
J.~Goldberg.
\newblock Invariant transformations and {N}ewman-{P}enrose constants.
\newblock {\em J. Math. Phys.}, 8:2161, 1967.

\bibitem{gn67}
J.~Goldberg, A.~Macfarlane, E.~Newman, F.~Rohrlich, and E.~Sudarshan.
\newblock Spin-$s$ spherical harmonics and $\eth$.
\newblock {\em J. Math. Phys.}, 8:2155, 1967.

\bibitem{grallafw05}
S.~Gralla, J.~Friedman, and A.~Wiseman.
\newblock Numerical radiation reaction for a scalar charge in {K}err circular
  orbit, February 2005.
\newblock gr-qc/0502123v1.

\bibitem{griffiths}
D.~Griffiths.
\newblock {\em Introduction to Quantum Mechanics}.
\newblock Prentice-Hall, New Jersey, 1995.

\bibitem{hampc}
A.~Hamilton.
\newblock Perturbation theory of spherically symmetric self-similar black
  holes, November 2007.
\newblock gr-qc/0706.3238v2 (personal communication).

\bibitem{hartle03}
J.~Hartle.
\newblock {\em Gravity}.
\newblock Addison-Wesley, San Francisco, 2003.

\bibitem{hawkhart72}
S.~Hawking and J.~Hartle.
\newblock Energy and angular momentum flow into a black hole.
\newblock {\em Commun. Math. Phys.}, 27:283, 1972.

\bibitem{hns05}
W.~Hikida, H.~Nakano, and M.~Sasaki.
\newblock Self-force regularization in the {S}chwarz\-schild spacetime.
\newblock {\em Class. Quantum Grav.}, 22:S753, 2005.

\bibitem{hobbs68}
J.~Hobbs.
\newblock A vierbein formalism of radiation damping.
\newblock {\em Ann. of Phys. (N.Y.)}, 47:141, 1968.

\bibitem{hs00}
S.~Hughes.
\newblock Computing radiation from {K}err black holes: {G}eneralization of the
  {S}asaki-{N}akamura equation.
\newblock {\em Phys. Rev. D}, 62:044029, 2000.
\newblock \emph{ibid.} 67:089902(E), 2003 (Erratum).

\bibitem{hsk00}
S.~Hughes.
\newblock Evolution of circular, nonequatorial orbits of {K}err black holes due
  to gravitational-wave emission.
\newblock {\em Phys. Rev. D}, 61:084004, 2000.
\newblock \emph{ibid.} 63:049902(E), 2001, \emph{ibid.} 65:069902(E), 2002,
  \emph{ibid.} 67:089901(E), 2003 (Errata).

\bibitem{hsdrff05}
S.~Hughes, S.~Drasco, \'E. Flanagan, and J.~Franklin.
\newblock Gravitational radiation reaction and inspiral waveforms in the
  adiabatic limit.
\newblock {\em Phys. Rev. Lett.}, 94:221101, 2005.

\bibitem{isaac167}
R.~Isaacson.
\newblock Gravitational radiation in the limit of high frequency. {I}. {T}he
  linear approximation and geometrical optics.
\newblock {\em Phys. Rev.}, 166:1263, 1968.

\bibitem{isaac267}
R.~Isaacson.
\newblock Gravitational radiation in the limit of high frequency. {II}.
  {N}onlinear terms and the effective stress tensor.
\newblock {\em Phys. Rev.}, 166:1272, 1968.

\bibitem{jack75}
J.~Jackson.
\newblock {\em Classical Electrodynamics}.
\newblock John Wiley, New York, second edition, 1975.

\bibitem{jt03}
S.~Jhingan and T.~Tanaka.
\newblock Improvement on the metric reconstruction scheme in the
  {R}egge-{W}heeler-{Z}erilli formalism.
\newblock {\em Phys. Rev. D}, 67:104018, 2003.

\bibitem{leavm86}
E.~Leaver.
\newblock Solutions to a generalized spheroidal wave equation.
\newblock {\em J. Math. Phys.}, 27:1238, 1986.

\bibitem{leav86}
E.~Leaver.
\newblock Spectral decomposition of the perturbation response of the
  {S}chwarz\-schild geometry.
\newblock {\em Phys. Rev. D}, 34:384, 1986.
\newblock \emph{ibid.} 38:725, 1988 (Erratum).

\bibitem{leopois97}
S.~Leonard and E.~Poisson.
\newblock Radiative multipole moments of integer-spin fields in curved
  spacetime.
\newblock {\em Phys. Rev. D}, 56:4789, 1997.

\bibitem{martel04}
K.~Martel.
\newblock Gravitational waveforms from a point particle orbiting a
  {S}chwarzschild black hole.
\newblock {\em Phys. Rev. D}, 69:044025, 2004.

\bibitem{martpois05}
K.~Martel and E.~Poisson.
\newblock Gravitational perturbations of the {S}chwarzschild spacetime: {A}
  practical covariant and gauge-invariant formalism.
\newblock {\em Phys. Rev. D}, 71:104003, 2005.

\bibitem{mandw}
J.~Mathews and R.~Walker.
\newblock {\em Mathematical Methods of Physics}.
\newblock Addison-Wesley, New York, second edition, 1970.

\bibitem{mns02}
Y.~Mino, H.~Nakano, and M.~Sasaki.
\newblock Covariant self-force regularization of a particle orbiting a
  {S}chwarzschild black hole.
\newblock {\em Prog. Theor. Phys.}, 108:1039, 2002.

\bibitem{progsupp97}
Y.~Mino, M.~Sasaki, M.~Shibata, H.~Tagoshi, and T.~Tanaka.
\newblock Black hole perturbation.
\newblock {\em Prog. Theor. Phys. Suppl.}, 128:1, 1997.

\bibitem{mst97}
Y.~Mino, M.~Sasaki, and T.~Tanaka.
\newblock Gravitational radiation reaction to a particle motion.
\newblock {\em Phys. Rev. D}, 55:3457, 1997.

\bibitem{mtw73}
C.~Misner, K.~Thorne, and J.~Wheeler.
\newblock {\em Gravitation}.
\newblock W. H. Freeman, New York, 1973.

\bibitem{monc74a}
V.~Moncrief.
\newblock Gravitational perturbations of spherically symmetric systems. {I}.
  {T}he exterior problem.
\newblock {\em Ann. of Phys. (N.Y.)}, 88:323, 1974.

\bibitem{monc74d}
V.~Moncrief.
\newblock Odd-parity stability of a {R}eissner-{N}ordstr\"om black hole.
\newblock {\em Phys. Rev. D}, 9:2707, 1974.

\bibitem{monc74e}
V.~Moncrief.
\newblock Stability of {R}eissner-{N}ordstr\"om black holes.
\newblock {\em Phys. Rev. D}, 10:1057, 1974.

\bibitem{nagar05}
A.~Nagar and L.~Rezzolla.
\newblock Gauge-invariant non-spherical metric perturbations of {S}chwarzschild
  black-hole spacetimes.
\newblock {\em Class. Quantum Grav.}, 22:R167, 2005.
\newblock \emph{ibid.} 23:4297, 2006 (Corrigendum).

\bibitem{naktag94}
T.~Nakamura and H.~Tagoshi.
\newblock Gravitational waves from a point particle in circular orbit around a
  black hole: {L}ogarithmic terms in the post-{N}ewtonian expansion.
\newblock {\em Phys. Rev. D}, 49:4016, 1994.

\bibitem{naka03}
H.~Nakano, N.~Sago, and M.~Sasaki.
\newblock Gauge problem in the gravitational self-force: First post-{N}ewtonian
  force in the {R}egge-{W}heeler gauge.
\newblock {\em Phys. Rev. D}, 68:124003, 2003.

\bibitem{np62}
E.~Newman and R.~Penrose.
\newblock An approach to gravitational radiation by a method of spin
  coefficients.
\newblock {\em J. Math. Phys.}, 3:566, 1962.
\newblock \emph{ibid.} 4:998, 1963 (Errata).

\bibitem{np66}
E.~Newman and R.~Penrose.
\newblock Note on the {B}ondi-{M}etzner-{S}achs group.
\newblock {\em J. Math. Phys.}, 7:863, 1966.

\bibitem{lisamiss}
LISA Mission~Science Office.
\newblock {\em LISA: Probing the Universe with Gravitational Waves}.
\newblock LISA Mission Science Office, 1.0 edition, 2007.

\bibitem{ori04}
A.~Ori.
\newblock Harmonic-gauge dipole metric perturbations for weak-field circular
  orbits in {S}chwarzschild spacetime.
\newblock {\em Phys. Rev. D}, 70:124027, 2004.

\bibitem{peters64}
P.~Peters.
\newblock Gravitational radiation and the motion of two point masses.
\newblock {\em Phys. Rev.}, 136:B1224, 1964.

\bibitem{peters63}
P.~Peters and J.~Mathews.
\newblock Gravitational radiation from point masses in a {K}eplerian orbit.
\newblock {\em Phys. Rev.}, 131:435, 1963.

\bibitem{pois93}
E.~Poisson.
\newblock Gravitational radiation from a particle in circular orbit around a
  black hole. {I}. {A}nalytical results for the nonrotating case.
\newblock {\em Phys. Rev. D}, 47:1497, 1993.

\bibitem{pois04}
E.~Poisson.
\newblock Absorption of mass and angular momentum by a black hole:
  {T}ime-domain formalisms for gravitational perturbations, and the small-hole
  or slow-motion approximation.
\newblock {\em Phys. Rev. D}, 70:084044, 2004.

\bibitem{poisslrr}
E.~Poisson.
\newblock The motion of point particles in curved spacetime.
\newblock {\em Living Rev. Relativity}, 7(6), May 2004.
\newblock Cited December 9, 2006. \urlstyle{same}
  \url{http://www.livingreviews.org/lrr-2004-6}.

\bibitem{poisrt04}
E.~Poisson.
\newblock {\em A Relativist's Toolkit}.
\newblock Cambridge University Press, New York, 2004.

\bibitem{odebook}
A.~Polyanin and V.~Zaitsev.
\newblock {\em Handbook of Exact Solutions for Ordinary Differential
  Equations}.
\newblock CRC Press, New York, second edition, 2003.

\bibitem{pound207}
A.~Pound and E.~Poisson.
\newblock Multi-scale analysis of the electromagnetic self-force in a weak
  gravitational field, August 2007.
\newblock gr-qc/0708.3037v1.

\bibitem{pound107}
A.~Pound and E.~Poisson.
\newblock Osculating orbits in {S}chwarzschild spacetime, with an application
  to extreme mass-ratio inspirals, August 2007.
\newblock gr-qc/0708.3033v1.

\bibitem{pound05}
A.~Pound, E.~Poisson, and B.~Nickel.
\newblock Limitations of the adiabatic approximation to the gravitational
  self-force.
\newblock {\em Phys. Rev. D}, 72:124001, 2005.

\bibitem{teukpr73}
W.~Press and S.~Teukolsky.
\newblock Perturbations of a rotating black hole. {II}. {D}ynamical stability
  of the {K}err metric.
\newblock {\em Astrophys. J.}, 185:649, 1973.

\bibitem{teukpr74}
W.~Press and S.~Teukolsky.
\newblock Perturbations of a rotating black hole. {III}. {I}nteraction of the
  hole with gravitational and electromagnetic radiation.
\newblock {\em Astrophys. J.}, 193:443, 1974.

\bibitem{numr}
W.~Press, S.~Teukolsky, W.~Vetterling, and B.~Flannery.
\newblock {\em Numerical Recipes in Fortran 77}.
\newblock Cambridge University Press, New York, second edition, 1992.

\bibitem{prth69}
R.~Price and K.~Thorne.
\newblock Non-radial pulsation of general-relativistic stellar models. {II}.
  {P}roperties of the gravitational waves.
\newblock {\em Astrophys. J.}, 155:163, 1969.

\bibitem{qwald97}
T.~Quinn and R.~Wald.
\newblock Axiomatic approach to electromagnetic and gravitational radiation
  reaction of particles in curved spacetime.
\newblock {\em Phys. Rev. D}, 56:3381, 1997.

\bibitem{qwald99}
T.~Quinn and R.~Wald.
\newblock Energy conservation for point particles undergoing radiation
  reaction.
\newblock {\em Phys. Rev. D}, 60:064009, 1999.

\bibitem{rw57}
T.~Regge and J.~Wheeler.
\newblock Stability of a {S}chwarzschild singularity.
\newblock {\em Phys. Rev.}, 108:1063, 1957.

\bibitem{rose05}
E.~Rosenthal.
\newblock Regularization of the second-order gravitational perturbations
  produced by a compact object.
\newblock {\em Phys. Rev. D}, 72:121503(R), 2005.

\bibitem{rose062}
E.~Rosenthal.
\newblock Construction of the second-order gravitational perturbations produced
  by a compact object.
\newblock {\em Phys. Rev. D}, 73:044034, 2006.

\bibitem{rose06}
E.~Rosenthal.
\newblock Second-order gravitational self-force.
\newblock {\em Phys. Rev. D}, 74:084018, 2006.

\bibitem{sago03}
N.~Sago, H.~Nakano, and M.~Sasaki.
\newblock Gauge problem in the gravitational self-force: Harmonic gauge
  approach in the {S}chwarzschild background.
\newblock {\em Phys. Rev. D}, 67:104017, 2003.

\bibitem{sakurai}
J.~Sakurai.
\newblock {\em Modern Quantum Mechanics}.
\newblock Addison-Wesley, New York, revised edition, 1994.

\bibitem{schutz90}
B.~Schutz.
\newblock {\em A First Course in General Relativity}.
\newblock Cambridge University Press, New York, 1990.

\bibitem{tsstn93}
T.~Tanaka, M.~Shibata, M.~Sasaki, H.~Tagoshi, and T.~Nakamura.
\newblock Gravitational wave induced by a particle orbiting around
  {S}chwarzschild black hole.
\newblock {\em Prog. Theor. Phys.}, 90:65, 1993.

\bibitem{teuk73}
S.~Teukolsky.
\newblock Perturbations of a rotating black hole. {I}. {F}undamental equations
  for gravitational, electromagnetic and neutrino-field perturbations.
\newblock {\em Astrophys. J.}, 185:635, 1973.

\bibitem{th80}
K.~Thorne.
\newblock Multipole expansions of gravitational radiation.
\newblock {\em Rev. Mod. Phys.}, 52:299, 1980.

\bibitem{vish70}
C.~Vishveshwara.
\newblock Stability of the {S}chwarzschild metric.
\newblock {\em Phys. Rev. D}, 1:2870, 1970.

\bibitem{wald73}
R.~Wald.
\newblock On perturbations of a {K}err black hole.
\newblock {\em J. Math. Phys.}, 14:1453, 1973.

\bibitem{wein72}
S.~Weinberg.
\newblock {\em Gravitation and Cosmology}.
\newblock John Wiley, New York, 1972.

\bibitem{wh55}
J.~Wheeler.
\newblock Geons.
\newblock {\em Phys. Rev.}, 97:511, 1955.

\bibitem{wolfbk}
S.~Wolfram.
\newblock {\em The Mathematica Book}.
\newblock Wolfram Research, 2005.
\newblock From the help browser for \textit{Mathematica}, version 5.2.

\bibitem{zare88}
R.~Zare.
\newblock {\em Angular Momentum}.
\newblock John Wiley, New York, 1988.

\bibitem{zert70}
F.~Zerilli.
\newblock {\em The Gravitational Field of a Particle Falling in a Schwarzschild
  Geometry Analyzed in Tensor Harmonics}.
\newblock PhD thesis, Princeton University, 1969.

\bibitem{zerp70}
F.~Zerilli.
\newblock Gravitational field of a particle falling in a {S}chwarzschild
  geometry analyzed in tensor harmonics.
\newblock {\em Phys. Rev. D}, 2:2141, 1970.

\bibitem{zerm70}
F.~Zerilli.
\newblock Tensor harmonics in canonical form for gravitational radiation and
  other applications.
\newblock {\em J. Math. Phys.}, 11:2203, 1970.

\bibitem{zill93}
D.~Zill.
\newblock {\em A First Course in Differential Equations}.
\newblock PWS-KENT Publishing Company, Boston, fifth edition, 1993.

\end{thebibliography}

\appendix

\chapter{\label{appa}Non-Zero Frequency Even Parity Solutions 
for $l\ge 2$}

Listed below are the non-zero frequency even parity solutions and 
their radial derivatives, for $l\ge 2$.  The derivation of these 
solutions is covered in subsection \ref{sec:nzevparge}.

\begin{equation}
\begin{split}\bhz=&-\frac{\lambda (1+\lambda) M
 (-3 M+(3+\lambda) r) \dptw}{3 (\iom)^2 r^3 (3 M+\lambda r)}+
\frac{(-M+r) \pz}{(2 M-r) r}\\&+\frac{2 \left(-2 M^2+M r+(\iom)^2 
r^4\right) \pzbf}{(2 M-r) r^4}+\frac{4 \iom (1+\lambda)\po}{2 M-r}
\\&-\frac{\lambda (1+\lambda)}{3 (\iom)^2 (2 M-r) r^4 (3 M+\lambda r)^2}
\left[18 M^4+3 (-3+4 \lambda) M^3 r+(\iom)^2 \lambda^2 r^6\right.\\&
\left.-3 \lambda M r^3 \left(1+\lambda-2 (\iom)^2
r^2\right)+M^2 \left(6 \lambda^2 r^2+9 (\iom)^2 r^4\right)\right] \ptw
\\&-\frac{8 \pi}{(\iom)^3 (2 M-r) r (3 M+\lambda r)^2}
\left[-2 (\iom)^2 \lambda^2 r^5+M^3 \left(9+4 \lambda+4 \lambda^2\right.\right.
\\&\left.+36 (\iom)^2 r^2\right)+2 M^2 r \left(\lambda-2 \lambda^2
-9 (\iom)^2 r^2+12 (\iom)^2 \lambda r^2\right)+\lambda M r^2
\\&\left.\times\left(1\!-\!12 (\iom)^2 r^2\!+\!\lambda\left(2\!+\!4 (\iom)^2
 r^2\right)\right)\right]Se_{01}+\frac{16 (1+\lambda) M \pi}{(\iom)^3 
(2 M\!-\!r) r^3 (3 M\!+\!\lambda r)^2}
\\&\times\left[48 M^3+15 (-1+2 \lambda) M^2 r
+\lambda (-7+6 \lambda) M r^2-\lambda (1+2 \lambda) r^3\right]Se_{02}
\\&-\frac{8 \pi \left((1-2 \lambda) M^2+2 (\iom)^2 \lambda r^4+M 
\left(r+2 \lambda r+6 (\iom)^2 r^3\right)\right) Se_{11}}{(\iom)^2
r (3 M+\lambda r)}
\\&+\frac{16 (1+\lambda) M \pi (8 M+(-1+2 \lambda) r) Se_{12}}
{(\iom)^2 r^2 (3 M+\lambda r)}+\dpz
\\&+\frac{2 M \dpzbf}{r^3}+\frac{4 (1+\lambda) M \dpo}{\iom r^3}
\end{split}
\end{equation}

\begin{equation}
\begin{split}\bho=&-\frac{\lambda(1+\lambda)\dptw}
{3 \iom r}-\frac{\left(6 M+4 \lambda M-3 r-2\lambda r-2(\iom)^2 r^3\right)\pz}
{4 \iom M r^2-2 \iom r^3}+\frac{(2+2 \lambda) \po}{2 M r-r^2}
\\&+\frac{2 \iom (-M+r)\pzbf}{(2 M-r) r^2}-\frac{\lambda (1+\lambda)
 \left(3 M^2+3 \lambda M r-\lambda r^2\right) \ptw}
{3\iom (2 M-r) r^2 (3 M+\lambda r)}
\\&+\frac{8 \pi \left(-(11+2 \lambda) M^2+2 (\iom)^2 \lambda r^4
+M \left(r-2 \lambda r+6 (\iom)^2 r^3\right)\right) Se_{01}}{(\iom)^2
(2 M-r) (3 M+\lambda r)}
\\&+\frac{16 (1+\lambda) \pi \left(16 M^2+(-5+6 \lambda) M r
-2 \lambda r^2\right) Se_{02}}{(\iom)^2 (2 M-r) r (3 M+\lambda
r)}-\frac{16 \pi r Se_{11}}{\iom}
\\&+\frac{32 (1+\lambda) \pi Se_{12}}{\iom}+\frac{\dpz}{2 \iom r}
+\frac{2 \iom \dpzbf}{r}+\frac{4 (1+\lambda) \dpo}{r}
\end{split}
\end{equation}

\begin{equation}
\begin{split}\bhtw=&\frac{\lambda (1+\lambda)
\left(-9 M^2+(3-5 \lambda) M r+2 \lambda r^2\right) \dptw}
{3 (\iom)^2 r^3 (3 M+\lambda r)}+\frac{(-3 M+2 r) \pz}{(2 M-r) r}
\\&+\frac{2 \left(6 M^2-(11+4 \lambda) M r+r^2 \left(4+2 \lambda
+(\iom)^2 r^2\right)\right) \pzbf}{(2 M-r) r^4}
\\&+\frac{4 (1+\lambda) \left(-4 (1+\lambda) M+r \left(2+2 \lambda
+(\iom)^2 r^2\right)\right) \po}{\iom (2 M-r) r^3}
\\&-\frac{\lambda (1+\lambda)}{3 (\iom)^2 (2 M-r) r^4 (3 M+\lambda r)^2}
\left[-54 M^4+3 (9-16 \lambda) M^3 r\right.\\&
+\lambda^2 r^4 \left(2+2 \lambda+(\iom)^2 r^2\right)+9 M^2 r^2
\left(2 \lambda-2 \lambda^2+(\iom)^2 r^2\right)\\&\left.+\lambda M r^3 
\left(3+5 \lambda-4 \lambda^2+6 (\iom)^2 r^2\right)\right] \ptw
+\frac{8 \pi}{(\iom)^3 (2 M-r) r^2 (3 M+\lambda r)^2}
\\&\times\left[192 (1+\lambda) M^4+2 \lambda^2 r^4\left(1+2\lambda
+3(\iom)^2 r^2\right)-3 M^3 r \left(43\!+\!8 \lambda\!-\!44 \lambda^2
\right.\right.\\&\left.+36 (\iom)^2 r^2\right)+\lambda M r^3 \left(9
-20 \lambda^2+36 (\iom)^2 r^2+\lambda \left(4-12 (\iom)^2 r^2\right)\right)
\\&\left.-6 M^2 r^2 \left(-2+14 \lambda^2-4 \lambda^3-9 (\iom)^2 r^2
+2 \lambda \left(5+6 (\iom)^2 r^2\right)\right)\right]Se_{01}
\\&-\frac{16 (1+\lambda) M \pi}{(\iom)^3 (2 M-r) r^3 (3 M+\lambda r)^2} 
\left[48 M^3+(-45+26 \lambda) M^2 r\right.\\&\left.
+\left(6-25 \lambda+2 \lambda^2\right)
 M r^2+(3-2 \lambda) \lambda r^3\right] Se_{02}
\\&-\frac{8 \pi}{(\iom)^2 r (3 M+\lambda r)}\left[-(35+26 \lambda) M^2
+2 \lambda r^2 \left(3+2 \lambda+(\iom)^2 r^2\right)\right.\\&
\left.+M r \left(19+2 \lambda-8 \lambda^2+6 (\iom)^2 r^2\right)\right] Se_{11}
\\&+\frac{16 (1+\lambda) \pi \left(-16 M^2+(11-6 \lambda) M r
+4 \lambda r^2\right) Se_{12}}{(\iom)^2 r^2 (3 M+\lambda r)}
\\&+\dpz+\frac{(-6 M+4 r) \dpzbf}{r^3}-
\frac{4 (1+\lambda) (M-r) \dpo}{\iom r^3}
\end{split}
\end{equation}

\begin{equation}
\begin{split}\bk=&\frac{\lambda (1+\lambda) (2 M-r) \dptw}
{3 (\iom)^2 r^3}+\frac{\pz}{r}+\frac{(-4 M+2 (2+\lambda) r) \pzbf}{r^4}
\\&+\frac{4 (1+\lambda)^2 \po}{\iom r^3}-\frac{\lambda (1+\lambda)
\left(6 M^2+3 \lambda M r+\lambda (1+\lambda) r^2\right) \ptw}
{3 (\iom)^2 r^4 (3 M+\lambda r)}\\&+\frac{8 \pi}{(\iom)^3 r^2 (3 M+\lambda r)}
\left[-16 (1+\lambda) M^2+\lambda r^2 \left(1+2 \lambda+2 (\iom)^2 r^2\right)
\right.\\&\left.+M r \left(2-\lambda-6 \lambda^2+6(\iom)^2
 r^2\right)\right]Se_{01}
-\frac{8 (3+2 \lambda) \pi (2 M-r) Se_{11}}{(\iom)^2 r}
\\&+\frac{16 (1+\lambda) M \pi (8 M+(-1+2 \lambda) r) Se_{02}}
{(\iom)^3 r^3 (3 M+\lambda r)}
+\frac{32 (1+\lambda) \pi (2 M-r) Se_{12}}{(\iom)^2 r^2}
\\&+\frac{(4 M-2 r) \dpzbf}{r^3}
+\frac{2 (1+\lambda) (2 M-r) \dpo}{\iom r^3}
\end{split}
\end{equation}

\begin{equation}
\begin{split}\hz=&\frac{\lambda (2 M-r) \dptw}{3 \iom r}
+\frac{(-2 M+r) \pz}{2 \iom r^2}+\frac{2 \iom \pzbf}{r}
+\frac{4 (1+\lambda) \po}{r}\\&-\frac{\lambda \left(6 M^2+3 \lambda M r
+\lambda (1+\lambda) r^2\right) \ptw}{3 \iom r^2 (3 M+\lambda r)}
-\frac{16 \pi (2 M-r) (2 M+\lambda r) Se_{02}}{(\iom)^2 r (3 M+\lambda r)}
\\&-\frac{8 \pi \left(16 M^2+(-5+6 \lambda) M r-2 \lambda r^2\right)
Se_{01}}{(\iom)^2 (3 M+\lambda r)}
+\frac{16 \pi r (-2 M+r) Se_{11}}{\iom}\\&-\frac{(-2 M+r) \dpz}{2 \iom r}
+\left(1-\frac{2 M}{r}\right) \dpo
\end{split}
\end{equation}

\begin{equation}
\begin{split}\ho=&\frac{\lambda ((3+\lambda) M
+\lambda (2+\lambda) r) \dptw}{3 (\iom)^2 r (3 M+\lambda r)}
-\frac{r \pz}{4 M-2 r}+\frac{4 \pzbf}{r^2}
\\&+\frac{\left(8 M\!+\!8 \lambda M\!-\!4 r\!-\!4 \lambda r\!
+\!(\iom)^2 r^3\right) \po}{2\iom M r^2-\iom r^3}-\frac{\lambda}
{3 (\iom)^2 (2 M\!-\!r) r^2 (3 M\!+\!\lambda
r)^2}\\&\times\left[9 (\iom)^2 M^2 r^3+2 \lambda^3 r^2 (-2 M\!+\!r)
\!+\!\lambda^2 r \left(-12 M^2\!+\!2 M r\!+\!2 r^2\!+\!(\iom)^2 r^4\right)
\right.\\&\left.+3 \lambda M \left(-4 M^2+r^2+2 (\iom)^2 r^4\right)\right]\ptw
-\frac{8 \pi}{(\iom)^3 (2 M-r) r (3 M+\lambda r)^2}
\\&\times\left[96 M^4+6 (-13+10 \lambda) M^3 r+2 (\iom)^2 \lambda^2 r^6+3
 \lambda M r^3 \left(1-2 \lambda\right.\right.\\&\left.\left.+4 (\iom)^2
r^2\right)+ 2 M^2 r^2 \left(3-24 \lambda+4 \lambda^2+9 (\iom)^2 r^2\right)
\right] Se_{01}\\&-\frac{16 \pi}{(\iom)^3 (2 M-r) r^2 (3 M+\lambda r)^2}
 \left[48 M^4+80 \lambda M^3 r+\left(-3-13 \lambda\right.\right.\\&
\left.\left.+38 \lambda^2\right) M^2 r^2+3 \lambda \left(-1-3 \lambda+2
\lambda^2\right) M r^3-\lambda^2 (1+2 \lambda) r^4\right] Se_{02}
\\&-\frac{8 \pi \left(16 M^2+(-11+6 \lambda) M r-4 \lambda r^2\right)
 Se_{11}}{(\iom)^2 (3 M+\lambda r)}
\\&-\frac{16 \pi \left(4 M^2+\lambda (1+2 \lambda) r^2
+4 M (r+2 \lambda r)\right) Se_{12}}{(\iom)^2 r (3 M+\lambda r)}
\\&-\frac{2 \dpzbf}{r}-\frac{2 (2 M+r+2 \lambda r) \dpo}{\iom r^2}
\end{split}
\end{equation}

\begin{equation}
\begin{split}\bg=&-\frac{\lambda (3+2 \lambda)(2 M-r)\dptw}
{6 (\iom)^2 r^2 (3 M+\lambda r)}-\frac{\pzbf}{r^3}-\frac{2 (1+\lambda) \po}
{\iom r^3}\\&+\frac{1}{6 (\iom)^2 r^3 (3 M+\lambda r)^2}
\left[4 \lambda^3 r^2+\lambda^4 r^2+27 (\iom)^2 M^2 r^2\right.\\&\left.
+9 \lambda M \left(M+2 (\iom)^2 r^3\right)+3 \lambda^2 \left(M^2+M
r+r^2+(\iom)^2 r^4\right)\right]\ptw
\\&-\frac{4 \pi \left(-48 M^3+15 (1-2 \lambda) M^2 r+(7-6 \lambda)
 \lambda M r^2+\lambda (1+2 \lambda) r^3\right) Se_{01}}{(\iom)^3
r^2 (3 M+\lambda r)^2}
\\&+\frac{8 \pi (2 M-r) \left(12 M^2+8 \lambda M r+\lambda
(1+2 \lambda) r^2\right) Se_{02}}{(\iom)^3 r^3 (3 M+\lambda r)^2}
\\&+\frac{4 \pi (2 M-r) (8 M+(-1+2 \lambda) r) Se_{11}}
{(\iom)^2 r (3 M+\lambda r)}+\frac{8 \pi (-2 M+r)^2 Se_{12}}
{(\iom)^2 r^2 (3 M+\lambda r)}\\&+\frac{(2 M-r) \dpo}{\iom r^3}
\end{split}
\end{equation}

\begin{equation}
\begin{split}
\dbhz=&-\frac{\lambda (1+\lambda) \left(18 M^3+2 (-9+\lambda) M^2 r
+(\iom)^2 \lambda r^5+3 M r^2 \left(2+(\iom)^2 r^2\right)\right)
\dptw}{3 (\iom)^2 (2 M-r) r^4 (3 M+\lambda r)}
\\&+\frac{\left(-6 M^2-2 M (r+2 \lambda r)+r^2 \left(3+2 \lambda
+(\iom)^2 r^2\right)\right) \pz}{r^2 (-2 M+r)^2}+\frac{2}{r^5 (-2 M+r)^2}
\\&\times\left[12 M^3-2 (9+2 \lambda) M^2 r+(\iom)^2 r^5+M r^2 
\left(6+2 \lambda+(\iom)^2 r^2\right)\right] \pzbf
\\&+\frac{4 (1+\lambda) \left(-4 (1+\lambda) M^2+(\iom)^2 r^4+M r
 \left(2+2 \lambda+(\iom)^2 r^2\right)\right) \po}{\iom r^4 (-2 M+r)^2}
\\&-\frac{\lambda (1+\lambda)}{3 (\iom)^2 r^5 (-2 M+r)^2 (3 M+\lambda r)^2}
\left[-108 M^5+(90-84 \lambda) M^4 r+(\iom)^2 \lambda^2 r^7\right.
\\&-3 M^3 r^2 \left(6-14 \lambda+12 \lambda^2+3 (\iom)^2 r^2\right)
+2 M^2 r^3 \left(6 \lambda+13 \lambda^2-2 \lambda^3\right.\\&\left.\left.
+9 (\iom)^2 r^2\right)+\lambda M r^4 \left(-6+2 \lambda^2+9 (\iom)^2 r^2
+\lambda \left(-4+(\iom)^2 r^2\right)\right)\right]\ptw
\\&+\frac{8 \pi}{(\iom)^3 r^3 (-2 M+r)^2 (3 M+\lambda r)^2}
\left[192 (1+\lambda) M^5+2 (\iom)^2 \lambda^2 r^7 \left(1
+(\iom)^2 r^2\right)\right.\\&-6 M^4 r \left(17+2 \lambda-24\lambda^2
+24 (\iom)^2 r^2\right)+\lambda M r^4 \left(-2+4 \lambda^2
+13 (\iom)^2 r^2\right.\\&\left.+12 (\iom)^4 r^4+2 \lambda \left(-1
+(\iom)^2 r^2\right)\right)+M^2 r^3 \left(-20 \lambda^3-18 \lambda^2
 \left(-1+(\iom)^2 r^2\right)\right.\\&\left.+3 (\iom)^2 r^2 
\left(7+6 (\iom)^2 r^2\right)+\lambda \left(8+7(\iom)^2 r^2\right)\right)
-M^3 r^2 \left(6+104 \lambda^2-24 \lambda^3\right.\\&\left.\left.
-3 (\iom)^2 r^2+2 \lambda \left(31+51 (\iom)^2 r^2\right)\right)\right]Se_{01}
-\frac{16 (1+\lambda) \pi}{(\iom)^3 r^4 (-2 M+r)^2 (3 M+\lambda r)^2} 
\\&\times\left[192 M^5+2 (-93+58 \lambda) M^4 r+2 (\iom)^2 \lambda^2 r^7
+2 M^3 r^2 \left(18-53 \lambda+10 \lambda^2\right.\right.
\\&\left.-24 (\iom)^2 r^2\right)+M^2 r^3 \left(-20 \lambda^2+15 (\iom)^2 r^2
+\lambda \left(14-34 (\iom)^2 r^2\right)\right)+\lambda M r^4 
\\&\left.\times\left(2+11 (\iom)^2 r^2+\lambda \left(4-6 (\iom)^2 r^2\right)
\right)\right] Se_{02}-\frac{16 \pi}{(\iom)^2 (2 M-r) r^2 (3 M+\lambda r)}
\\&\times\left[-(19+10 \lambda) M^3+(\iom)^2 \lambda r^5+M^2 r \left(9
-4 \lambda-4 \lambda^2+6 (\iom)^2 r^2\right)+M r^2 \left(1
\right.\right.\\&\left.\left.+2 \lambda^2+3 (\iom)^2 r^2+\lambda
 \left(5+2 (\iom)^2 r^2\right)\right)\right] Se_{11}
+\frac{32 (1+\lambda) \pi}{(\iom)^2 (2 M-r) r^3 (3 M+\lambda r)}
\\&\times\left[-20 M^3+3 (5-2 \lambda) M^2 r+(\iom)^2 \lambda r^5+M r^2 
\left(-1+4 \lambda+3 (\iom)^2 r^2\right)\right] Se_{12}
\\&+\frac{(M+r) \dpz}{2 M r-r^2}+\frac{2 \left(-6 M^2+4 M r
+(\iom)^2 r^4\right) \dpzbf}{(2 M-r) r^4}
\\&+\frac{4 (1+\lambda) \left(-4 M^2+3 M r+(\iom)^2 r^4\right) \dpo}
{\iom (2 M-r) r^4}
\end{split}
\end{equation}

\begin{equation}
\begin{split}\dbho=&-\frac{\lambda (1+\lambda)
M (M+r+\lambda r) \dptw}{\iom (2 M-r) r^2 (3 M+\lambda r)}
+\frac{1}{2 \iom r^3 (-2 M+r)^2}\left[4 (5+4 \lambda) M^2
\right.\\&\left.-2 M r \left(13+10 \lambda+2 (\iom)^2 r^2\right)
+r^2 \left(8+6 \lambda+5 (\iom)^2 r^2\right)\right]\pz
\\&+\frac{2 \iom \left(-(7+4 \lambda) M+r \left(4+2 \lambda
+(\iom)^2 r^2\right)\right) \pzbf}{r^2 (-2 M+r)^2}
\\&+\frac{4 (1+\lambda) \left(-(5+4 \lambda) M+r \left(3+2 \lambda
+(\iom)^2 r^2\right)\right) \po}{r^2 (-2 M+r)^2}
\\&+\frac{\lambda (1+\lambda)}{3 \iom r^3 (-2 M+r)^2 (3 M+\lambda r)^2}
\left[72 M^4+9 (-5+8 \lambda) M^3 r+\lambda M r^3 \left(3-13 \lambda
\right.\right.\\&\left.\left.+4 \lambda^2-6 (\iom)^2 r^2\right)
-\lambda^2 r^4 \left(2 \lambda+(\iom)^2 r^2\right)-3 M^2 r^2 
\left(16 \lambda\!-\!8 \lambda^2\!+\!3 (\iom)^2 r^2\right)\right]\ptw
\\&+\frac{8 \pi}{(\iom)^2 r (-2 M+r)^2 (3 M+\lambda r)^2}\left[6 (43
+34 \lambda) M^4+2 \lambda^2 r^4 \left(2\!+\!2 \lambda\!+\!3 (\iom)^2 r^2
\right)\right.\\&-M^3 r \left(237+2 \lambda-136 \lambda^2
+72 (\iom)^2 r^2\right)+\lambda M r^3 \left(23-20 \lambda^2+36 (\iom)^2 r^2
\right.\\&\left.-4 \lambda \left(1+2 (\iom)^2 r^2\right)\right)
-6 M^2 r^2 \left(-6+14 \lambda^2-4 \lambda^3-9 (\iom)^2 r^2\right.\\&
\left.\left.+4 \lambda \left(5+2 (\iom)^2 r^2\right)\right)\right]Se_{01}
+\frac{16 (1+\lambda) \pi}{(\iom)^2 r (-2 M+r)^2 (3 M+\lambda r)^2}
\left[(63+2 \lambda) M^3\right.\\&\left.+\left(-18+43 \lambda
+2 \lambda^2\right) M^2 r+\lambda (-13+6 \lambda) M r^2
-2 \lambda^2 r^3\right] Se_{02}
\\&-\frac{8 \pi}{\iom (2 M-r) (3 M+\lambda r)}\left[-(35+26 \lambda) M^2
+2 \lambda r^2 \left(4+2 \lambda+(\iom)^2 r^2\right)\right.\\&\left.
+M r \left(25+2 \lambda-8 \lambda^2+6(\iom)^2 r^2\right)\right] Se_{11}
-\frac{2 \iom (M-2 r) \dpzbf}{(2 M-r) r^2}
\\&-\frac{16 (1+\lambda) \pi \left(4 M^2+(-11+2 \lambda) M r
-4 \lambda r^2\right) Se_{12}}{\iom (2 M-r) r (3 M+\lambda r)}
\\&+\frac{\left(-(3+2 \lambda) M+r \left(2+\lambda+(\iom)^2 r^2\right)
\right) \dpz}{\iom (2 M-r) r^2}+\frac{(6+6 \lambda) \dpo}{2 M r-r^2}
\end{split}
\end{equation}
\vspace*{-18pt}
\begin{equation}
\setlength{\jot}{0.9pt}
\begin{split}\dbhtw=&\frac{\lambda (1+\lambda)}
{3 (\iom)^2 (2 M-r) r^4 (3 M+\lambda r)}\left[54 M^3+2(-21+13 \lambda)M^2 r
+M r^2 \left(6-18 \lambda\right.\right.\\&\left.\left.+4 \lambda^2
-3 (\iom)^2 r^2\right)-\lambda r^3 \left(-2+2 \lambda+(\iom)^2 r^2\right)
\right]\dptw+\frac{1}{r^2 (-2 M+r)^2}\left[14 M^2\right.\\&\left.
-2 (11+2 \lambda) M r+r^2 \left(8+2 \lambda+(\iom)^2 r^2\right)\right]\pz
+\frac{1}{r^5 (-2 M+r)^2}\left[-72 M^3\right.\\&\left.+4(47+18\lambda)M^2 r
+2 r^3 \left(16+10 \lambda+3 (\iom)^2 r^2\right)
-2 M r^2 \left(70+38 \lambda\right.\right.\\&\left.\left.
+3 (\iom)^2 r^2\right)\right]\pzbf+\frac{4 (1+\lambda)}{\iom r^4 (-2 M+r)^2}
\left[28 (1+\lambda) M^2+2 r^2 \left(4+4 \lambda\right.\right.\\&\left.\left.
+(\iom)^2 r^2\right)-M r \left(30+30 \lambda+(\iom)^2 r^2\right)\right]\po
-\frac{\lambda (1+\lambda)}{3 (\iom)^2 r^5 (-2 M+r)^2 (3 M\!+\!\lambda r)^2}
\\&\times\left[324 M^5+114 (-3+2 \lambda) M^4 r+\lambda^2 r^5 
\left(2+2 \lambda-(\iom)^2 r^2\right)+3 M^3 r^2 \left(30\!-\!74 \lambda
\right.\right.\\&\left.\left.+20 \lambda^2
+9 (\iom)^2 r^2\right)+2 \lambda M^2 r^3 \left(21-31 \lambda+2 \lambda^2
+12 (\iom)^2 r^2\right)+\lambda M r^4 \left(6\right.\right.\\&\left.\left.
-6 \lambda^2-3 (\iom)^2 r^2+\lambda \left(12+5 (\iom)^2 r^2\right)
\right)\right]\ptw+\frac{8 \pi}{(\iom)^3 r^3 (-2 M+r)^2 (3 M+\lambda r)^2}
\\&\times\left[-1344 (1+\lambda) M^5+6 M^4 r \left(283+158 \lambda
-152 \lambda^2+72 (\iom)^2 r^2\right)+2 \lambda^2 r^5 \left(3
\right.\right.\\&\left.+10 (\iom)^2 r^2+(\iom)^4 r^4+2 \lambda \left(3
+(\iom)^2 r^2\right)\right)+\lambda M r^4 \left(34+121 (\iom)^2 r^2
\right.\\&\left.+12 (\iom)^4 r^4-4 \lambda^2 \left(23
+2 (\iom)^2 r^2\right)-2 \lambda \left(1+21 (\iom)^2 r^2\right)\right)
+M^3 r^2 \left(-618\right.\\&\left.+1112 \lambda^2-152 \lambda^3
-609 (\iom)^2 r^2+\lambda \left(358+210 (\iom)^2 r^2\right)\right)
+M^2 r^3 \left(212 \lambda^3\right.\\&-2 \lambda^2 \left(175
+(\iom)^2 r^2\right)-\lambda \left(322+365 (\iom)^2 r^2\right)
+3 \left(16+61 (\iom)^2 r^2\right.\\&\left.\left.\left.
+6 (\iom)^4 r^4\right)\right)\right]Se_{01}
-\frac{16 (1+\lambda) \pi}{(\iom)^3 r^4 (-2 M+r)^2 (3 M+\lambda r)^2}
\left[-192 M^5\right.\\&+2 (183+50 \lambda) M^4 r
+2 \lambda^2 r^5 \left(1+2 \lambda+(\iom)^2 r^2\right)
-2 M^3 r^2 \left(96-21 \lambda\right.\\&\left.-62 \lambda^2
+24 (\iom)^2 r^2\right)+M^2 r^3 \left(24-80 \lambda-84 \lambda^2+24 
\lambda^3+15 (\iom)^2 r^2\right.\\&\left.\left.-34 (\iom)^2\lambda r^2\right)
+\lambda M r^4 \left(14-20 \lambda^2+11 (\iom)^2 r^2
+\lambda \left(6-6 (\iom)^2 r^2\right)\right)\right]Se_{02}
\\&-\frac{16 \pi}{(\iom)^2 (2 M-r) r^2 (3 M+\lambda r)}
\left[(121+94 \lambda) M^3+\lambda r^3 \left(13+10 \lambda
+3 (\iom)^2 r^2\right)\right.\\&-M^2 r \left(137+66 \lambda-32 \lambda^2
+6 (\iom)^2 r^2\right)+M r^2 \left(39-36 \lambda^2
+9 (\iom)^2 r^2-2 \lambda \left(8\right.\right.\\&\left.\left.\left.
+(\iom)^2 r^2\right)\right)\right]Se_{11}
+\frac{32 (1+\lambda) \pi}{(\iom)^2 (2 M-r) r^3 (3 M+\lambda r)}
\left[44 M^3+(-61+2 \lambda) M^2 r\right.\\&\left.
+\lambda r^3 \left(7+2 \lambda+(\iom)^2 r^2\right)
+M r^2 \left(21-14 \lambda-4 \lambda^2+3 (\iom)^2 r^2\right)\right]Se_{12}
\\&-\frac{(M-2 r) \dpz}{2 M r-r^2}
+\frac{4 (1+\lambda) \left(4 M^2-(9+4 \lambda) M r
+r^2 \left(4+2 \lambda+(\iom)^2 r^2\right)\right) \dpo}{\iom (2 M-r)r^4}
\raisetag{27.5pt}\\&+\frac{2 \left(18 M^2-4 (6+\lambda) M r
+r^2 \left(8+2 \lambda+(\iom)^2 r^2\right)\right) \dpzbf}{(2 M-r) r^4}
\end{split}
\end{equation}

\begin{equation}
\begin{split}\dbk=&-\frac{\lambda (1+\lambda)
\left(18 M^2+(-6+7 \lambda) M r+(-1+\lambda) \lambda r^2\right) \dptw}
{3 (\iom)^2 r^4 (3 M+\lambda r)}-\frac{3 \pz}{r^2}
\\&+\frac{2 \left(12 M^2-2 (11+5 \lambda) M r+r^2 \left(8
+5 \lambda+(\iom)^2 r^2\right)\right) \pzbf}{(2 M-r) r^5}
\\&+\frac{2 (1+\lambda) \left(-16 (1+\lambda) M+r \left(8+8 \lambda
+(\iom)^2 r^2\right)\right) \po}{\iom (2 M-r) r^4}
\\&+\frac{\lambda (1+\lambda)}{3 (\iom)^2 (2 M-r) r^5 (3 M+\lambda r)^2}
\left[108 M^4+6 (-9+13 \lambda) M^3 r-\lambda^2 r^4 \left(1+\lambda
\right.\right.\\&\left.\left.-(\iom)^2 r^2\right)+2 \lambda M r^3 
\left(-3-5 \lambda+\lambda^2+3 (\iom)^2 r^2\right)+3 M^2 r^2 
\left(-9 \lambda+8 \lambda^2\right.\right.\\&\left.\left.
+3 (\iom)^2 r^2\right)\right]\ptw
+\frac{8 \pi}{(\iom)^3 (2 M-r) r^3 (3 M+\lambda r)^2}
\left[384 (1+\lambda) M^4-6 M^3 r\right.\\&\times\left(43+8 \lambda
-44 \lambda^2+24 (\iom)^2 r^2\right)+2 \lambda M r^3 
\left(9-18 \lambda^2+27 (\iom)^2 r^2+\lambda \left(3\right.\right.
\\&\left.\left.-2 (\iom)^2 r^2\right)\right)+\lambda^2 r^4 
\left(3+9 (\iom)^2 r^2+2 \lambda\left(3+(\iom)^2 r^2\right)\right)
+M^2 r^2 \left(24-172 \lambda^2\right.\\&\left.\left.+44 \lambda^3
+81(\iom)^2 r^2-3\lambda\left(43+26 (\iom)^2 r^2\right)\right)\right]Se_{01}
\\&-\frac{16 (1+\lambda) \pi}{(\iom)^3 (2 M-r) r^4 (3 M+\lambda r)^2}
\left[96 M^4+2 (-45+2 \lambda) M^3 r+\left(12-35 \lambda
\right.\right.\\&\left.\left.-26 \lambda^2\right) M^2 r^2
+3 \lambda \left(2+\lambda-2 \lambda^2\right) M r^3
+\lambda^2 (1+2 \lambda) r^4\right]Se_{02}
\\&-\frac{8 \pi}{(\iom)^2 r^2 (3 M+\lambda r)}\left[-2 (35+26 \lambda)
M^2+\lambda r^2 \left(13+10 \lambda+2 (\iom)^2 r^2\right)\right.\\&\left.
+M r \left(38+5 \lambda-18 \lambda^2+6 (\iom)^2 r^2\right)\right]Se_{11}
\\&+\frac{16 (1+\lambda) \pi \left(-32 M^2+(22-4 \lambda) M r
+\lambda (7+2 \lambda) r^2\right) Se_{12}}{(\iom)^2 r^3(3 M+\lambda r)}
+\frac{\dpz}{r}\\&+\frac{2 (-6 M+(4+\lambda) r) \dpzbf}{r^4}
+\frac{4 (1+\lambda) (-2 M+(2+\lambda) r) \dpo}{\iom r^4}
\end{split}
\end{equation}

\begin{equation}
\begin{split}\dhz=&-\frac{\lambda \left(6 M^2
+3 \lambda M r+\lambda (1+\lambda) r^2\right) \dptw}
{3 \iom r^2 (3 M+\lambda r)}+\frac{(-2 M+r) \dpz}{2 \iom r^2}
+\frac{2 \iom \dpzbf}{r}
\\&+\frac{\left(4 M^2-4 (2+\lambda) M r+r^2 \left(3+2 \lambda
+(\iom)^2 r^2\right)\right) \pz}{2\iom (2 M-r) r^3}-\frac{2 \iom \pzbf}{r^2}
\\&+\frac{\left(4 (1+\lambda) M+r \left(-2-2\lambda+(\iom)^2 r^2\right)
\right) \po}{r^2 (-2 M+r)}
+\frac{\lambda}{3 \iom (2 M-r) r^3 (3 M+\lambda r)^2}
\\&\times\left[36 M^4+18 (-1+\lambda) M^3 r-2 \lambda M r^3 
\left(\lambda+\lambda^2-3 (\iom)^2 r^2\right)+\lambda^2 r^4\left(1+\lambda
\right.\right.\\&\left.\left.+(\iom)^2 r^2\right)+M^2 \left(-9 \lambda r^2
+9 (\iom)^2 r^4\right)\right]\ptw
+\frac{8 \pi}{(\iom)^2 (2 M-r) r (3 M+\lambda r)^2}
\\&\times\left[48 M^4+16 (-3+2 \lambda) M^3 r+2 (\iom)^2 \lambda^2 r^6
+2 \lambda M r^3 \left(1-2 \lambda+6 (\iom)^2 r^2\right)\right.\\&\left.
+M^2 r^2 \left(3-32 \lambda+4 \lambda^2+18 (\iom)^2 r^2\right)\right]Se_{01}
+\frac{16 \pi}{(\iom)^2 (2 M-r) r (3 M+\lambda r)^2}
\\&\times\left[(48+52 \lambda) M^3+\left(-15+15 \lambda+34 \lambda^2\right)
M^2 r+\lambda \left(-10-5 \lambda+6 \lambda^2\right)M r^2
\right.\\&\left.-2 \lambda^2 (1+\lambda) r^3\right] Se_{02}
+\frac{8 \pi \left(4 M^2+(-5+2 \lambda) M r-2 \lambda r^2\right) Se_{11}}
{\iom (3 M+\lambda r)}\\&+
\frac{16 \pi \left(-2 M^2+(7+6 \lambda) M r+2 \lambda (1+\lambda)
 r^2\right) Se_{12}}{\iom r (3 M+\lambda r)}+\frac{4 (1+\lambda) \dpo}{r}
\end{split}
\end{equation}

\begin{equation}
\begin{split}\dho=&\frac{\lambda \left(\lambda^2 (4 M-r) r
-3 M r \left(-1+(\iom)^2 r^2\right)+\lambda \left(4 M^2+3 M r
-(\iom)^2 r^4\right)\right)\dptw}{3 (\iom)^2 (2 M-r) r^2 (3 M+\lambda r)}
\\&+\frac{(3 M-2 r) \pz}{(-2 M+r)^2}-\frac{2}{r^3(-2 M+r)^2}\left[12 M^2
-2 (9+2 \lambda) M r+r^2 \left(6+2 \lambda\right.\right.\\&\left.\left.
+(\iom)^2 r^2\right)\right]\pzbf-\frac{2}{\iom r^3 (-2 M+r)^2}
\left[8 (1+\lambda) M^2+M r \left(-16\right.\right.\\&\left.\left.
-24 \lambda-8 \lambda^2+(\iom)^2 r^2\right)+r^2 \left(6+4 \lambda^2
+(\iom)^2 r^2+2 \lambda \left(5+(\iom)^2 r^2\right)\right)\right] \po
\\&+\frac{\lambda}{3 (\iom)^2 r^3 (-2 M+r)^2 (3 M+\lambda r)^2}
\left[-12 (3+5 \lambda) M^4-6 M^3 r \left(-3+\lambda\right.\right.\\&
\left.+10 \lambda^2+3 (\iom)^2 r^2\right)+M^2 r^2 \left(22 \lambda^2-20
\lambda^3+9 (\iom)^2 r^2+\lambda \left(24-9 (\iom)^2 r^2\right)\right)
\\&+\lambda^2 r^4 \left(2 \lambda^2+2 (\iom)^2 r^2+\lambda \left(2+(\iom)^2
r^2\right)\right)+\lambda M r^3 \left(-3+6 \lambda^2-4 \lambda^3\right.
\\&\left.\left.+9 (\iom)^2 r^2+2 \lambda \left(2+(\iom)^2 r^2\right)\right)
\right]\ptw+\frac{8 \pi}{(\iom)^3 r^2 (-2 M+r)^2 (3 M+\lambda r)^2}
\\&\times\left[192 M^5-12 (37+6 \lambda) M^4 r-2 \lambda^2 r^5 \left(1
+2 \lambda+4 (\iom)^2 r^2\right)+4 M^3 r^2 \left(57\right.\right.\\&\left.
-32 \lambda-28 \lambda^2+30 (\iom)^2 r^2\right)+M^2 r^3
\left(-18+116 \lambda+60 \lambda^2-24 \lambda^3-69 (\iom)^2 r^2
\right.\\&\left.\left.+82 (\iom)^2 \lambda r^2\right)+\lambda M r^4
\left(-11+20 \lambda^2-47 (\iom)^2 r^2+2 \lambda \left(2+7 (\iom)^2
 r^2\right)\right)\right] Se_{01}
\\&+\frac{16 \pi}{(\iom)^3 r^3 (-2 M+r)^2 (3 M+\lambda r)^2}
\left[96 M^5+16 (-3+10 \lambda) M^4 r+\lambda^2 r^5 \left(1\right.\right.
\\&\left.+2 \lambda-(\iom)^2 r^2\right)+4 M^3 r^2 \left(-9-35 \lambda
+18 \lambda^2+3 (\iom)^2 r^2\right)+M^2 r^3 \left(9-78 \lambda^2\right.
\\&\left.+8 \lambda^3-6 (\iom)^2 r^2+5 \lambda \left(-1
+2 (\iom)^2 r^2\right)\right)+\lambda M r^4 \left(7-10 \lambda^2
-5 (\iom)^2 r^2\right.
\\&\left.\left.+\lambda \left(11+2 (\iom)^2 r^2\right)\right)\right]Se_{02}
+\frac{8 \pi}{(\iom)^2 (2 M-r) r (3 M+\lambda r)}\left[32 M^3
-8 (9+2 \lambda)\right.\\&\left.\times M^2 r+2\lambda r^3\left(5
+2\lambda+(\iom)^2 r^2\right)+M r^2 \left(31-10 \lambda-8 \lambda^2
+6 (\iom)^2 r^2\right)\right] Se_{11}
\\&+\frac{16 \pi}{(\iom)^2 (2 M-r) r^2 (3 M+\lambda r)}
\left[8 M^3+2 \left(-7-6 \lambda+4 \lambda^2\right) M r^2-\lambda (5
+6 \lambda) r^3\right.\\&\left.+12 M^2 (r+2 \lambda r)\right]Se_{12}
-\frac{r \dpz}{4 M-2 r}+\frac{2 (-4 M+3 r) \dpzbf}{r^2 (-2 M+r)}
\\&+\frac{\left(8 M^2+8 \lambda M r+r^2 \left(-6-8 \lambda
+(\iom)^2 r^2\right)\right) \dpo}{\iom (2 M-r) r^3}
\end{split}
\end{equation}

\begin{equation}
\begin{split}\dbg=&\frac{\left(5 \lambda^2 M+\lambda^3 r
+9 (\iom)^2 M r^2+3 \lambda \left(3 M-r+(\iom)^2 r^3\right)\right) \dptw}
{6(\iom)^2 r^3 (3 M+\lambda r)}+\frac{3 \pzbf}{r^4}
\\&+\left(\frac{4 (1+\lambda)}{\iom r^4}+\frac{\iom}{2 M r-r^2}\right) \po
+\frac{1}{6 (\iom)^2 (2 M-r) r^4 (3 M+\lambda r)^2}
\\&\times\left[\lambda^4(2 M-r)r^2+27 (\iom)^2 M^2 r^2 (-2 M+r)
+2 \lambda^3 r\left(6 M^2-5 M r+r^2\right.\right.\\&\left.\left.
-(\iom)^2 r^4\right)-9 \lambda M \left(2 M^2-(\iom)^2 r^4
+M r\left(-1+4 (\iom)^2 r^2\right)\right)
+3 \lambda^2 \left(2 M^3\right.\right.\\&\left.\left.
-M^2 r+r^3-2 M \left(r^2+2(\iom)^2 r^4\right)\right)\right]\ptw
+\frac{4 \pi}{(\iom)^3 r^3 (-2 M+r) (3 M+\lambda r)^2}
\\&\times\left[192 M^4+12 (-13+10 \lambda) M^3 r+M^2 r^2 \left(21-92 \lambda
+20 \lambda^2+24 (\iom)^2 r^2\right)\right.\\&\left.
+\lambda r^4 \left(1-(\iom)^2 r^2+2 \lambda\left(1+(\iom)^2 r^2\right)\right)
+M r^3 \left(-16 \lambda^2-3 (\iom)^2 r^2+2\lambda \left(4\right.\right.\right.
\\&\left.\left.\left.+7 (\iom)^2 r^2\right)\right)\right]Se_{01}
-\frac{8 \pi}{(\iom)^3 (2 M-r) r^4 (3 M+\lambda r)^2}
\left[96 M^4+16 (-3+7 \lambda) M^3 r\right.\\&\left.+\lambda r^4 
\left(1+\lambda-2 \lambda^2-(\iom)^2 r^2\right)+M^2 r^2 \left(9
-41 \lambda+46 \lambda^2+6 (\iom)^2 r^2\right)\right.\\&
\left.+M r^3 \left(\lambda-17 \lambda^2+6 \lambda^3-3 (\iom)^2 r^2+2 (\iom)^2
\lambda r^2\right)\right]Se_{02}
\\&-\frac{4 \pi \left(32 M^2+(-21+10 \lambda) M r+(1-6 \lambda)
r^2\right) Se_{11}}{(\iom)^2 r^2 (3 M+\lambda r)}
\\&-\frac{8 \pi \left(8 M^2+8 \lambda M r+\left(1+\lambda+2 \lambda^2\right)
r^2\right) Se_{12}}{(\iom)^2 r^3 (3 M+\lambda r)}
\\&-\frac{\dpzbf}{r^3}-\frac{2 (2 M+\lambda r) \dpo}{\iom r^4}
\end{split}
\end{equation}
\chapter{\label{appb}Five Zero Frequency Even Parity Solutions 
for $l\ge 2$}

Listed below are five zero frequency even parity solutions and 
their radial derivatives, for $l\ge 2$.  The five are $\bhz$, $\bhtw$, 
$\ho$, $\bk$ and $\bg$.  The derivation of these functions and the 
expressions for $\bho$ and $\hz$ are given in subsection~\ref{sec:zevparge}.

\begin{equation}
\begin{split}
\bhz=&\frac{\left((3+2 \lambda) M^2-2 \lambda (1+\lambda) M r
+\lambda (1+\lambda) r^2\right) \dptw}{(1+\lambda)r (3 M+\lambda r)}
-\frac{2 M \bmtwa}{r^4}+\frac{M \pz}{2 (1+\lambda) r^2}
\\&+\frac{\left(3 (3+4 \lambda) M^3+15 \lambda (1+\lambda) M^2 r+4 
\lambda^2 (1+\lambda) M r^2+\lambda^2 (1+\lambda)^2 r^3\right)
\ptw}{(1+\lambda) r^2 (3 M+\lambda r)^2}
\\&+\frac{8 \pi r^3 \left((3+2 \lambda) M^2-2 \lambda (1+\lambda) M r
+\lambda (1+\lambda) r^2\right) Se_{00}}{(1+\lambda)^2
(2 M-r) (3 M+\lambda r)^2}+\frac{2 M \dbmtwa}{r^3}
-\frac{M\dpz}{2 r+2 \lambda r}
\\&+\frac{8 \pi (2 M-r) r (M-(1+\lambda) r) Se_{11}}{(1+\lambda) 
(3 M+\lambda r)}
-\frac{16 \pi \left(M^2+3 (1+\lambda) M r-(1+\lambda) r^2\right) Se_{12}}
{(1+\lambda) (3 M+\lambda r)}
\end{split}
\end{equation}
\begin{equation}
\begin{split}
\bhtw=&\frac{\left(-3 (3+2 \lambda) M^2+2 \left(3+\lambda
-\lambda^2\right) M r+\lambda (1+\lambda) r^2\right) \dptw}{(1+\lambda)
r (3 M+\lambda r)}+\frac{(6 M-4 (2+\lambda) r) \bmtwa}{r^4}
\\&+\frac{(-3 M+2 (2+\lambda) r) \pz}{2 (1+\lambda) r^2}
-\frac{1}{(1+\lambda) r^2 (3 M+\lambda r)^2}\left[9 (3+4 \lambda)M^3
\right.\\&\left.+3 \lambda (11+13 \lambda) M^2 r
+6 \lambda \left(-1+\lambda+2 \lambda^2\right) M r^2
+\lambda^2\left(-1+\lambda^2\right) r^3\right] \ptw
\\&+\frac{8 \pi r^3 \left(-3 (3+2 \lambda) M^2+2 \left(3+\lambda
-\lambda^2\right) M r+\lambda (1+\lambda) r^2\right) Se_{00}}
{(1+\lambda)^2(2 M-r) (3 M+\lambda r)^2}
\\&-\frac{8 \pi (2 M-r) r (3 M+(-1+\lambda) r) Se_{11}}{(1+\lambda)
(3 M+\lambda r)}+\frac{(-6 M+4 r) \dbmtwa}{r^3}
\\&+\frac{16 \pi \left(3 M^2+(-1+\lambda) M r-(1+\lambda) r^2\right) 
Se_{12}}{(1+\lambda) (3 M+\lambda r)}
+\frac{(3 M-2 r) \dpz}{2 r+2 \lambda r}
\end{split}
\end{equation}

\begin{equation}
\begin{split}
\bk=&\frac{(3+2 \lambda) M (2 M-r) \dptw}
{(1+\lambda)r (3 M+\lambda r)}+\frac{(-4 M+2 (2+\lambda) r) \bmtwa}{r^4}
-\frac{(-2 M+r) \pz}{2 (1+\lambda) r^2}
\\&+\frac{1}{(1+\lambda) r^2 (3 M+\lambda r)^2}\left[6 (3+4 \lambda) M^3
+3 \lambda (8+9 \lambda) M^2 r+\lambda \left(-3+5 \lambda
\right.\right.\\&\left.\left.+8 \lambda^2\right)M r^2
+\lambda^3(1+\lambda) r^3\right] \ptw
+\frac{8 (3+2 \lambda) M \pi r^3 Se_{00}}{(1+\lambda)^2 (3 M+\lambda r)^2}
+\frac{8 \pi r (-2 M+r)^2 Se_{11}}{(1+\lambda)(3 M+\lambda r)}
\\&+\frac{16 \pi (-2 M+r) (M+r+\lambda r) Se_{12}}{(1+\lambda)
(3 M+\lambda r)}+\frac{(4 M-2 r) \dbmtwa}{r^3}+\frac{(-2 M+r) \dpz}
{2 (1+\lambda) r}
\end{split}
\end{equation}
\begin{equation}
\begin{split}
\ho=&-\frac{r ((6+5 \lambda) M+\lambda (1+\lambda) r)\dptw}
{2 (1+\lambda) (3 M+\lambda r)}+\frac{4 \bmtwa}{r^2}-\frac{\pz}{4+4\lambda}
\\&-\frac{\lambda \left(3 M^2+6 (1+\lambda) M r+\lambda(1+\lambda) r^2\right)
\ptw}{2 (1+\lambda) (3 M+\lambda r)^2}-\frac{4\pi r^5 ((6+5 \lambda) M
+\lambda (1+\lambda) r) Se_{00}}{(1+\lambda)^2 (2 M-r) (3 M+\lambda r)^2}
\\&-\frac{4 \pi (2 M-r) r^3 Se_{11}}{(1+\lambda) (3 M+\lambda r)}
+\frac{8 \pi r^2 (M+r+\lambda r) Se_{12}}{(1+\lambda)
(3 M+\lambda r)}-\frac{2 \dbmtwa}{r}+\frac{r \dpz}{4+4 \lambda}
\end{split}
\end{equation}
\begin{equation}
\bg=-\frac{\bmtwa}{r^3}
\end{equation}

\begin{equation}
\begin{split}
\dbhz=&\frac{1}{(1+\lambda) (2 M-r) r^2 (3 M
+\lambda r)}\left[-6 (3+2 \lambda) M^3-2 \left(-6-3 \lambda
+\lambda^2\right) M^2 r\right.\\&\left.+\lambda \left(3+5 \lambda
+2\lambda^2\right) M r^2-\lambda (1+\lambda)^2 r^3\right] \dptw
-\frac{4 M (-3 M+(3+\lambda) r) \bmtwa}{(2 M-r) r^5}
\\&+\frac{M (-3 M+(3+\lambda) r) \pz}{(1+\lambda) (2 M-r) r^3}
-\frac{1}{(1+\lambda)(2 M-r) r^3 (3 M+\lambda r)^2}
\left[18 (3+4 \lambda) M^4\right.\\&\left.+6 \left(-3+6 \lambda
+11\lambda^2\right)M^3 r+3 \lambda \left(-9-5 \lambda+4\lambda^2\right)
M^2 r^2-2 \lambda^2 \left(3+4\lambda+\lambda^2\right) M r^3
\right.\\&\left.+\lambda^2 (1+\lambda)^2 r^4\right]\ptw
-\frac{8 \pi r^2}{(1+\lambda)^2 (-2 M+r)^2 (3 M+\lambda r)^2}
\left[6 (3+2 \lambda) M^3\right.\\&\left.+2 \left(-6-3 \lambda
+\lambda^2\right) M^2 r-\lambda \left(3+5 \lambda+2 \lambda^2\right)
M r^2+\lambda (1+\lambda)^2 r^3\right]Se_{00}
\\&+\frac{8 \pi \left(-6 M^2-4 \lambda M r+(1+\lambda)
r^2\right) Se_{11}}{(1+\lambda) (3 M+\lambda r)}
+\frac{16 \pi }{(1+\lambda)(2 M-r) r (3 M+\lambda r)}
\left[6 M^3\right.\\&\left.+4 \lambda M^2 r+\left(-1+\lambda+2 
\lambda^2\right) M r^2-(1+\lambda)^2 r^3\right]Se_{12}
+\frac{4 M (-3 M+2 r) \dbmtwa}{(2 M-r) r^4}
\\&+\frac{M (3 M-2 r) \dpz}{(1+\lambda) (2 M-r) r^2}
\end{split}
\end{equation}

\begin{equation}
\begin{split}
\dbhtw=&\frac{1}{(1+\lambda) (2 M-r) r^2 (3 M+\lambda r)}
\left[18 (3+2 \lambda) M^3-2 \left(36+29 \lambda+3\lambda^2\right) M^2 r
\right.\\&\left.+\left(24+15 \lambda-5 \lambda^2
-2 \lambda^3\right)M r^2+\lambda \left(3+4 \lambda+\lambda^2\right)
r^3\right]\dptw+\frac{4 }{r^5 (-2 M+r)}\left[9 M^2\right.\\&\left.
-(19+9 \lambda) M r+(8+5 \lambda) r^2\right] \bmtwa+\frac{\left(9 M^2
-5 (3+\lambda) M r+3 (2+\lambda)r^2\right) \pz}{(1+\lambda) (2 M-r) r^3}
\\&+\frac{1}{(1+\lambda) (2 M-r) r^3 (3 M+\lambda r)^2}
\left[54 (3+4 \lambda) M^4+6 \left(-15+14 \lambda+39 \lambda^2\right)M^3 r
\right.\\&\left.+3 \lambda \left(-51-37 \lambda+20 \lambda^2\right)
M^2 r^2+2 \lambda \left(12-7 \lambda-16 \lambda^2+3 \lambda^3\right) M r^3
-3 \lambda^2 \left(-1\right.\right.\\&\left.\left.+\lambda^2\right)
 r^4\right]\ptw+\frac{8 \pi r^2}{(1+\lambda)^2 (-2 M+r)^2 (3 M+\lambda r)^2}
\left[18 (3+2 \lambda) M^3-2 \left(36+29 \lambda\right.\right.\\&
\left.\left.+3 \lambda^2\right)M^2 r+\left(24+15 \lambda-5 \lambda^2
-2 \lambda^3\right) M r^2+\lambda \left(3+4 \lambda+\lambda^2\right)
 r^3\right]Se_{00}\\&-\frac{8\pi\left(-18 M^2+20 M r+(-5+\lambda) r^2
\right) Se_{11}}{(1+\lambda)(3 M+\lambda r)}
-\frac{16 \pi }{(1+\lambda) (2 M-r) r (3 M+\lambda r)}
\\&\times\left[18 M^3+4 (-2+3 \lambda) M^2 r-\left(9+17 \lambda
+2 \lambda^2\right) M r^2+\left(5+6 \lambda+\lambda^2\right)r^3\right]Se_{12}
\\&+\frac{4 \left(9 M^2-2 (6+\lambda) M r
+(4+\lambda) r^2\right)\dbmtwa}{(2 M-r) r^4}
\\&+\frac{\left(-9 M^2+2 (6+\lambda) M r-(4+\lambda)r^2\right) 
\dpz}{(1+\lambda) (2 M-r) r^2}
\end{split}
\end{equation}

\begin{equation}
\begin{split}
\dbk=&\frac{\left(-6 (3+2 \lambda) M^2+\left(12+10 \lambda
+\lambda^2\right) M r+\lambda \left(2+3 \lambda+\lambda^2\right)
r^2\right) \dptw}{(1+\lambda) r^2 (3 M+\lambda r)}
-\frac{2 }{r^5}\left[-6 M\right.\\&\left.+(8+5 \lambda) r\right] \bmtwa
+\frac{(-6 M+(5+2 \lambda) r) \pz}{2 (1+\lambda) r^3}
+\frac{1}{(1+\lambda) r^3 (3 M+\lambda r)^2}\left[-18 (3
\right.\\&\left.+4 \lambda) M^3-3 \lambda (22+25 \lambda) M^2 r-6 \lambda 
\left(-2+\lambda+3\lambda^2\right) M r^2+\lambda^2\left(2+\lambda
-\lambda^2\right) r^3\right] \ptw
\\&+\frac{8 \pi r^2 \left(-6 (3+2 \lambda) M^2+\left(12+10 \lambda
+\lambda^2\right) M r+\lambda \left(2+3 \lambda+\lambda^2\right)
r^2\right) Se_{00}}{(1+\lambda)^2 (2 M-r) (3 M+\lambda r)^2}
\\&-\frac{8 \pi (2 M-r) (6 M+(-2+\lambda) r) Se_{11}}{(1+\lambda) 
(3 M+\lambda r)}+\frac{16 \pi (2 M-r) (3 M+2 (1+\lambda)
r) Se_{12}}{(1+\lambda) r (3 M+\lambda r)}
\\&+\frac{2 (-6 M+(4+\lambda) r) \dbmtwa}{r^4}-\frac{3 (-2 M+r) \dpz}
{2 (1+\lambda) r^2}
\end{split}
\end{equation}

\begin{equation}
\begin{split}
\dho=&\frac{\left(3 (4+3 \lambda) M^2+\left(-9-6 \lambda
+\lambda^2\right) M r-\lambda (1+\lambda) r^2\right) \dptw}{(1+\lambda)
(2 M-r) (3 M+\lambda r)}-\frac{4 (-3 M+(3+\lambda) r) \bmtwa}{r^3 (-2 M+r)}
\\&+\frac{(3 M-(3+\lambda) r) \pz}{2 (1+\lambda) (2 M-r) r}
+\frac{2 (-4 M+3 r) \dbmtwa}{r^2 (-2 M+r)}
+\frac{(-3 M+2 r) \dpz}{2 (1+\lambda) (2 M-r)}
\\&+\frac{\left(9 (2+3 \lambda) M^3+3 \lambda (9+11 \lambda) M^2 r
+9 \lambda \left(-1+\lambda^2\right) M r^2+\lambda^2 \left(-1
+\lambda^2\right)r^3\right) \ptw}{(1+\lambda) (2 M-r) r (3 M+\lambda r)^2}
\\&-\frac{8 \pi r^4 \left(-3 (4+3 \lambda) M^2+\left(9+6 \lambda
-\lambda^2\right) M r+\lambda (1+\lambda) r^2\right) Se_{00}}{(1+\lambda)^2
(-2 M+r)^2 (3 M+\lambda r)^2}
\\&+\frac{8 \pi (3 M-2 r) r^2 Se_{11}}{(1+\lambda) (3 M+\lambda r)}
-\frac{16 \pi (3 M-2 r) r (M+r+\lambda r) Se_{12}}{(1+\lambda)
(2 M-r) (3 M+\lambda r)}
\end{split}
\end{equation}
\begin{equation}
\dbg=\frac{3 \bmtwa}{r^4}-\frac{\dbmtwa}{r^3}
\end{equation}
\chapter{\label{appc}Zero Frequency Even Parity Solution for $l=1$}

Below are zero frequency $\bk$ and $\dbk$ for $l=1$, as derived in 
subsection~\ref{sec:zevpareqo}.
\begin{equation}
\bk=\bk_{a}+\bk_{b}\;,
\end{equation}
where
\begin{equation}
\begin{split}
\bk_{a}=&\frac{4 M \left(2 M \left(4 M^2-13 M r+6 r^2\right)+3 r 
\left(2 M^2-5 M r+2 r^2\right)\lnff\right) \pza}{r^4}
+\frac{1}{60 r^4}\\&\times\bigg\{\!\!-\!16 M^3+2 M^2 r+13 M r^2+6 r^3
\!+8 \bigg(2 M \left(M^2-4 M r+2 r^2\right)\!+r \!\left(2 M^2
-5 M r\right.\\&\left.+2 r^2\right)\lnff\bigg) \lnr+8 r \left(2 M^2-5 M r
+2 r^2\right) \plog\bigg\}\pz
\\&+\frac{\pi  r \left(2 M (2 M-3 r)+3 (M-r) r\lnff\right)}{30 M^5} 
\bigg\{\!\!-\!2 M \bigg(\!-\!42 M^3\!+56 M^2 r-9 M r^2
\\&+18 r^3+8 M^2 (M-r) \lnr\bigg)+(M-r)\lnff \bigg(3 r \left(20 M^2-3 M r
\right.\\&\left.+6 r^2\right)+8 M^2(M-r)\lnr\bigg)
+32 M^2(M-r)^2\plog\bigg\}Se_{00}
\\&-\frac{\pi(-2 M+r)^2\left(2 M (2 M-3 r)+3 (M-r) r\lnff\right)}{30 M^6 r} 
\bigg\{-2 M \bigg(-42 M^4\\&+56 M^3 r+111 M^2 r^2-222 M r^3+90 r^4
+8 M^3 (M-r) \lnr\bigg)+(M-r)\\&\times \lnff\left(3 r \left(20 M^3-3 M^2 r
-54 M r^2+30 r^3\right)+8 M^3 (M-r) \lnr\right)
\\&+32 M^3 (M-r)^2 \plog\bigg\}Se_{11}\\&\!+\!\frac{\pi(M-2 r) (-2 M+r)^2}{M^6}
\bigg(2 M (2 M\!-\!3 r)+3 (M-r) r\lnff\bigg)^2 Se_{12}\;,
\end{split}
\end{equation}

\begin{equation}
\begin{split}
\bk_{b}=&\frac{\pi  (2 M-r) \left(2 M (2 M-3 r)+3 (M-r) r\lnff\right)}
{15 M^5 r^2} \bigg\{2 M \bigg(42 M^3-56 M^2 r
\\&-21 M r^2+27 r^3-8 M^2 (M-r) \lnr\bigg)+(M-r)\lnff \bigg(-3 r 
\\&\times\left(-20 M^2+3 M r+9 r^2\right)
+8 M^2 (M-r) \lnr\bigg)+32 M^2 (M-r)^2 \\&\times\plog\bigg\}Ue_{22}
+\frac{(-2 M+r) }{60 r^3}\bigg\{r (M+6 r)+8 \bigg(M (M-2 r)
\\&+(M-r)r\lnff\bigg)\lnr+8 (M-r)r \plog\bigg\}\dpz
\\&-\frac{4 M (2 M-r) \left(2 M (2 M-3 r)+3 (M-r) r\lnff\right)}{r^3}\dpza\;.
\end{split}
\end{equation}
\begin{equation}
\dbk=\dbk_{c}+\dbk_{d}\;,
\end{equation}
where
\begin{equation}
\begin{split}
\dbk_{c}=&-\frac{8 M \left(2 M \left(6 M^2-17 M r+6 r^2\right)
+3 r \left(2 M^2-5 M r+2 r^2\right)\lnff\right)}{r^5}\pza
\\&+\frac{1}{30 r^5}\bigg\{24 M^3+30 M^2 r-22 M r^2+3 r^3
-8 \bigg(M \left(3 M^2-10 M r+4 r^2\right)
\\&+r \left(2 M^2-5 M r+2 r^2\right)\lnff\bigg) \lnr
-8 r \left(2 M^2-5 M r+2 r^2\right)\\&\times\plog\bigg\}\pz
-\frac{\pi}{15 M^5 (2 M-r)}\bigg(2 M \left(6 M^2-10 M r+3 r^2\right)+3 r
\\&\times\left(2 M^2-3 M r+r^2\right)\lnff\bigg)
\bigg\{\!-2 M \bigg(\!-42 M^3+56 M^2 r-9 M r^2+18 r^3
\\&+8 M^2 (M-r) \lnr\bigg)+(M-r)\lnff\bigg(3 r\left(20 M^2-3 M r+6 r^2\right)
\\&+8 M^2 (M-r) \lnr\bigg)+32 M^2 (M-r)^2 \plog\bigg\}  Se_{00}
+\frac{\pi(2 M\!-\!r)}{15 M^6 r^2}\\&\times\bigg(2 M \left(6 M^2-10 M r
+3 r^2\right)+3 r \left(2 M^2-3 M r+r^2\right) \lnff\bigg)\bigg\{-2 M
\\& \times\left(-42 M^4+56 M^3 r+111 M^2 r^2-222 M r^3+90 r^4
+8 M^3 (M-r) \lnr\right)\\&+(M-r) \lnff 
\bigg(3 r \left(20 M^3-3 M^2 r-54 M r^2+30 r^3\right)
\\&+8 M^3 (M-r) \lnr\bigg)+32 M^3 (M-r)^2 \plog\bigg\}Se_{11}\;,
\end{split}
\end{equation}

\begin{equation}
\begin{split}
\dbk_{d}=&-\frac{2 \pi  \left(2 M^2-5 M r+2 r^2\right)}{M^6 r} 
\bigg\{4 M^2 \left(12 M^3-38 M^2 r+36 M r^2-9 r^3\right)+12 M r
\\& \times\left(5 M^3-14 M^2 r+12 M r^2-3 r^3\right)\lnff+
9 (M-r)^2 (2 M-r) r^2\\&\times\lnff^2\bigg\}  Se_{12}
-\frac{2 \pi}{15 M^5 r^3}\bigg(2 M \left(6 M^2-10 M r+3 r^2\right)
+3 r \left(2 M^2-3 M r\right.\\&\left.+r^2\right)\lnff\bigg) 
\bigg\{2 M \bigg(42 M^3-56 M^2 r-21 M r^2+27 r^3-8 M^2 (M-r) 
\\&\times\lnr\bigg)+(M-r)\lnff \bigg(-3 r \left(-20 M^2+3 M r+9 r^2\right)
+8 M^2 \\&\times(M-r) \lnr\bigg)+32 M^2 (M-r)^2 \plog\bigg\}  Ue_{22}
+\frac{1}{30 r^4}\\&\times\bigg\{r \left(-18 M^2+14 M r+3 r^2\right)
+8 \bigg(M \left(3 M^2-6 M r+2 r^2\right)+r \left(2 M^2-3 M r
\right.\\&\left.+r^2\right) \lnff\bigg) \lnr
+8 r \left(2 M^2-3 M r+r^2\right) \plog\bigg\} \dpz
\\&+\frac{8 M \left(2 M \left(6 M^2-10 M r+3 r^2\right)+3 r 
\left(2 M^2-3 M r+r^2\right)\lnff\right)}{r^4}\dpza\;.
\end{split}
\end{equation} 
The function $\plog$ is defined in equation \eqref{eq:polylog}.
\chapter{\label{appd}Zero Frequency Even Parity Solutions for $l=0$}

Listed below are the zero frequency even parity solutions 
$\bhz$, $\bhtw$ and $\bk$, and their radial derivatives, for $l=0$.  
Also for this mode, we have $\bho=0$ and $\dbho=0$.  The 
derivation of these solutions is in subsection~\ref{sec:zevpareqz}.

\begin{equation}
\begin{split}
\bhz=&\frac{\left(-16 M^3-8 M^2 r-3 M r^2+3 r^3+8 M^3 
\lnr\right) \pz}{6 r^4}+\frac{1}{3 r^4}\bigg\{-2 M \left(52 M^3
\right.\\&\left.+24 M^2 r+12 M r^2-9 r^3\right)
+\left(96 M^4-30 M r^3+9 r^4\right) \lnff\bigg\} \pza
\\&+\frac{\pi r}{9 M^3 (-2 M+r)^2}\bigg\{\left(-8 M^4
+\left(96 M^4-30 M r^3+9 r^4\right) \lnff\right) 
\\&\times\left(-4 M^2+3 \lnff \left(8 M^2+r^2+4 M (2 M-r) 
\lnr\right)\right)\bigg\} Se_{00}
\\&-\frac{\pi}{9M^4 r}\bigg\{-8 \left(4 M^7+9 M^4 r^3\right)
+9 \left(32 M^4-10 M r^3+3 r^4\right) \lnff^2 
\\&\times\left(-24 M^3+M r^2+r^3+4 M^2 (2 M-r) \lnr\right)-12 M^3 \lnff 
\\&\times\left(-80 M^4+2 M^2 r^2+12 M r^3-3 r^4+8 M^3 (2 M-r) 
\lnr\right)\bigg\}Se_{11}
\\&+\frac{\pi \lnff}{3 M^4 r^2}\bigg\{\left(-8 M^4
+\left(96 M^4-30 M r^3+9 r^4\right) \lnff\right)
\\&\times\left(-8 M^2-4 M r-r^2+8 M^2 \lnr\right)\bigg\} Ue_{22}
\\&+\frac{\left(8 M^3+4 M^2 r+M r^2-3 r^3-8 M^3 \lnr\right) \dpz}{6 r^3}
\\&+\frac{\left(8 M^4+\left(-96 M^4+30 M r^3-9 r^4\right) 
\lnff\right) \dpza}{3 r^3}
\end{split}
\end{equation}

\begin{equation}
\begin{split}
\bhtw=&\frac{\left(48 M^3-8 M^2 r-7 M r^2+3 r^3
-8 M^2 (3 M-2 r) \lnr\right) \pz}{6 r^4}
+\frac{1}{3 r^4}\bigg\{2 M \left(156 M^3\right.\\&\left.
-32 M^2 r-12 M r^2+3 r^3\right)
-3 \left(96 M^4-64 M^3 r+6 M r^3-r^4\right) \lnff\bigg\}\pza
\\&+\frac{\pi r}{9 M^3 (-2 M+r)^2}\bigg\{\bigg(8 M^3 (3 M-2 r)
-3 \left(96 M^4-64 M^3 r+6 M r^3-r^4\right)
\\&\times\lnff\bigg)\bigg(-4 M^2+3 \lnff \bigg(8 M^2+r^2+4 M (2 M-r) 
\\&\times\lnr\bigg)\bigg)\bigg\}Se_{00}
+\frac{\pi}{9 M^4 r}\bigg\{8 \left(4 M^7+9 M^4 r^3\right)
-9 \left(32 M^4-10 M r^3+3 r^4\right)
\\&\times\lnff^2 \left(-24 M^3+M r^2+r^3+4 M^2 (2 M-r) \lnr\right)+12 M^3
\\&\times\lnff \left(-80 M^4+2 M^2 r^2+12 M r^3
-3 r^4+8 M^3 (2 M-r) \lnr\right)
\\&+2(2 M-r)\left(-8 M^3+\left(96 M^3-3 r^3\right) \lnff\right)
\bigg(4 M^3+3 \lnff \\&\times\left(-24 M^3+M r^2+r^3
+4 M^2(2 M-r)\lnr\right)\bigg)\bigg\}Se_{11}
\\&-\frac{\pi \lnff}{3 M^4 r^2}\bigg\{ \bigg(8 M^3 (-3 M+2 r)
+3 \left(96 M^4-64 M^3 r+6 Mr^3-r^4\right)
\\&\times\lnff\bigg)\left(-8 M^2-4 M r-r^2
+8 M^2 \lnr\right)\bigg\} Ue_{22}
\\&+\frac{\left(-24 M^3+4 M^2 r+5 M r^2-r^3+8 M^2
(3 M-2 r) \lnr\right) \dpz}{6 r^3}
\\&+\frac{\left(8 M^3 (-3 M+2 r)+3 \left(96 M^4-64 M^3 r
+6 M r^3-r^4\right) \lnff\right) \dpza}{3 r^3}
\end{split}
\end{equation}

\begin{equation}
\begin{split}
\bk=&\frac{(2 M-r) \left(-16 M^2-8 M r-3 r^2
+8 M^2 \lnr\right)\pz}{6 r^4}+\frac{1}{3 r^4}\bigg\{-208 M^4
\\&+8 M^3 r+6 M r^3+3 \left(64 M^4
-32 M^3 r-2 M r^3+r^4\right) \lnff\bigg\} \pza
\\&+\frac{\pi r}{9 M^3 (2 M-r)}\bigg\{\left(-8 M^3+\left(96 M^3
-3 r^3\right) \lnff\right) 
\\&\times\left(-4 M^2+3 \lnff \left(8 M^2+r^2+4 M (2 M-r) \lnr\right)
\right)\bigg\} Se_{00}
\\&-\frac{\pi (2 M-r)}{9 M^4 r}\bigg\{ \left(-8 M^3
+\left(96 M^3-3 r^3\right) \lnff\right) \bigg(4 M^3
\\&+3 \lnff \left(-24 M^3+M r^2+r^3+4 M^2 (2 M-r)
\lnr\right)\bigg)\bigg\}Se_{11}
\\&+\frac{\pi (2 M-r) \lnff }{3 M^4 r^2}\bigg\{\left(-8 M^3
+\left(96 M^3-3 r^3\right) \lnff\right)
\\&\times\left(-8 M^2-4 M r-r^2+8 M^2 \lnr\right)\bigg\} Ue_{22}
\\&-\frac{(-2 M+r) \left(8 M^2+4 M r+r^2-8 M^2\lnr\right)\dpz}{6 r^3}
\\&+\frac{(-2 M+r) \left(-8 M^3+\left(96 M^3-3 r^3\right) 
\lnff\right) \dpza}{3 r^3}
\end{split}
\end{equation}

\begin{equation}
\begin{split}
\dbhz=&\frac{\left(96 M^4-16 M^3 r-14 M^2 r^2-6 M r^3+3 r^4
+16 M^3 (-3 M+2 r) \lnr\right) \pz}{6(2 M-r) r^5}
\\&-\frac{4 M}{3 (2 M-r) r^5}\bigg\{2 M \left(-78 M^3+16 M^2 r
+6 M r^2+3 r^3\right)+3 \left(48 M^4-32 M^3 r
\right.\\&\left.+r^4\right) \lnff\bigg\}\pza
-\frac{4 \pi}{9 M^2 (2 M-r)^3}\bigg\{\bigg(4 M^3 (-3 M+2 r)
+3 \left(48 M^4\right.\\&\left.-32 M^3 r+r^4\right) \lnff\bigg)
\bigg(-4 M^2+3 \lnff \bigg(8 M^2+r^2+4 M
\\&\times (2 M-r) \lnr\bigg)\bigg)\bigg\}Se_{00}
+\frac{4 \pi}{9 M^3 (2 M-r) r^2}\bigg\{4 M^3 \left(-12 M^4+8 M^3 r
\right.\\&\left.-9 M r^3+9 r^4\right)+9 (-2 M+r)^2 \left(12 M^2
+4 M r+r^2\right) \lnff^2 \bigg(-24 M^3
\\&+M r^2+r^3+4 M^2 (2 M-r) \lnr\bigg)-12 M^3 \lnff \bigg(-120 M^4
\\&+80 M^3 r+3 M^2 r^2+M r^3-3 r^4+4 M^2 \left(6 M^2-7 M r
+2 r^2\right) \lnr\bigg)\bigg\} Se_{11}
\\&-\frac{4 \pi \lnff}{3 M^3 (2 M-r) r^3}\bigg\{\left(4 M^3 
(-3 M+2 r)+3 \bigg(48 M^4-32 M^3 r+r^4\right)
\\&\times\lnff\bigg)\left(-8 M^2-4 M r-r^2+8 M^2\lnr\right)\bigg\}Ue_{22}
\\&+\frac{\left(-48 M^4+8 M^3 r+10 M^2 r^2+4 M r^3-3 r^4
+16 M^3 (3 M-2 r) \lnr\right) \dpz}{6 (2 M-r) r^4}
\\&+\frac{4 \left(4 M^4 (-3 M+2 r)+3 \left(48 M^5-32 M^4 r
+M r^4\right) \lnff\right) \dpza}{3 (2 M-r) r^4}
\end{split}
\end{equation}

\begin{equation}
\begin{split}
\dbhtw=&\frac{1}{6 (2 M-r) r^5}\bigg\{-288 M^4+176 M^3 r
+10 M^2 r^2-18 M r^3+3 r^4+16 M^2 \left(9 M^2\right.
\\&\left.-10 M r+3 r^2\right) \lnr\bigg\}\pz
+\frac{4 M}{3 (2 M-r) r^5}\bigg\{-2 M \left(234 M^3-152 M^2 r
\right.\\&\left.+12 M r^2+3 r^3\right)+3 (-2 M+r)^2 \left(36 M^2
-4 M r-r^2\right) \lnff\bigg\}\pza
\\&+\frac{4 \pi}{9 M^2 (2 M-r)^3}\bigg\{\bigg(-4 M^2 \left(9 M^2
-10 M r+3 r^2\right)+3(-2 M+r)^2\left(36 M^2
\right.\\&\left.-4 M r-r^2\right)\lnff\bigg)\bigg(-4 M^2+3 \lnff
\bigg(8 M^2+r^2+4 M (2 M-r) 
\\&\times\lnr\bigg)\bigg)\bigg\}Se_{00}
-\frac{4 \pi}{9 M^3 (2 M-r) r^2}\bigg\{-4 M^3 \left(36 M^4-40 M^3 r
+12 M^2 r^2\right.\\&\left.-9 M r^3+9 r^4\right)+9 (-2 M+r)^2 
\left(36 M^2-4 M r-r^2\right)\lnff^2 \bigg(-24 M^3
\\&+M r^2+r^3+4 M^2 (2 M-r) \lnr\bigg)-12 M^2 \lnff \bigg(-360 M^5
\\&+400 M^4 r-111 M^3 r^2-M^2 r^3-6 M r^4+3 r^5+4 M^2 
\left(18 M^3-29 M^2 r+16 M r^2
\right.\\&\left.-3 r^3\right) \lnr\bigg)\bigg\}Se_{11}
+\frac{4 \pi \lnff}{3 M^3 (2 M-r) r^3}\bigg\{\bigg(-4 M^2 
\left(9 M^2-10 M r+3 r^2\right)
\\&+3 (-2 M+r)^2\left(36 M^2-4 M r-r^2\right) \lnff\bigg)
\bigg(-8 M^2-4 M r-r^2+8 M^2
\\&\times\lnr\bigg)\bigg\} Ue_{22}
-\frac{1}{6(2 M-r) r^4}\bigg\{-144 M^4+88 M^3 r+14 M^2 r^2-16 M r^3
\\&+3 r^4+16 M^2 \left(9 M^2-10 M r+3 r^2\right) \lnr\bigg\} \dpz
-\frac{4 }{3(2 M-r) r^4}\bigg\{-4 M^3 \left(9 M^2
\right.\\&\left.-10 M r+3 r^2\right)+3 M (-2 M+r)^2 
\left(36 M^2-4 M r-r^2\right) \lnff\bigg\} \dpza
\end{split}
\end{equation}

\begin{equation}
\begin{split}
\dbk=&-\frac{\left(-32 M^3+2 M r^2+r^3-8 M^2(-2 M+r)
\lnr\right)\pz}{2 r^5}-\frac{8 M^3}{r^5}\bigg\{-26 M+r
\\&+12(2 M-r) \lnff\bigg\}\pza
-\frac{8 \pi }{6 M-3 r}\bigg\{\left(-1+12 \lnff\right)
\\&\times\bigg(-4 M^2+3 \lnff\left(8 M^2+r^2+4 M (2 M-r) 
\lnr\right)\bigg)\bigg\} Se_{00}
\\&+\frac{8 \pi (2 M-r)}{3 M r^2}\bigg\{\left(-1+12 \lnff\right)
\bigg(4 M^3+3\lnff\bigg(-24 M^3
\\&+M r^2+r^3+4 M^2 (2 M-r)\lnr\bigg)\bigg)\bigg\}Se_{11}
-\frac{8 \pi (2 M-r) \lnff }{M r^3}
\\&\times\bigg\{\bigg(-1+12\lnff\bigg) \left(-8 M^2-4 M r-r^2
+8 M^2 \lnr\right)\bigg\}Ue_{22}
\\&+\frac{(2 M-r) \left(-8 M^2-4 M r-r^2+8 M^2 \lnr\right) \dpz}{2 r^4}
\\&+\frac{8 M^3 (2 M-r) \left(-1+12 \lnff\right) \dpza}{r^4}
\end{split}
\end{equation}

\end{document}